\documentclass[%
 aip,
 amsmath,amssymb,
 reprint,%
]{revtex4-1}

\usepackage{graphicx}
\usepackage{dcolumn}
\usepackage{bm}

\usepackage[utf8]{inputenc}
\usepackage[T1]{fontenc}
\usepackage{mathptmx}
\usepackage{etoolbox}
\usepackage{xcolor}
\usepackage{float}
\usepackage{orcidlink}
\usepackage{adjustbox}
\usepackage{tikz}
\usepackage{fontawesome}

\usepackage{placeins}
\usepackage[breakable]{tcolorbox}
\usepackage{caption}
\usepackage[normalem]{ulem} 
\usepackage{verbatim}

\usepackage{titletoc}
\titlecontents{section}
  [0em]
  {\normalfont\addvspace{0.5ex}}
  {\contentslabel{2em}\bfseries}
  {\bfseries}
  {\normalfont\titlerule*[0.7em]{.}\contentspage}
\titlecontents{subsection}
  [2em]
  {\normalfont}
  {\contentslabel{2em}}
  {}
  {\titlerule*[0.7em]{.}\contentspage}
\titlecontents{subsubsection}
  [4em]
  {\normalfont}
  {\contentslabel{2em}}
  {}
  {\titlerule*[0.7em]{.}\contentspage}

\makeatletter
\makeatletter
\def\FM@note{\frontmatter@thefootnote}
\def\frontmatter@makefnmark{%
  \@textsuperscript{\normalfont
    \ifx\@thefnmark\FM@note
      \ifnum\csname c@\@mpfn\endcsname=1
        \faTree
      \else
        \@thefnmark)%
      \fi
    \else
      \@thefnmark)%
    \fi}}
\makeatother

\makeatletter
\let\AIP@contactemails\@empty
\patchcmd{\titleblock@produce}
  {\frontmatter@RRAPformat}
  {\AIP@contactemails\frontmatter@RRAPformat}
  {}{}
\def\@email#1#2{%
  \endgroup
  \gappto\AIP@contactemails{%
    \frontmatter@RRAPformat{%
      \produce@RRAP{*#1\href{mailto:#2}{#2}}%
    }%
  }%
}
\makeatother
\makeatletter
\newcommand{\startconstantfooter}[1]{%
  \clearpage
  \gdef\@oddfoot{%
    \hfil
    \normalfont\scriptsize\itshape\color{gray}#1%
    \hfil
  }%
  \global\let\@evenfoot\@oddfoot
}
\makeatother

\begin{document}



\title{The Trajectory Statistics of Biological Exploratory Dynamics}

\author{Sara D. Mahdavi \orcidlink{0009-0004-4540-4268}}
\thanks{These authors contributed equally to this work.}
\email[Contact author: ]{smahdavi@caltech.edu}
\affiliation{Division of Biology and Biological Engineering, California Institute of Technology, Pasadena, CA 91125}

\author{Gabriel L. Salmon \orcidlink{0000-0003-2163-8399}}
\email[Contact author: ]{gsalmon@alumni.caltech.edu}
\thanks{These authors contributed equally to this work.}
\affiliation{NSF-Simons National Institute for Theory and Mathematics in Biology, Chicago, IL 60611, USA}

\author{Minakshi Ashok \orcidlink{0009-0007-0504-3826}}
\affiliation{Division of Biology and Biological Engineering, California Institute of Technology, Pasadena, CA 91125}

\author{Madhav Mani \orcidlink{0000-0002-5812-4167}}
\affiliation{Department of Mathematics, Imperial College, London, UK}

\author{Marc Kirschner \orcidlink{0000-0001-6540-6130}}
\affiliation{Department of Systems Biology, Harvard Medical School, Boston, MA 02115}

\author{Jane~Kondev \orcidlink{0000-0001-7522-7144}}
\affiliation{Department of Physics, Brandeis University, Waltham, MA 02453}

\author{Rob~Phillips \orcidlink{0000-0003-3082-2809}}
\affiliation{Department of Physics, California Institute of Technology, Pasadena, CA 91125}
\affiliation{Division of Biology and Biological Engineering, California Institute of Technology, Pasadena, CA 91125}
\email[Contact author: ]{phillips@pboc.caltech.edu}

\date{August 31, 2026}
\begin{abstract}

Biological processes, from molecules diffusing to their regulatory destinations to whales pursuing food or mates, are widely exploratory. Such behaviors are effective when some verifiable functional state or outcome can be reached regardless of initial conditions. In these cases, systems repeatedly undergo distinct and abortive trajectories; the process only ends when the system finds ``right'' outcomes. This dominance of the terminal condition (and indifference to the initial state) provokes a very different perspective than the conventional ``dynamical systems'' framework that has been a centerpiece of the quantitative sciences for centuries. We hypothesize that many of these problems defy the initial-condition driven or gradients on landscapes so useful in problems ranging from mechanics to electrodynamics to chemical kinetics to mass and heat transport. We examine several mathematical frameworks that capture and unify key aspects of exploratory dynamics. One powerful way of thinking of such processes is the geometric distribution, where repeated failures are punctuated by a successful trajectory. We develop intuitions for how random walks with resets accelerate search processes. We highlight fresh and surprising behaviors of search under drift, cues, and checkpoints. Last, appreciating the probability of trajectories conditioned on satisfying macroscopic final outcomes reveals a language for the potency of variation sculpted by selection in their broadest forms. This can give the fictitious appearance of the future making itself known in the present, but we view this as conceptually similar to the way ``fictitious forces'' arise in non-inertial reference frames. These approaches stress unity, open questions, and applications across a range of biological phenomena.
\end{abstract}

\maketitle

\section{Introduction}
\label{sec:intro}

It is our very sad privilege to participate in this special issue dedicated to the life and work of Prof. Rudi Podgornik.  Two of us (RP and JK) got our start in the fascinating world of biology by thinking about the packing of DNA in viruses and that led us to a wonderful encounter with Rudi Podgornik and Adrian Parsegian~\cite{Podgornik1989,Podgornik1995,Strey1999,Zandi2020}.  We learned many things about the mechanics of DNA from Rudi.  He was particularly proud of the fact that one of the earliest measurements of the DNA persistence length was done by a Slovenian scientist~\cite{Peterlin1953}, announced in a paper in \emph{Nature} several months before the paper of Watson and Crick.  One of the striking features of Rudi's work was its mathematical elegance and approachability.   In this paper, we attempt to strike a similar tone in the context of a broad class of biological processes that can be thought of as examples of the widespread phenomena of exploratory dynamics.  Indeed, we go further in the sense that we have actively imitated one of Rudi's outstanding papers, his article entitled ``On virus growth and form'' \cite{Zandi2020}.

In the nearly 350 years since the 1687 publication of Newton's {\it Principia}, 
we have had an incredibly 
powerful and beautiful description of the temporal 
evolution of systems of all kinds~\cite{newton1687,Chandrasekhar1995,Pask2013,arnold1978,lin1988,strogatz2015}. In most of these cases, if dissected 
carefully, we see a common paradigm described by the update rule 
\begin{equation}
\text{stuff}(t+\Delta t) = \text{stuff}(t) + f(\text{stuff},t)\, \Delta t.
\label{eqn:DynamicalSystems}
\end{equation}
To find the value of ``stuff'' now, we take its value a short time ago and 
add to it some ``update'' 
that
 is computed using laws such as $F=ma$ or 
the law of mass action, for example.
Differential equations like these describe 
planetary motion, the flow of heat, the conversion of atoms and molecules into 
different kinds of molecules, the time evolution of the electromagnetic 
field according to Maxwell's equations, the changes in the sizes of 
populations and much, much more.
  And yet, we argue that there are huge swathes of biological phenomena that are not described in this way.

In their provocative 1997 book  {\it Cells, Embryos and Evolution}, Gerhart and Kirschner  provided an unusual vista on
some of the unique processes of living organisms.   Rather than an impressive compendium of the many great achievements of modern
biology,  each of their chapters expounds tentative principles that help explain the facts of the living world.     One of those chapters introduces the notion of ``the exploratory behavior of biological systems,'' which paints a completely different picture of biological dynamics in which the trajectories of the system are a repeated series of trials which end upon achieving some target function.
Following on the discovery of the dynamic instability of microtubules and its role in chromosome search and capture,  they provided a plethora of examples from across scales of 
biological exploratory behavior, whether the foraging of animals to find 
food~\cite{Bennison2018}, the development of the cells of the immune system~\cite{victora2012}, the active 
dynamics of microtubules as they search for chromosomes as a prerequisite 
to cell division~\cite{mitchison1984}, the plasticity of nervous connections~\cite{lowery2009} or the development 
of vasculature~\cite{gerhardt2003}. 
However, their exposition was verbal, not mathematical.
Our goal in this paper dedicated to Rudi Podgornik is to examine to what extent the qualitative statement of the exploratory behavior of biological systems can be restated mathematically.

\begin{figure*}[t]
    \centering
    \includegraphics[width=\textwidth]{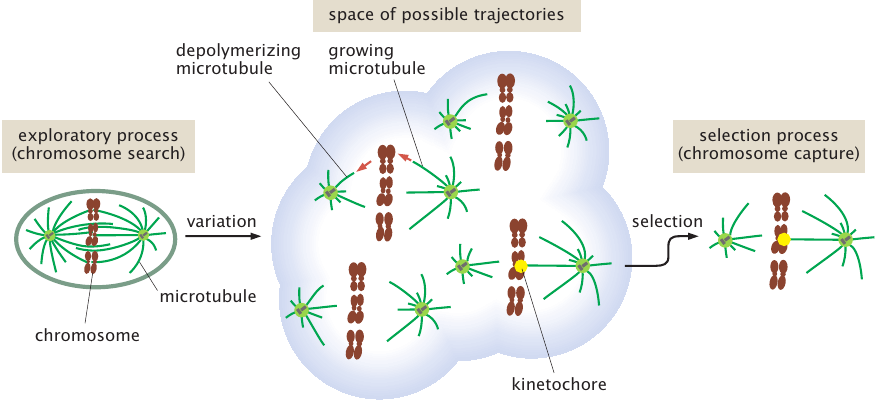}
\caption{\raggedright The variation and selection of exploratory dynamics. A cell or an organism needs to perform some function such as chromosome segregation.  To that end, it generates a collection of possible microscopic trajectories (variation) and when some function such as attachment to the kinetochore is achieved, that trajectory is stabilized (selection).}
    \label{fig:VariationSelection}
\end{figure*}

One of the most powerful facets of formulating our thinking in mathematical terms is that it reveals that phenomena or processes that were outwardly not the same, can fruitfully be viewed as the same thing.  For example, the use of statistical mechanical models to treat allosteric molecules reveals the unity of ligand-gated ion channels, membrane receptors, transcription factors, hemoglobin and beyond~\cite{Martins2011,Phillips2020}.  The aim of the present paper is to examine several different mathematical frameworks that describe a myriad of exploratory processes.  What unites these cases is a common 
architecture shown schematically
in Figure~\ref{fig:VariationSelection}: some mechanism generates variation, that variation 
samples a space of possibilities, and only a small subset of 
outcomes is selectively stabilized.
A central idea running through these ostensibly distinct thrusts is the distinction between micromanagement and verification. One way to build a system is to specify every step in advance (i.e. micromanagement) so that the desired outcome is guaranteed by construction. Biology often uses a different strategy: generate many possible trajectories, let them explore, and stabilize the rare successful outcome when it appears.

Another more general question raised but not answered by our paper is the extent to which the analysis of these fascinatingly diverse biological processes might teach us much deeper lessons and principles about the physical description of the natural world.  We find it a strange irony that in his famous book {\it Chance and Necessity} ~\cite{monod1974chance}, Jacques Monod came down strongly against the idea that living systems would teach us any new physical principles, noting:

``Biology occupies, among the sciences, a place that is at once marginal and central. Marginal in that the living world constitutes only an infinitesimal and very special part of the known universe, so that the study of living beings does not seem likely ever to reveal general laws applicable outside the biosphere. But if the ultimate ambition of all science is, as I believe, to elucidate the relation of man to the universe, then one must recognize that biology occupies a central place, since it is, of all the disciplines, the one that attempts to go most directly to the heart of the problems that must be resolved before one can even pose the question of human nature in terms other than metaphysical.''

Given Monod's own enormous contributions to biology 
 we find his position ironic. The {\it lac} operon is arguably one of the most apparently  specific, contingent, bacteria-in-a-test-tube-with-lactose thing you could imagine. And what did it reveal? The logic of regulatory circuits, the distinction between structural and regulatory genes, the concept of the operon as a unit of coordinated expression, the molecular basis of induction and repression~\cite{Jacob1961,Monod1963,Monod1965,Muller-Hill1996}.  So moved was Monod by his work on these topics that he named it the ``second secret of life''~\cite{Ullmann2011}.

Physics is full of such examples where the highly specific illuminates the very general.  The specific heat of solids at low temperatures occupies a tiny slice of physical phenomena, yet Einstein and Debye's treatment of it revealed deep insights into the quantum theory of matter~\cite{pais1979einstein,Rogers2005}. The photoelectric effect is a specialized laboratory observation about light impinging on surfaces that unlocked an entirely new picture of the nature of light and matter ~\cite{pais1979einstein,Rigden2005}. Spectral lines from hydrogen helped establish the entire architecture of quantum mechanics and with it the periodic table and all of chemistry~\cite{Rigden2002}.
In fact one could argue that life seems more likely to reveal deep principles precisely because it is so apparently improbable and so structured. A system that maintains itself far from equilibrium, replicates with variation, and evolves  is not a sign that it is parochial. Rather, we hypothesize that it is a sign that there is more physical insight to be gained.  And it is our hope that some item from  the long list of exploratory processes highlighted above might give rise to deeper insights.  

\begin{figure*}[t]
    \centering
    \includegraphics[width=\textwidth]{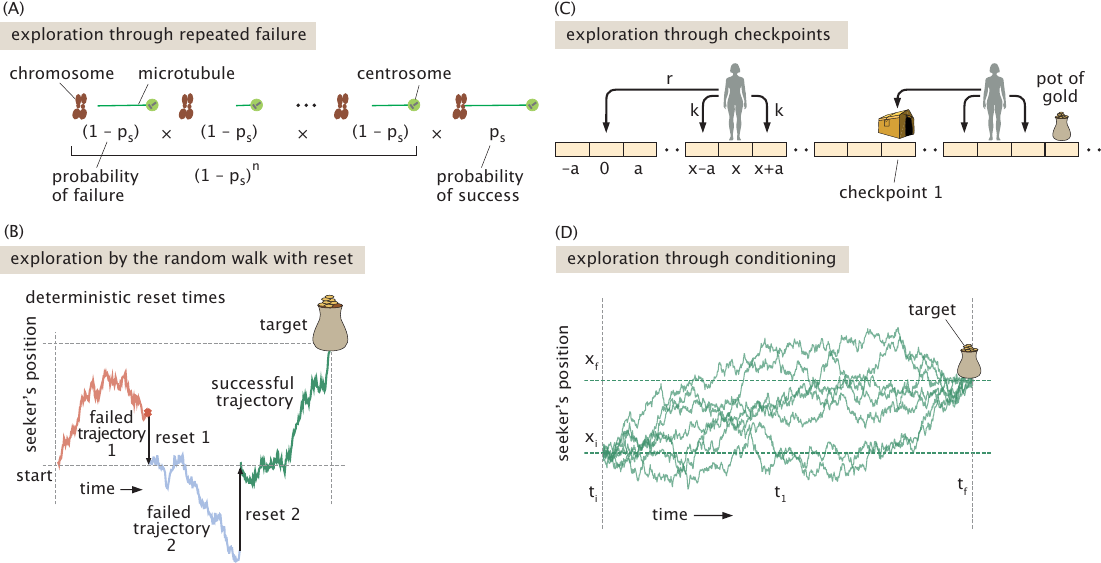}
\caption{\raggedright The statistical physics of exploratory trajectories. (A) Repeated attempts for microtubules to reach chromosomes are described by the geometric distribution. $p_s$ is the probability of success on a given trial.  This distribution applies much more broadly than to just the problem of chromosome search and capture. (B) The random walk with reset is a type of exploratory dynamics in which the seeker returns back to the origin after a certain number of steps known as the reset time. (C) Exploratory behavior can be codified by checkpoints in which the seeker no longer has to reset to the origin, but rather can reset only to some intermediate point.
(D) Exploration through conditioning considers only the subset of trajectories that reach the target. 
}
    \label{fig:DifferentMath}
\end{figure*}

We are also amused and inspired by a second unexpected connection to another biological luminary, the eminent developmental biologist Conrad Waddington. Towards the very end of his life, he wrote a book entitled, ``OR in World War 2 --- Operational Research Against the U Boat,'' in which he articulates the science underlying what we might call search with a purpose \cite{waddington1973or}. Waddington notes that, ``There are many more places in which a similar approach would be valuable. It is one of the main purposes of this book, by expounding what operational research did in one particular field, to suggest what kinds of things it might be called upon to do in others.''  Waddington's perspective argues that this field facilitates ``the general method of science employed to study any problem which may be of importance to an executive.''  Our view is that biological exploratory dynamics calls for ``executive action'' in the sense of achieving some function such as the high fidelity copying of genetic information or the attachment of microtubules to chromosomes.

Despite biology's many impressive ``executive actions,'' natural history reminds us that biology need not be ``optimal.'' This is true due to our poor human guesses about what actually matters to organisms over evolutionary time, or because inherited accidents also rule over living systems. Nonetheless, understanding how the success and timing of searches vary with natural parameters, including with respect to possible optima in minimal models, is revealing even without assuming that biological systems operate exactly at such optima. Characterizing the space of possible dynamics in simple models can identify when a particular strategy is beneficial, quantify the largest improvements that it can provide, and crystallize tradeoffs that could apply to a wide array of systems.

\begin{figure*}[t]
\centering
\includegraphics[width=\textwidth]{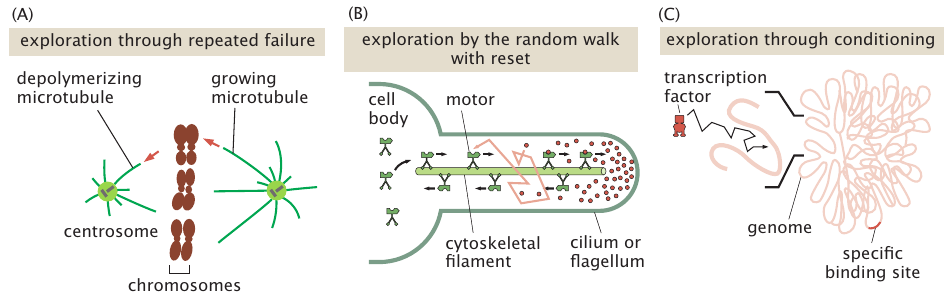}
\caption{\raggedright Gallery of biological examples illustrating the ways in which the different mathematical strategies are instructive. (A) Chromosome search and capture can be examined using the geometric distribution~\cite{Holy1994,Wollman2005,kondev2025}. (B) Random walks with resets are a convenient mathematical formalism for thinking about gradient formation in structures such as cilia and flagella. (C) The search of transcription factors for their target sites on DNA can be considered as a problem in conditional probability~\cite{Hachmo2023}.}
\label{fig:BiologicalCaseStudies}
\end{figure*}

 The remainder of the paper traces several different mathematical perspectives offered in Figure~\ref{fig:DifferentMath} with corresponding biological case studies illustrated in Figure~\ref{fig:BiologicalCaseStudies} and is organized as follows.  In section~\ref{section:GeometricDistribution}, we set up one of the simplest and most revealing of mathematical structures for studying exploratory processes, namely, the geometric distribution.   In this section, we will see that this distribution provides a broad umbrella for thinking about exploratory processes. Section~A, ``Quixotic exploration through random walks,'' of the Supplemental Material reviews 
 how the simplest exploratory motions, bare random walks, suffer from fundamental mathematical limitations in their speeds or probability of success. This basis motivates why biology amends  random walks with surprising strategies such as resets to the start, ballistic drift, and chemical cues; these dynamical strategies are explored in section~\ref{section:RandomWalkReset}. Within this framework, we also examine how a spatially localized chemical
cue can guide the search, and identify a combination of cue strength and
spatial extent that determines when resetting remains beneficial. Section~\ref{section:RWCheckpoints} studies how exploratory random walks change under resets to intermediate checkpoints rather than to the origin, and section~\ref{sec:severalSearchers} investigates parallel searches accomplished by multiple searchers. Such parallel exploration is common in biology; for example, multiple
microtubules act in parallel during chromosome capture and segregation.
More generally, completion may depend on the first, several, or all
successful arrivals.
 Finally, in section~\ref{sec:ExplorationConditioning}, we consider the language of conditional probability as an insightful and unifying approach to a wide variety of search problems.  
 These various approaches are summarized in Table~\ref{tab:exploratory} which will serve as a road map  and reference point for the remainder of the paper.

\begin{table*}[t]
\centering
\small
\setlength{\tabcolsep}{3pt}
\renewcommand{\arraystretch}{1.08}

\begin{adjustbox}{max width=\textwidth}
\begin{tabular}{@{}llll@{}}
\textbf{framework} & \textbf{key equation} & \textbf{resulting behavior} & \parbox[t]{2.4in}{\textbf{exploratory dynamics story}}\\
\hline
geometric distribution &
$P(k\text{~steps needed}) = (1-p_s)^{k-1}~p_s$ &
$T = t_{\text{step}}/p_s$ (eqn.~\ref{eq:geometricMeanTime}) &
\parbox[t]{2 in}{\raggedright searcher (e.g. microtubule) explores at each timestep $t_{\text{step}}$ with success probability $p_s$ of reaching target (e.g. chromosome).} \\\noalign{\vskip 3pt}

\parbox[t]{1.25 in}{\raggedright random walk with Poissonian reset (with or without drift)}& \parbox[t]{2.2 in}{\raggedright$0 = 1 + D\, \frac{d^2 T}{dx^2} + v\, \frac{dT}{dx} - r\, T(x) + r\, T(0)$ (eqns.~\ref{eq:backward_equation_finalmain},~\ref{eq:backward_drift_reset})} & \parbox[t]{2in}{\raggedright $T = T_R \left(\exp\left[\frac{x_T}{\lambda(D,v,T_R)}\right] -1 \right)$ \\ (eqns.~\ref{eqn:searchtimeresets},~\ref{eq:mfpt_drift_reset})} & \parbox[t]{2 in}{\raggedright searchers diffusively search for target at $x_T$ with diffusion coefficient $D$ and drift $v$, resetting to origin on average every $T_R$ steps. An emergent length scale $\lambda$ governs mean time to success $T$.}\\\noalign{\vskip 3pt}

\parbox[t]{1.25 in}{\raggedright random walk with Poissonian reset and spatially dependent drift due to cues}& \parbox[t]{2.2 in}{\raggedright$0 = 1 + D\, \frac{d^2 T}{dx^2} + v(x)\, \frac{dT}{dx} - r\, T(x) + r\, T(0)$  (eqn.~\ref{eq:backward_equation_position_dependent_drift})} & \parbox[t]{2in}{\raggedright organized by Pe$\equiv \frac{v x_T}{D}$\\ (see e.g. eqns.~\ref{eq:pecletDescriptionOfEq100}~and \ref{eq:cue_blind_optimal_ratios})} & \parbox[t]{2in}{\raggedright biochemical cues induce spatially-varying drift field; prudence of reset and cue admit surprising regime transitions in P{\'e}clet number (see especially Fig.~\ref{fig:CueOptimization})}\\\noalign{\vskip 3pt}

\parbox[t]{1.25 in}{\raggedright random walk with intermediate checkpoints} & \parbox[t]{2in}{\raggedright $n$ independent subtrajectories: $\displaystyle T = \sum_{k=0}^{n} T_k$ \\ (eqn.~\ref{eq:indepSubCheckpoints})} & \parbox[t]{2in}{\raggedright $ T = \frac{n+1}{r}\left(e^{x_T/(n+1)\lambda_D} - 1\right)$ \\ (see eqn.~\ref{eq:general_mean})} & \parbox[t]{2in}{\raggedright $n$ checkpoints preserve progress between a searcher's starting location and a target at $x_T$, subject to a diffusive search length scale $\lambda_D=\sqrt{\frac{D}{r}}$.}\\\noalign{\vskip 3pt}

\parbox[t]{1.25 in}{\raggedright first success among multiple independent searchers} & $S^{\text{tot}}(t) = S(t)^N$ & \parbox[t]{2in}{\raggedright $T_{\min}\sim \dfrac{x_T^2}{4D\ln N}$ for $N\gg1$ (eqn.~\ref{eq:Tmin_leading_largeN})} & \parbox[t]{2 in}{\raggedright $N$ independent searchers, each with survival probability $S(t)$ of not yet finding the target by time $t$, collectively explore more efficiently, but only logarithmically, with the search time scaling as $1/\ln N$.}\\\noalign{\vskip 3pt}

\parbox[t]{1.25 in}{\raggedright $M$ successes among multiple independent searchers}&\parbox[t]{2.2in}{\raggedright
$S_{M:N}(t)=
\displaystyle\sum_{\ell=0}^{M-1}
\binom{N}{\ell}
[1-S(t)]^\ell S(t)^{N-\ell}$
\\
(eqn.~\ref{eq:MN_survival})
}&\parbox[t]{2in}{\raggedright
$T_{M:N}\sim \dfrac{x_T^2}{4D\ln(N/M)}$
for fixed $M\ll N$;
$T_{N:N}\propto \ln N$ for $M=N$ and $N\gg1$
\\
(eqns.~\ref{eq:TMN_largeN},~\ref{eq:last_arrival_largeN})
}&\parbox[t]{2 in}{\raggedright 
$N$ independent searchers explore in parallel; successful completion can require $M=$ one, several, or all searchers to reach the target.}
\\\noalign{\vskip 3pt}

\parbox[t]{1.25 in}{\raggedright conditioning search trajectories on final and initial states: Schr{\"o}dinger bridges on graphs} & \parbox[t]{2 in}{\raggedright $\hat{p}(\ell,t)\equiv  p\left(\ell ~\text{at }t|(i\text{ at } t=0, j\text{ at }t=T ) \right) = \frac{\left(\exp\left[t\mathbf{L}\right]\right)_{\ell i}~\left(\exp\left[(T-t)\mathbf{L}\right]\right)_{j \ell}}{\left(\exp\left[T\mathbf{L}\right]\right)_{j i}} $}  & \parbox[t]{1.25 in}{\raggedright  new marginal distributions $\hat{\mathbf{p}}(t)$ and new, effective dynamics $\frac{d\hat{\mathbf{p}}}{dt}=\hat{\mathbf{L}}(t)\hat{\mathbf{p}}(t) $}& \parbox[t]{2 in}{\raggedright given transitions between states of a system at rates stored in a graph Laplacian matrix $\mathbf{L}$, conditioning trajectories to start at state $i$ and end in state $j$ at time $T$ induces effective dynamics over states at intermediate times, described by a new time-dependent $\hat{\mathbf{L}}(t)$} \\\noalign{\vskip 3pt}

\end{tabular}
\end{adjustbox}

\caption{Mathematical frameworks for exploratory dynamics.}
\label{tab:exploratory}
\end{table*}

\section{Exploration Through Repeated Failure}
\label{section:GeometricDistribution}

We begin by building the mathematical language of
the statistical mechanics of exploratory trajectories by considering 
trajectories built up of ``wrong'' and ``right'' moves.  This structure appears in many guises~\cite{kondev2025}. In one biological system, an attempt may be a microtubule growth episode that either finds a kinetochore or collapses~\cite{mitchison1984,mitchison1985,Holy1994,Gundersen2002,Wollman2005,Heald2015}. In another, it may be an excursion of a transcription factor away from DNA followed by rebinding~\cite{vonHippel1979,Winter1981a,Winter1981b,Hu2006,Hachmo2023}. The microscopic meaning of an attempt differs from system to system, but the statistical structure is the same.
  A series of wrong moves followed by a final right
move.

Suppose that each attempt succeeds with probability $p_s$ and fails with probability $p_f=1-p_s$ ($f$ for failure and $s$ for success). If success occurs after $i$ failed attempts, then the probability of that trajectory is
\begin{equation}
p_i=\underbrace{p_f p_f \cdots p_f}_{i\ \text{factors}} p_s=p_f^i p_s.
\end{equation}
Equivalently, we can write this more informatively as
\begin{equation}
p_i=(1-p_s)^i p_s,
\label{eq:geomPi}
\end{equation}
which is the geometric distribution~\cite{Blitzstein2019}. This is the first foundational model for exploratory dynamics. It says that the central quantity is the probability of success  per attempt.  This fundamental equation has a structure that we will use over
and over in thinking about the statistical mechanics of exploratory
trajectories since it highlights the importance of the final state.

In particular, we ask how long does it take
on average for the exploratory dynamics to terminate due to success in finding its
target?
The answer to that question is obtained by averaging over the time spent in all
the possible trajectories and given by
\begin{equation}
\langle t\rangle=\sum_{i=0}^{\infty} \underbrace{\left(i \, t_{\text {step }}\right.}_{\text {failures }}+\underbrace{t_{\text {step}}}_{\text {success}}) p_i,
\label{eq:geomTsuccForm}
\end{equation}
where we have introduced the quantity $t_{\text{step}}$ as the duration of each instance where the system makes an unsuccessful excursion (e.g. a microtubule polymerizes without finding the chromosome).
The second term in the sum immediately yields $t_{\text{step}}$ since the distribution is normalized
leaving us with
\begin{equation}
\langle t\rangle=\sum_{i=0}^{\infty}i \, t_{\text{step}}\, p_i+t_{\text{step}}.
\label{eq:simpleGeomTsteppedAvgTime}
\end{equation}
Next, we need to evaluate
\begin{equation}
\langle t \rangle = t_{\text{step}}\sum_{i=0}^{\infty} i \, p_i+t_{\text{step}}=  t_{\text{step}} \, p_s \sum_{i=0}^{\infty} i  (1-p_s)^{i}+t_{\text{step}},
\label{eqn:AverageTime1}
\end{equation}
which we recognize as having a term of the form
\begin{equation}
\sum_{i=0}^{\infty} i x^i = x {d \over dx} \sum_{i=1}^{\infty} x^i  = x {d \over dx}  {1 \over 1-x}= {x \over (1-x)^2}
= {1- p_s \over p_s},
\label{eqn:DerivativeTrick}
\end{equation}
where we have used the definition $x=1-p_s$.
If we now invoke this result in the context of eqn.~\ref{eqn:AverageTime1}, we find
the very intuitive result that 
\begin{equation}
\langle t \rangle =t_{\text{step}}\left(1 + {1-p_s \over p_s}\right)= {t_{\text{step}} \over p_s}.
\label{eq:geometricMeanTime}
\end{equation}
The intuition for this result is best served by concretely imagining that $p_s=1/M$ which means
that $1$ out of every $M$ trials is successful.  This implies in turn that $\langle t \rangle=M \, t_{\text{step}}$,
stating that on average we need to carry out $M$ trials for the exploration
to be a success when the probability of success on a given trajectory is $p_s=1/M$.

This example showcases the structure of a first model for many problems in exploratory dynamics that both we and others have described elsewhere~\cite{Holy1994,Wollman2005,kondev2025}
We will often characterize the exploratory
process as a statistical mechanics of trajectories, with all such trajectories 
sharing the feature that they involve repeated ``failures'' until they succeed on
the final trajectory.  Interestingly, for the examples considered throughout
the paper, we will see many different realizations of the space of trajectories,
but as noted above, they are often of the form $p_i=\underbrace{p_f p_f \cdots p_f}_{\text i} p_s$, and perhaps surprisingly, even when our mathematical analysis does not start with the geometric distribution, it will often end there. 
In an earlier paper, we showed a number of results in exploratory dynamics that can be described precisely in terms of the geometric distribution~\cite{Holy1994,Wollman2005,kondev2025}. As a result, we turn to the next class of mathematical approaches shown in
Figure~\ref{fig:DifferentMath}, random walks---and, a realization of them particularly practical in biology, namely random walks with resets.

\section{Exploration Through Random Walks with Reset}
\label{section:RandomWalkReset}
\begin{figure}[h]
\centering
\includegraphics[width=\linewidth]{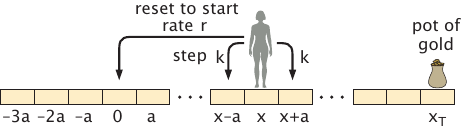}
\caption{\raggedright The random seeker with reset. Schematic of the steps available to the random seeker during a time step. During each discrete time step $\Delta t$, the walker can step to the right or left with probability $k \Delta t$ or reset to the origin with probability $r \Delta t$. }
\label{fig:RandomWalkReset}
\end{figure}
\FloatBarrier

\begin{figure*}[t]
\centering
\includegraphics[width=\textwidth]{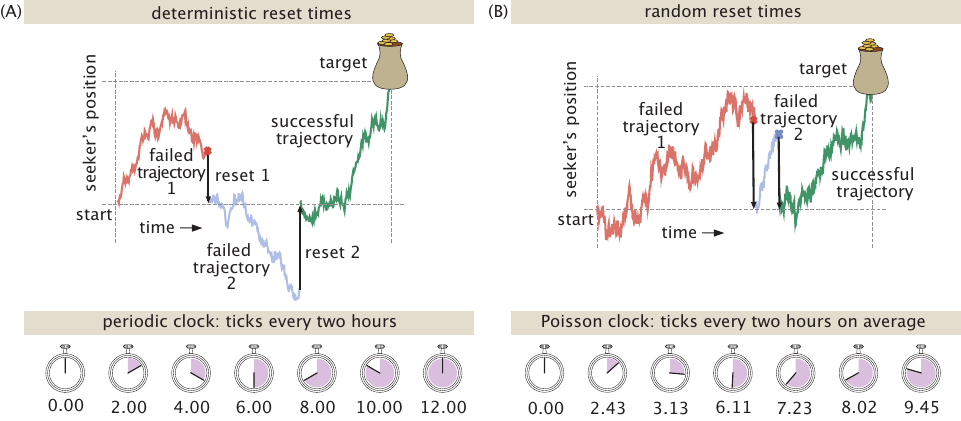}
\caption{\raggedright Random walks with resets. (A) Top: Deterministic resets in which after a fixed number of steps $T_R$, the walker resets to the origin. Bottom: This corresponds to resets performed with a periodic clock. (B) Random resets, where resets to the origin occur at a stochastic rate $r = 1/T_R$, making the time between resets match $T_R$ only on average. Bottom: This protocol corresponds to a Poissonian clock instead of a perfectly periodic clock. }
\label{fig:RandomWalkResets}
\end{figure*}

\begin{figure}
    \centering
    \includegraphics[width=\linewidth]{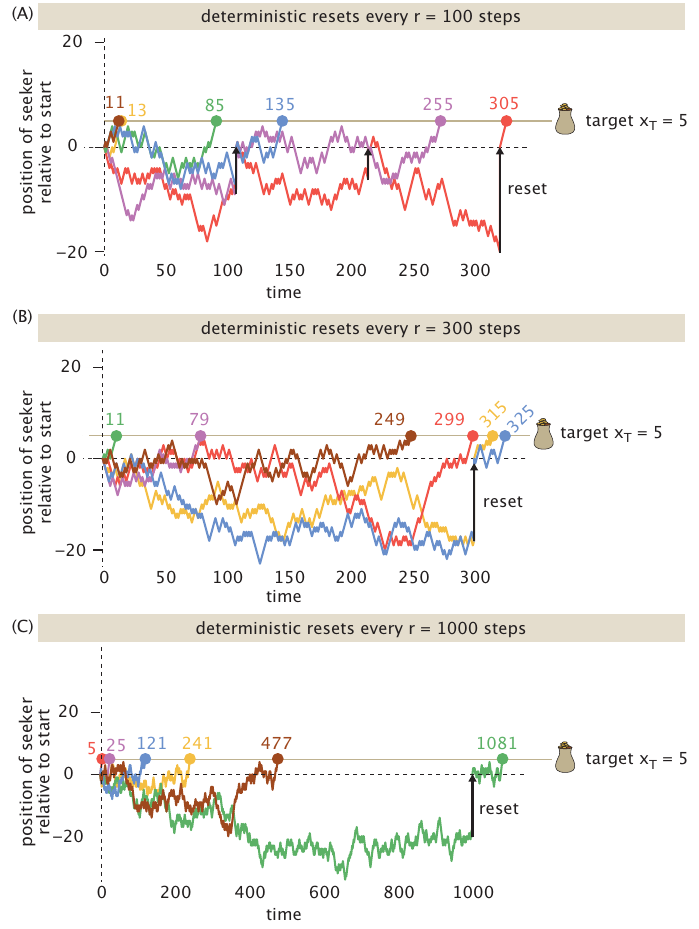}
\caption{\raggedright Time to reach a target in the presence of reset. Every $T_R$ steps, each walker deterministically returns to the origin, no matter its present proximity to the target. The resets are indicated by vertical black arrows. The three examples show representative trajectories reaching the target at position $x_T=5$, under deterministic reset times of (A) $T_R=$ 100, (B) $T_R=$ 300, and (C) $T_R=$ 1000 steps. The times at which each stochastic trajectory first reaches the target are marked by dots.}
    \label{fig:TimeToReachTarget2}
\end{figure}

A second foundational model is the random walk with reset~\cite{evans2011,Reuveni2016,Evans2020a,Bressloff2020a,Bressloff2020b}, as shown in Figure~\ref{fig:RandomWalkReset}. In this model, a ``seeker'' begins at the origin, $x=0$, and searches for a target at position $x_T$. The seeker walks randomly (e.g. diffuses), but at a given rate $r$ it is returned to the origin and begins again. As shown by Figure~\ref{fig:RandomWalkResets}, the seeker suffers either deterministic resets at a fixed time $T_R=1/r$, or else resets occur stochastically at an average rate $r$. Figure~\ref{fig:TimeToReachTarget2} gives further examples of trajectories under deterministic resets at varying rates to visualize the  arrivals to a target at an example location of $x_T = 5$. This is a deliberately stripped-down model of repeated exploration. It is inspired by systems such as microtubule search for kinetochores in the process of spindle assembly. There, failed growth episodes are erased by catastrophes, and new attempts begin from a preferred location~\cite{mitchison1984,Holy1994,Wollman2005}. Why should such resets or structural decorations on random walks be deployed by biological systems?
In Sec.~A of the Supplemental Material, we examine the properties of random walks without resets with an eye to why they might not suffice for some biological exploratory processes on generic mathematical grounds.

\subsection{Simulating the Simple Random Walk With Reset}
\begin{figure*}[t]
    \centering
    \includegraphics[width=\textwidth]{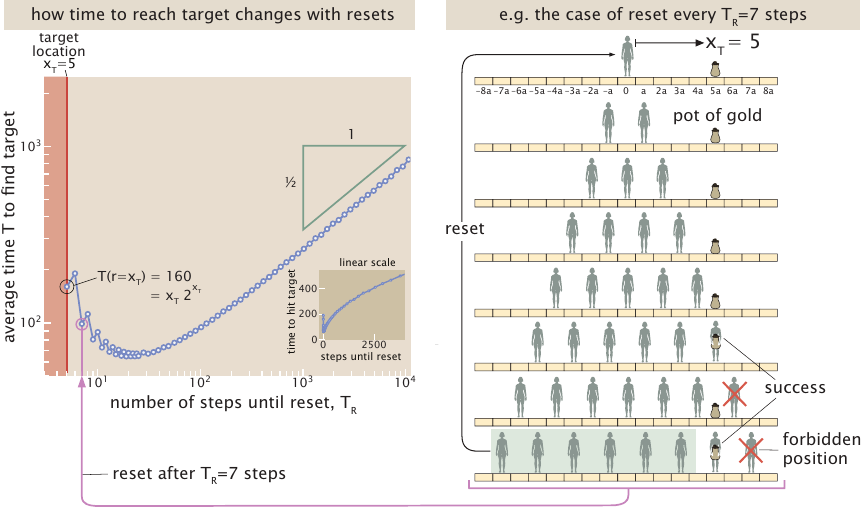}
\caption{\raggedright Simulation of random walk with reset in one-dimension with a target at position $x_T$.  The graph at left shows the average time to find the target, $T$, as a function of the number of steps until a hard reset to the origin, $T_R$. The inset shows a zoom of the same curve on a linear scale. Note sawtooth oscillations which reflect parity structure in the discrete counting search problem (not sampling artifacts); 
see also Sec.~B, ``First-passage properties of ordinary random walks,'' of the Supplemental Material. The schematic on the right illustrates the possible positions of the seeker after consecutive steps, subject to reset after $T_R=7$ steps.  The seekers shaded in green show the endpoints of trajectories that did not find the target and are hence sent back to the origin to begin their target search again.}
    \label{fig:TimeToReachTarget}
\end{figure*}

How do resets change the average time a seeker needs to reach a target? Direct numerical simulations give an immediate feel for this question. Figure~\ref{fig:TimeToReachTarget} averages the required length of many simulated trajectories (such as those highlighted in Fig.~\ref{fig:TimeToReachTarget2}) to arrive at a target located at $x_T=5$ for the first time, as a function of the reset time $T_R$, in $d=1$ dimension. Intriguingly, as the graph in Figure~\ref{fig:TimeToReachTarget} makes clear, these resets can intensify the exploration process: prudently choosing a reset rate can minimize the average time to a target. This acceleration of typical searches is initially surprising, given that  resets in fact abort many trajectories on the precipice of success. 
The graph in Figure~\ref{fig:TimeToReachTarget} showing how success time varies with reset rate has rich qualitative and numerical structure, and now our goal is to understand the different parts of the graph.

Before proceeding with a quantitative analysis, it is useful to ask why the curve in Figure~\ref{fig:TimeToReachTarget} should exhibit a minimum at all. If the reset time $T_R$ is too short, the searcher is repeatedly returned to the origin before it has a reasonable chance of reaching the target, and successful trajectories become rare. (If the time between resets $T_R$ is shorter than the target location (when both are measured as a number of steps), as in the red shaded region in Fig.~\ref{fig:TimeToReachTarget}, then the target is never reached.) On the other hand, if
$T_R$ is very large, resets occur infrequently and the search approaches unconstrained diffusion. In that regime, the searcher can spend a long time on unproductive excursions far from the target, leading again to large search times; this is the original quixotic character discussed in Sec.~A of the Supplemental Material.
These two competing effects suggest that an optimal finite reset
time should exist.

We can also anticipate the scale of this optimum using a simple diffusive argument. In a time $t$, a walker diffusing with diffusion coefficient $D$ typically explores a distance of order
$\sqrt{Dt}$. Reaching a target located a distance $x_T$ from the origin
therefore requires a characteristic time of order
\begin{equation}
T_{\rm diff} \thicksim \frac{x_T^2}{D}\thicksim  \frac{1}{k}\frac{x_T^2}{a^2},
\end{equation}
where $k$ is the stepping rate and $a$ is the lattice spacing (as schematized in Fig.~\ref{fig:RandomWalkReset}), and $D=ka^2$ is the emergent diffusion coefficient.
Resetting on a timescale much shorter than $T_{\rm diff}$ prevents the
searcher from reaching the target, while resetting on a much longer
timescale allows rare but costly excursions to dominate the search.
Consequently, we expect the optimal reset time to be of the same order as
the diffusive first-passage time,
\begin{equation}
T_R^* \sim \frac{x_T^2}{D},
\end{equation}
a prediction that will emerge naturally from the calculations that follow. (This is also the only timescale defineable by the physical parameters entering the problem, anticipating the optimal time will vary with this quantity.)

First, to appreciate the curve to the left of the optimal reset rate in Figure~\ref{fig:TimeToReachTarget}, can we understand the search time for the special case in which the reset time is equal to the distance of the target from the origin? Note that here we use dimensionless units of ``steps'' for both distance and time.  
For the case study considered here, the target is at $x_T=5$, thus any reset time shorter than $T_R=5$ can never find the target.  If we allow 5 steps before reset, there is only one successful trajectory  which is $RRRRR$ with probability $1/32$.  The seeker must make 5 successive steps to the right, each with probability 1/2.  This means that out of the $2^5 =32$ possible trajectories, 31 of them do {\it not} reach target and they are characterized by probability
 $31/32$.  Hence, if we want to sum over all the possible trajectories to find the average time to reach the target, we can write 
\begin{equation}
\langle \#  \text { steps}\rangle=\sum_{n=1}^{\infty} 5n \left({31 \over 32}\right)^{n-1} {1 \over 32},
\end{equation}
where $n-1$ is once again the number of ``failures'' before success.  The factor of $5n$ measures the fact that for each of the $n-1$ failed attempts, they were of length 5 before the reset and then the final trajectory is itself of length 5 since that one is of the form $RRRRR$.
Using the geometric distribution as highlighted in  Table~\ref{tab:exploratory} and described in detail in section~\ref{section:GeometricDistribution}, we find that the average number of steps until hitting the target is 160, precisely what we see for the point on the vertical red line in Figure~\ref{fig:TimeToReachTarget}.

Let us now solve this problem for the more general case in which $x_T=n$ and we reset after n steps.  
 We assume that each step is either $+1$ or $-1$ with probability $1/2$.
In this special case there is a very simple observation. Since the walker has only $n$ steps available before reset, the largest position it can possibly attain is $x=n$. That can happen only if every one of the $n$ steps is a step to the right. Thus there is exactly one successful trajectory in each attempt,
\begin{equation}
(+1,+1,\ldots,+1),
\end{equation}
and therefore the probability of success in a single attempt is
\begin{equation}
p_s=\left(\frac{1}{2}\right)^n.
\end{equation}
The probability of failure in a single attempt is hence
\begin{equation}
1-p_s=1-\left(\frac{1}{2}\right)^n.
\end{equation}

Now let $i$ be the number of the attempt on which success first occurs. Then the probability that the first $i-1$ attempts fail and the $i$th attempt succeeds is
\begin{equation}
p(i)=\left[1-\left(\frac{1}{2}\right)^n\right]^{i-1}\left(\frac{1}{2}\right)^n,
\qquad i=1,2,3,\ldots
\end{equation}
This is once again a geometric distribution, specifically an application of eqn.~\ref{eq:geomPi}, revealing that our work in the previous section sheds light on the problem of random seeker with resets as well. Since each attempt lasts exactly $i$ steps, the total number of steps taken before success on the $n$th attempt is
\begin{equation}
N=in.
\end{equation}
Hence, the mean number of steps is given by,
\begin{equation}
\langle N\rangle
=
\sum_{i=1}^{\infty}
in
\left[1-\left(\frac{1}{2}\right)^n\right]^{i-1}
\left(\frac{1}{2}\right)^n,
\end{equation}
which is an application of eqn.~\ref{eq:geomTsuccForm}.
Using the standard identity for the mean of the geometric distribution, this becomes
\begin{equation}
\langle N\rangle
=
n\,\frac{1}{(1/2)^n}
=
n\,2^n.
\end{equation}
It is worth checking that this reproduces the case $n=5$. In that case we find
\begin{equation}
\langle N\rangle = 5\cdot 2^5 = 160,
\end{equation}
as already determined above, via eqn.~\ref{eq:simpleGeomTsteppedAvgTime}, and shown by the point circled in black in Figure~\ref{fig:TimeToReachTarget}.

Now that we have explained the numerical value of the time to find the target for the special case in which the reset time is equal to the distance of the target from the origin, we look to the opposite limit of the graph where the time to find the target is an increasing function of the reset time. 

\subsection{Approximate Description of The Simple Random Walk With Reset}
\label{sec:Approximate Description of The Simple Random Walk With Reset}

An approximate approach to the seeker-with-reset problem that we find particularly appealing is to replace the explicit reset dynamics by random walk in a finite spatial domain. The central idea is simple. In the original problem, the seeker begins at the origin and repeatedly explores until reset returns it to the starting point. Rather than tracking the complicated history of many interrupted excursions, we ask whether there is a simpler description that captures the same physics at the level of scaling behavior. The approximation is to imagine that reset limits how far the seeker can typically wander and to replace that limitation by a reflecting boundary. This idea is justified when the target location is small compared to the resetting interval, $x_T \ll T_R$.

\begin{figure*}[t]
\centering
\includegraphics[width=\textwidth]{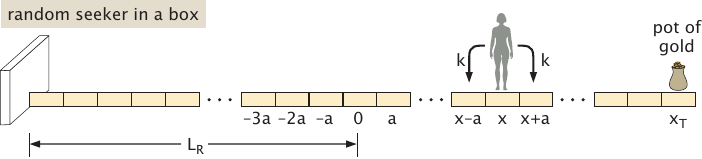}
\caption{\raggedright Replacing the random walker with reset by a walker in a finite box. The width of the box in the negative $x$-direction is set by the average number of steps $L_R$ that the seeker walks before being sent back to the origin through reset. A naive estimate for the characteristic exploration length under reset is $L_R = \sqrt{D T_R}$.}
\label{fig:BoxModel}
\end{figure*}

Hence, we keep the geometry of the original problem intact as shown in Figure~\ref{fig:BoxModel}. The seeker begins at the origin, the treasure remains at position $x=x_T$, and the effect of reset is represented by a reflecting wall placed a distance $L_R$ to the left of the origin. The resulting finite-box problem plays out on the interval
\begin{equation}
-L_R\le x\le x_T .
\end{equation}
The point $x=x_T$ is absorbing because reaching the treasure terminates the search, while the boundary at $x=-L_R$ is reflecting because the seeker is not allowed to wander indefinitely far away from the target. 
A natural estimate for the scale of this confinement length comes from ordinary diffusion. A random walker diffuses a characteristic distance of order $\sqrt{DT_R}$ during the reset time $T_R$, where the diffusion coefficient $D \equiv \frac{a^2}{2 \Delta t}$ relates to the hopping rate $k$. The simplest guess would then take the effective distance of the reflective wall $L_R$ as $L_R=\sqrt{DT_R}$. This naive estimates already captures the essential scaling. 
However, more sophisticated treatments of this problem are tractable and interesting; these more precisely specify the appropriate lengthscale sampled by walks under reset. These analyses give numerical prefactors for the lengthscale $L_R$ that accords more closely with the dynamics of reset as we explore in Sec.~C, ``Analytics of random walks with deterministic reset,'' of the Supplemental Material.
Accordingly, we acknowledge such a dimensionless constant $C$ entering $L_R$ as,
\begin{equation}
L_R=C\sqrt{DT_R}.
\label{eq:constrainLRlengthscale}
\end{equation}
 In effect, we have replaced a difficult dynamical problem involving repeated resets by a much simpler first-passage problem in a finite domain.

\begin{figure*}
    \centering
    \includegraphics[width=\linewidth]{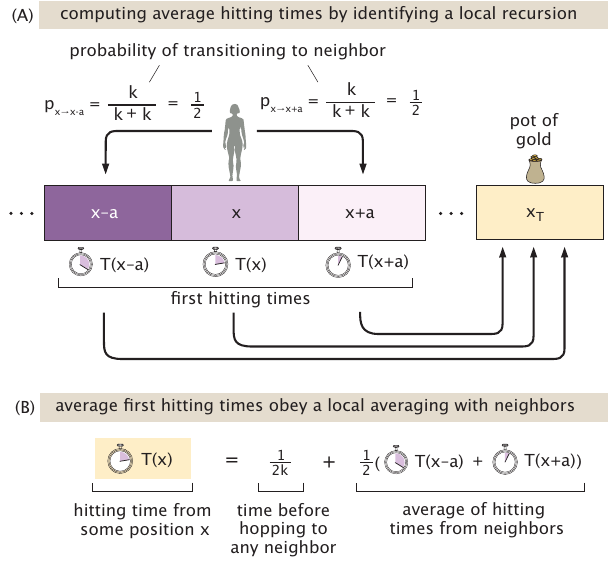}
\caption{\raggedright Origins of the local averaging property of a hitting time to a target. (A) The average  time $T(x)$ for a seeker to hit a target starting from position $x$ is consistent with the analogous times $T(x-a)$ and $T(x+a)$ starting from neighboring positions. The purple shading of each site schematizes its local first hitting time; the self-consistency between adjacent sites means a site's time (shading) smooths between those of its neighbors. The explicit mention of $k/(k+k_)$ is to highlight that there is a branching ratio between two processes. In this case, those two processes have the same rate and hence the step probabilities are the same. (B) This self-consistency is a local average.} \label{fig:explainSimpleHitTimeRecurse}
\end{figure*}

Armed with this picture of motion in a domain effectively bounded by resets, how long on average will it take for a seeker to reach the treasure at $x_T$? To approach this question, which inquires about the mean first hitting time from the walker's position to the target position $x_T$ under diffusive motion, we use the following gambit as illustrated in Figure~\ref{fig:explainSimpleHitTimeRecurse}. This strategy is well established~\cite{Norris1998}, mechanically simple, but conceptually subtle. It will recur again and again throughout our studies of complex exploratory trajectories, rewarding some discussion now.

While we care about the time $T(x)$ a particular walker released at a \emph{particular} position $x$ (say, the origin) will take to reach the target, we could easily ask the same question of any other starting position. For example,  if we had started instead one site to the left at position $x-a$. The key idea is that once a walker arrives at a position, its propensity for where to step next, and crucially, the average time it takes to later wander anywhere else in the domain, does not depend at all on its past, and depends only on its current position. This property (the so-called Markov property of the walk) means there is a set of well-defined numbers $T(x-a), T(x), T(x+a)$, and so on, that specify the average times that a walker (released at $x-a$, $x$, $x+a$) will take to arrive at any target, no matter the paths they took to arrive at those starting points. Crucially, this requires that the time $T(x)$ to hit a target from $x$ must be internally consistent with those of its neighbors, $T(x-a)$ and $T(x+a)$, according to a very specific and natural pattern. Specifically, as shown in Fig.~\ref{fig:explainSimpleHitTimeRecurse}, the average time $T(x)$ to reach position $x_T$ satisfies a sort of local average with those of its neighbors, according to,
\begin{equation}
T(x)=\frac{1}{2k}+ \frac{1}{2}T(x-a)+\frac{1}{2}T(x+a),
\label{eq:recursionBasic}
\end{equation}
where the first constant term $\frac{1}{2k} \equiv \Delta t$ represents the average time before a jump occurs to either neighboring position, where $k$ is the rate at which the walker steps right or left.
 This recursion relation eqn.~\ref{eq:recursionBasic} quantifies the intuition that if neighboring positions can quickly reach a target, and a walker hops between neighboring sites quickly, then the time to reach the target from the present position should also be rapid. The logic of this insightful procedure was developed by Kolmogorov \cite{dynkin1989kolmogorov, kolmogorov2019selected} and such recurrence relations are often referred to as ``backwards equations.'' Another way to see the emergence of such equations is via explicit averages over trajectories; see Sec.~D, ``Average hitting times as explicit averages over trajectories,'' of the Supplemental Material.

This recursion can be simplified by resorting to a
Taylor expansion of the form
\begin{equation}
\begin{aligned}
T(x)=\Delta t & +\frac{1}{2}\left(T(x)-a T^{\prime}(x)+\frac{a^2}{2} T^{\prime \prime}(x)\right) \\
& +\frac{1}{2}\left(T(x)+a T^{\prime}(x)+\frac{a^2}{2} T^{\prime \prime}(x)\right)
\end{aligned}
\end{equation}
After performing the relevant cancellations we are left with
\begin{equation}
\frac{d^2 T}{d x^2}=-\frac{1}{D} 
\end{equation}
where we define the diffusion coefficient
$D=a^2/2\Delta t$.
The general solution of the differential equation is
\begin{equation}
T(x)=-{x^2\over 2D}+Ax+B.
\end{equation}
To determine the coefficients $A$ and $B$ we need to impose the boundary conditions.
Of course, if we start at $x_T$, we don't have to make any steps and hence the boundary condition there is that the time to reach the target is zero, written mathematically as
\begin{equation}
T(x_T) = 0.
\end{equation}
For the boundary condition at $-L_R$, we note that 
reaching the point $-L_R$ means that the seeker has only one option which is to step back to the right.
This condition is equivalent to
\begin{equation}
\left.{dT \over dx}\right|_{x=-L_R}=0.
\end{equation}
To understand why this reflecting boundary condition materializes, consider the relation for the first hitting time at the boundary, where only one move is possible, enforcing,
\begin{align}
    T(-L_R) = \frac{1}{2k} + T(-L_R + a).
\end{align}
Taylor expanding the right-most term near $x=-L_R$ gives,
\begin{align}
    T(-L_R) = \frac{1}{2k} + T(-L_R) + \left.\frac{dT}{dx}\right|_{x=-L_R} a
\end{align}
Substituting $1/2k = a^2/2D$ yields
\begin{align}
    \left.\frac{dT}{dx}\right|_{x=-L_R} = - \frac{a}{2D}.
\end{align}
In the continuum limit of relatively small lattice spacing $a/L_R\rightarrow 0$, this gives indeed, 
\begin{align}
    \left.\frac{dT}{dx}\right|_{x=-L_R} = 0.
\end{align}

Imposing the reflecting condition gives
\begin{equation}
A=-{L_R\over D},
\end{equation}
and imposing the absorbing condition at the treasure gives
\begin{equation}
B={x_T^2\over 2D}+{L_Rx_T\over D}.
\end{equation}
As a result, the mean first-passage time becomes
\begin{equation}
T(x)=-{x^2\over 2D}-{L_Rx\over D}+{x_T^2\over 2D}+{L_Rx_T\over D}.
\label{eq:firstHitFiniteRegionResult}
\end{equation}

Since the seeker begins at the origin, the quantity of interest is simply
\begin{equation}
T(0)={x_T^2\over 2D}+{L_Rx_T\over D}.
\label{eq:mainHitTimeFromZeroIn1dReflectingBox}
\end{equation}
This expression has a simple interpretation. The first term is the ordinary diffusive contribution associated with reaching a target at distance $x_T$, while the second term reflects the penalty imposed by reset through the effective confinement length $L_R$. When the reset length is large compared with the target distance, so that $L_R\gg x_T$, the dominant contribution becomes
\begin{equation}
\label{eqn: MFPT under confinement if LR >> xT}
T(0)\approx {L_Rx_T\over D}.
\end{equation}
Substituting the heuristic $L_R = C\sqrt{DT_R}$ of eqn.~\ref{eq:constrainLRlengthscale} gives,
\begin{equation}
T(0)\approx x_T \, C \, \sqrt{{T_R\over D}},
\label{eq:approxSeekerBoxFinal}
\end{equation}
showing that the search time grows as the square root of the reset time as shown in Figure~\ref{fig:TimeToReachTarget}.

How does the above compare to an exact analytical expression for the mean first passage time of a random searcher subject to deterministic reset? In the large reset time limit $T_R\rightarrow \infty$, 
we find in Sec.~E.2, ``Equivalence of asymptotics of mean first passage times,'' of the Supplemental Material that the mean first passage time is given by 
\begin{align}
    T_{\text{exact}}(0) = x_T  \frac{2}{\sqrt{\pi}} \sqrt{\frac{T_R}{D}}.
\end{align}
Comparing to eqn.~\ref{eq:approxSeekerBoxFinal} identifies the prefactor $C = \frac{2}{\sqrt{\pi}}$, resulting in the reflecting boundary being placed at
\begin{align}
    \label{eqn: location of reflecting boundary following extremal value arguments}
    L_R = \frac{2}{\sqrt{\pi}}\sqrt{D T_R}.
\end{align}
Interestingly, this prefactor may be physically rationalized by understanding that the presiding length scale is not precisely the typical diffusive lengthscale $\sqrt{D T_R}$, but rather the expectation of the largest distance explored to the left  $\left\langle \max_{0\leq t\leq T_R} x(t) \right \rangle = \left(2 / \sqrt{\pi}\right) \sqrt{D T_R}$ between resets \cite{HUANG2024129389}. Indeed, the confining box must be large enough to reproduce the full spatial region visited by the walker.

In addition, the result of eqn.~\ref{eq:firstHitFiniteRegionResult} is also illuminating in the limit of no resets, namely $L_R\rightarrow \infty$: the fact that $T(0)$ grows to infinity in this limit is another explicit affirmation of the fact that the mean time to reach a target in 1D with pure diffusion is infinite, as presaged by Sec.~A of the Supplemental Material. The action of the reset is to introduce the finite lengthscale $L_R$ and makes this time finite.

Thus a seemingly complicated problem involving repeated interruption and restart admits a remarkably simple approximation in which reset merely confines the seeker to a finite region of exploration.

\subsection{Exact Description of The Simple Random Walk With Reset}

Having earned a qualitative feeling
for the time for the seeker to find
the target as a function of reset time,
we resort to a full calculation to appreciate the structure of the whole curve shown in Figure~\ref{fig:TimeToReachTarget}. Thus far,
we have spoken of deterministic resets as shown in 
Figure~\ref{fig:RandomWalkResets}(A). However, as shown in Figure~\ref{fig:RandomWalkResets}(B), a close and mathematically tractable alternative is to consider the case where the seeker suffers a stochastic reset with an average rate $r=1/T_R$ (instead of a deterministic periodic reset). 

Once again, we appeal to the idea of making a recursive calculation that recognizes that the time  to reach the target given the seeker is currently at position $x$ can be obtained by adding the times it would take had it been at adjacent positions.
In particular, we can write
\begin{equation}
\begin{aligned}
& T(x)=\frac{1}{2 k+r}+\frac{k}{2 k+r} T(x+a)+\frac{k}{2 k+r} T(x-a) \\
&+\frac{r}{2 k+r} T(0),
\end{aligned}
\end{equation}
where the average rate of steps is by $2k+r$. The first term on the right, $1/(2k+r)$, is the length of a time step accomplishing any jump; it is the analog of the $1/2k$ term in the earlier more basic example of eqn.~\ref{eq:recursionBasic}. The weights for each of the remaining terms are given by branching ratios. The two terms with factors $k/(2k+r)$ correspond to the seeker jumping left or right. Last, the term with factor $r/(2k+r)$ gives the branching probability of reset to the origin.  In light of this recursive structure, we can rewrite our equation as,
\begin{equation}
\begin{aligned}
T(x)(2 k+r)= & 1+k T(x+a)+k T(x-a) \\
& +r T(0)
\end{aligned}
\end{equation}
which invites us to perform Taylor expansions of $T(x+a)$ and $T(x-a)$ with
the result
\begin{equation}
0=1-r T(x)+r T(0)+\frac{k a^2}{2} T^{\prime \prime}(x).
\end{equation}
Hence, in the absence of drift, the differential equation for the mean first passage time is given by
\begin{equation}
0=1+D\,T''(x)-r\,T(x)+r\,T(0).
\label{eq:backward_equation_finalmain}
\end{equation}
We can rewrite this as 
\begin{equation}
0=D T''(x)-r\left(T(x)-T(0)- {1 \over r}\right).
\end{equation}
This differential equation holds for all $x<x_T$, and must be supplemented with the absorbing boundary condition
\begin{equation}
T(x_T)=0,
\end{equation}
which is the statement that the target is at $x=x_T$ as shown in Figure~\ref{fig:TimeToReachTarget}.
We now introduce a transformed function
\begin{equation}
    y(x) = T(x) - T(0) - \frac{1}{r}
\end{equation}
which leads to the transformed differential equation
\begin{equation}
    y''(x) - \frac{1}{\lambda_D^2} y(x) = 0,
\end{equation}
where we define  $\lambda_D = \sqrt{D/r}$ as the length scale over which the seeker diffuses before the reset occurs.
The general solution for this equation is
\begin{equation}
    y(x)
    =
    A e^{-x/\lambda_D}
    +
    B e^{x/\lambda_D},
\end{equation}
with $A$ and $B$ real constants.
This gives us
\begin{equation}
    T(x) = Ae^{-x/\lambda_D} + B e^{x/\lambda_D} + T(0)+\frac{1}{r}.
    \label{eq:generalFirstHitTimeFromReset}
\end{equation}
Because of the resets, we know that even when trajectories wander to positions $x$ very far away, $x\to -\infty$, the waiting time to reach the target $x_T$ will not be infinite. This is because searchers will eventually reset to the start, and then go from there. This yields the additional required boundary condition,
\begin{align}
    \lim_{x\rightarrow -\infty} T(x) =\text{finite}.
    \label{eq:finite1DresetRandomWalkBC}
\end{align}
This condition requires we take $A=0$. Next, we enforce the additional boundary condition $T(x_T)=0$, namely that the time to success is zero once the searcher has already reached the target. This condition applied to eq.~\ref{eq:generalFirstHitTimeFromReset} specifies the value of the remaining  coefficient $B$ as,
\begin{equation}
    B= -e^{-x_T/\lambda_D} \left(T(0) + \frac{1}{r}\right)
\end{equation}
which results in the expression
\begin{equation}
    T(x) = \left(T(0) + \frac{1}{r}\right)\left(1-e^{(x-x_T)/\lambda_D}\right).
\end{equation}
When we start at the origin, this simplifies to
\begin{equation} 
    T(0) = T(0)\left(1-e^{-x_T/\lambda_D}\right) + \frac{1}{r}\left(1-e^{-x_T/\lambda_D}\right),
\end{equation}
which can be simplified in turn to the form
\begin{equation}
    T(0)=  \frac{e^{x_T/\lambda_D}}{r}\left(1-e^{-x_T/\lambda_D}\right).
\end{equation}
We can rewrite this as the extremely informative final expression, giving a theoretical prediction for the behavior of stochastic reset, 
\begin{equation}
\label{eqn:searchtimeresets}
    T(0)=  \frac{1}{r}\left(e^{x_T/\lambda_D}-1\right).
\end{equation}

\begin{figure}
    \centering
    \includegraphics[width=0.5\textwidth]{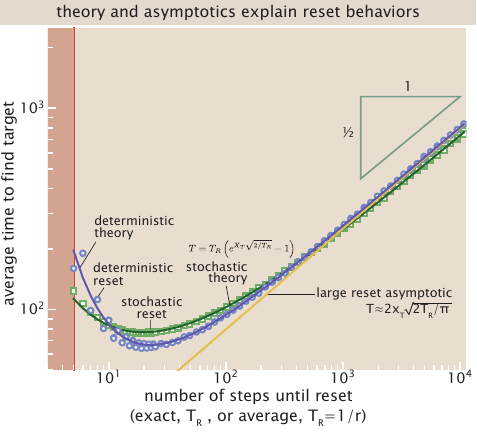}
\caption{\raggedright Agreement of theoretical descriptions of random walk under reset with direct numerical simulations. The blue-edged circle markers are the same numerically simulated behaviors visualized in Fig.~\ref{fig:TimeToReachTarget}. The green-edged square markers show the analog of these direct simulations, but where the resets are executed stochastically with an average rate, instead of deterministically with the period of the reset. The gold straight line visualizes the large reset time asymptotic in the case of deterministic reset, $T\approx 2x_T \sqrt{2 T_R/\pi}$. The dark green and dark purple lines visualize the theoretical expressions for stochastic resets (eqn.~\ref{eqn:SearchTimeFinal}) and deterministic resets (derived in Sec.~C.3, ``Mean first passage time of random searcher,'' of the Supplemental Material) 
respectively, showing that despite the technical difference in executions of these resets, this description describes the quantitative features of both reset scenarios with excellent agreement. (The red shaded region where $T_R < x_T$ delimits the physical region where resets are too fast to allow any successful encounters of the target.)}
    \label{fig:TimeToReachTarget_stochasticAndFits}
\end{figure}

How well does the stochastic model prediction of eqn.~\ref{eqn:searchtimeresets} compare to exact numerical simulations of the sort we set out to understand? Fig.~\ref{fig:TimeToReachTarget_stochasticAndFits} reports in green square markers the results of simulations with stochastic reset summarized by their average times to hit a target, analogous to the original deterministically-reset curve of interest in Fig.~\ref{fig:TimeToReachTarget} (also reproduced for comparison as blue circles). The theoretical prediction of eqn.~\ref{eqn:searchtimeresets} for stochastic reset is shown as a black line (informed by the knowledge that $D=1/2$ in units of time and space steps in this discrete simulation, and recalling $1/r = T_R$). The theoretical prediction accords with the stochastic reset simulations, and also closely reproduces the major features of the related deterministic reset problem in blue. We note that the distinctions and comparative performances between deterministic resetting and stochastic resetting support an active arena of mathematical inquiry; for instance,  investigations show that a deterministic resetting strategy can reach targets in smaller optimal times than stochastic resetting strategies \cite{chechkin2018random}, but stochastic resetting strategies can show performance benefits for randomly-distributed targets \cite{evans2025stochastic}.

The next question we might ask in our qualitative attempt to understand the graph is  how the position of the minimum depends upon the distance to the target. The distance to the target ($x_T$) and the diffusion constant $D$ combine to give the only time scale in the problem $x_T^2/D$. This implies that the minimum time to target must scale as  $T_{min} \thicksim  x_T^2/D$.  
To formalize that result and estimate the prefactors, we consider the average search time as a function of the reset time $T_R=1/r$ reformulating 
eqn.~\ref{eqn:searchtimeresets} as
\begin{equation}
\label{eqn:SearchTimeFinal}
T(T_R)=T_R\left(e^{x_T/\sqrt{D T_R}}-1\right),
\end{equation}
and seek the value of $T_R$ that minimizes it.
Here, as mentioned earlier, we approximated the original problem shaped by a deterministic reset time using a closely related problem steered by stochastic resets with rate $ r = 1/T_R$. 
To that end, we define a dimensionless average search time, $\tilde{T} = T D/x_T^2$, and a dimensionless reset time $\tilde{T_R}= T_R D/x_T^2$, which transforms eqn.~\ref{eqn:SearchTimeFinal} into 
\begin{equation}
\label{eqn:SearchTimeFinalDimensionless}
\tilde{T}(\tilde{T_R})=\tilde{T_R}\left(e^{1/\sqrt{ \tilde{T_R}}}-1\right).
\end{equation}
Minimization leads to the condition
\begin{equation}
\label{eqn:SearchTimeMinimizer}
e^{1/\sqrt{ \tilde{T_R}}} \left(1 - {1 \over 2\sqrt{\tilde{T_r}}} \right) = 1.
\end{equation}
This transcendental equation has a unique positive solution
\begin{equation}
\tilde{T_R}^* \approx 0.394,
\end{equation}
corresponding to an optimal reset time in real units of,
\begin{equation}
T_R^* \approx 0.394\,\frac{x_T^2}{D}.
\label{eq:optimalReset}
\end{equation}
The corresponding minimal value for the average search time is
\begin{equation}
T_{\min}  \approx 1.54\,\frac{x_T^2}{D}.
\end{equation}

\begin{figure}[t!]
    \centering
    \includegraphics[width=\linewidth]{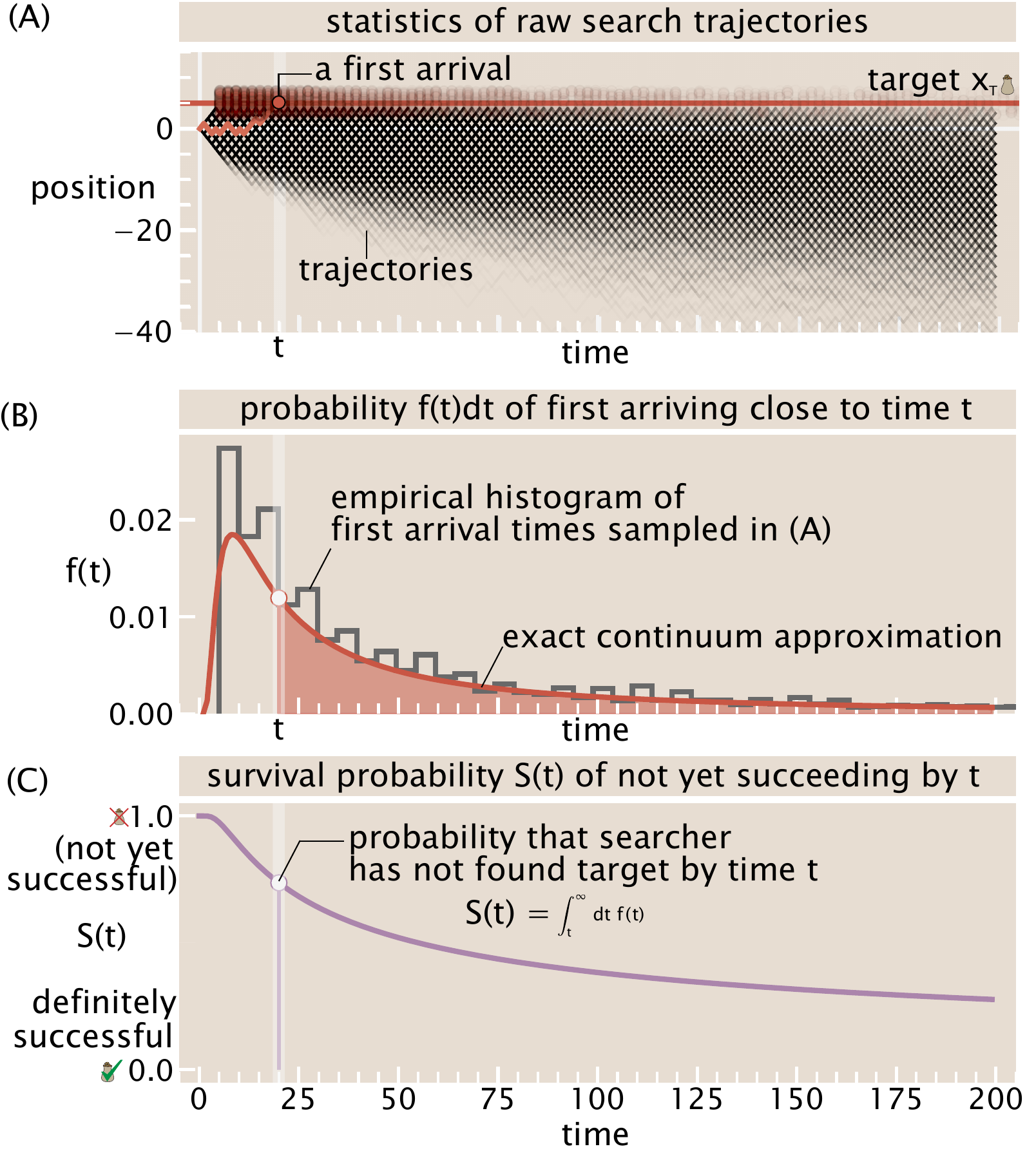}
    \caption{\raggedright The first hitting time probability density and survival probability from trajectories. (A) Many random trajectories (in black lines) wander until they first hit a target (here, shown at $x_T=5$, a location emphasized by the red horizontal line). The stochastic first arrival times $\tau_i$ of each trajectory $i$ are marked by red circles. Clearly, some first arrival times are much more common than others, as stressed by the gradation in density of these first arrivals. (B) The histogram of first arrival times to the target, shown as a grey empirical histogram for the example trajectories in (A), approach a continuum probability density $f(t)$ of first arriving near time $t$, shown in the red curve. Specifically, $f(t)dt$ gives the probability of first arriving at the target within an interval $dt$ of time $t$. (C) The total ``survival probability'' $S(t)$ that a searcher has not yet arrived at the target by time $t$, given by $S(t)=\int_t^\infty dt~f(t)$, gives another handle on the statistics of search; such a survival probability is explored further in the context of Fig.~\ref{fig:explainNotionOfSurvivalProbability_andNsearchIdea}. (The light vertical lines in all plots illustrate a particular time $t\approx T_R^*$, for explicit visual reference, and comparison to Fig.~\ref{fig:contrastWithNaive}.)}
    \label{fig:explainFirstHitDensityAndSurv}
\end{figure}

This optimal reset policy is interesting. It does not correspond to perfunctory heuristics one might guess before performing the calculations we just discussed. To appreciate these real subtleties, it is revealing to return to the beginning and ask, what would be appealing guesses for the optimal rate of resets? Given just the specified physics of the problem (the target location $x_T$ and the statistics of raw search trajectories without reset), one immediate guess might be to choose the time $T_R$ between resets that matches the most common time that trajectories otherwise ordinarily first hit the target. More precisely, say we run the unrestricted search many times, as in Fig.~\ref{fig:explainFirstHitDensityAndSurv}(A), and keep track of the times $\tau_i$ that each trajectory $i$ first hits the target (shown by the set of red circles first hitting the red target line). This would yield a distribution of first arrival times, which could be summarized by an empirical histogram (shown by the grey histogram of Fig.~\ref{fig:explainFirstHitDensityAndSurv}(B)). In the continuum limit of (many) long searches, this can be described by a probability density function $f(t)$ shown as the red curve in Fig.~\ref{fig:explainFirstHitDensityAndSurv}(c)), which encodes the probability $f(t)dt$ of a searcher arriving at the target within a time interval of $[t, t+dt]$. Our initial na{\"i}ve guess of taking the most common time that a searcher reaches the target for the reset corresponds to the mode of this first passage time probability distribution, which turns out to be $x_T^2/3$ (with $D=1$); see Sec.~B.2, ``Interpretations and continuum approximations,'' of the Supplemental Material. 
As stressed by Fig.~ \ref{fig:contrastWithNaive}, this guess conspicuously disagrees with the actual optimal reset rate identified by our more careful calculations. Giving trajectories more time before reset than the modal arrival time $x_T^2/3$ is needed to minimize the time to arrive at the target. 
Another tempting guess at the outset might be to choose some reset policy that nonetheless allows each search trajectory to have a decent chance of hitting the target before being compelled to reset. However, yet again, interestingly this expectation is not borne out. To appreciate this, it is instructive to consider the so-called ``survival probability'' $S(t)$ of a searcher not yet having succeeded at reaching the target by time $t$, as illustrated by Fig~\ref{fig:explainFirstHitDensityAndSurv}(C). This is a quantity worth appreciating as we will regularly refer to such behavior throughout calculations in this paper. As shown by Fig.~\ref{fig:contrastWithNaive},  the actual optimal reset time remains surprisingly aggressive in pruning trajectories: the probability that the seeker reaches the target within the optimal reset time is only $S(T_R^*) = \int_0^{T_R*} dt~ f(t) \approx 0.14$, meaning that $\approx$ 86\% of attempts are reset before finding the target. In general, these features would be obscure were it not for the understanding given by the exact calculation of eqn.~\ref{eq:optimalReset}. 
We are persuaded that these trajectory-level metrics are a powerful way to gauge the distinct character of microtrajectories, following on the idea of the {\it potency} of such trajectories we described elsewhere.

\begin{figure}[h!]
    \centering
    \includegraphics[width=\linewidth]{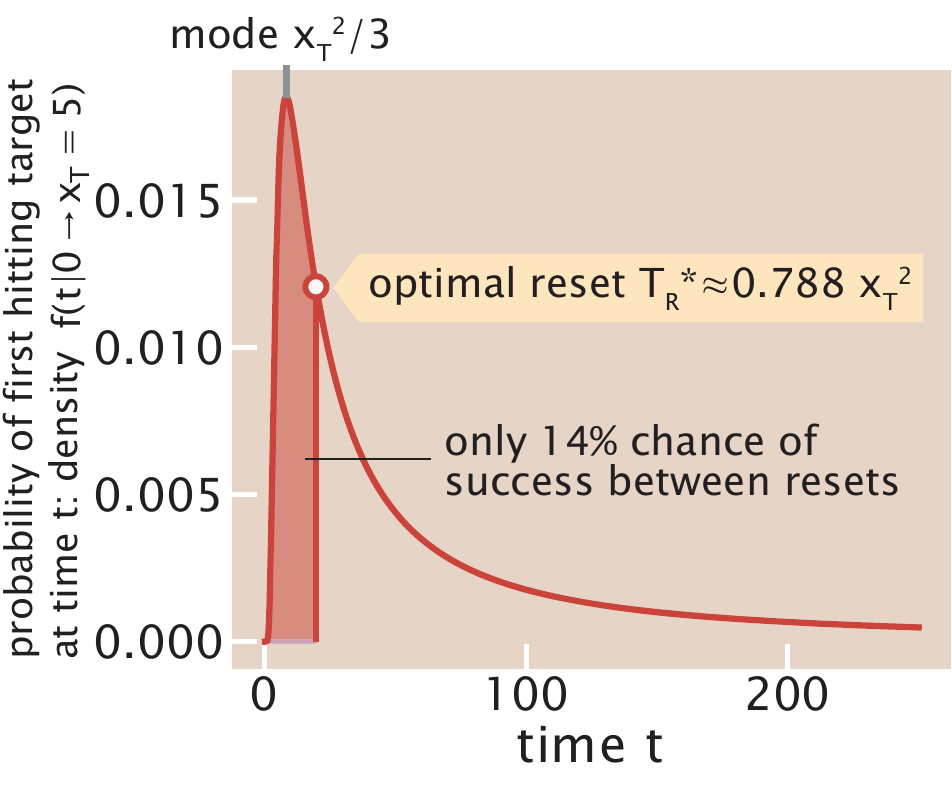}
    \caption{\raggedright The optimal reset time revealed by eqn.~\ref{eq:optimalReset}, shown as the teal vertical line, in the context of the first-passage time density (of a random walk), shown as the purple curve. The most-common time to reach the target for the first time is given by the mode $x_T^2/3$ marked by the vertical grey line which notably is quite distinct from the (longer) optimal reset policy specified by eqn.~\ref{eq:optimalReset}. The integral of the first-passage time density up to an allotted time $t$ gives the net probability of the walk reaching the target by this time; integrating up to the optimal reset time (denoted by the purple shading) suggests a $\approx$14\% probability of reaching the target in a bout between two resets according to the optimal policy.}
    \label{fig:contrastWithNaive}
\end{figure}

Eqn.~\ref{eqn:SearchTimeFinal} gives revealing limiting asymptotics that fortify intuition. First, consider the case of large reset time $T_R$: Taylor expanding eqn.~\ref{eqn:SearchTimeFinal} near $T_R\rightarrow \infty$ gives $T\approx x_T \sqrt{\frac{T_R}{D}} + \cdots$ to first order, explicitly recovering the square-root dependence of the mean time to find the target, emphasized by the teal triangle in Fig.~\ref{fig:TimeToReachTarget_stochasticAndFits}, that we reasoned about in the limiting physical scenario of the walker in a finite box in the preceding subsection and eqn.~\ref{eq:approxSeekerBoxFinal}. Next, consider the opposite extreme of a small reset time $T_R$ in the vicinity of the target location itself, namely $T_R = x_T + \epsilon$ for small $\epsilon$. Substituting this and expanding near $\epsilon \rightarrow 0$ gives $T\approx x_T \left(\exp\left[\sqrt{\frac{x_T}{D}}\right] - 1 \right) + \left(\exp\left[\sqrt{\frac{x_T}{D}}\right]\left(1-\sqrt{\frac{x_T}{D}} \right) - 1 \right) \epsilon + \cdots$ where $x_T$, $T_R$, and $\epsilon$ are all measured in units of steps; the precise form is less important than the takeaway of a local linear behavior in the departure $\epsilon$ from the $T_R = x_T$ case. These behaviors are visible in Fig.~\ref{fig:TimeToReachTarget_stochasticAndFits} as eqn.~\ref{eqn:SearchTimeFinal} stitches together. 

For additional concreteness and intuition, Sec.~F.1, ``Additional sweep of behavior of random walk under resets from simulation,'' of the Supplemental Material, shows additional direct simulation sweeps reflecting how reset rates and target locations change the mean time to successfully reach targets, and optimal reset rates. Further interesting features visible in the behavior of Figure~\ref{fig:TimeToReachTarget}, including the sawtooth oscillations especially visible at small reset times (which are not sampling artifacts but reflect parity structure in the discrete search problem), are also ratifiable in light of lattice path counting, which we detail in Sec.~B of the Supplemental Material. Together, all of these results provide a coherent explanatory framework for the rich competing behaviors shown in Figure~\ref{fig:TimeToReachTarget}.

\subsection{Case Study: Intraflagellar Transport as a Random Walk with Reset}
\label{sec:iftMainText}
\begin{figure}
\centering
\includegraphics[width=\linewidth]{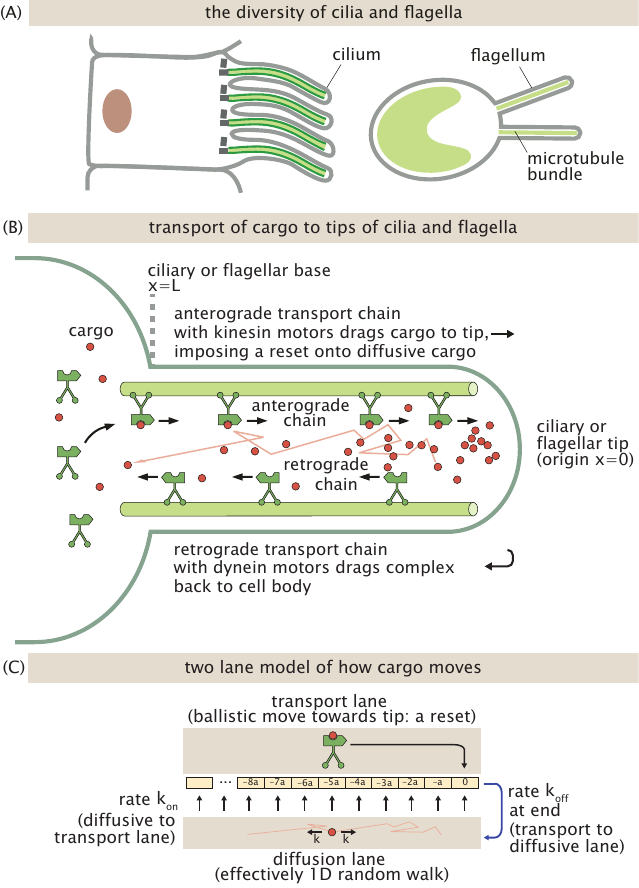}
\caption{\raggedright Intraflagellar transport (IFT) spatially organizes cargo in cilia and flagella. (A) Cilia and flagella are ubiquitous structures in many different cell types. (B) Mechanisms of intraflagellar transport, where cargo interact with motorized complexes  moving ballistically in the anterograde direction towards the ciliary or flagellar tip, and otherwise diffuse in the lumen. Retrograde transport chains also replenish complexes impoverished by cargo back to the cell body, visualized by the empty motors returning to the cytosolic body. (C) A one dimensional model of cargo transport, organized by two lanes, yields an effective resetting mechanism.}
\label{fig:ift_Overview}
\end{figure}

To what extent might these mathematical approaches shed light on biological phenomena?  Reframing the problem in terms of concentration fields rather than the trajectory statistics of individual seekers provides a direct path to interesting biological case studies.  One interesting example is offered by intraflagellar transport (IFT), directed motion of proteins in eukaryotic cilia driven by kinesin and dynein motors as indicated schematically in Figure~\ref{fig:ift_Overview}(A). These structures are built up of about one thousand proteins of a few dozen types, which then assemble into a large protein complex with kinesin motors to form the anterograde IFT train~\cite{van2020ciliary}, as depicted in Figure~\ref{fig:ift_Overview}(B). Powered by ATP and the action of kinesin motors, this protein complex moves along ciliary microtubules toward the cilium tip at a speed of about $v\approx 1~\mu$m/s. When it reaches the tip, the train disassembles and then reforms to make the retrograde IFT train, which contains a similar number of proteins and is transported by dynein motors toward the cell body at a similar speed. In addition to this directed transport by IFT trains, proteins in the lumen of the cilium also move diffusively, with a diffusion constant of about $D\approx 1~\mu\mathrm{m}^2/s$. This means that in a typical cilium about $L\approx 5~ \mu$m in length, transport and diffusion across the length of the cilium take anywhere between a few and few tens of seconds. Though our focus on random walks with resets has thus far emphasized the individual walkers, here we examine how random walks with resets shapes not only individual search trajectories but achieves interesting concentration distributions \cite{naoz2008protein} within cells that are crucial for exquisite and downstream functions for organisms.

The motion of a protein in a cilium can be described by a two-lane model ~\cite{naoz2008protein}, as illustrated by Figure~\ref{fig:ift_Overview}(C). (Specifically, this is true when anterograde transport dominates retrograde transport (such that the latter can be neglected). A two-lane model behaves qualitatively similarly to a fuller three-lane model acknowledging both anterograde and reterograde motion, as long as the net motion of cargo is towards the tip.) 

Each lane in this two-lane model is described by a one dimensional lattice, with a lattice spacing $a$, like in Figure~\ref{fig:ift_Overview}(C). In the transport lane the protein undergoes directed motion toward the tip of the cilium (the positive-$x$ direction) with a rate $v$. In the diffusion lane the protein hops left and right with rate $k$. The protein can switch between the two lanes, with rate $k_{\text{on}}$ describing the transition from the diffusion lane to the transport lane, and rate $k_{\text{off}}$ dictating transitions from the transport to the diffusion lane. In the limit when the time spent in the transport lane, which on average is $1/k_{\text{off}}$, is much larger than the time it takes to transport the protein to the tip of the cilium (a few seconds), the two lane model becomes a random walk with resets. In particular, every time a protein diffusing in the lumen of the cilium is captured by an anterograde IFT train, its fate is sealed and it is deposited at the tip before it has a chance to disassociate from the train, effectively sending the walker back to the origin. Direct experiments tracking the motion of a fluorescently-labeled sodium receptor protein cargo inside the cilium of a \emph{C. elegans} nerve cell explicitly identified and measured this type of motion, with the protein either diffusing or being primarily transported by the anterograde IFT \cite{van2020ciliary}. The result is the accumulation of this protein at the tip of the cilium, which was shown to be required for sensitive sensing of salt in the environment. In a very different setting, similar accumulation of a depolymerizing kinesin, which also rides on the IFT trains, was observed in the flagella of \emph{Giardia} \cite{dawson2010life}. Other examples of this sort of protein localization by intraflagellar transport have been reported, making the cilium tip into a hub for signaling, sensing, and regulation \cite{ott2025design}. 

The spatial localization of proteins to the cilium tip observed in experiments can be described using the mathematics of diffusion with resets. Along the length of the cilium, the concentration at a location $x$ changes either by diffusion or by molecules being whisked away by the IFT at rate $k_{\text{on}}$. The tip enjoys an additional source term from the accumulation of delivered cargos, accumulated over possible intial arrival points all along the microtubule. 

These deliveries contribute a total positive arrival rate given by,
\begin{align}
    \text{additional source term at tip $x=0$:  }\hspace{5pt}k_{\text{on}}\displaystyle \int_0^L dx' c(x',t).
\end{align}
The integral accounts for the delivery of all the captured proteins along the length of the cilium, to the tip of the cilium at $x=0$.
Taken together, these effects make the continuum equation for these dynamics adopt the form,

{\footnotesize
\begin{align}
  \frac{\partial c(x,t)}{\partial t}
  & = \begin{cases}
      D\,\frac{\partial^2 c}{\partial x^2}
  - k_{on} \,c & \text{if $x\neq 0$ (not at the tip),}\\
   D\,\frac{\partial^2 c}{\partial x^2}
  - k_{on} \,c + k_{\text{on}} \!\displaystyle \int_0^L dx' c(x',t) & \text{if $x=0$ (at the tip)}.
  \end{cases}
 \label{eq:pdeConcentration}
\end{align} 
}

\begin{figure}[t]
\centering
\includegraphics[width=\linewidth]{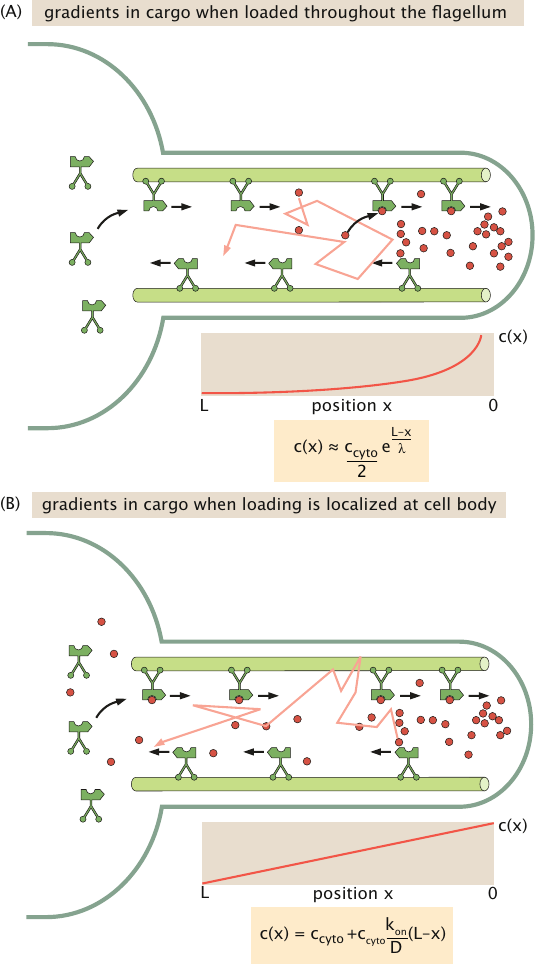}
\caption{\raggedright Gradients in cilia and flagella.  (A) Intraflagellar transport in which the cargo can jump onto motors within the flagellum, thus serving as a reset to the flagellar tip. (B) Intraflagellar transport (IFT) in which the cargo can only be loaded at the cell body.}
\label{fig:ift_gradients}
\end{figure}
The steady-state concentration is described by the balance between diffusion and resets (to the tip), given by\sout{,}
\begin{equation}
  D\,\frac{\partial^2 c}{\partial x^2} = k_{\text{on}} \,c. 
\label{eq:pdeConcentrationSteadyState}
\end{equation}
This equation has the general solution,
\begin{align}
    c(x) & =  A \exp\left[\frac{x}{\sqrt{D/k_{\text{on}}}} \right] + B\exp\left[-\frac{x}{\sqrt{D/k_{\text{on}}}} \right]\\
    & \equiv  A \exp\left[x/\lambda \right] + B\exp\left[-x/\lambda \right],
\end{align}
where we have defined the emergent length scale $\lambda \equiv \sqrt{D/k_{\text{on}}}$, the same length scale we encountered when analyzing the first passage time for random walks with resets. 

To find the constants $A$ and $B$ in this solution, we consider the boundary conditions set by the physics of the process. At the cytosolic end of the cilium ($x=L$), the ambient concentration of cargo matches that of the cytoplasm, $c_{\text{cyto}}$; this delivers the boundary condition $c(L) = c_{\text{cyto}}$.This fact, which expresses the reasonable assumption that the cytoplasm is not depleted of cargo, imposes the behavior,
\begin{align}
    c_{\text{cyto}}= c(L)=A \exp\left[L/\lambda \right] + B\exp\left[-L/\lambda \right].
    \label{eq:cytosolicBoundCond}
\end{align}
In addition, the diffusive flux $-\left.D\frac{\partial c}{\partial x}\right|_{x=0}$ experienced at the tip must balance the accumulated influx from resets along the length, $k_{\text{on}}\displaystyle \int_0^L dx' c(x',t)$. This gives the additional boundary condition,
\begin{align}
    -\left.D\frac{\partial c}{\partial x}\right|_{x=0} & = k_{\text{on}}\displaystyle \int_0^L dx' \left(A \exp\left[x'/\lambda \right] + B\exp\left[-x'/\lambda \right]\right),
\end{align}
or,
\begin{align}
     -D \left( \frac{A}{\lambda} - \frac{B}{\lambda}\right) & = k_{
\text{on}} ~ \lambda ~\left(A\left(\exp\left[L/\lambda\right]-1\right) - B\left(\exp[-L/\lambda]-1 \right) \right).
\end{align}
Recalling that $k_{\text{on}} = D/\lambda^2$, this yields,
\begin{align}
    -D \left( \frac{A}{\lambda} - \frac{B}{\lambda}\right) & = \frac{D}{\lambda} ~\left(A\left(\exp\left[L/\lambda\right]-1\right) - B\left(\exp[-L/\lambda]-1 \right) \right),
\end{align}
or,
\begin{align}
    B-A = A\left(\exp\left[L/\lambda\right]-1\right) - B\left(\exp[-L/\lambda]-1 \right),
\end{align}
and subtracting $B-A$ from both sides finally gives,
\begin{align}
A\exp\left[L/\lambda\right]=B\exp[-L/\lambda]. \label{eq:boundaryCondVerb}
\end{align}
This consequence of the second boundary condition eqn.~\ref{eq:boundaryCondVerb} combines with the first cytosolic boundary condition eqn.~\ref{eq:cytosolicBoundCond} to require that 
\begin{align}
    A\exp\left[L/\lambda\right]=B\exp[-L/\lambda]= \frac{c_{\text{cyto}}}{2},
\end{align}
so,
\begin{align}
    c(x) & = \frac{c_{\text{cyto}}}{2}\left(\exp\left[\frac{L-x}{\lambda} \right] +  \exp\left[-\frac{L-x}{\lambda} \right]\right)\\
    & =c_{\text{cyto}} \cosh \left(\frac{L-x}{\lambda} \right).
\end{align}

This profile is perhaps of greatest interest in positions close to the ciliary tip, namely $L\gg x$ (in units of the presiding length scale $\lambda$). More specifically, when $L \gg \lambda$, so that $\exp\left[- \frac{L - x}{\lambda}\right] \rightarrow 0$ in the limit $\frac{\lambda}{L} \rightarrow 0$ at the tip $x = 0$, we enjoy the approximation,
\begin{align}
    c^{\text{near tip}}(x) \approx  \frac{c_{\text{cyto}}}{2}\exp\left[\frac{L-x}{\lambda} \right] = \frac{c_{\text{cyto}}}{2}\exp\left[\frac{L-x}{\sqrt{D/k_{\text{on}}}} \right].
\end{align}
Such an exponential concentration profile is illustrated in Fig.~\ref{fig:ift_gradients}(A).

Does this quantitative picture accord with real gradients of cargo observed in cilia? To address this question, we first must estimate the on rate $k_{\text{on}}$ for cargo arriving to the cytoskeletal filaments where the transport (and thus resetting) occurs. To anticipate the scale of how quickly molecules arrive from the cytoplasm to the central cytoskeletal filament, we recall that the time that a molecule takes to transversely diffuse across the radius of the cilium is simply of order $\tau_{\text{diff}} \approx R^2 /D$, implicating an on rate of order $k_{\text{on}} \approx 1/{\tau_\text{diff}} \approx D/R^2$. 
If we substitute this approximation to $k_{\text{on}}$ into the anticipated exponential length scale $\lambda \equiv \sqrt{D/k_{\text{on}}}$, we find $\lambda \approx R$, namely, the length scale of the gradient is roughly equal to the width of the cilium. Strikingly, measurements of the exponential profile of the concentration of kinesin-13 cargo in the flagella of \emph{Giardia} are consistent with that expectation \cite{McInally2019}. For each of the four pairs of flagella in this organism, the exponential profile gave the same $\lambda \approx 350$~nm length scale, which is consistent with the 200-300~nm range for the widths of  microtubule-based flagella and cilia. 

Intriguingly, this description linking the width of a cilium $R$ with the longitudinal gradient length scale $\lambda$ (via the arrival rate $k_{\text{on}}$ of cargo to central cytoskeletal filaments) anticipates a relationship potentially under cellular control, or else testable in a synthetic engineered setting. This description, albeit simplified, anticipates that the transverse geometry of a cilium or flagellum should couple tightly with the shape of the longitudinal gradient, with wider cilia or flagella eliciting shallower gradients (modulo other regulation or saturation effects) \cite{datta2022assemble}. This intriguing speculation invites further inquiry both experimentally and theoretically.

Note that another mechanism under cellular control concerns where in space the cargo can load onto the machinery of the intraflagellar transport chain's complexes. If cargo can load only at the base of the cilium (at the cell body), instead (as assumed earlier) occurring anywhere along the flagellum or cilium, the resulting dynamics and concentration gradient adopt a very different form. Specifically, except at the cell body at $x=L$ where loading occurs at a rate $k_{\text{on}}c_{\text{cyto}}$, the sink terms disappear throughout the domain, making the only way that concentration changes in time changes from the dynamics specified earlier in eqn.~\ref{eq:pdeConcentration} to simply,
\begin{align}
    \frac{\partial c(x,t)}{\partial t} = D\frac{\partial^2 c}{\partial x^2}.
\end{align}
The solution to this equation now has a simple linear profile in space,
\begin{align}
c(x) 
& = c_{\text{cyto}} \left(1  + \frac{k_{\text{on}}}{D}\left(L-x \right)\right),
\end{align}
where this solution enforces the boundary conditions $-D \left.\frac{dc}{dx}\right|_{x=0} = k_{\text{on}} c_{\text{cyto}}$ and $c(L) = c_{\text{cyto}}$. In other words, merely by adjusting which spatial positions enjoy replenishment of cargo (just the base of the cilium, or all of the sites along the cilium), the concentration profile of cargo over space can be tuned to change from an exponential behavior to a linear gradient, as shown by Fig.~\ref{fig:ift_gradients}(B). In fact, distinct examples of such a linear profile have been observed for kinesin-2 motors in experimental works \cite{chien2017dynamics}.

In summary, the calculations we just performed show how resets and diffusion conspire to set the typical concentration profiles of molecules along cilia and flagella, in tunable ways. Such concentration profiles reflect the probability of encountering molecules at any point in space, no matter whether they have visited that point in space before. This analysis gives a natural informative complement to the separate mathematics in the other sections of this paper that attend to first arrivals of searchers like molecules, organelle structures, cells, or organisms to specified targets of interest.

\subsection{Random Walk With Resets and Drift}
\label{subsec:RandomWalkResetsConstantDrift}
\begin{figure}
    \centering
    \includegraphics[width=\columnwidth]{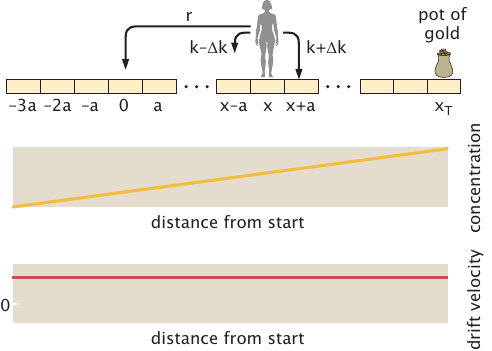}
    \caption{\raggedright Schematic of the steps available to the random seeker during a time step when there is a bias in one direction. The top graph posits a linear concentration gradient of some cue that leads to a constant drift velocity as shown in the lower graph.}
    \label{fig:DynamicsModelDrift}
\end{figure}
We now generalize the calculation of random walk with reset to include a bias, or drift, in the 
direction of the target.  The physical motivation comes from the fact that in many 
biological contexts the seeker is not an unbiased random walker but 
moves preferentially toward its target, whether because of a chemical 
gradient, a mechanical bias, explicit flow, motor-assisted transport, or an asymmetry in the dynamics of growth 
and shrinkage.  We model this by allowing the rate of stepping toward 
the target to differ from the rate of stepping away, as shown in Fig.~\ref{fig:DynamicsModelDrift}.  Specifically, 
we assign rate $k_1 = k + \Delta k$ to steps in the positive direction 
(toward the target) and rate $k_2 = k - \Delta k$ to steps in the 
negative direction (away from the target), where $\Delta k \geq 0$ 
measures the degree of bias.  The reset rate $r$ remains unchanged.

The recursive equation for the mean first-passage time $T(x)$ now 
takes the form
\begin{equation}
T(x)(2k + r) = 1 + k_1\, T(x+a) + k_2\, T(x-a) + r\, T(0),
\end{equation}
where as before the factor $2k + r$ is the total rate of leaving 
position $x$, and the three terms on the right involving $T(x)$ account for stepping 
right, stepping left, and resetting to the origin, respectively.  
Substituting $k_1 = k + \Delta k$ and $k_2 = k - \Delta k$ gives
\begin{equation}
T(x)(2k + r) = 1 + (k+\Delta k)\, T(x+a) + (k-\Delta k)\, T(x-a) + r\, T(0).
\end{equation}
We now perform a Taylor expansion of $T(x \pm a)$ to second order in 
the step size $a$,
leaving
\begin{equation}
r\, T(x) = 1 + 2k \cdot \frac{a^2}{2}\, T''(x) 
+ 2\Delta k \cdot a\, T'(x) + r\, T(0).
\end{equation}
Rearranging and identifying the diffusion coefficient 
$D = ka^2$ and the drift velocity $v = 2\Delta k \cdot a$ in the continuum limit, we arrive at the ordinary differential equation
\begin{equation}
0 = 1 + D\, T''(x) + v\, T'(x) - r\, T(x) + r\, T(0),
\label{eq:backward_drift_reset}
\end{equation}
which holds for all $x < x_T$, supplemented by the absorbing boundary 
condition $T(x_T) = 0$.  This is the natural generalization of 
eqn.~\ref{eq:backward_equation_finalmain}: the only new term is 
$v\, T'(x)$, which encodes the drift toward the target.  When 
$v = 0$ we recover the unbiased result exactly.

To solve eqn.~\ref{eq:backward_drift_reset}, we again define a helper function given by,
\begin{equation}
y(x) = T(x) - T(0) - \frac{1}{r},
\end{equation}
which transforms eqn.~\ref{eq:backward_drift_reset} into,
\begin{equation}
D\, y''(x) + v\, y'(x) - r\, y(x) = 0.
\end{equation}
This is a second-order linear ODE with constant coefficients.  
We seek solutions of the form $y(x) = e^{\sigma x}$, which leads 
to the characteristic equation
\begin{equation}
D\sigma^2 + v\sigma - r = 0.
\end{equation}
The two roots are
\begin{equation}
\sigma_{\pm} = \frac{-v \pm \sqrt{v^2 + 4Dr}}{2D}.
\end{equation}
Note that $\sigma_+ > 0$ and $\sigma_- < 0$ for all positive 
values of $v$, $D$, and $r$.  The general solution is hence
\begin{equation}
y(x) = A\, e^{\sigma_- x} + B\, e^{\sigma_+ x},
\end{equation}
which results in 
\begin{equation}
T(x) = A\, e^{\sigma_- x} + B\, e^{\sigma_+ x} + T(0) + \frac{1}{r}.
\end{equation}
As in the unbiased case, we require the mean first-passage time to remain finite as $x \to -\infty$. This is again because resets will eventually return even the most eccentrically waywards seeker back to the origin and give another chance at success. Since $\sigma_- < 0$, the term $e^{\sigma_- x}$ grows without bound as $x \to -\infty$, whereas $e^{\sigma_+ x} \to 0$ because $\sigma_+ > 0$. As a result, this condition imposes $A = 0$ to guarantee that $T(x)$ is finite. This leaves
\begin{equation}
T(x) = B\, e^{\sigma_+ x} + T(0) + \frac{1}{r}.
\end{equation}
Imposing the absorbing boundary condition $T(x_T) = 0$ (that the time to reach the target is zero once the seeker is already there) gives
\begin{equation}
B = -e^{-\sigma_+ x_T}\left(T(0) + \frac{1}{r}\right),
\end{equation}
and hence
\begin{equation}
T(x) = \left(T(0) + \frac{1}{r}\right)
\left(1 - e^{\sigma_+(x - x_T)}\right).
\end{equation}
Setting $x = 0$ and solving for $T(0)$ yields
\begin{equation}
T(0)
=
\frac{1 - e^{-\sigma_+ x_T}}
{r\,e^{-\sigma_+ x_T}}
=
\frac{1}{r}\left(e^{\sigma_+ x_T} - 1\right).
\end{equation}
We can write this compactly as
\begin{equation}
T(0) = \frac{1}{r}\left(e^{x_T/\lambda} - 1\right).
\label{eq:mfpt_drift_reset}
\end{equation}
where the characteristic length scale is
\begin{equation}
\lambda
=
\frac{1}{\sigma_+}
=
\frac{2D}{\sqrt{v^2 + 4Dr} - v}.
\label{eq:xstar_drift}
\end{equation}

Here, it is worth noting that we have taken $v>0$ throughout, corresponding to a productive drift towards the target. 
If instead the drift points away from the target, $v<0$, the same algebraic calculation 
applies with $v=-|v|$. In this case the characteristic length becomes
\begin{equation}
\lambda_-=
\frac{2D}{\sqrt{v^2+4Dr}+|v|},
\end{equation}
and the mean first-passage time retains the form
\begin{equation}
T(0)=\frac{1}{r}\left(e^{x_T/\lambda_-}-1\right).
\end{equation}
Unlike the case of drift toward the target, however, the mean first-passage 
time diverges in the no-reset limit $r\to0$, since a walker drifting away from 
the target has a finite probability of never reaching it.

\begin{figure}[t!]
    \centering
    \includegraphics[width=\linewidth]{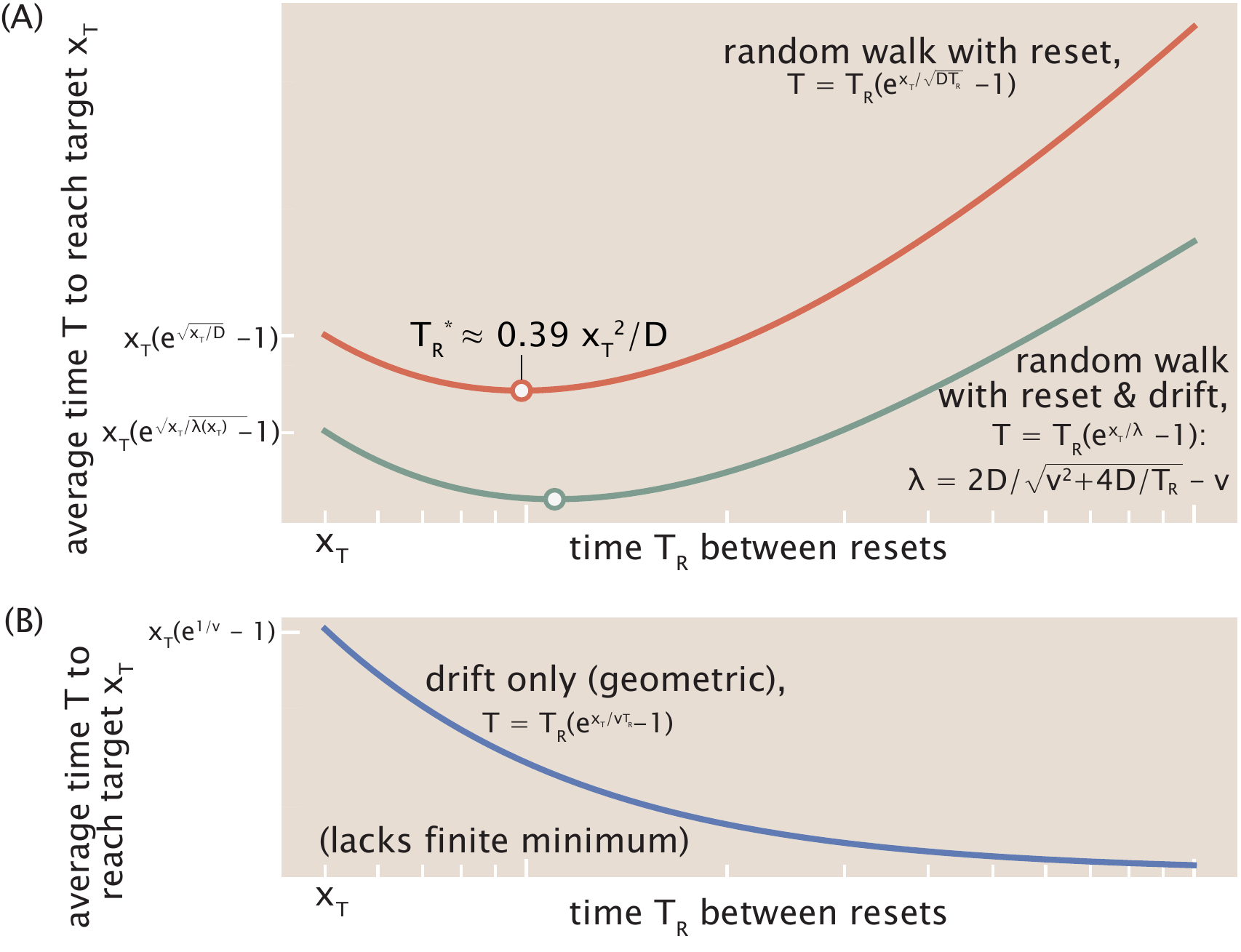}
    \caption{\raggedright The typical time to reach a target adopts a common mathematical form for a surprising variety of search dynamics. (A) A simple random walk with reset shows a mean time to find a target given by eqn.~\ref{eqn:SearchTimeFinal}  and is shown by the red curve. The additional complexity of a drift towards the target is given by eqn.~\ref{eq:mfpt_drift_reset} and shown by the (faster)  teal curve. Both curves show that the time to reach a target can be minimized by using some finite reset rate. These curves display the evocative case of a target located at $x_T=5$, diffusion coefficient $D=1$, and velocity $v=0.05$. (B) Searches by coarse-grained drift alone (as considered in Ref. \cite{kondev2025}'s eqn.~27, taking the catastrophe rate indicated there as $k_c \equiv 1/T_R$ and recognizing $d$ as $x_T$) shows a similar algebraic form of the time to reach the target. Here, without diffusive motion, however, no finite optimal reset rate minimizes the average time to arrive at the target. These disparate cases can be understood by recognizing that the mean time to reach the target always displays an exponential dependence on the target location $x_T$, only with respect to different effective length scales $\lambda(T_R)$ for each search dynamic. The appearance of a finite optimal reset rate depends on how aggressively this effective length scale changes with the reset interval $T_R$.}
    \label{fig:quasiUniversalExponentialTimeToSuccess}
\end{figure}

The behavior of the first-passage time eqn.~\ref{eq:mfpt_drift_reset} (and its length scale eqn.~\ref{eq:xstar_drift}) is the central result.  The mean first-passage time with 
drift and reset has exactly the same exponential form as the 
unbiased (diffusion with reset) case seen in eqn.~\ref{eqn:SearchTimeFinal}, but with a modified length scale $\lambda$ that depends jointly on the drift velocity $v$, the diffusion 
coefficient $D$, and the reset rate $r$. Fig.~\ref{fig:quasiUniversalExponentialTimeToSuccess} summarizes and emphasizes the commonality of an exponential form of the mean time to reach a target, with different effective length scales, across disparate search dynamics. 

Interestingly, as shown in Fig.~\ref{fig:quasiUniversalExponentialTimeToSuccess}(B), a quite separate ``drift-only'' search behavior analyzed by Ref.~\cite{kondev2025}'s eqn.~27 in the context of collapsing microtubule catastrophes also shows a similar algebraic form. The microscopic dynamics are nevertheless distinct: whereas the drift direction is fixed in the present model, in the microtubule problem the direction of motion after a reset can be either toward or away from the target. Despite these differences, the resulting mean search time can be written as
\begin{align}
T(T_R) & = T_R\left(e^{\frac{x_T}{vT_R}} - 1\right) \label{eq:basalDriftEquationThatAttachesToPeclet2}
\\
& = T_R\left(e^{x_T/\lambda(T_R)} - 1\right),
\end{align}
where $T_R=1/r$ and $\lambda(T_R)\equiv vT_R$. Because this effective length scale grows linearly with $T_R$, the mean search time decreases monotonically with $T_R$, and the optimum is therefore the no-reset limit, $T_R\to\infty$. Thus, the same mathematical form can arise from quite distinct search dynamics, with the dependence of the effective length scale on resetting determining the resulting search behavior.

In addition to the diffusive reset length $\lambda_D=\sqrt{\frac{D}{r}}$ defined above, it is useful to introduce the ballistic reset length
\begin{equation}
\lambda_B = \frac{v}{r},
\end{equation}
corresponding to the distance traveled by drift during the mean time interval between resets. The effective length scale for the case in which both diffusion and drift are in play can then be written as
\begin{equation}
\lambda =
\frac{\lambda_B+\sqrt{\lambda_B^2+4\lambda_D^2}}{2}.
\end{equation}

Several limits are instructive. When the drift vanishes, $v=0$, we have $\lambda_B=0$ and recover the unbiased result exactly,
\begin{equation}
\lambda=\lambda_D=\sqrt{\frac{D}{r}}.
\end{equation}
More generally, in the diffusion-dominated regime $\lambda_B\ll \lambda_D$, or equivalently $v\ll \sqrt{Dr}$, the effective length scale reduces to the diffusive lengthscale,
\begin{equation}
\lambda \approx \lambda_D.
\end{equation}
In the opposite limit, when drift dominates over diffusion, 
$\lambda_B\gg \lambda_D$, or equivalently $v\gg\sqrt{Dr}$, we write
\begin{equation}
\lambda
=
\frac{\lambda_B}{2}
\left(
1+\sqrt{1+4\frac{\lambda_D^2}{\lambda_B^2}}
\right).
\end{equation}
Expanding the square root for $\lambda_D/\lambda_B\ll 1$ gives
\begin{equation}
\lambda
\approx
\lambda_B+\frac{\lambda_D^2}{\lambda_B}
\approx
\lambda_B.
\end{equation}
In this ballistic regime, the mean first-passage time becomes
\begin{equation}
T(0) \approx \frac{1}{r}\left(e^{x_T/\lambda_B}-1\right).
\end{equation}
Here the relevant length scales are the distance traveled between resets, $\lambda_B=v/r$ and the distance to the target $x_T$.

In our present context, whether or not a finite reset rate can improve the search can be expressed in terms of this Péclet number 
(see Sec.~G, ``Dependence of the mean first-passage time on the reset rate $r$ in 1D with diffusion and drift,'' of the Supplemental Material)
through the condition
\begin{equation}
\mathrm{Pe}<2.
\end{equation}
This condition has been derived previously for diffusion with drift and stochastic resetting~\cite{Ray2019Peclet}.

This condition exposes a central takeaway: resetting is therefore beneficial only when diffusion is sufficiently important relative to drift. In this regime, trajectories can wander away from the target, and resets provide a useful way to restart the search from the origin. When $\mathrm{Pe}>2$, drift carries the seeker toward the target efficiently enough that resets mostly interrupt productive trajectories. The optimal reset rate then tends to zero, and the best strategy is not to reset.

We can understand where this condition comes from by looking near the no-reset limit. In the limit of very frequent resets, $T_R\to0$, the walker is returned to the origin before it has time to reach the target, and the mean first-passage time diverges. In the opposite limit of no resetting, $T_R\to\infty$, the mean first-passage time approaches the ballistic time $x_T/v$. Expanding the search time via eqn.~\ref{eq:basalDriftEquationThatAttachesToPeclet2} about this no-reset limit gives,
\begin{equation}
\tilde T
=
\frac{1}{\mathrm{Pe}}
+
\frac{1}{\tilde T_R}
\left(
\frac{\mathrm{Pe}-2}{2\mathrm{Pe}^3}
\right)
+\cdots,
\label{eq:pecletExpandClever}
\end{equation}
where $\tilde T_R=T_RD/x_T^2$. The sign of the first correction determines whether introducing a small amount of resetting improves the search. For $\mathrm{Pe}<2$ this correction is negative, so resetting initially lowers the mean first-passage time; since the search time diverges again as $T_R\to0$, a finite optimum must occur. For $\mathrm{Pe}>2$, introducing resets instead increases the search time, and the optimum remains the no-reset limit. This core behavior is justified more completely and rigorously 
in Sec.~G of the Supplemental Material,
but the essence is already well captured by this expansion.

\subsection{Random Seeker with an Exponential Cue}
\label{subsec:exponentialCue}
\begin{figure}
    \centering
    \includegraphics[width=\columnwidth]{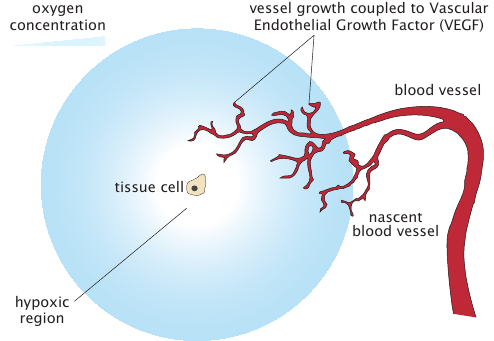}
    \caption{\raggedright Many biological settings steer exploratory motion using cues of biochemicals, such as the exquisitely sensitive development of new blood vessels and capillaries across the circulatory system that respond to gradients of oxygen and associated signaling molecules such as vascular endothelial growth factor (VEGF).}
    \label{fig:cueCartoon}
\end{figure}

The random walk models considered so far treat the explorer as having, at most, a compass: a constant drift can point it toward the target, but gives no indication of how close the target is. Many biological seekers have access to richer information, responding to cues that grow stronger or weaker as they move through space. The next generalization is to allow the drift to vary with position.

\begin{figure}[t]
    \centering
    \includegraphics[width=\columnwidth]{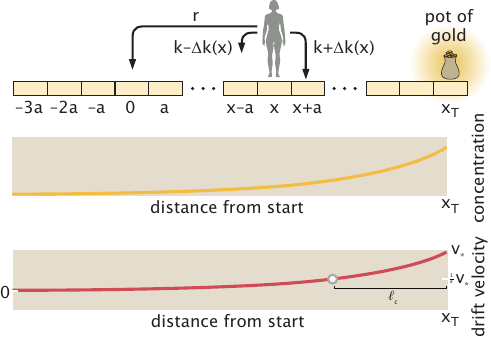}
    \caption{\raggedright Schematic of the steps available to the random seeker
    in the presence of a cue. 
    A concentration gradient of an underlying cue field, such as the growth factor VEGF during angiogenic sprouting or the extracellular protein netrin guiding axon growth, results in a gradient in the effective drift velocity experienced by the random seeker.}
    \label{fig:DynamicsModelCue}
\end{figure}

For example, as hinted in Fig.~\ref{fig:cueCartoon}, endothelial tip cells respond to the growth factor VEGF during angiogenic sprouting. The exquisite accomplishment of such a cue-guided search process is to position almost all cells in vertebrate bodies within a hundred microns of some capillary \cite[ch. 23, p. 1448]{alberts2008molecular}, \cite{kelch2015organ}. Further examples abound. Axons are guided by the extracellular protein netrin, and bacteria such as \textit{E. coli} bias their run-and-tumble motion in chemical gradients~\cite{Kennedy1994,gerhardt2003,Berg1972}. These examples should not be identified with a single microscopic sensing rule. They motivate a family of models indicated schematically in Figure~\ref{fig:DynamicsModelCue} in which cue detection is summarized by an effective drift field $v(x)$. This class of models is closely related to the literature on stochastic resetting in external potentials, where spatially varying drift fields are due to effective forces or guiding landscapes~\cite{Ahmad2019,Ray2021}.

Here, we assume that the cue acts by changing the relative rates of stepping toward or away from the target. Generalizing the continuum-limit derivation of eqn.~\ref{eq:backward_drift_reset} to a position-dependent bias $\Delta k(x)$ (and hence position-dependent drift velocity $v(x)$) gives
\begin{equation}
0=1+D T''(x)+v(x)T'(x)-rT(x)+rT(0),
\label{eq:backward_equation_position_dependent_drift}
\end{equation}
where $D=ka^2$, and $v(x)=2a\Delta k(x)$. 
This is the ``backward equation'' for the first hitting time $T$ derived using the same scheme outlined in Figure~\ref{fig:explainSimpleHitTimeRecurse}. Because the two stepping rates are written as $k\pm \Delta k(x) $, their sum remains fixed at $2k$. The cue therefore changes the local directional bias, and hence the local drift velocity, while leaving the diffusion coefficient unchanged. 
As a concrete example, consider an exponential cue profile
\begin{equation}
c(x)=c_0 e^{-(x_T-x)/\ell_c},
\end{equation}
where $c_0$ is the concentration at the target and $\ell_c$ is the cue decay length. This profile is defined for all $x<x_T$. If the drift velocity is proportional to the local gradient of the cue, namely,
\begin{align}
v(x)&=\chi {dc\over dx},
\end{align}
where $\chi$ is a mobility that characterizes the relation between the gradient and the drift speed, then
\begin{align}
v(x)&={\chi c_0\over \ell_c}e^{-(x_T-x)/\ell_c}\\
&\equiv v_*e^{-(x_T-x)/\ell_c},
\end{align}
where $v_*=\chi c_0/\ell_c$ is the drift speed at the target. Larger values of the constant mobility $\chi$ encode stronger ``sensing'' of the underlying cue field by the random seeker. (In general, of course, the mobility $\chi$ can adopt more complex functional forms; the sweep of  physics describing these couplings of velocity to cue is the majestic field of chemotaxis. Here, we sweep all of these complexities into a single effective and constant mobility.)

Armed with this expression for the velocity over space, the  eqn.~\ref{eq:backward_equation_position_dependent_drift} for the mean first-passage time becomes,
\begin{equation}
0=1+D T''(x)+v_*e^{-(x_T-x)/\ell_c}T'(x)-rT(x)+rT(0).
\label{eq:backward_exp_cue}
\end{equation}
  The absorbing boundary condition is $T(x_T)=0$; the solution's behavior is further specified by requiring $T(x)$ to remain finite as $x\to -\infty$. (Namely, no matter what $x$ the seeker begins at, it can arrive at the target in a finite amount of time.) Indeed, when $x$ becomes a highly negative number, the exponential cue becomes negligible, and the seeker no longer experiences the drift. Hence, eqn.~\ref{eq:backward_equation_position_dependent_drift} reduces to the pure diffusion equation with resets, eqn.~\ref{eq:backward_equation_finalmain}, corresponding to the case $v(x)=0$. In this far-field limit, the mean first-passage time must therefore recover the exponential behavior with respect to the position obtained previously in eqn.~\ref{eqn:searchtimeresets}.

\begin{figure*}[t!]
    \centering
    \includegraphics[width=0.9\linewidth]{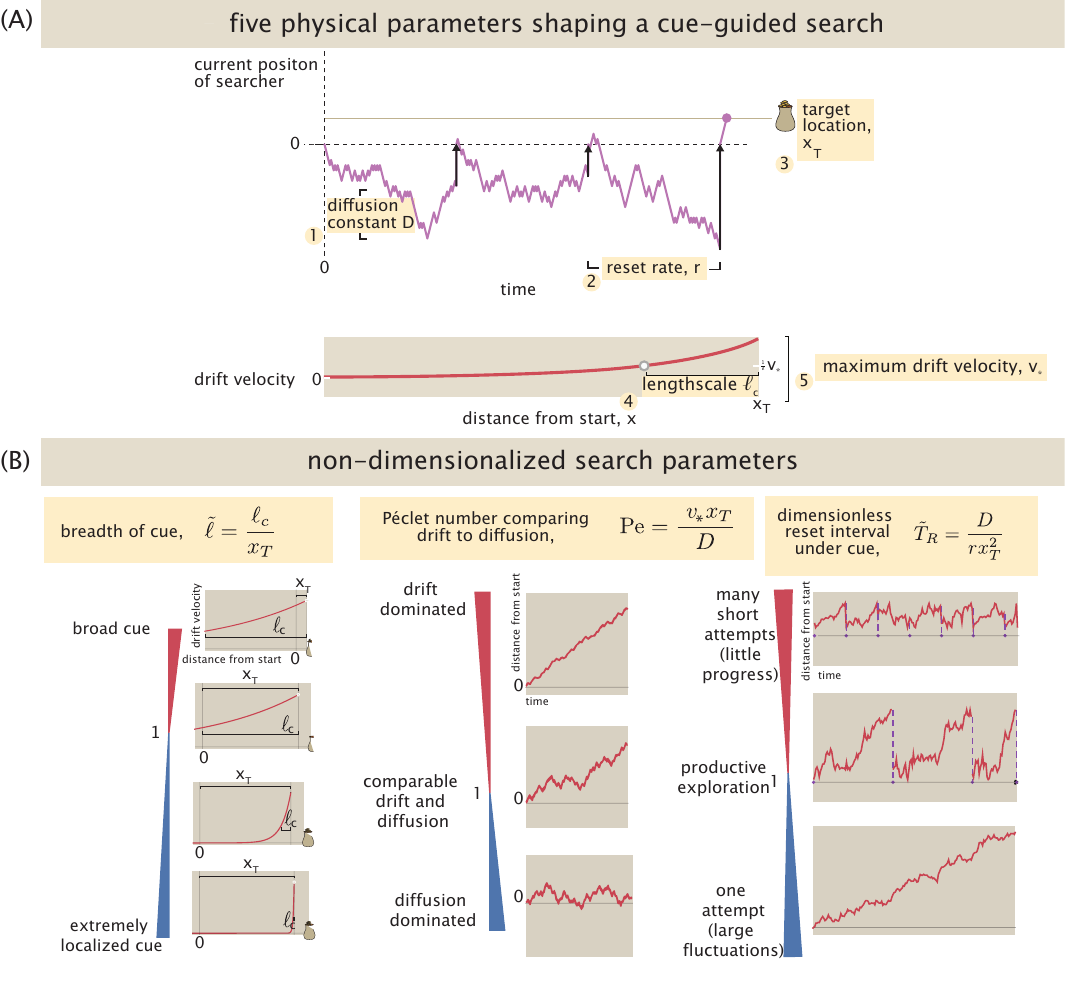}
    \caption{\raggedright Physical parameters and their nondimensionalized representations for search with cue. (A) Five physical parameters in the raw description of the cue problem, entering eqn.~\ref{eq:backward_exp_cue}. (B) Three non-dimensional parameters economize and organize the study of the search problem.}
    \label{fig:possibleCartoonForCueSearch}
\end{figure*}

\begin{figure}
    \centering
    \includegraphics[width=\linewidth]{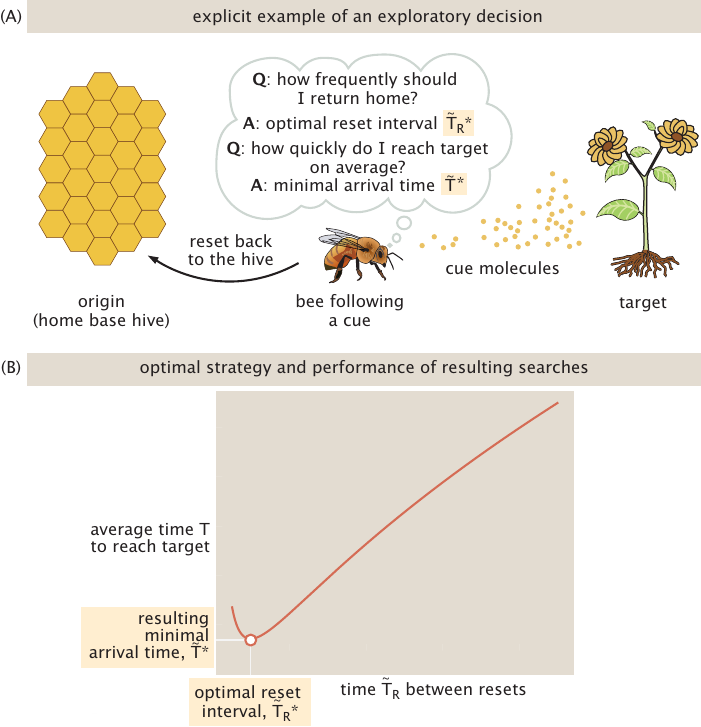}
     \caption{\raggedright Biological importance of search strategy and resulting arrival time for search under cue. (A) A schematic example of a biological search where a decision about the optimal reset parameters for a search matters is a bee executing a one dimensional search for a flower target, but subject to deciding the optimal rate at which it returns home to a hive to rest. Bees can use long-range cues, including odors, to seek and orient towards flowers~\cite{chaffiol2005}.
     (B) Two features of the resulting search behavior, namely the optimal reset interval $\tilde{T}_R^*$ that minimizes the average time to encounter the target, and the resulting minimum time $\tilde{T}^*$ itself, are meaningful metrics of performance, and soon analyzed in Fig.~\ref{fig:CueOptimization}. }\label{fig:beeAndMetricsOfPerformance}
\end{figure}

Eqn.~\ref{eq:backward_exp_cue} states how a full set of five physical parameters shape the search process under such a cue, as highlighted by Fig.~\ref{fig:possibleCartoonForCueSearch}(A). To make progress despite this profusion of parameters, it is useful to nondimensionalize positions by $x_T$ and time by $x_T^2/D$. This leads us to define the dimensionless variables
\begin{equation}
\tilde x=\frac{x}{x_T},
\qquad
\tilde T=\frac{D T}{x_T^2},
\end{equation}
and the three dimensionless parameters
\begin{equation}
\tilde \ell=\frac{\ell_c}{x_T},
\qquad
\mathrm{Pe}=\frac{v_*x_T}{D},
\qquad
\tilde T_R=\frac{D}{r x_T^2}.
\end{equation}
Thus, the five physical parameters collapse into three dimensionless control
parameters, $\tilde \ell$, $\mathrm{Pe}$, and $\tilde T_R$, as illustrated in
Fig.~\ref{fig:possibleCartoonForCueSearch}(B). The parameter $\tilde\ell=\ell_c/x_T$ measures the range of the cue relative to the target distance: small $\tilde\ell$ corresponds to a localized cue, while large $\tilde\ell$ corresponds to a broad cue. The Péclet number $\mathrm{Pe}=v_*x_T/D$ measures the strength of drift relative to diffusion: small $\mathrm{Pe}$ corresponds to diffusion-dominated motion, while large $\mathrm{Pe}$ corresponds to drift-dominated motion. The dimensionless reset interval
$\tilde T_R=D/(r x_T^2)$ compares the mean time between resets, $1/r$, to the
diffusive time $x_T^2/D$ needed to explore the target distance: small
$\tilde T_R$ corresponds to frequent resetting, while large $\tilde T_R$
corresponds to rare resetting.

Using these dimensionless variables, the earlier equation on the mean first passage time, eqn. \ref{eq:backward_exp_cue},
can be rewritten in the cleaner and more expressive form,
\begin{equation}
0=
1
+
\tilde T''(\tilde x)
+
\mathrm{Pe} \,
e^{-(1-\tilde x)/\tilde\ell}
\tilde T'(\tilde x)
-
\frac{1}{\tilde T_R}\tilde T(\tilde x)
+
\frac{1}{\tilde T_R}\tilde T(0),
\label{eq:pecletDescriptionOfEq100}
\end{equation}
as derived in Sec.~H.1, ``Nondimensionalization of the exponential-cue backward equation,'' of the Supplemental Material.

Even after nondimensionalization, the mean first-passage time depends on three parameters: $\mathrm{Pe}$, $\tilde\ell$, and $\tilde T_R$. For each given cue, we ask which reset time gives the fastest search. We denote this optimal reset time by $\tilde T_{R,\rm cue}^*$ and the corresponding minimum search time by $\tilde T_{\rm cue}^*$. These values reflect interesting aspects of exploratory decisions important to organisms. For instance, Fig.~\ref{fig:possibleCartoonForCueSearch} schematizes how a bee searching for an aromatic flower decides how frequently to reset its position back to the hive. Remarkably, bees follow odor, visual, and social (waggle dance) cues on lengthscales of centimeters to tens of meters \cite{sprayberry2018prevalence, chaffiol2005, dong2023social, spaethe2001visual, riley2005flight}. In the language of the present mathematics, how quickly they can arrive at flower targets is subject to behavioral decisions captured by $\tilde T_{R,\rm cue}^*$ and $\tilde T_{\rm cue}^*$ under different types of cue. Ultimately, asking these questions reflects sensitivity to the two remaining parameters $\mathrm{Pe}$ and $\tilde\ell$. 

\begin{figure}[t]
    \centering
    \includegraphics[width=\columnwidth]
    {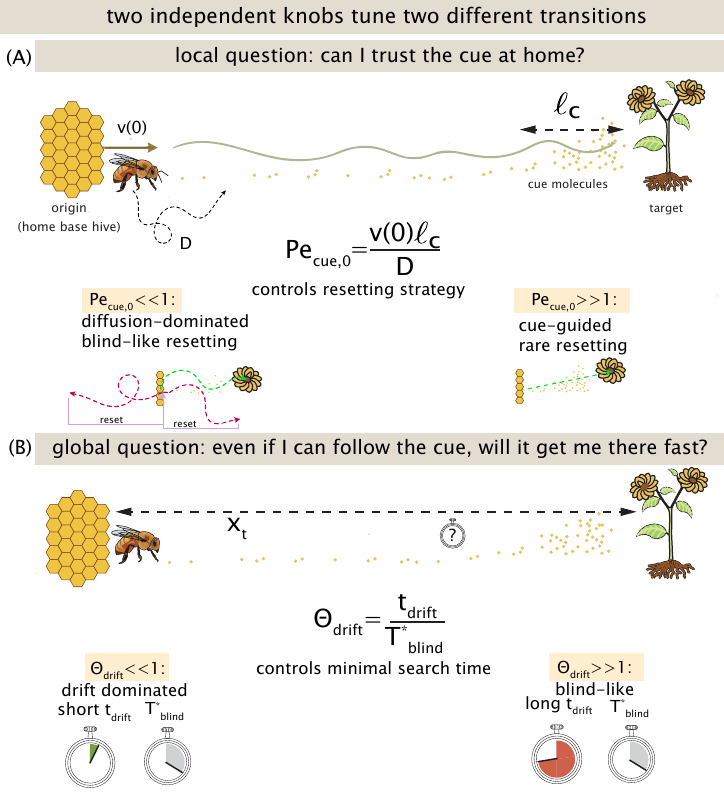}
    \caption{\raggedright
    Two dimensionless quantities control cue-guided search.
    (A) $\mathrm{Pe}_{\rm cue,0}=v(0)\ell_c/D$ controls the resetting strategy:
    small values give blind-like resetting, while large values make resetting rare.
    (B) $\Theta_{\rm drift}=\tilde t_{\rm drift}/\tilde T_{\rm blind}^*$
    controls the optimized search time: small values are drift dominated, while
    large values remain close to the blind-search limit.
    }
    \label{fig:twoCueKnobs}
\end{figure}

As we show below, two distinct dimensionless combinations organize cue-guided searches. One controls the optimized search time, while the other
controls the resetting strategy. These two knobs are illustrated
schematically in Fig.~\ref{fig:twoCueKnobs}.

How different are the statistics and optimal timing of cue-guided search compared to blind searches without cues? 
Two natural limits guide the comparison. For a negligible cue, we recover
blind search, for which the dimensionless optimal reset time and optimal search
time are fixed constants. As derived from eqn.~\ref{eqn:SearchTimeFinalDimensionless}, the
smallest time to arrive at a target without any cues occurs at
\begin{equation}
\tilde T_{R,\rm blind}^*\simeq 0.394,
\qquad
\tilde T_{\rm blind}^*\simeq 1.54.
\end{equation}
These values are useful constant reference points to understand the performance of cue-guided searches. We therefore normalize by these blind-search values. We compare the two problems through the ratios
\begin{equation}
\frac{\tilde T_{\rm cue}^*}{\tilde T_{\rm blind}^*},
\qquad
\frac{\tilde T_{R,\rm cue}^*}{\tilde T_{R,\rm blind}^*}.
\label{eq:cue_blind_optimal_ratios}
\end{equation}
Both ratios approach one when the cue becomes negligible. The meanings of these metrics of performance are visualized in Fig.~\ref{fig:beeAndMetricsOfPerformance}(B).

\begin{figure}
\centering
\includegraphics[width=\columnwidth]{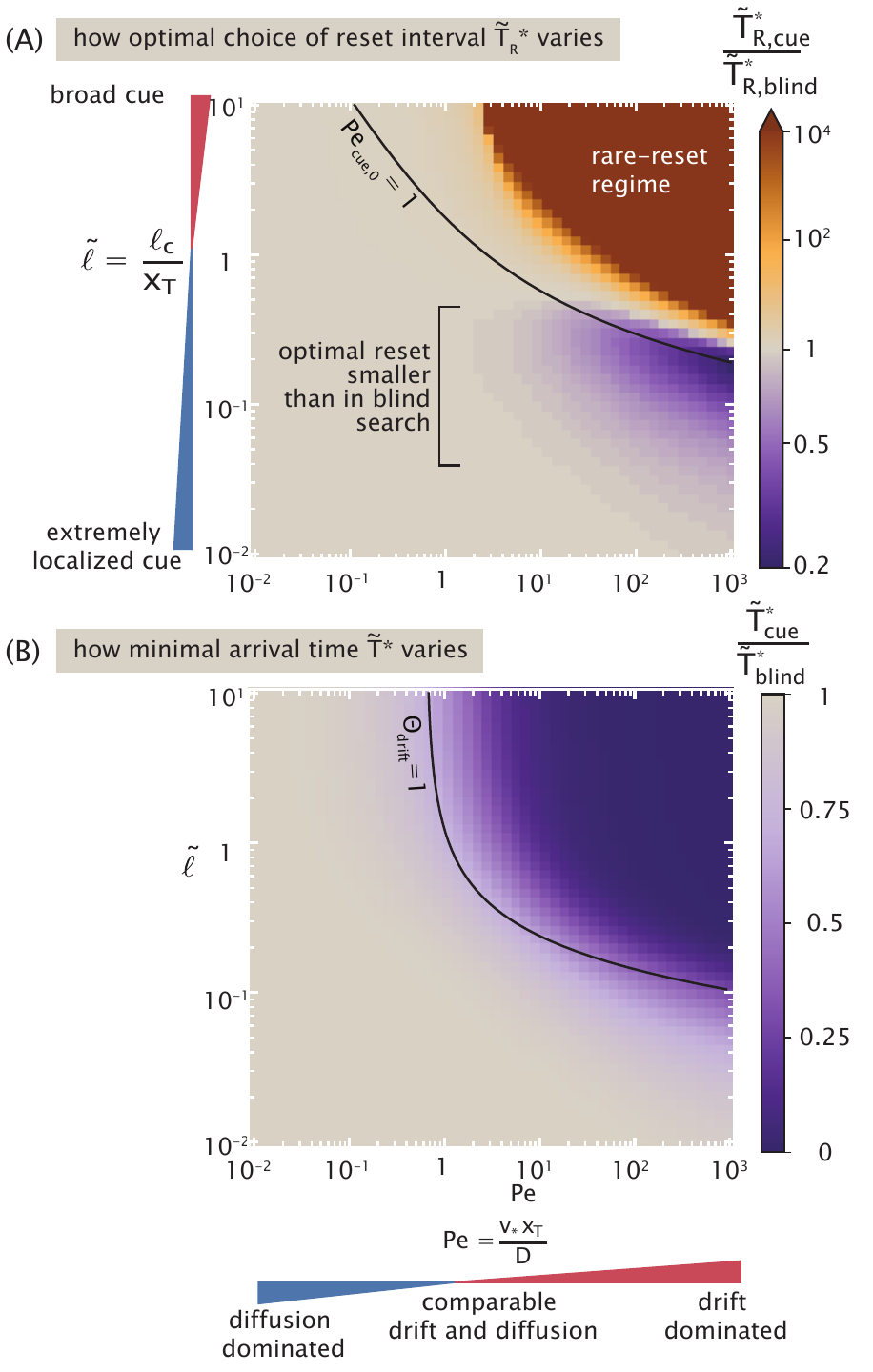}
\caption{
\raggedright
Effect of an exponential cue on optimized search.
The searcher experiences the drift
$v(x)=v_{*}\exp[-(x_{T}-x)/\ell_{c}]$,
with $\tilde{\ell}=\ell_{c}/x_{T}$ and
$\mathrm{Pe}=v_{*}x_{T}/D$.
(A) Optimal reset time with the cue, normalized by the blind-search optimum,
$\tilde T^{*}_{R,\mathrm{cue}}/\tilde T^{*}_{R,\mathrm{blind}}$. For visualization, values of $\frac{\tilde T_{R,\rm cue}^*}{\tilde T_{R,\rm blind}^*}$ exceeding $10^4$ are clipped to the same fixed dark brown color, to accommodate the divergence of the optimal reset time in the strong-cue limit. The black curve marks the informative boundary $\mathrm{Pe}_{\rm cue,0}=1$.
(B) Optimized mean first-passage time, normalized by the blind-search optimum,
$\tilde T^{*}_{\mathrm{cue}}/\tilde T^{*}_{\mathrm{blind}}$.
The black curve marks the informative boundary $\Theta_{\rm drift}=1$.
}
\label{fig:CueOptimization}
\end{figure}

Next, to appreciate the behavior of this larger physical problem, we
evaluate the solution of eqn.~\ref{eq:pecletDescriptionOfEq100} for the mean first-passage time (see Sec.~H.2, ``Solving the backward equation for the mean first-passage time,'' of the Supplemental Material).
Figure~\ref{fig:CueOptimization} shows how these two quantities vary across the $(\mathrm{Pe},\tilde\ell)$ plane. When the cue is weak, both ratios remain close to one and the optimal strategy
approaches that of blind search. As the cue becomes stronger and extends farther from the target, the search becomes increasingly drift dominated. The optimal search time decreases, while the optimal reset time increases. In this regime, resetting becomes less useful: once the cue reliably guides the searcher toward the target, a reset only sends it back to a less favorable position. For any finite cue strength, the optimal reset time remains finite, but in the strong-cue regime it becomes very large, corresponding to an effectively rare-reset search, as shown in Sec.~H.3.b, ``Large $\tilde T_R$ for fixed $\tilde\ell$ and $\mathrm{Pe}$,'' of the Supplemental Material.

In this rare-reset regime, the optimized search time approaches the deterministic time needed to follow the cue from the starting point to the target,
\begin{equation}
\tilde t_{\mathrm{drift}}
= \int_0^1
\frac{d\tilde x}{
\tilde{v}(\tilde{x})} =
\int_0^1
\frac{d\tilde x}{
\mathrm{Pe}\,e^{-(1-\tilde x)/\tilde\ell}
}
=
\frac{\tilde\ell\left(e^{1/\tilde\ell}-1\right)}{\mathrm{Pe}}.
\end{equation}
This asymptotic result is derived in Sec.~H.3.a, ``Large $\mathrm{Pe}$ for fixed $\tilde\ell$ and $\tilde T_R$,'' of the Supplemental Material,
where we also show that the optimized search time approaches this drift-time limit.

To understand what controls the optimal search time itself, we compare the deterministic
drift time with the optimal blind-search time and define the normalized drift time
\begin{equation}
\Theta_{\rm drift}
=
\frac{\tilde t_{\rm drift}}{\tilde T_{\rm blind}^*}
=
\frac{\tilde\ell\left(e^{1/\tilde\ell}-1\right)}
{\mathrm{Pe}\,\tilde T_{\rm blind}^*},
\end{equation}
where $\tilde T_{\rm blind}^*\simeq 1.54$ is the dimensionless optimal
blind-search time defined above.
When $\Theta_{\rm drift}\gg1$, the optimal search time remains close to the blind-search limit. When $\Theta_{\rm drift}\ll1$, the optimal search time approaches the deterministic drift time. Replotting the search-time ratio in the
$(\Theta_{\rm drift},\tilde\ell)$ plane produces nearly vertical contours, showing
that much of its dependence on $\mathrm{Pe}$ and $\tilde\ell$ collapses onto
$\Theta_{\rm drift}$, as shown in Sec.~H.3.a of the Supplemental Material.

A different combination of parameters controls the resetting strategy. The relevant
question is whether, at the starting position, drift is strong enough to carry the
searcher through one cue length before diffusion moves it away. A central result of
our analysis is that this behavior is organized by the Péclet number evaluated at the scale of the cue,
\begin{equation}
\mathrm{Pe}_{\rm cue,0}
=
\frac{v(0)\ell_c}{D}
=
\mathrm{Pe}\,\tilde\ell\,e^{-1/\tilde\ell}.
\end{equation}
This single combination of cue strength and cue length controls the transition
from blind-like resetting to a regime in which resetting becomes increasingly
rare.

When $\mathrm{Pe}_{\rm cue,0}\ll1$, diffusion dominates over one cue length at the
starting point, and the optimal reset time remains close to that of blind search.
When $\mathrm{Pe}_{\rm cue,0}\gg1$, the cue already provides reliable directional
motion from the starting position, so resetting becomes rare and the optimal
reset time becomes very large.
For broad cues, $\tilde\ell\gtrsim1$, curves of constant
$\mathrm{Pe}_{\rm cue,0}$ approximately align with the transition in
Fig.~\ref{fig:CueOptimization}(A), showing that the reset-time dependence on
$\mathrm{Pe}$ and $\tilde\ell$ largely collapses onto $\mathrm{Pe}_{\rm cue,0}$.
The localized-cue regime discussed below is the main exception.

Thus, the cue presents two distinct knobs. The normalized drift time
$\Theta_{\rm drift}$ controls the optimal time to reach the target, whereas
$\mathrm{Pe}_{\rm cue,0}$ controls the optimal resetting strategy. Importantly,
these are independent dimensionless coordinates: the transition in search time and
the transition in reset frequency need not occur together. These two regimes are also visible directly in Fig.~\ref{fig:CueOptimization}.
In panel (A), $\mathrm{Pe}_{\rm cue,0}\ll1$ corresponds to blind-like resetting,
whereas $\mathrm{Pe}_{\rm cue,0}\gtrsim1$ marks the onset of increasingly rare
resetting. Similarly, in panel (B), $\Theta_{\rm drift}\gg1$ corresponds to
blind-like search times, while $\Theta_{\rm drift}\ll1$ corresponds to the
drift-dominated regime.

Finally, the purple region marked in Fig.~\ref{fig:CueOptimization}(A) reveals a less intuitive regime: for sufficiently localized cues, the optimal reset time can be smaller than in blind search. In this regime, resetting gives the searcher repeated opportunities to reach and exploit the localized cue. Thus, while broad and strong cues suppress resetting, localized cues can instead enhance its usefulness. The onset of this regime and the corresponding threshold cue length are analyzed in Sec.~H.4, ``Existence of a minimum for the optimal reset time,'' of the Supplemental Material.

\section{Random Walk with Resets to Checkpoints}
\label{section:RWCheckpoints}
\begin{figure}[h!]
    \centering
    \includegraphics[width=\linewidth]{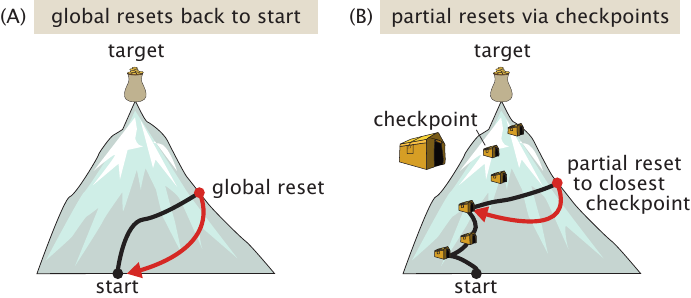}
    \caption{\raggedright Random walk with resets to checkpoints.  (A) Resets  globally to the beginning of an exploratory journey.   (B) Partial resets to the closest checkpoints (shown as orange camp tents), thus saving intermediate progress along a journey.}
    \label{fig:mountainCheckpointsGraphic}
\end{figure}
\begin{figure*}
\centering
\includegraphics[width=\textwidth]{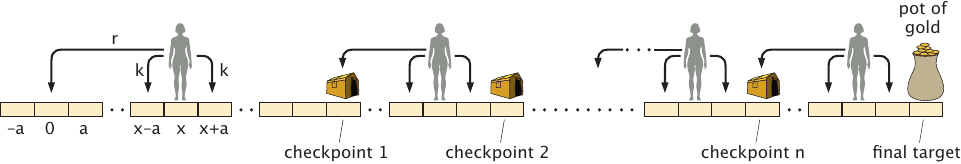}
\caption{\raggedright The seeker that resets to checkpoints (shown as orange camp tents). Once the seeker arrives at the intermediate treasure at checkpoint 1, it commences a new search to arrive at checkpoint 2.  The process repeats until arriving at the giant pot of gold at the target.}
\label{fig:SeekerCheckpoints}
\end{figure*}

The reset models considered in the previous section share a common 
feature: every failed attempt returns the seeker to the same starting point after reset. 
In contrast, however, many natural exploratory processes consolidate partial progress. Progress that has already been consolidated is not lost, and only the most recent exploratory episode must be repeated as indicated schematically in Figs.~\ref{fig:mountainCheckpointsGraphic}~and~\ref{fig:SeekerCheckpoints}. 
These are the biological analog of a mountain summit expedition in which one stops for the night at a series of refuges: if the weather forces a retreat the following morning, one returns to the last refuge rather than all the way  to the base camp. The mathematics of such a climb is the subject of this section. 

\begin{figure*}
    \centering
    \includegraphics[width=0.82\linewidth]{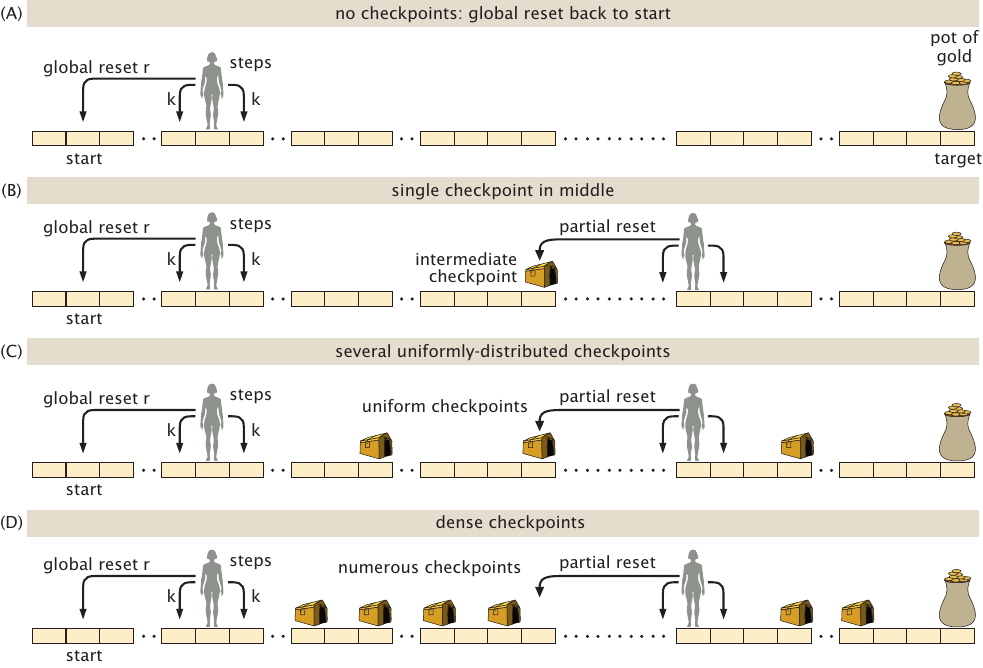}
    \caption{\raggedright Different implementations of checkpoint architectures. (A) Global reset back to the origin without any intermediate checkpoints. (B) A single intermediate checkpoint receiving partial resets once the seeker has progressed past it, instead of being subject to resets all the way back to the origin. (C) Multiple, uniformly-spaced checkpoints between the start and the target location. (D) The limit of many checkpoints.}
    \label{fig:allCheckpointCases}
\end{figure*}

Biological implementations of such checkpoints reflect rich and potentially-useful behaviors; Ravasio and colleagues give a creative contemporary synthesis \cite{Ravasio2026b}. Other examples include a microtubule that has laid down a GTP island mid-filament, which does not collapse all the way to the centrosome when catastrophe strikes; instead,  it retreats only to the most recent island. An axon that has already reached a waypoint does not retract to the cell body when its growth cone stalls. Instead, it resets only to the last stable branch point. A B cell that has already achieved an intermediate affinity does not lose all prior progress when a hypermutation fails: it reverts only to the last positively selected clone.

We will show that intermediate checkpoints reduce the mean 
first-passage time from a quantity that grows exponentially in 
the target distance to one that grows only linearly, and that 
the optimal checkpoint spacing is simply the characteristic 
exploration length $\lambda_D$ that appeared in the single-reset 
problem.  These results are exact, and they follow from a single 
key observation: once a checkpoint has been reached, the 
future problem is statistically identical to a fresh single-segment 
problem starting from the origin.  The segments are therefore 
independent, and everything follows from that independence.
Previewing our approach, we consider the cases shown in Fig.~\ref{fig:allCheckpointCases} in turn.

\subsection{The Single Intermediate Checkpoint}

We begin with the simplest case that illustrates the principle: 
one intermediate checkpoint placed at the midpoint of the journey as shown in Fig.~\ref{fig:allCheckpointCases}(B).  
The target is at position $x_T$ and the checkpoint is at 
$x_1 = x_T/2$.  The seeker starts at the origin.  While the seeker 
is between the origin and $x_1$, any reset returns it to the 
origin.  When the seeker reaches $x_1$ for the first time, that 
position is consolidated and becomes the new reset point.  The 
seeker then attempts to reach $x_T$, with any subsequent reset 
returning it to $x_1$ rather than to the origin.  We work with 
diffusion coefficient $D$ and reset rate $r$, and we 
recall from the previous section that the mean first-passage time 
for a single segment of length $L$ with resets to the start of 
that segment is
\begin{equation}
\langle T \rangle_L = \frac{1}{r}\left(e^{L/\lambda_D} - 1\right),
\qquad \lambda_D = \sqrt{\frac{D}{r}}.
\label{eq:single_segment}
\end{equation}

As noted above, the key observation is that the two exploratory segments are independent.  Once the seeker has reached $x_1$, the future evolution depends only on the distance to the next target, $x_T - x_1 = x_T/2$, and not on how long it took to reach $x_1$ or how many resets occurred along the way. This is simply the Markov property: the seeker starts fresh from $x_1$. Hence, the total first-passage time is the sum of two independent first passage times and is given by,
\begin{equation}
T_{\text{total}} = T_0 + T_1,
\end{equation}
where $T_0$ is the time to cross the first segment from $0$ to $x_1 = x_T/2$ (with resets to $0$), and $T_1$ is the time to cross the second segment from $x_1$ to $x_T$ with resets to $x_1$. Since both segments have the same length $L = x_T/2$, 
both $T_0$ and $T_1$ have the same statistical distribution, and their means are equal. The mean total time is thus given by
\begin{equation}
\langle T_{\text{one checkpoint}} \rangle 
= \langle T_0 \rangle + \langle T_1 \rangle 
= \frac{2}{r}\left(e^{x_T/2\lambda_D} - 1\right).
\label{eq:one_checkpoint_mean}
\end{equation}
Compare this to the mean first-passage time without any 
checkpoint, which from eqn.~\ref{eq:single_segment} with 
$L = x_T$ is
\begin{equation}
\langle T_{\text{no checkpoint}} \rangle 
= \frac{1}{r}\left(e^{x_T/\lambda_D} - 1\right).
\label{eq:no_checkpoint_mean}
\end{equation}
The ratio of these two times is
\begin{equation}
\frac{\langle T_{\text{one checkpoint}} \rangle}
{\langle T_{\text{no checkpoint}} \rangle}
= \frac{2\left(e^{x_T/2\lambda_D} - 1\right)}
{e^{x_T/\lambda_D} - 1}.
\label{eq:ratio_one_checkpoint}
\end{equation}
To understand what this ratio means, we introduce the 
dimensionless distance $\Lambda = x_T/\lambda_D$ permitting us to write
\begin{equation}
\frac{\langle T_{\text{one checkpoint}} \rangle}
{\langle T_{\text{no checkpoint}} \rangle}
= \frac{2\left(e^{\Lambda/2} - 1\right)}{e^{\Lambda} - 1}.
\end{equation}
For large $\Lambda$, the denominator grows as $e^\Lambda$ and the 
numerator grows only as $2e^{\Lambda/2}$, so the ratio decays as
\begin{equation}
\frac{\langle T_{\text{one checkpoint}} \rangle}
{\langle T_{\text{no checkpoint}} \rangle}
\approx 2\,e^{-\Lambda/2} = 2\,e^{-x_T/2\lambda_D},
\qquad x_T \gg \lambda_D.
\label{eq:ratio_asymptotic}
\end{equation}
 One 
intermediate checkpoint reduces the mean first-passage time by a 
factor that is exponentially small in $x_T/2\lambda_D$. This is often the biologically relevant regime, since the distance to the target can be much larger than the typical distance explored between reset events. For a 
target that is ten exploration lengths away, a single checkpoint 
at the midpoint reduces the mean search time by a factor of 
roughly $2e^{-5} \approx 0.013$, a reduction of almost two 
orders of magnitude. In the opposite limit, $x_T \ll \lambda_D$, checkpoints still provide
a speedup, but the exponential behavior of
eqn.~\ref{eq:ratio_asymptotic} no longer applies. Instead, the
improvement becomes weaker and scales linearly with
$x_T/\lambda_D$, as shown in Sec.~I, ``The Near-Target Limit for a Single Checkpoint,'' of the Supplemental Material.

Because $T_0$ and $T_1$ are independent, the variance of the 
total time is simply the sum of the variances of the two 
segment times given by
\begin{equation}
\mathrm{Var}[T_{\text{one checkpoint}}] 
= \mathrm{Var}[T_0] + \mathrm{Var}[T_1]
= 2\,\mathrm{Var}[T_k],
\end{equation}
where $\mathrm{Var}[T_k]$ is the variance of a single segment 
of length $x_T/2$.  To measure noise per unit signal, we calculate the coefficient of variation, given by the ratio of 
the standard deviation to the mean is given by
\begin{align}
\frac{\sqrt{\mathrm{Var}[T_{\text{one checkpoint}}]}}
{\langle T_{\text{one checkpoint}} \rangle}
&= \frac{\sqrt{2\,\mathrm{Var}[T_k]}}{2\,\langle T_k \rangle}\\
& = \frac{1}{\sqrt{2}}
\frac{\sqrt{\mathrm{Var}[T_k]}}{\langle T_k \rangle}.
\end{align}
The coefficient of variation of the full journey is reduced by 
a factor of $1/\sqrt{2}$ relative to that of a single segment.  
The checkpoint not only reduces the mean search time exponentially 
but also makes the arrival time more reproducible, which might matter 
biologically if we hypothesize that developmental and cell-division processes 
operate on tight schedules.

\subsection{The General Case: $n$ Uniform Checkpoints}
\begin{table*}
\setlength{\tabcolsep}{18pt}
\centering
\renewcommand{\arraystretch}{1.6}
\begin{tabular}{llll}
 & \textbf{case} & $\langle T \rangle$ & \textbf{scaling with $x_T$} \\
\hline
\includegraphics[height=12pt]{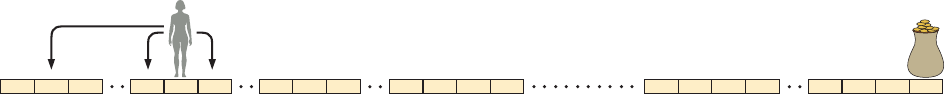} & no checkpoints & 
$r^{-1}(e^{\Lambda} - 1)$ &
exponential \\

\includegraphics[height=12pt]{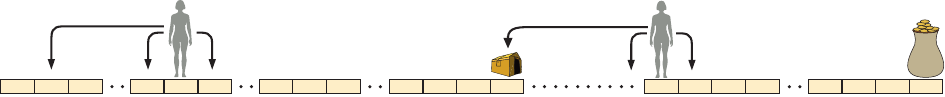} & one checkpoint at midpoint &
$2r^{-1}(e^{\Lambda/2} - 1)$ &
sub-exponential \\

\includegraphics[height=12pt]{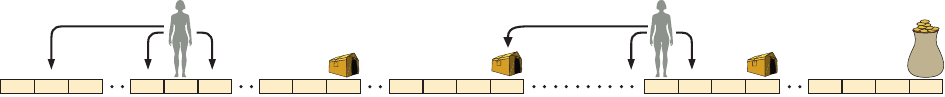}  & $n$ uniform checkpoints &
$(n+1)r^{-1}(e^{\Lambda/(n+1)} - 1)$ &
sub-exponential \\

\includegraphics[height=12pt]{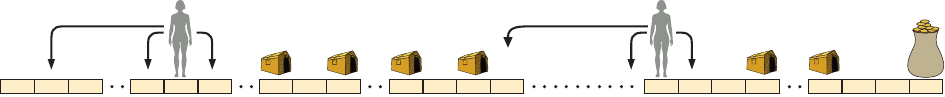} & dense checkpoints ($n\gg \Lambda$) &
$x_T/\sqrt{Dr}$ &
linear \\
\end{tabular}
\caption{
\raggedright Summary of scaling results for the mean first-passage time to a target located at $x_T$ as a
function of the number of checkpoints $n$, where $\Lambda \equiv \frac{x_T}{D}$. 
}
\label{tab:ScalingSummary}
\end{table*}

We now generalize to $n$ uniformly spaced checkpoints, dividing 
the full distance $x_T$ into $n+1$ segments of equal length
\begin{equation}
L = \frac{x_T}{n+1},
\end{equation}
as shown in Fig.~\ref{fig:allCheckpointCases}(C).
The checkpoints are at positions $c_k = kL$ for 
$k = 0, 1, \ldots, n$, with $c_0 = 0$ the origin and 
$c_{n+1} = x_T$ the target.  When the seeker is in the interval 
$(c_k, c_{k+1})$, any reset returns it to $c_k$.  Upon reaching 
$c_{k+1}$, that position is consolidated and becomes the new 
reset point.

By the same Markov argument as before, the time to traverse 
each segment is independent of all other segments.  The total 
first-passage time is
\begin{equation}
T_{\text{total}} = \sum_{k=0}^{n} T_k,
\label{eq:indepSubCheckpoints}
\end{equation}
where each $T_k$ has the same distribution as the first-passage 
time for a segment of length $L$ with resets to the start of 
that segment.  Taking the mean and using 
eqn.~\ref{eq:single_segment}, we find
\begin{equation}
\langle T_n \rangle 
= (n+1)\cdot\frac{1}{r}\left(e^{L/\lambda_D} - 1\right)
= \frac{n+1}{r}\left(e^{x_T/(n+1)\lambda_D} - 1\right).
\label{eq:general_mean}
\end{equation}
The no-checkpoint result is recovered by setting $n = 0$, 
giving $\langle T_0 \rangle = r^{-1}(e^{x_T/\lambda_D}-1)$ as 
expected.  The variance is equally simple and given by
\begin{equation}
\mathrm{Var}[T_n] = (n+1)\,\mathrm{Var}[T_k],
\end{equation}
meaning the coefficient of variation (the mean divided by the standard deviation) scales as $1/\sqrt{n+1}$, 
decreasing with the number of checkpoints. Furthermore, since the total first-passage time is the sum of
$n+1$ independent segment times, the central limit theorem implies
that, for large $n$, the distribution of $T_n$ approaches a
Gaussian. Thus, increasing the number of checkpoints not only
decreases the mean search time but also makes the completion time
progressively more reproducible.

\subsection{The Limit of Many Checkpoints}

It is instructive to ask what happens as $n \gg x_T/\lambda_D$ with 
$x_T$ fixed.  As the segment length $L = x_T/(n+1) \to 0$, 
we expand the exponential in 
eqn.~\ref{eq:general_mean} to first order,
\begin{equation}
e^{L/\lambda_D} = 1 + \frac{L}{\lambda_D} + O\!\left(\frac{L^2}{\ell^{*2}}\right),
\end{equation}
and substitute to obtain
\begin{equation}
\langle T_n \rangle \approx \frac{n+1}{r}\cdot\frac{x_T}{(n+1)\lambda_D}
= \frac{x_T}{r\lambda_D} = \frac{x_T}{\sqrt{Dr}}.
\label{eq:dense_checkpoints}
\end{equation}
In this limit the mean first-passage time grows only linearly 
in $x_T$, in stark contrast to the exponential growth of the 
no-checkpoint result.  Dense checkpoints completely eliminate 
the exponential cost of long-distance search, replacing it 
with a cost proportional to the distance divided by the 
geometric mean of the diffusion coefficient and the reset rate.

\subsection{The Optimal Checkpoint Spacing}

Given fixed $x_T$, $D$, and $r$, how many checkpoints minimize 
the mean first-passage time?  We treat $n+1$ as a continuous 
variable $m$ and minimize $\langle T \rangle = m\,r^{-1}(e^{x_T/m\lambda_D}-1)$ 
with respect to $m$.  Differentiating and setting to zero gives 
the condition
\begin{equation}
e^{x_T/m^*\lambda_D}\left(1 - \frac{x_T}{m^*\lambda_D}\right) = 1,
\label{eq:optimal_condition}
\end{equation}
which must be solved numerically in general.  In the biologically 
relevant limit $x_T \gg \lambda_D$, the solution satisfies
\begin{equation}
\frac{x_T}{m^*} \approx \lambda_D,
\end{equation}
so the optimal segment length is approximately $\lambda_D$.  The 
optimal strategy places one checkpoint per characteristic 
exploration length.  Fewer checkpoints and the exponential cost 
of crossing long segments dominates.  More checkpoints and the 
overhead of repeatedly consolidating short segments dominates.  
The optimal spacing is set entirely by the physics of the reset process.

\subsection{Summary of Scaling Results}

Table~\ref{tab:ScalingSummary} summarizes these key results for the mean first-passage time as a function of the number of checkpoints, writing $\Lambda \equiv x_T/\lambda_D$ for the target's distance in units of the diffusive exploration distance.

The progression from exponential to linear scaling as the number 
of checkpoints increases is the main message of this section.  
The biological implication is direct: a system that can 
consolidate partial progress 
is exponentially more efficient at long-distance search than 
one that must start over completely after every failure.  The 
checkpoint spacing that achieves this efficiency is not a fine-tuned 
parameter: it is simply the characteristic exploration length 
$\lambda_D = \sqrt{D/r}$, which is already determined by the 
diffusion and reset properties of the system.

A biological example that is conceptually similar to the random seeker with intermediate checkpoints is the translocation of polymers across membranes.
This mechanism is often described as a translocation ratchet~\cite{Simon1992,Peskin1993,Lubensky1999,Muthukumar2011}.
Concretely, the problem of DNA viruses ejecting their DNA into bacterial hosts is one in which tens of thousands of basepairs have to get from within the viral capsid to the cellular interior.
The checkpoint idea is relevant here because the binding of proteins such as RNA polymerase to the translocated viral genome can inhibit backward motion of the DNA, thereby making later stages of translocation less likely to undo earlier progress~\cite{Zandi2003,Evilevitch2003,Inamdar2006,Hepp2016}. More generally, similar ideas arise whenever a multistep process
passes through relatively stable intermediates that make reverse
transitions less likely.

\section{Several searchers and several targets}
\label{sec:severalSearchers}

\begin{figure*}[t]
     \centering
     \includegraphics[width=0.75\textwidth]{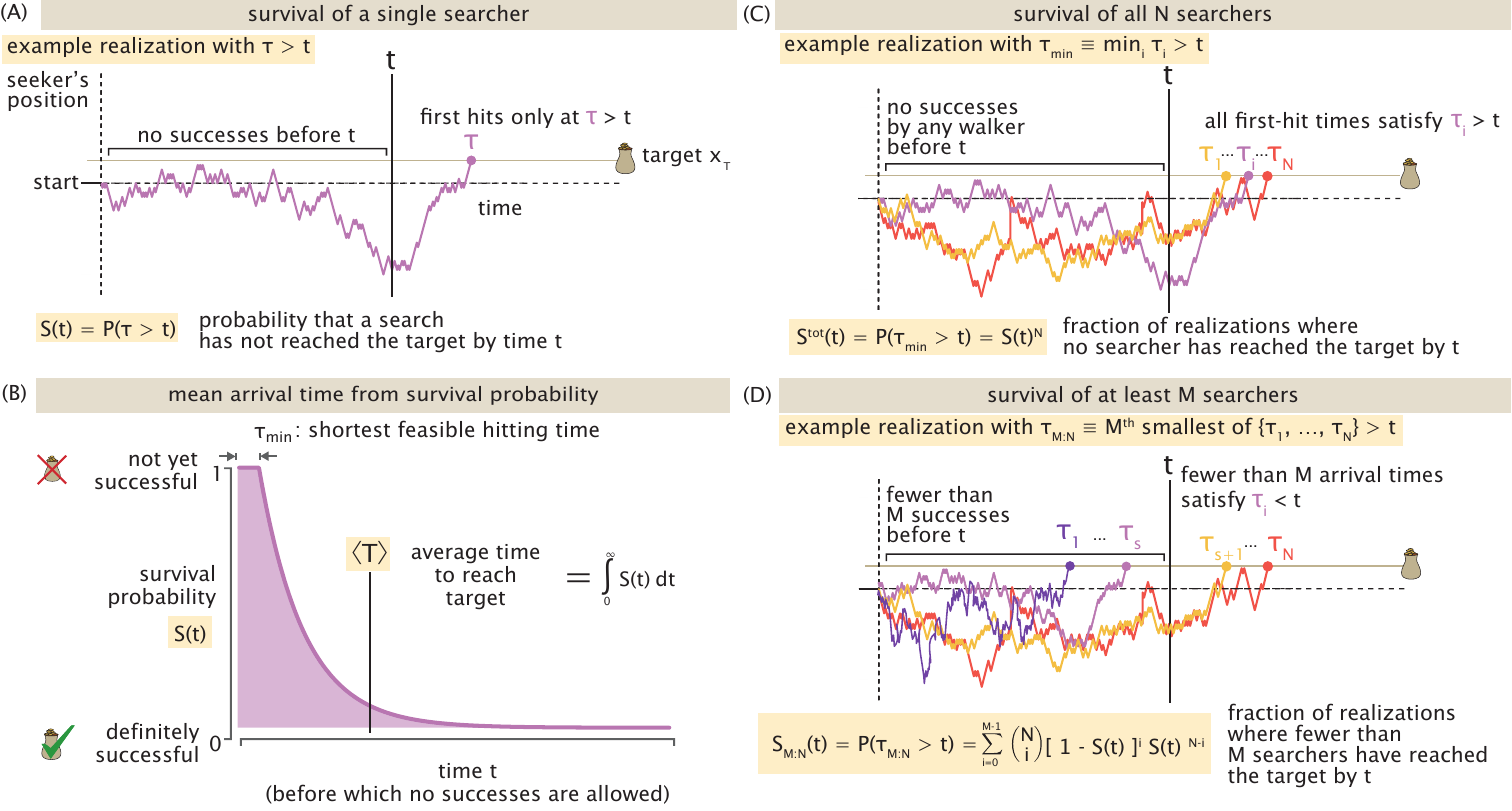}
     \caption{\raggedright Many biological functions are organized by many searchers pursuing many targets. (A) The canonical example of dynamically-unstable microtubules exploring space while setting up the kinetochore is a manifestly parallel search problem. (B) Multiple filopodia in the growth cones of neurons extend and contract in various directions during synaptogenesis, ultimately underlying the exquisite specificity of neuronal connections. (C) At a much more macroscopic scale, the first root (of many searching) to encounter water or nutrients often sets the destiny of plants. (D) Slime molds such as the charismatic \emph{Physarum polycephalum} use many individual extensions of their plasmodia  to perform parallel searches for food in their environments.}
     \label{fig:multipleTargets_asBiologicalExamples}
 \end{figure*}

 \begin{figure*}[!t]
    \centering
    \includegraphics[width=\linewidth]{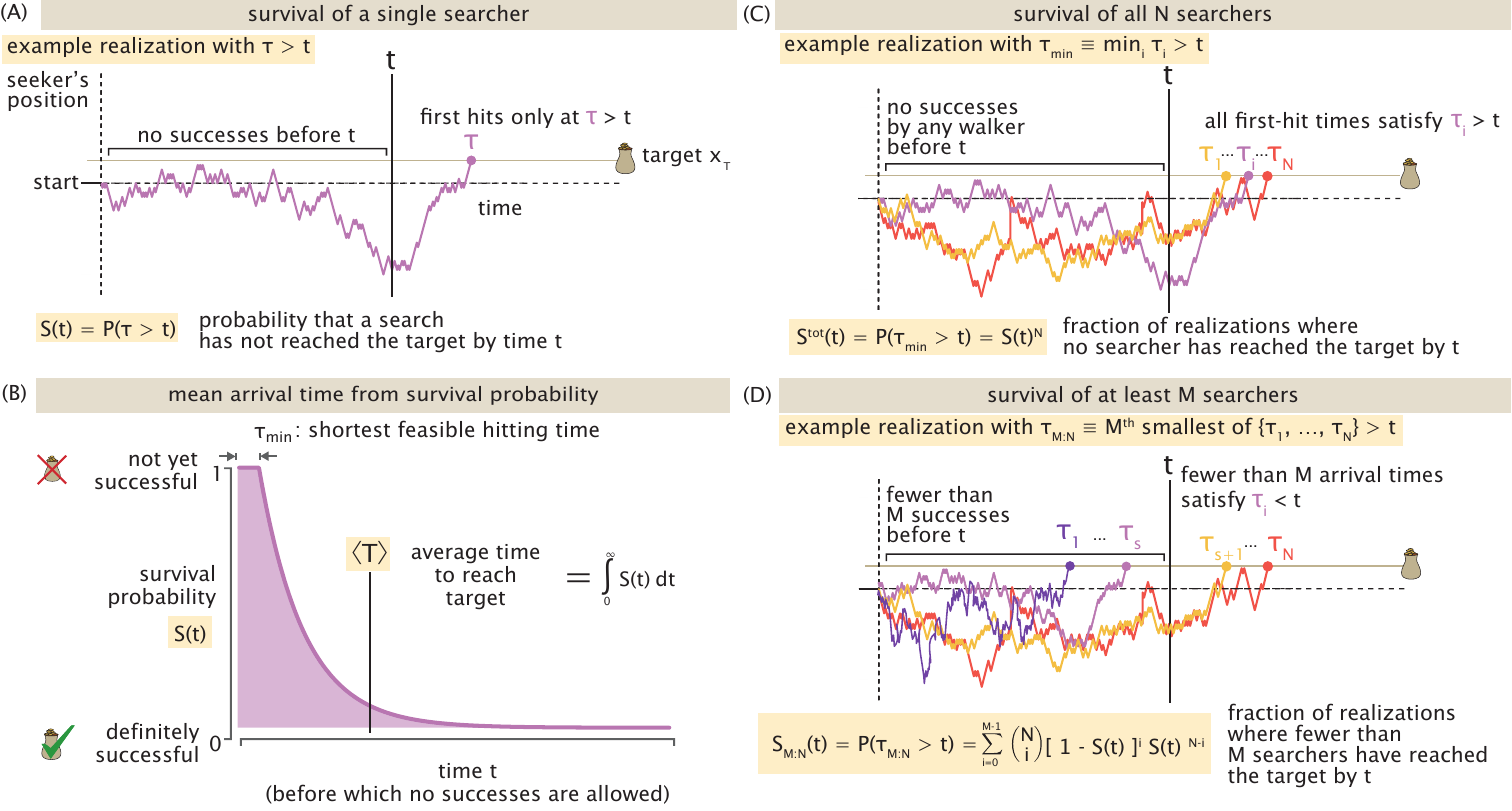}
    \caption{\raggedright Survival probabilities quantify the likelihood that a specified success criterion has not yet been met by time $t$.
(A) A representative single-searcher trajectory that has not yet reached the target by time $t$, corresponding to the event $\tau>t$. Over an ensemble of trajectories, the fraction satisfying this condition is the survival probability $S(t)=\mathbb{P}(\tau>t)$.
(B) The area under a survival-probability curve gives the corresponding mean arrival time, $\langle T\rangle$.
(C) A representative realization of $N$ independent searchers in which none has reached the target by time $t$. The probability of this event is the collective survival probability $S^{\mathrm{tot}}(t)=\mathbb{P}(\tau_{\min}>t)=S(t)^N$.
(D) A representative realization in which fewer than $M$ of the $N$ searchers have reached the target by time $t$. The probability of this event defines the $M$th-arrival survival probability $S_{M:N}(t)$.}
    \label{fig:explainNotionOfSurvivalProbability_andNsearchIdea}
\end{figure*}

\begin{figure*}[t]
    \centering
    \includegraphics[width=\linewidth]{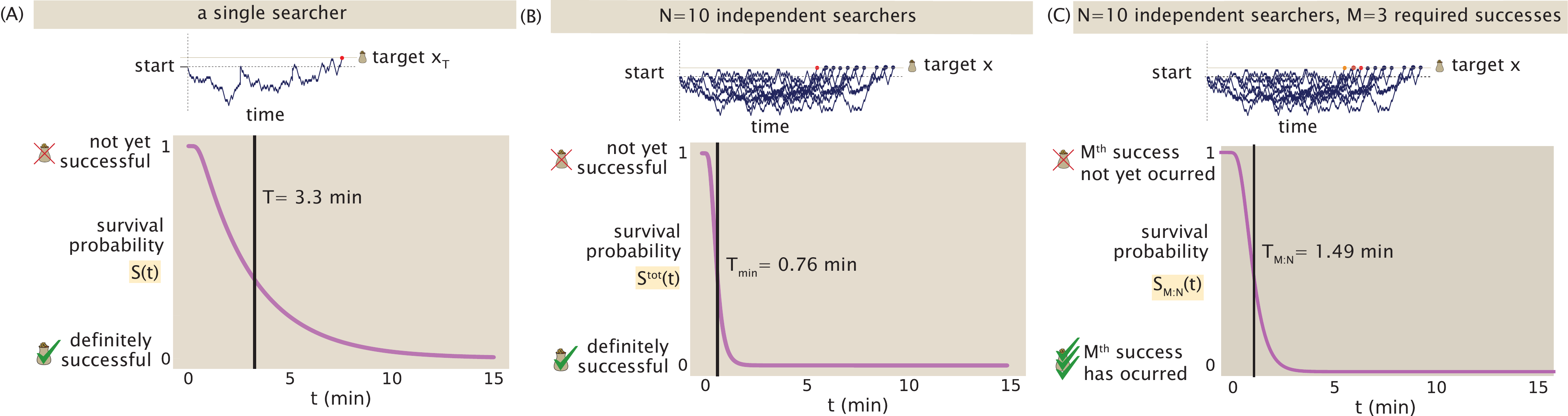}
    \caption{\raggedright 
Survival probabilities and arrival statistics for the 1D
drift--diffusion--reset model.
(A) Single-searcher survival probability $S(t)$, giving the probability
an individual searcher has not yet reached the target by time $t$. The
mean time to reach the target is indicated by the black vertical line.
(B) The collective survival probability $S(t)^N$ for $N=10$ independent
searchers, showing a reduction in the mean time for at least one
of the $N$ searchers to reach the target.
(C) The survival probability $S_{M:N}(t)$ for $N=10$ independent searchers
with $M=3$ required successes, with the mean $M$th-arrival time indicated
by the black vertical line.
Parameters used in these simulations are
$D=0.5\,\mu\mathrm{m}^2/\mathrm{min}$,
$v=1\,\mu\mathrm{m}/\mathrm{min}$, $r=1\,\mathrm{min}^{-1}$,
and $x_T=2\,\mu\mathrm{m}$.}
    \label{fig:singlevsmultiple1d}
\end{figure*}

\subsection{Parallel search with $N$ independent agents}
\label{subsubsec:parallelSearchNindep}

Many biological search processes do not rely on a single seeker. Instead, multiple agents explore the same environment in parallel. Examples pervade biology, as hinted at in Fig.~\ref{fig:multipleTargets_asBiologicalExamples}; these include many microtubules searching for chromosomes ~\cite{Heald2015}, many transcription factors searching out their binding sites on DNA ~\cite{sokolov2005target}, or many cellular protrusions exploring the extracellular environment ~\cite{bornschlogl2013filopodia}. In such situations, the relevant timescale is no longer the first-passage time of a single searcher, but rather the time at which the first member of the population reaches the target. 

Consider $N$ independent searchers evolving under identical dynamics and starting from the same initial condition. Denoting by $\tau_i$ the first-passage time of the $i$th searcher to a target located at $x_T$, we define the earliest-arrival time
\begin{equation}
\tau_{\min}
=
\min(\tau_1,\ldots,\tau_N).
\end{equation}

This quantity describes the time of first successful encounter among the population of searchers. We first study the earliest arrival. Then, in \S\ref{subsubsec:mofnsuccesses}, we generalize this construction to appreciate the timing of at least multiple arrivals (in particular, the $M$-th arrival, with the last arrival recovered by setting $M=N$).

We first introduce the survival probability $S(t)$ of a single searcher, namely the probability that a single searcher has not yet reached the target by time $t$ starting at the origin, or
\begin{equation}
S(t)
=
\mathbb{P}(\tau_i>t),
\end{equation}
where $\mathbb{P}$ denotes probability. This $S(t)$ is illustrated in
Fig.~\ref{fig:explainNotionOfSurvivalProbability_andNsearchIdea}(A)-(B).
The corresponding survival probability for a single searcher is shown in
Fig.~\ref{fig:singlevsmultiple1d}(A), illustrated for the case of a constant drift random walk under reset in one dimension. The event $\tau_{\min}>t$ means that none of the $N$ searchers has reached the target by time $t$. Since each searcher survives independently with probability $S(t)$, the probability that all $N$ searchers survive up to time $t$ is
\begin{equation}
\mathbb{P}(\tau_{\min}>t)
=
S(t)^N.
\label{eq:collective_survival_first_arrival}
\end{equation}
This collective survival probability is illustrated schematically in
Fig.~\ref{fig:explainNotionOfSurvivalProbability_andNsearchIdea}(B),
and the corresponding survival curve for $N=10$ independent searchers is
shown in Fig.~\ref{fig:singlevsmultiple1d}(B).

The mean earliest-arrival time is
\begin{equation}
T_{\min}
=
\left\langle \tau_{\min}\right\rangle
=
\int_0^\infty S(t)^N\,dt,
\label{eq:Tmin_parallel_search}
\end{equation}
as derived in Sec.~J.1, ``Mean earliest-arrival time from the collective survival probability,'' of the Supplemental Material and illustrated geometrically in
Fig.~\ref{fig:explainNotionOfSurvivalProbability_andNsearchIdea}(C). This reduces the parallel-search problem to determining the survival probability $S(t)$ of a single searcher.  These relations are standard results for the extreme first-passage time
of independent searchers
\cite{weiss1983order,yuste2001order,lawley2020universal}.

To determine the survival probability of a searcher initialized at the origin,
we introduce the more general position-dependent survival probability
$S(x,t)$, defined as the probability that a searcher initialized at position
$x$ at time $t=0$ has not yet reached the target at $x_T$ by time $t$. The
single-searcher survival probability introduced above is then
\begin{equation}
S(t)=S(0,t).
\end{equation}

To determine $S(x,t)$, we use the same backward-equation approach introduced
previously for the mean first-passage time, now applying the local averaging
argument to survival probabilities rather than times. The derivation is given 
in Sec.~J.2 of the Supplemental Material.
For diffusion with constant drift $v$
and Poissonian resetting to the origin at rate $r$, this gives ~\cite{pal2019local}
\begin{equation}
\frac{\partial S(x,t)}{\partial t}
=
v\,\frac{\partial S(x,t)}{\partial x}
+
D\,\frac{\partial^2 S(x,t)}{\partial x^2}
-rS(x,t)
+rS(0,t),
\label{eq:backward_survival_drift_reset}
\end{equation}
with the absorbing boundary condition
\begin{equation}
S(x_T,t)=0,
\end{equation}
and initial condition
\begin{equation}
S(x,0)=1,
\qquad x<x_T.
\end{equation}

We now turn to the large-$N$ asymptotic behavior. When many searchers explore in parallel under diffusion and drift, what does the earliest-arrival time approach as $N\to\infty$? Once this asymptotic limit is established, can we determine a threshold number of searchers beyond which the system is effectively in this large-$N$ regime?

For large $N$, the first successful searcher arrives increasingly early.
To understand which part of the single-searcher dynamics controls this
limit, let
\begin{equation}
F(t)
=
1-S(t)
\end{equation}
denote the probability that a single searcher \emph{has} reached the target by time
$t$. At sufficiently early times, $F(t)\ll1$. Among $N$ independent
searchers, the expected number that have reached the target by time $t$ is
$NF(t)$. If this number is much smaller than one, it is unlikely that any
searcher has yet succeeded; if it is much larger than one, at least one
successful trajectory is overwhelmingly likely. Intuitively, the first
arrival therefore occurs when the expected number of successful searchers
becomes of order one,
\begin{equation}
NF(t)=O(1).
\label{eq:NF_order_one}
\end{equation}
A more formal derivation of this criterion is given in
Sec.~J.3, ``Characteristic earliest-arrival time and relation to the mean,'' of the Supplemental Material.

This observation is useful because the large-$N$ problem is therefore
controlled by exceptionally fast single-searcher trajectories. In this
short-time regime, the first-passage probability has the asymptotic form
\begin{equation}
F(t)
\sim
\frac{2\sqrt{Dt}}{x_T\sqrt{\pi}}
\exp\left[
-\frac{x_T^2}{4Dt}
+
\frac{v x_T}{2D}
\right],
\qquad
t\to0^+,
\label{eq:F_short_time_parallel}
\end{equation}
as derived in Sec.~J.4, ``Short-time and large-$N$ asymptotics of the earliest-arrival time,'' of the Supplemental Material.

The full order-statistics calculation, including the relation between a
characteristic first-arrival time and the mean $T_{\min}$, is also given
there.

Using this short-time form, the mean earliest-arrival time of a single searcher among $N$ admits the
large-$N$ expansion (in increasing powers of $1/\ln N$) of
\begin{equation}
T_{\min}
=
T_{\min}^{(0)}
+
T_{\min}^{(1)}
+
T_{\min}^{(2)}
+\cdots,
\label{eq:Tmin_largeN_expansion}
\end{equation}
with
\begin{equation}
T_{\min}^{(0)}
=
\frac{x_T^2}{4D\ln N},
\end{equation}
\begin{equation}
T_{\min}^{(1)}
=
\frac{x_T^2}{8D}
\frac{\ln\ln N}{(\ln N)^2},
\end{equation}
and
\begin{equation}
T_{\min}^{(2)}
=
\frac{x_T^2}{4D(\ln N)^2}
\left[
\frac{1}{2}\ln\pi
-\gamma_E
-\frac{v x_T}{2D}
\right].
\label{eq:Tmin_second_correction}
\end{equation}

The leading behavior is therefore
\begin{equation}
T_{\min}
\sim
\frac{x_T^2}{4D\ln N},
\qquad
N\to\infty.
\label{eq:Tmin_leading_largeN}
\end{equation}
This inverse-logarithmic scaling is a classical result of extreme
first-passage statistics
~\cite{weiss1983order,yuste1996order,yuste2001order,lawley2020universal}. Parallel search therefore accelerates the first encounter only
logarithmically with the number of searchers. Because
$T_{\min}\propto 1/\ln N$, halving $T_{\min}$ requires increasing the
population from $N$ to $N^2$.

\begin{figure}[t]
    \centering
    \includegraphics[width=\columnwidth]{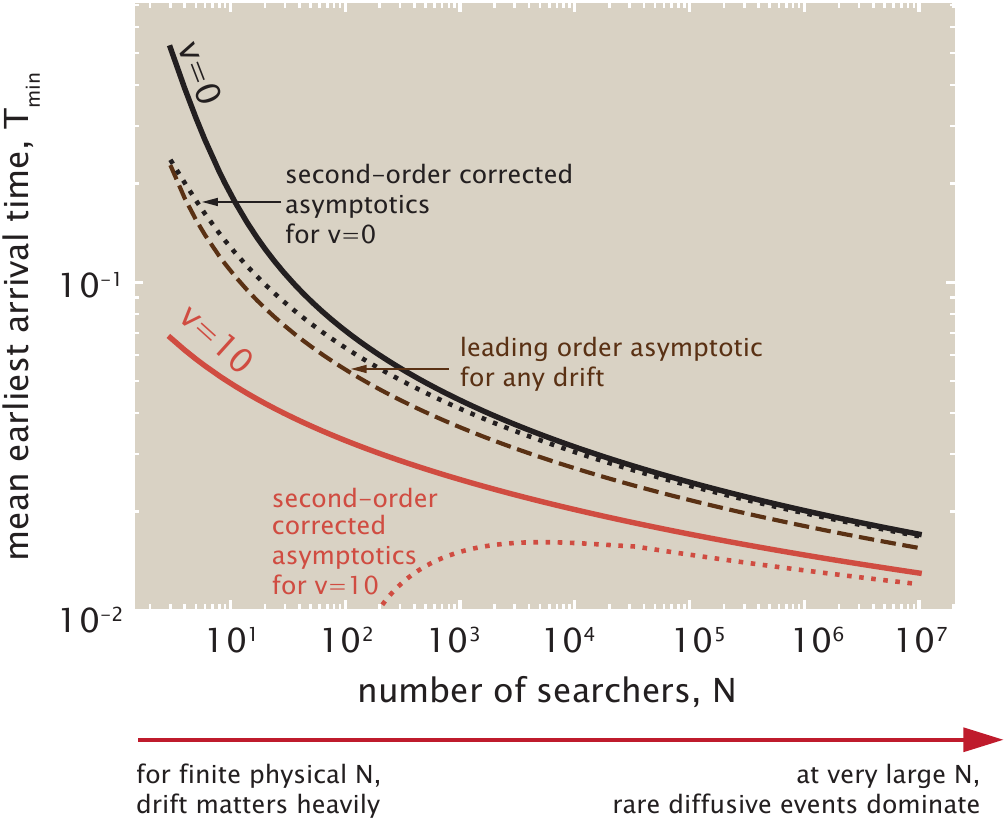}
    \caption{
    \raggedright
    Mean earliest-arrival time as a function of the number of parallel
    searchers $N$ for diffusion without drift ($v=0$) and with drift toward
    the target ($v=10$). Solid curves show the exact mean
    $T_{\min}$, while the corresponding corrected asymptotic curves include
    $T_{\min}^{(0)}+T_{\min}^{(1)}+T_{\min}^{(2)}$.
    The common leading-order result
    $T_{\min}^{(0)}=x_T^2/(4D\ln N)$, which applies for any drift, is shown separately as a brown dashed curve.
    }
    \label{fig:Tmin_largeN_drift}
\end{figure}

The structure of this expansion is illustrated in
Fig.~\ref{fig:Tmin_largeN_drift}. The leading term
$T_{\min}^{(0)}$ is independent of drift ~\cite{lawley2020universal}. The first correction,
$T_{\min}^{(1)}$, is also drift independent and decays only
logarithmically with $N$, producing a slow approach to the leading
large-$N$ behavior. Drift first appears in $T_{\min}^{(2)}$.
Consequently, although all fixed drift velocities ultimately approach
the same leading $1/\ln N$ scaling, their earliest-arrival times can
remain substantially different over a broad range of finite $N$.

The physical origin of this universal behavior can be understood from
eqn.~\ref{eq:F_short_time_parallel}. Because the first arrival occurs when
$NF(t)=O(1)$ according to the heuristic eqn.~\ref{eq:NF_order_one}, increasing $N$ selects progressively earlier arrival times.
At these short times, the diffusive factor
$\exp\left[-x_T^2/(4Dt)\right]$
dominates the first-passage probability. Accordingly, the requirement $NF(t)\approx N\exp\left[-x_T^2/(4Dt)\right] = \text{constant}$ suggests that among a sufficiently large
population, one searcher realizes the exceptionally rare and fast
diffusive trajectory required to reach the target, giving rise to the
universal $1/\ln N$ behavior in
eqn.~\ref{eq:Tmin_leading_largeN}.

Drift affects these rare trajectories differently. Its contribution to
the short-time first-passage probability, $\exp[vx_T/(2D)]$, remains finite as $t\to0$. Drift therefore changes the
probability of observing a fast target-reaching trajectory, but does not
change the dominant short-time scaling that determines
$T_{\min}^{(0)}$. This distinction is visible in
Fig.~\ref{fig:Tmin_largeN_drift}: for finite $N$, drift can
substantially accelerate the first arrival, while at sufficiently large
$N$ rare diffusive events dominate the leading behavior. Importantly,
this asymptotic regime may require extremely large populations, so drift
can remain quantitatively important over the range of searcher numbers
relevant to biological systems.

\begin{figure}[t]
    \centering
    \includegraphics[width=\columnwidth]{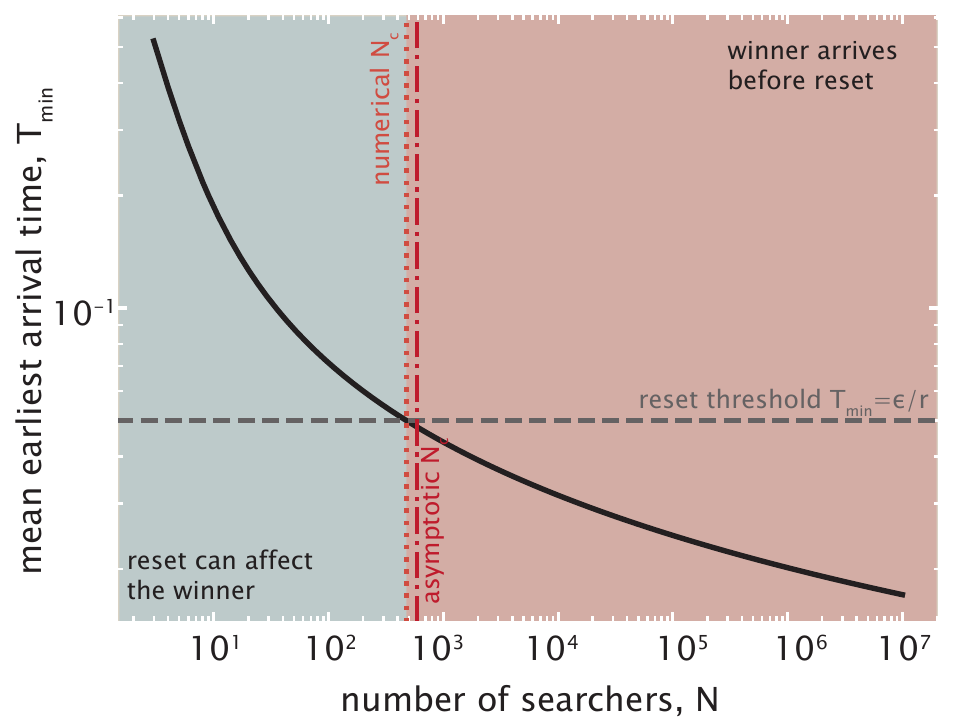}
    \caption{
    \raggedright
Approach to the reset-independent regime as the number of parallel
searchers increases. The mean earliest-arrival time $T_{\min}$ decreases
with $N$, and the characteristic population size $N_c$ is defined by
$T_{\min}=\varepsilon/r$. The numerical value of $N_c$ is compared with the
large-$N$ asymptotic prediction from
eqn.~\ref{eq:Nc_parallel_search}.
    }
    \label{fig:Nc_multiple_searchers}
\end{figure}

Given the parameters of a particular search process, we can now ask how
many parallel searchers are ``enough.'' A complementary version of this
question has been considered by asking how the optimal reset rate changes
with the number of searchers \cite{biroli2023critical}. Here, instead, we
introduce a criterion based on the timescale of the winning trajectory:
how many searchers are required for the first successful trajectory to
occur before resetting typically has time to act? Since the mean reset
time is $1/r$, we define a characteristic population $N_c(\varepsilon)$
by requiring the first-arrival time to be a small fraction $\varepsilon$
of this timescale. Using the large-$N$ condition $NF(t)=O(1)$ at
$t=\varepsilon/r$ gives
\begin{equation}
N_c(\varepsilon)
\sim
\sqrt{\frac{\pi\rho}{\varepsilon}}
\exp\left[
\frac{\rho}{\varepsilon}
-\frac{v x_T}{2D}
\right],
\qquad
\rho=\frac{r x_T^2}{4D},
\label{eq:Nc_parallel_search}
\end{equation}
with the derivation given in Sec.~J.5, ``Approach to the reset-independent large-$N$ regime,'' of the Supplemental Material 
This characteristic population
size is illustrated in Fig.~\ref{fig:Nc_multiple_searchers}.

For $N<N_c$, resetting can still influence the winning trajectory, whereas for $N>N_c$ the first arrival typically occurs before a reset has time to act. The figure also compares the characteristic population size obtained from the full numerical calculation with the large-$N$ asymptotic prediction in eqn.~\ref{eq:Nc_parallel_search}.

A second notion of ``enough'' arises when each additional searcher carries
a cost~\cite{meyer2025optimal}. We consider an effective cost
\begin{equation}
\mathcal C(N)
=
T_{\min}(N)+\gamma N
\simeq
\frac{x_T^2}{4D\ln N}+\gamma N,
\end{equation}
where $\gamma$ is the cost associated with each additional searcher,
measured in units of time, and could represent, for example, the time
required to create or deploy a searcher. More generally, the linear cost $\gamma N$ could be
replaced by a population-dependent cost $G(N)$. For the linear case
considered here, the diminishing gain from parallelization competes with
the cost of additional searchers, selecting a finite optimal population
$N^*$, whose derivation is given in Sec.~J.6, ``Optimal number of searchers in the presence of a resource cost,'' of the Supplemental Material.

The comparison between $N^*$ and $N_c$ then distinguishes two regimes.
If $N^*<N_c$, adding searchers becomes too costly before the system reaches
the reset-independent regime, and resetting still influences the winning
trajectory at the resource optimum. If $N^*>N_c$, the system can afford
enough searchers that the first arrival typically outruns resetting before
the resource optimum is reached.

\begin{figure*}[t]
     \centering
     \includegraphics[width=0.65\textwidth]{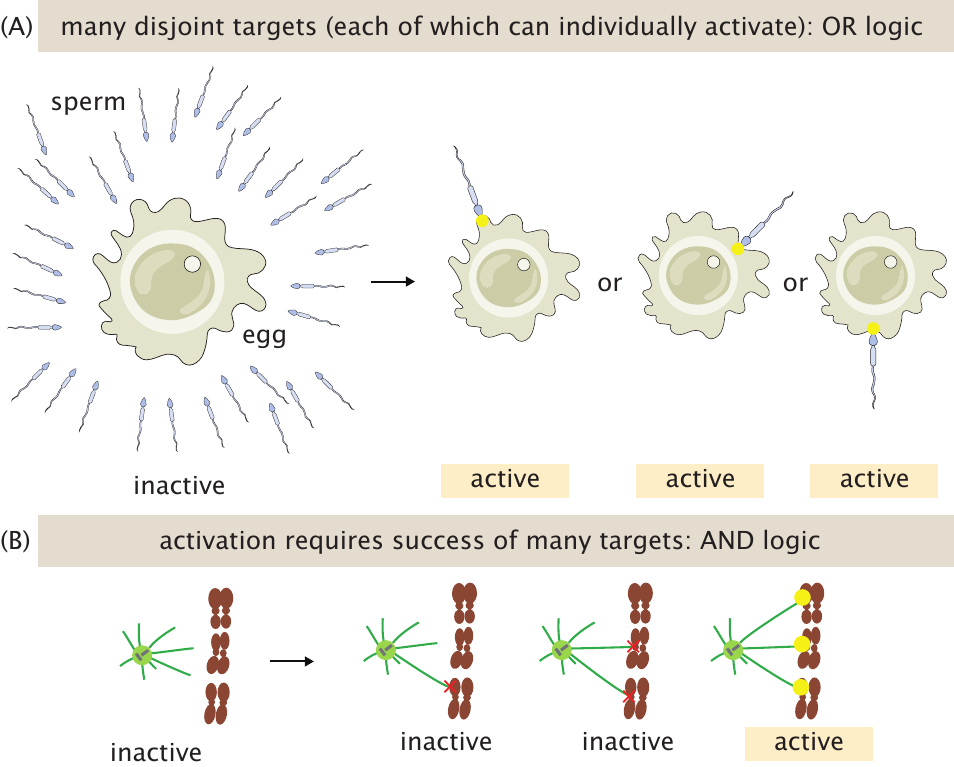}
     \caption{\raggedright Multiple targets for exploratory dynamics can sculpt biological activity. (A) Some biological trajectories terminate and unlock downstream activity when the first of any searchers encounters a target, effectively implementing ``OR'' logic among the search trajectories. For instance, an egg can be fertilized by any of arriving sperm and the first arrival among a group is often salient. A sea urchin egg being fertilized is a particularly direct example of such a dynamic. (In reality, fertilizations do not occur deterministically once sperm encounter eggs but rather at some finite probability estimable by measurements and modeling \cite{levitan1991kinetics, levitan1993importance}, but high fertilization rates motivate the abstraction depicted here.) (B) Other biological contexts do not permit just one encountered target to elicit downstream consequences. Instead, they can require at least several targets to be met before terminating and prompting downstream activity, effectively implementing ``AND'' logic. For instance,  the division of the mitotic spindle is under tight error regulation to effectively execute division only when exactly twenty three chromosomes are captured \cite{ha2024measuring}, an example of aspired AND logic.}
     \label{fig:multipleTargetsSchematic}
 \end{figure*}

\subsection{Multiple searchers and multiple required successes}
\label{subsubsec:mofnsuccesses}

So far, we have focused on the first success among $N$ parallel searchers. We now consider a different completion criterion in which a single target must be reached by multiple searchers before the process is considered complete. This differs from searching for multiple distinct targets, although both involve multiple successful search events.

One suggestive biological setting is kinetochore capture during cell division. Multiple dynamic microtubules can search for and attach to the same kinetochore, with the number of microtubule attachments supported by a single kinetochore varying substantially across organisms: a budding-yeast kinetochore binds a single microtubule, fission-yeast kinetochores bind a few, whereas mammalian kinetochores typically bind on the order of tens of microtubules \cite{winey1995three,joglekar2008molecular,long2019mammalian}. These attachment numbers do not imply that cell division waits for a
prescribed $M$th attachment before proceeding. Rather, they show that a single biological target can receive multiple successful exploratory
arrivals, motivating the more general question of how the time to the $M$th success depends on the number of parallel searchers.

Figure~\ref{fig:multipleTargetsSchematic} illustrates more generally how
biological processes can distinguish between ``OR''-like logic, in which a
single successful encounter is sufficient, and threshold or ``AND''-like
logic, in which several successful encounters are required. Search
processes with stochastic resetting have also been extended to situations
in which a searcher can encounter several possible targets \cite{Bressloff2020a}. Motivated by this broader logic, we generalize the earliest-arrival problem by asking for the time at which $M$ of the $N$ searchers have reached the target.

Denoting by $\tau_i$ the first-passage time of the $i$th searcher, we define
$\tau_{M:N}$ as the $M$th smallest value among
$(\tau_1,\ldots,\tau_N)$. Equivalently, $\tau_{M:N}$ is the first time at
which $M$ distinct searchers have successfully reached the target.

By time $t$, a single searcher has arrived with probability $1-S(t)$ and
remains unsuccessful with probability $S(t)$. Among the $N$ independent searchers, the number of successful arrivals by time $t$ is binomially
distributed, with ``success'' corresponding to arrival by time $t$ and success probability $1-S(t)$. For the $M$th arrival not to have occurred yet, there may have
been no successful arrivals, exactly one successful arrival, exactly two,
and so on, up to at most $M-1$ successful arrivals. Summing over these
possibilities gives the survival probability of the $M$th arrival \cite{weiss1983order,yuste2001order},
\begin{equation}
S_{M:N}(t)=
\mathbb P(\tau_{M:N}>t)
=
\sum_{i=0}^{M-1}
\binom{N}{i}
\bigl[1-S(t)\bigr]^i
S(t)^{N-i}.
\label{eq:MN_survival}
\end{equation}
Figure~\ref{fig:explainNotionOfSurvivalProbability_andNsearchIdea}(D)
illustrates this construction: the $M$th arrival has not yet occurred at
time $t$ whenever fewer than $M$ searchers have reached the target.

The corresponding mean completion time is
\begin{equation}
T_{M:N}
=
\left\langle\tau_{M:N}\right\rangle
=
\int_0^\infty
S_{M:N}(t)\,dt.
\label{eq:TMN_general}
\end{equation}
Figure~\ref{fig:singlevsmultiple1d}(C) shows the corresponding
$M$th-arrival survival probability and mean completion time for
$N=10$ independent searchers with $M=3$ required successes.

The two limiting cases are particularly informative.
For $M=1$, only one searcher needs to succeed and we recover the
earliest-arrival problem, $T_{1:N}=T_{\min}$. At the opposite extreme,
$M=N$ requires every searcher to succeed, so completion is set by the
slowest member of the population. In this case,
\begin{equation}
\mathbb P(\tau_{N:N}>t)
=
1-\bigl[1-S(t)\bigr]^N,
\label{eq:last_arrival_survival}
\end{equation}
since the last arrival has not yet occurred whenever at least one searcher
remains unsuccessful.

We first consider the regime $N\gg M$, with $M$ fixed. By time $t$, the
expected number of successful searchers is $NF(t)$, where
$F(t)=1-S(t)$. If $NF(t)\ll M$, obtaining $M$ successes is unlikely,
whereas once $NF(t)\gg M$, the required number of arrivals has typically
been exceeded. The $M$th arrival therefore occures in the regime
\begin{equation}
NF(t)=O(M),
\end{equation}
or equivalently $F(t)=O(M/N)$. For $N\gg M$, this remains in the
short-time tail of the single-searcher first-passage distribution.

\begin{figure}[t]
\centering
\includegraphics[width=\columnwidth]{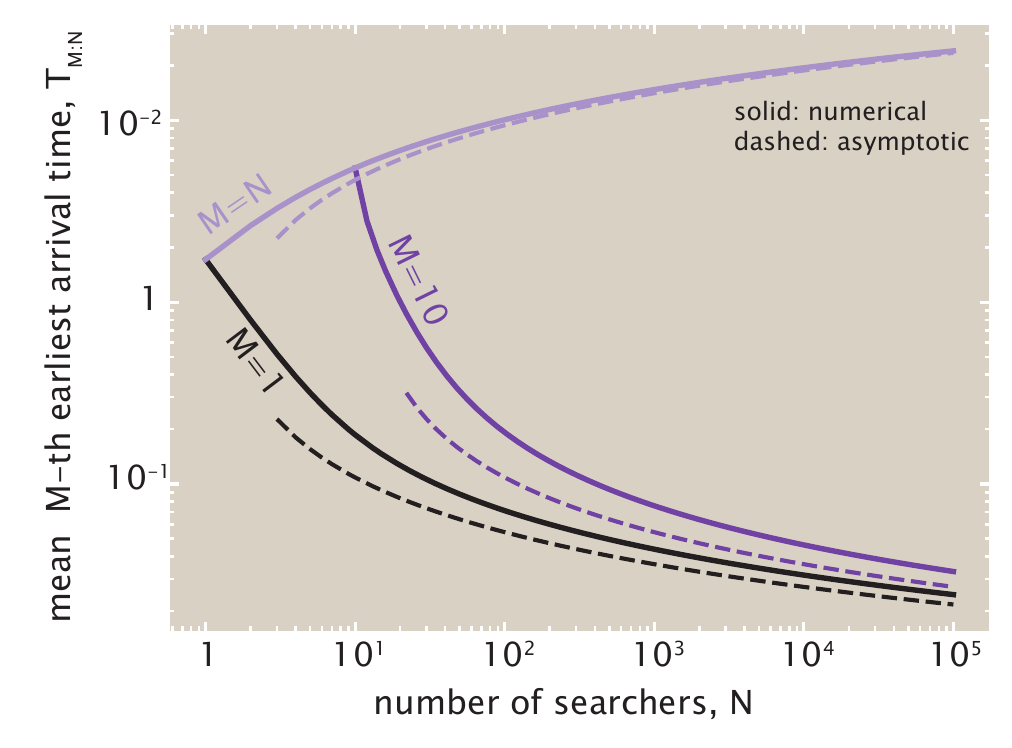}
\caption{\raggedright
Mean $M$th-arrival time $T_{M:N}$ as a function of the number of independent
searchers $N$ for $M=1$, $M=10$, and $M=N$. Solid curves show the numerical
results, while dashed curves show the corresponding leading large-$N$
asymptotic predictions.}
\label{fig:MN_completion_times}
\end{figure}

Using the short-time form of $F(t)$ derived above, the mean $M$th-arrival
time obeys
\begin{equation}
T_{M:N}
\sim
\frac{x_T^2}
{4D\ln(N/M)},
\qquad
N\to\infty
\quad\text{with fixed }M.
\label{eq:TMN_largeN}
\end{equation}
As shown in Fig.~\ref{fig:MN_completion_times}, this large-$N$ prediction
captures the behavior of the mean completion time for fixed $M$, while also
highlighting the contrasting behavior of the last-arrival case $M=N$.
A derivation of eqn.~\ref{eq:TMN_largeN} is given in Sec.~K.1, ``Large-$N$ behavior of the $M$th arrival,'' of the Supplemental Material. 
  The first-arrival result is recovered by
setting $M=1$, and this behavior is consistent with the general large-$N$
theory of fixed first-passage order statistics
\cite{yuste2001order,lawley2020universal}. More generally, requiring more
successful searchers increases the completion time, since the process must
wait for progressively later arrivals among the $N$ searchers; equivalently,
the benefit of parallel search is reduced from $N$ effective searchers to
roughly $N/M$.

We next ask how many parallel searchers are needed for $M$ successful
arrivals to typically occur before resetting becomes important. The
corresponding characteristic number of searchers is
\begin{equation}
N_c(M,\varepsilon)
\sim
M\,N_c(\varepsilon)
=
M
\sqrt{\frac{\pi\rho}{\varepsilon}}
\exp\!\left[
\frac{\rho}{\varepsilon}
-\frac{v x_T}{2D}
\right],
\qquad
\rho=\frac{r x_T^2}{4D},
\label{eq:Nc_M}
\end{equation}
where $\varepsilon = t/r$ is again the time measured in units of the reset time. 
The number of searchers required to outrun resetting therefore increases
linearly with $M$. This scaling follows naturally from the same effective
parallelization that controls the $M$th-arrival time: requiring $M$
successes effectively reduces the number of parallel searchers from $N$ to
$N/M$. Maintaining the same short-time search advantage therefore requires
increasing $N$ by a factor of $M$. We derive this result in
Sec.~K.2, ``Approach to the reset-independent regime for the $M$th arrival,'' of the Supplemental Material.
The opposite limit, $M=N$, corresponds to requiring every searcher to
succeed before the process is complete. Closely related last-arrival
logic arises in ovarian aging, where menopause timing is controlled by
the last small fraction of primordial follicles to leave the ovarian
reserve \cite{lawley2023slowest}.
The probability that all $N$ searchers have reached the target by time $t$ is
$[1-S(t)]^N$. At the late times relevant for the final arrival,
$S(t)\ll1$, so
\begin{equation}
[1-S(t)]^N
\simeq
e^{-NS(t)}.
\end{equation}
If $NS(t)\gg1$, it is therefore very unlikely that every searcher has
succeeded, whereas if $NS(t)\ll1$, essentially all searchers have reached
the target. The final arrival consequently occurs in the regime
$NS(t)=O(1)$, or equivalently $S(t)=O(1/N)$. For large $N$, the last arrival is therefore controlled by the long-time tail of $S(t)$.

For the drift--diffusion process with Poisson resetting considered here,
this long-time behavior can be understood intuitively in terms of repeated
attempts. Each reset returns the searcher to its initial position and begins
a new attempt to reach the target. Let $q$ denote the probability that a
single attempt reaches the target before the next reset. The probability
of failing in $n$ successive attempts is then
\begin{equation}
(1-q)^n
=
\exp\left[n\ln(1-q)\right].
\end{equation}
Since resets occur at a constant rate $r$, the typical time between resets
is $1/r$, so the number of attempts grows proportionally with the duration
of a long trajectory, $n=O(rt)$. The probability of remaining unsuccessful
therefore decreases exponentially with time. This repeated-attempt argument
suggests the long-time form
\begin{equation}
S(t)\sim C e^{-\alpha t},
\qquad
t\to\infty,
\end{equation}
where $\alpha>0$ is the long-time decay rate ~\cite{pal2019local}. The precise value of $\alpha$
depends on the distribution of the attempt durations and is derived more
formally in Sec.~K.3, ``Long-time survival and the last arrival,'' of the Supplemental Material.
Combining this exponential decay with the condition $NS(t)=O(1)$ gives
$NCe^{-\alpha t}=O(1)$, and hence the leading large-$N$ behavior of the
mean last-arrival time ~\cite{lawley2023slowest},
\begin{equation}
T_{N:N}
\sim
\frac{\ln N}{\alpha},
\qquad
N\gg1.
\label{eq:last_arrival_largeN}
\end{equation}
Unlike the fixed-$M$ case, adding more searchers now increases the
completion time, because it also increases the chance that one searcher
will take unusually long to reach the target.

These different regimes are illustrated in
Figure~\ref{fig:MN_completion_times}, which shows the mean $M$th-arrival
time as a function of $N$ for the first arrival ($M=1$), a fixed
intermediate number of required successes ($M=10$), and the last arrival
($M=N$). For fixed $M$, increasing the number of searchers decreases the
completion time, whereas requiring all $N$ searchers to succeed produces
the opposite behavior. The dashed curves show the corresponding leading
large-$N$ asymptotic predictions.

This contrast highlights the difference between processes controlled by
the fastest successful trajectories and those that must wait for the
slowest member of the population.

\section{Exploration Through Conditioning}
\label{sec:ExplorationConditioning}

So far in our journey through exploratory dynamics in sections \S\ref{sec:intro}-\S\ref{sec:severalSearchers}, we contemplated trajectories united only by the condition that they eventually successfully terminate at a target state (no matter their particular approaches or delays in arriving there). Yet biology may impose and experience broader flavors of conditioning than mere eventual success. For instance, trajectories may only be relevant to an organism if they terminate by (or at) an allotted time, set by physiological, environmental, or reproductive demands. 

Accordingly, to appreciate how these wider senses of conditioning sculpt biological exploration, our final foray into mathematical representations of exploratory dynamics focuses on revealing incarnations of the powerful idea of conditional probability~\cite{Jaynes2003ProbabilityTheory,Amir2021ThinkingProbabilistically,Hachmo2023}.  
Though our discussion will focus primarily on the easily visualizable case of one-dimensional diffusion, we note that many biological processes can be described as random walks between the nodes of graphs with
the coin flips of a honest coin used to describe the simplest random walk replaced by rolls of a multisided and dishonest die that allow the transition rates between different nodes to be unequal.

At the heart of this viewpoint is Bayes' theorem. In words, it states that the probability of the present state, given a desired future outcome, is obtained by reweighting the probability of the present state according to how likely that future outcome is from the present state. Symbolically, we write this as
\begin{equation}
\begin{split}
P(\hbox{present state}
\mid
\hbox{future outcome})
=
\\[3pt]
\frac{
P(\hbox{future outcome}
\mid
\hbox{present state})
}
{
P(\hbox{future outcome})
}
\,
P(\hbox{present state}).
\end{split}
\label{eqn:BayesTrajectories}
\end{equation}
A more compact way to write the same idea is
\begin{equation}
P(X_t \mid \mathcal O)
=
\frac{
P(\mathcal O \mid X_t)
}
{
P(\mathcal O)
}
\,P(X_t),
\end{equation}
where $X_t$ denotes the current state of the system and $\mathcal O$ denotes a specified future outcome. The significance of this equation is that it provides a recipe for constructing dynamical laws. The primary
example we will use to illustrate those ideas is what is sometimes called
the Schr\"{o}dinger or Brownian bridge~\cite{schrodinger1931,Chetrite2021}.
Rather than merely analyzing trajectories after they have succeeded, Bayes' theorem tells us how a desired future outcome reweights the present state of the system. The Brownian bridge provides the canonical example. Ordinary diffusion is transformed into a new stochastic process whose trajectories are generated according to the condition that they arrive at a specified endpoint. Our goal is to explore how this same logic can be extended beyond diffusion to more general exploratory processes.

One of the very interesting facets of the problems we describe below is that
in almost all of the cases, we will construct our conditional probabilities
using Bayes-like ratios of constrained microscopic trajectories.
Many of the building blocks for our analysis were written down succinctly
and clearly long ago in a famed 1943 article by S. Chandrasekhar entitled ``Stochastic Problems in Physics and Astronomy'' \cite{chandrasekhar1943stochastic}. In pages 3--7 of that extremely impressive tome, Chandrasekhar demonstrates how to solve
the generic random walk, the problem with both reflecting and absorbing
boundary conditions as well as the first-passage properties needed to figure out
the probability of arriving at position $x$ at time $t$ for the first time. He carries
out these analyses both in discrete and continuous language.
In the pages that follow, we try to develop a concise but explicit language for
describing these various conditional probabilities in both the discrete and continuous
frameworks and then assembling these pieces using Bayes-like ratios.

The goal of this part of the paper is to illustrate how to  construct and analyze such conditioned ensembles of trajectories, and to see how the conditioning process induces a new dynamics. In particular, we will study the statistics of diffusive paths conditioned on future outcomes, such as arriving at a prescribed position at a fixed final time, or reaching a target for the first time at that final time~\cite{Bertoin2003}. 
These constrained trajectory ensembles let us ask how apparently
purposeful dynamics can emerge from purely diffusive motion once
successful outcomes are imposed as constraints.
Throughout, we consider a one-dimensional Brownian particle with diffusion coefficient $D$, starting at position $x_i$ at time $t_i$.
 \begin{figure*}[t]
\centering
\includegraphics[width=\textwidth]{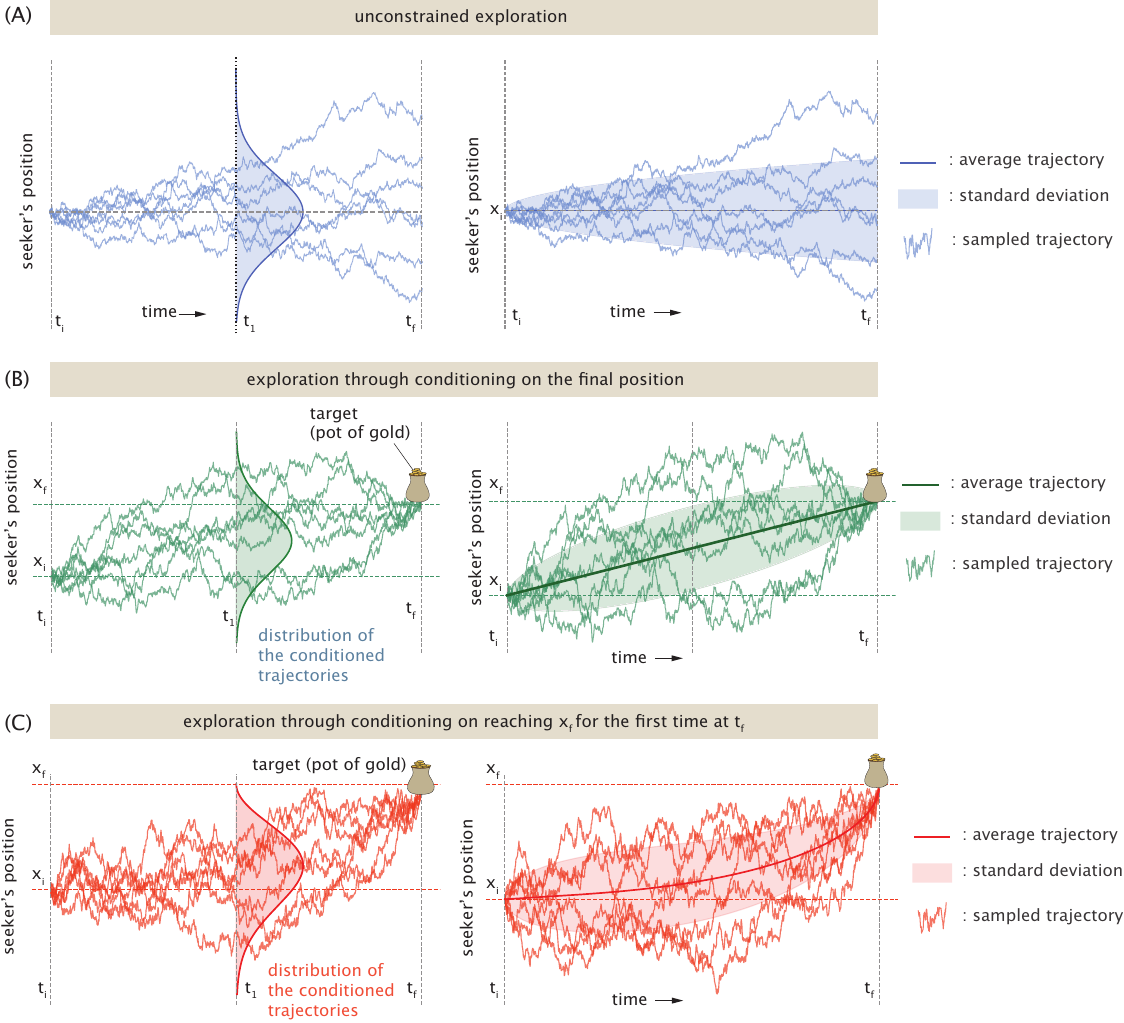}
\caption{\raggedright Conditioning on the trajectories of the random seeker. (A) Representative trajectories for a random seeker with no constraints. (B) Representative trajectories for a random seeker that leaves position $x_i$ at time $t_i$ and arrives at the target at position $x_T$ at time $t_f$. (C) Representative trajectories for a random seeker that leaves position $x_i$ at time $t_i$ and arrives at the target at position $x_T$ at time $t_f$ but with the added constraint that it never arrives at $x_T$ at any time before time $t_f$.}
\label{fig:BrownianBridgeTrajectories}
\end{figure*}

\subsection{Free diffusion}

The elementary building block for all of the conditioned processes considered below is free Brownian diffusion shown in Figure~\ref{fig:BrownianBridgeTrajectories}(A). An intuitive way to think about this process~\cite{Amir2021ThinkingProbabilistically} is that over a small time interval $\Delta t$, the displacement $\Delta x$ is drawn from a Gaussian distribution with mean zero and variance $2D\Delta t$,
\begin{equation}
\Delta x
\sim
\mathcal N(0,2D\Delta t).
\end{equation}
As a result, the transition probability density for a particle diffusing freely from position $x$ at time $t$ to position $y$ at time $t+\Delta t$ is the Gaussian propagator~\cite{Feynman2010,Wiegel1986}
\begin{equation}
p_0(y,t+\Delta t \mid x,t)
=
\frac{1}{\sqrt{4\pi D\Delta t}}
\exp\left[
-\frac{(y-x)^2}{4D\Delta t}
\right].
\end{equation}
Here and throughout, $p_0$ denotes the free-diffusion probability density and we will often use the $p_0$ shorthand rather than writing the whole formula. In particular, the probability density of finding the particle at position $x$ at time $t$, given that it started at $x_i$ at time $t_i$, is
\begin{equation}
p_0(x,t \mid x_i,t_i)
=
\frac{1}{\sqrt{4\pi D(t-t_i)}}
\exp\left[
-\frac{(x-x_i)^2}{4D(t-t_i)}
\right].
\end{equation}

Another important quantity associated with free diffusion is the first-passage-time density~\cite{redner2001,Redner2022}. Consider a particle starting at position $x<x_T$ at time $t$ and diffusing freely until it reaches the target position $x_T$ for the first time. We introduce the notation $f_{\mathrm{FP}}(\tau \mid x,t;x_T)$ for the probability density that this first arrival occurs after an elapsed time $\tau$. For free one-dimensional diffusion, this density is
\begin{equation}
f_{\mathrm{FP}}(\tau \mid x,t;x_T)
=
\frac{x_T-x}{\sqrt{4\pi D\tau^3}}
\exp\left[
-\frac{(x_T-x)^2}{4D\tau}
\right].
\end{equation}
This quantity gives the probability density that the particle first reaches the target position $x_T$ at time $t+\tau$, given that it was at position $x$ at time $t$.
The corresponding survival probability is obtained by integrating over all the seekers who have arrived at the target before time $t$ as
\begin{equation}
S_0(x,t)
=
1-\int_0^t f_{\mathrm{FP}}(x,\tau)\,d\tau .
\end{equation}
For one-dimensional Brownian motion, this integral can be evaluated explicitly \cite{redner2001} as 
\begin{equation}
S_0(x,t)
=
\operatorname{erf}\left(
\frac{x_T-x}{\sqrt{4Dt}}
\right).
\end{equation}

\subsection{Brownian bridge}

We next consider the Brownian bridge, the simplest example of a
diffusive trajectory conditioned on its endpoint. This idea goes back at
least to Schr\"odinger's formulation of conditioned diffusions, now often
called the Schr\"odinger bridge problem~\cite{schrodinger1931,
Chetrite2021}, and is treated systematically in references on
Brownian motion and stochastic processes~\cite{Borodin2012}. In the
Brownian bridge considered here, trajectories are conditioned only to
arrive at the target position $x_f$ at a prescribed final time $t_f$, as
seen in Figure~\ref{fig:BrownianBridgeTrajectories}(B). Importantly,
trajectories are still allowed to cross $x_f$ before the final time.

We introduce the notation
\begin{equation}
p_{\mathrm{bb}}(x,t \mid x_i,t_i;x_T,t_f)
\end{equation}
for the conditional probability density of being at position $x$ at time $t$, given that the trajectory starts at position $x_i$ at time $t_i$ and ends at position $x_T$ at time $t_f$. For $t_i<t<t_f$, Bayes' theorem and the Markov property give
\begin{equation}
p_{\mathrm{bb}}(x,t \mid x_i,t_i;x_T,t_f)
=
\frac{
p_0(x,t \mid x_i,t_i)
\,p_0(x_T,t_f \mid x,t)
}{
p_0(x_T,t_f \mid x_i,t_i)
}.
\end{equation}
The numerator counts all trajectories that start at position $x_i$ at
time $t_i$, pass through position $x$ at the intermediate time $t$, and
subsequently arrive at the final position $x_f$ at time $t_f$. The
denominator counts all trajectories that start at $x_i$ at time $t_i$
and arrive at $x_f$ at time $t_f$. The conditional probability density
therefore corresponds to the ratio between these two trajectory
ensembles. Note that our use of the word ``count'' is imprecise because there are an infinite number of such trajectories.

Substituting the Gaussian diffusion kernels and simplifying yields a Gaussian conditional distribution,
\begin{equation}
\begin{split}
p_{\mathrm{bb}}(x,t \mid x_i,t_i;x_f,t_f)
&=\frac{1}{\sqrt{2 \pi \mathrm{Var}[x(t)]}}
\exp\Bigg[ -
\frac{\left(
x-\langle x(t)\rangle
\right)^2}{2 \mathrm{Var}[x(t)]}
\Bigg],
\end{split}
\end{equation}
with mean trajectory
\begin{equation}
\langle x(t)\rangle
=
x_i+\frac{t-t_i}{t_f-t_i}(x_T-x_i),
\end{equation}
and variance
\begin{equation}
\mathrm{Var}[x(t)]
=
2D\frac{(t-t_i)(t_f-t)}{t_f-t_i}.
\end{equation}
The variance vanishes at both endpoints because the trajectory is constrained to begin at $x_i$ at time $t_i$ and end at $x_T$ at time $t_f$.

The same construction can be carried out locally in time to obtain the one-step transition kernel for the Brownian bridge,
\begin{equation}
p_{\mathrm{bb}}(y,t+\Delta t \mid x,t;x_T,t_f)
=
p_0(y,t+\Delta t \mid x,t)
\frac{
p_0(x_T,t_f \mid y,t+\Delta t)
}{ 
p_0(x_T,t_f \mid x,t)
}.
\end{equation}
This one-step transition was used to generate the trajectories shown in Figure~\ref{fig:BrownianBridgeTrajectories}(B).

\subsection{First-passage Brownian bridge}
\label{sec:first_passage_brownian_bridge}

We now introduce a more restrictive conditioned ensemble. Rather than
merely requiring the trajectory to arrive at the target position $x_T$
at time $t_f$, as in the Brownian bridge, we require that $t_f$ be the
first time at which the target is reached, as illustrated in
Figure~\ref{fig:BrownianBridgeTrajectories}(C). Trajectories that reach
$x_T$ at any earlier time are therefore excluded.

We denote the corresponding conditional probability density by
\begin{equation}
p_{\mathrm{FPbb}}(x,t\mid x_i,t_i;x_T,t_f),
\end{equation}
which gives the probability density of finding the trajectory at
position $x$ at an intermediate time $t$, given that it starts at
$(x_i,t_i)$ and first reaches $x_T$ at time $t_f$
\cite{Bertoin2003,Pitman2006}.

The construction again follows the Bayes-like logic introduced above.
For a trajectory to pass through position $x$ at time $t$, it must
satisfy two requirements. First, it must travel from $(x_i,t_i)$ to
$(x,t)$ without having reached the target at any earlier time. Second,
starting from $x$ at time $t$, it must first reach $x_T$ after exactly
the remaining time $t_f-t$.

The first requirement is described by the absorbing-boundary propagator
\begin{equation}
p_{\mathrm{abs}}(x,t\mid x_i,t_i;x_T),
\end{equation}
which gives the probability density of finding the particle at $x<x_T$
at time $t$, given that it started at $x_i<x_T$ at time $t_i$ and has
remained below the absorbing boundary at $x_T$ at all earlier times. For free
one-dimensional diffusion, the method of images gives
\begin{equation}
p_{\mathrm{abs}}(x,t\mid x_i,t_i;x_T)
=
p_0(x,t\mid x_i,t_i)
-
p_0(x,t\mid 2x_T-x_i,t_i),
\label{eqn:absorbing_propagator}
\end{equation}
for $x<x_T$,
where $p_0$ is the free-diffusion propagator introduced above.

The second requirement is described by the first-passage-time density
$f_{\mathrm{FP}}$ introduced in the preceding subsection. The conditional probability density
is therefore obtained by counting trajectories that satisfy both
requirements and dividing by the total probability density of first
passage from the initial to final state and given by
\begin{equation}
p_{\mathrm{FPbb}}
(x,t\mid x_i,t_i;x_T,t_f)
=
\frac{
p_{\mathrm{abs}}(x,t\mid x_i,t_i;x_T)
\,f_{\mathrm{FP}}(t_f-t\mid x,t;x_T)
}{
f_{\mathrm{FP}}(t_f-t_i\mid x_i,t_i;x_T)
},
\label{eqn:FPbb_density_factorized}
\end{equation}
for $x<x_T$. The numerator represents trajectories that begin at $(x_i,t_i)$,
survive below the target up to time $t$, pass through $x$ at that time,
and subsequently first reach $x_T$ exactly at $t_f$. The denominator
represents all trajectories beginning at $(x_i,t_i)$ that first reach
$x_T$ at $t_f$. 
For the remainder of this subsection, we set $x_i=0$ and $t_i=0$ for
readability and introduce the remaining distance to the target,
\begin{equation}
z=x_T-x.
\end{equation}
It is also convenient to define
\begin{equation}
\mu(t)
=
\frac{x_T(t_f-t)}{t_f},
\qquad
\sigma(t)^2
=
\frac{2Dt(t_f-t)}{t_f}.
\label{eqn:FPbb_mu_sigma}
\end{equation}
Substituting the absorbing propagator and the first-passage densities
into eqn.~\ref{eqn:FPbb_density_factorized} gives the compact expression
\begin{widetext}
\begin{equation}
p_{\mathrm{FPbb}}
(x_T-z,t\mid 0,0;x_T,t_f)
=
\frac{z}
{\mu(t)\sqrt{2\pi\sigma(t)^2}}
\left[
\exp\left(
-\frac{[z-\mu(t)]^2}{2\sigma(t)^2}
\right)
-
\exp\left(
-\frac{[z+\mu(t)]^2}{2\sigma(t)^2}
\right)
\right].
\label{eqn:FPbb_density_compact}
\end{equation}
\end{widetext}
The derivation of eqn.~\ref{eqn:FPbb_density_compact} and the moments
quoted below are given in Sec.~L, ``Derivation of the first-passage Brownian bridge statistics,'' of the Supplemental Material.

The conditioned density can be used to determine both the average
trajectory and the fluctuations around it. The mean position is
\begin{equation}
\begin{split}
\langle x(t)\rangle
=
x_T
&-
\left(
\mu(t)
+
\frac{\sigma(t)^2}{\mu(t)}
\right)
\operatorname{erf}\left(
\frac{\mu(t)}{\sqrt{2}\sigma(t)}
\right)
\\
&-
\sqrt{\frac{2}{\pi}}
\,\sigma(t)
\exp\left[
-\frac{\mu(t)^2}{2\sigma(t)^2}
\right],
\end{split}
\label{eqn:FPbb_mean}
\end{equation}
while the variance is
\begin{equation}
\mathrm{Var}[x(t)]
=
\mu(t)^2
+
3\sigma(t)^2
-
\left(
x_T-\langle x(t)\rangle
\right)^2.
\label{eqn:FPbb_variance}
\end{equation}
The mean and standard deviation predicted by these expressions are shown
together with representative conditioned trajectories in
Figure~\ref{fig:BrownianBridgeTrajectories}(C).

Whether the average first-passage bridge moves monotonically toward the target depends on the competition between the distance to the target, $x_T$, and the characteristic distance explored by diffusion over the allotted time $t_f$ to reach the target.
When
\begin{equation}
\sqrt{2Dt_f}<x_T,
\end{equation}
the average first-passage bridge moves monotonically toward the target.
In this regime, reaching $x_T$ by time $t_f$ already requires an unusually
large displacement toward the target, while arriving much earlier is unlikely.

In contrast, when
\begin{equation}
\sqrt{2Dt_f}>x_T,
\end{equation}
the mean trajectory becomes non-monotonic: it initially moves away from
the target before turning back toward $x_T$. In this regime, reaching the
target too early is likely enough that the first-passage constraint
selects against trajectories that approach $x_T$ too quickly.

In the strongly non-monotonic regime,
$\sqrt{2Dt_f}\gg x_T$, the maximal displacement of the average
trajectory away from the target is itself set by the diffusive length
scale. If $t_*$ denotes the time at which the mean trajectory is farthest
from $x_T$, then
\begin{equation}
x_T-\langle x(t_*)\rangle
\simeq
2\sqrt{\frac{Dt_f}{\pi}}.
\label{eqn:FPbb_max_excursion}
\end{equation}
The derivation of the transition between the monotonic and
non-monotonic regimes and of
eqn.~\ref{eqn:FPbb_max_excursion} is given in
Sec.~L of the Supplemental Material.

The contrast between the three rows of
Figure~\ref{fig:BrownianBridgeTrajectories} summarizes how knowledge of
a future outcome reshapes the dynamics. For unconstrained diffusion,
there is no preferred direction of motion and the distribution simply
broadens with time. Conditioning only on the final position produces a
Brownian bridge whose mean trajectory is directed toward the prescribed
endpoint. Conditioning instead on the first arrival at that
endpoint imposes an additional requirement: the trajectory must both
arrive at $x_T$ at time $t_f$ and avoid arriving there at any earlier
time. 
As for the Brownian bridge, this conditioned ensemble can be expressed
locally as a new stochastic dynamics. Applying the same construction
over a short time interval $\Delta t$ gives the transition kernel
\begin{widetext}
\begin{equation}
p_{\mathrm{FPbb}}
(y,t+\Delta t\mid x,t;x_T,t_f)
=
p_{\mathrm{abs}}
(y,t+\Delta t\mid x,t;x_T)
\frac{
f_{\mathrm{FP}}
(t_f-t-\Delta t\mid y,t+\Delta t;x_T)
}{
f_{\mathrm{FP}}
(t_f-t\mid x,t;x_T)
},
\qquad y<x_T.
\label{eqn:FPbb_local_kernel}
\end{equation}
\end{widetext}
The absorbing propagator appears here rather than the free propagator
because the first-passage condition must be enforced during the short
interval from $t$ to $t+\Delta t$ itself. This local transition kernel
provides a direct prescription for generating first-passage bridge
trajectories and was used to generate the sampled paths shown in
Figure~\ref{fig:BrownianBridgeTrajectories}(C).

The Brownian bridge and first-passage Brownian bridge illustrate how conditioning on a prescribed future outcome can induce new effective dynamics through the reweighting of microscopic trajectories.
This viewpoint also provides a complementary interpretation of the resetting processes considered earlier. A search with resetting can be viewed as a sequence of unsuccessful excursions followed by a final successful trajectory. Conditioning allows us to isolate this last trajectory and ask what paths are typical given that success occurs at a prescribed time. In this sense, resetting organizes exploration through repeated attempts, whereas bridges describe the statistics of the successful attempt itself. To our knowledge, the corresponding first-passage Brownian bridge with
stochastic resetting has not previously been characterized. 
In Sec.~M, ``First-passage Brownian bridges with stochastic resetting,'' of the Supplemental Material, we construct this process explicitly.
We then decompose conditioned trajectories into failed excursions
followed by a final successful excursion and derive analytical results
for the number of resets in the conditioned ensemble.

\subsection{A Conditioned Bridge on a Graph}

The Brownian and first-passage bridges considered above offer two ways
of thinking about conditioning: retrospectively, as selecting a subset
of trajectories, and dynamically, as reweighting the local transitions
that generate them. We now ask what this same construction looks like
beyond diffusion in continuous space.

As noted at the beginning of this section, ``diffusion'' can be interpreted so much more broadly than the simple coin flips of traditional Brownian motion or the continuum version as embodied in Fick's law. Indeed, as we argue here, many of the most fundamental biochemical networks can be thought of as random walks on the nodes of graphs~\cite{Gunawardena2012,Gunawardena2013}, with transition rates that depend upon which node of the graph the random walker occupies.

\begin{figure*}[t]
\centering
\includegraphics[width=\linewidth]{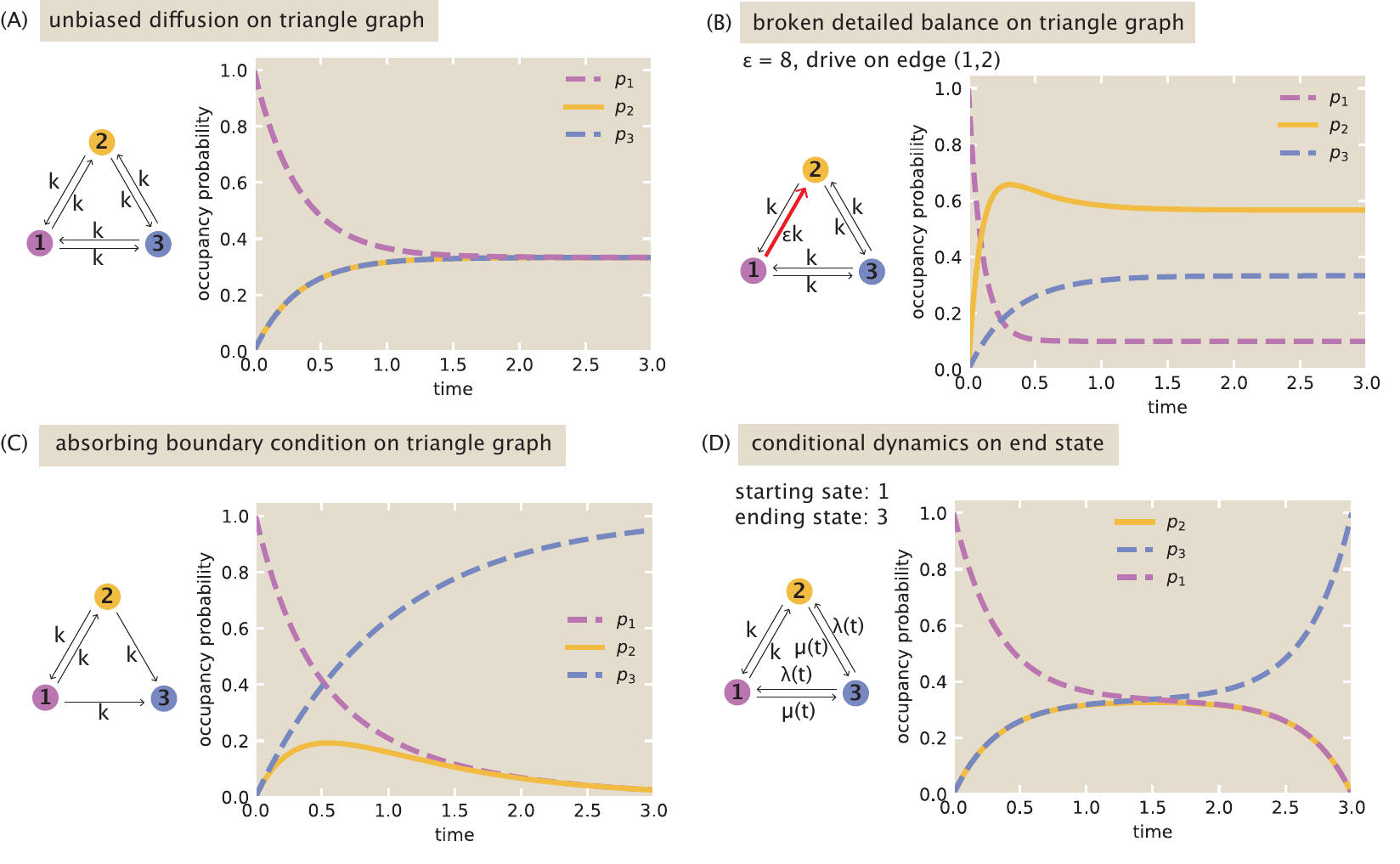}
\caption{\raggedright
Conditioning on the trajectories of the triangular graph. Throughout, we set $k=1$, so that time is measured in units of $1/k$. (A) Relaxation to equilibrium on the three-noded graph with all rate constants equal. (B) Broken detailed balance on the three-noded graph. (C) The three-noded graph with an absorbing boundary condition at node 3. (D) The three-noded graph for a random seeker that leaves node 1 at time $t_i=0$ and arrives at the target at node 3 at time $t_f=3$.
}
\label{fig:TriangleGraph}
\end{figure*}

To illustrate how conditioning induces new dynamics on biochemical networks, we adopt the highly simplified three-nodedgraph shown in Figure~\ref{fig:TriangleGraph}. We consider four instructive versions of dynamics on this graph, culminating in what we will call a ``graph bridge,'' conditioned to begin at node 1 at $t=0$ and arrive at node 3 at a prescribed final time $t_f$.
Case 1: Unbiased diffusion on the three-noded graph.  As shown in Figure~\ref{fig:TriangleGraph}(A), we first consider the triangular graph in which  every allowed transition has the same rate $k$, with arrows in both directions between every pair of nodes. For all of our examples, we will consider the initial condition of releasing the graph walker at node 1, corresponding to the initial condition 
\begin{equation}
{\bf p}(0)=
\left(
\begin{array}{c}
1\\
0\\
0
\end{array}
\right),
\end{equation}
associated with the master equation
\begin{equation}
{d\over dt}
\left(
\begin{array}{c}
p_1\\
p_2\\
p_3
\end{array}
\right)
=
k
\left(
\begin{array}{ccc}
-2 & 1 & 1\\
1 & -2 & 1\\
1 & 1 & -2
\end{array}
\right)
\left(
\begin{array}{c}
p_1\\
p_2\\
p_3
\end{array}
\right).
\end{equation}
We can solve this set of coupled equations directly by elimination.  
As shown in Figure~\ref{fig:TriangleGraph}(A), 
the solution is given by
\begin{equation}
p_1(t)={1\over 3}+{2\over 3}e^{-3kt},
\end{equation}
and
\begin{equation}
p_2(t)=p_3(t)={1\over 3}-{1\over 3}e^{-3kt}.
\end{equation}

This is diffusion on a graph. Probability starts localized at node 1 and then spreads until all three nodes have equal occupancy of
1/3. Interestingly, when we turn to conditioning in Case 4, we will find these solutions serve as the propagators for solving the conditioned problem.

Case 2: Broken detailed balance on the three-noded graph. In earlier work, we have shown how broken detailed balance greatly enriches the behaviors available to biochemical networks~\cite{Mahdavi2024}.  The simplest implementation of broken detailed balance is to imagine that
one of the edges is driven as shown in
Figure~\ref{fig:TriangleGraph}(B).

To be concrete, let the transition $1\to 2$ have rate $\varepsilon k$, while all other rates remain equal to $k$. In this case, the master equation becomes
\begin{equation}
{d\over dt}
\left(
\begin{array}{c}
p_1\\
p_2\\
p_3
\end{array}
\right)
=
k
\left(
\begin{array}{ccc}
-(\varepsilon+1) & 1 & 1\\
\varepsilon & -2 & 1\\
1 & 1 & -2
\end{array}
\right)
\left(
\begin{array}{c}
p_1\\
p_2\\
p_3
\end{array}
\right).
\end{equation}
Using the fact that $p_1+p_2=1-p_3$,
we can solve for $p_3(t)$ immediately yielding
\begin{equation}
p_3(t)=\tfrac{1}{3}\left(1-e^{-3kt}\right).
\end{equation}
Next, we can solve for $p_1(t)$
using the same trick by noting that 
$p_2+p_3=1-p_1$ which yields
\begin{equation}
{d p_1 \over dt}=k\,\big[\,1-(2+\varepsilon)p_1\,\big],
\end{equation}
whose steady state value is $1/(2+\varepsilon)$ and whose rate of approach to steady state is
$(2+\varepsilon)k$, so that
\begin{equation}
p_1(t)=\frac{1}{2+\varepsilon}+\frac{1+\varepsilon}{2+\varepsilon}\,
e^{-(2+\varepsilon)kt}.
\end{equation}
The last probability $p_2$ follows from conservation, $p_2=1-p_1-p_3$, resulting in
\begin{equation}
p_2(t)=\frac{1+2\varepsilon}{3(2+\varepsilon)}
+\tfrac{1}{3}e^{-3kt}
-\frac{1+\varepsilon}{2+\varepsilon}\,e^{-(2+\varepsilon)kt}.
\end{equation}
The system now forgets its initial condition at two distinct rates, the old $3k$ carried
by the bystander at node 3 and a new relaxation rate $(2+\varepsilon)k$ carried by the altered edge.
The steady state it reaches is no longer
uniform and given by 
\begin{equation}
p_1^{\rm ss}={1\over \varepsilon+2},
\end{equation}
\begin{equation}
p_2^{\rm ss}={2\varepsilon+1\over 3(\varepsilon+2)},
\end{equation}
and
\begin{equation}
p_3^{\rm ss}={\varepsilon+2\over 3(\varepsilon+2)}={1\over 3}.
\end{equation}
When $\varepsilon=1$, all three probabilities are $1/3$, as in the unbiased case. But when $\varepsilon\ne 1$, the system has a nonequilibrium steady state with a net circulating flux. For example, the net flux from node 1 to node 2 is
\begin{equation}
J_{12}
=
\varepsilon k p_1^{\rm ss}-k p_2^{\rm ss}
=
{k(\varepsilon-1)\over 3(\varepsilon+2)}.
\end{equation}
The same net flux circulates around the other edges. Thus the system reaches a time-independent distribution, but it is not an equilibrium distribution. The random walkers have a circulation.

Case 3: Absorbing boundary condition on the three-noded triangle.  As shown in 
Figure~\ref{fig:TriangleGraph}(C), in this case, the fact that node 3 is absorbing is captured through the fact that there are no outgoing arrows from this node.  This assumption results in
a master equation of the form
\begin{equation}
{d\over dt}
\left(
\begin{array}{c}
p_1\\
p_2\\
p_3
\end{array}
\right)
=
k
\left(
\begin{array}{ccc}
-2 & 1 & 0\\
1 & -2 & 0\\
1 & 1 & 0
\end{array}
\right)
\left(
\begin{array}{c}
p_1\\
p_2\\
p_3
\end{array}
\right).
\end{equation}
As usual, we have the initial condition $p_1(0)=1$ implying that 
 the solution is
\begin{equation}
p_1(t)={1\over 2}e^{-kt}+{1\over 2}e^{-3kt},
\end{equation}
\begin{equation}
p_2(t)={1\over 2}e^{-kt}-{1\over 2}e^{-3kt},
\end{equation}
and
\begin{equation}
p_3(t)=1-e^{-kt}.
\end{equation}
with the  long-time limit given by
\begin{equation}
p_1(t)\to 0,\qquad p_2(t)\to 0,\qquad p_3(t)\to 1 
\end{equation}
as shown in Figure~\ref{fig:TriangleGraph}(C).

Case 4: The conditioned bridge on the three-noded graph. In the final and
most important case we neither let the walker diffuse freely on the graph, nor do
we subject it to absorbing boundary conditions at node 3. Instead, we condition on arrival at a particular place (node 3) at a particular time ($t_f$), as shown in
Figure~\ref{fig:TriangleGraph}(D). We ask what dynamics the
trajectories that respect this constraint must follow.

We invoke Bayes' theorem to determine the reweighted rates, exactly as we did for
the Brownian bridge. Imagine the walker sits at node $i$ at time $t$ and consider
its next step to a neighboring node $j$. With no constraint, that step is taken at
the bare rate $k$. However, we keep only those histories that end at node 3 at time
$t_f$, and a step to $j$ is consistent with that demand in proportion to the chance
of completing the trip once we are at $j$. That chance is captured by the graph
propagator $P_{ij}(\tau)$, the probability that a walker initially at node $i$ is
found at node $j$ a time $\tau$ later under the unbiased dynamics of Case 1. It is the exact counterpart of the diffusion propagator $p_0(x_f,t_f\mid x,t)$, with positions replaced by nodes on a graph, and describes propagation under the unconditioned, unbiased dynamics of Case 1. The chance of completing the trip from $j$
is then $P_{j3}(t_f-t)$, the propagator carrying node $j$ to the target over the
time that remains. Bayes' theorem tells us that the conditioned probability of the
step $i\to j$ is its bare probability, multiplied by the propagator to the target
from $j$ and divided by the propagator to the target from $i$, so that the bare
rate is reweighted to
\begin{equation}
k_{ij}^{\ast}(t)=k\,\frac{P_{j3}(t_f-t)}{P_{i3}(t_f-t)}.
\end{equation}
The numerator measures the chance of ultimately reaching the target after taking the step to node $j$, while the denominator measures the chance of reaching the target before the step is taken. We can formalize this by writing
\begin{equation}
\frac{P_{33}(t_f-t)}{P_{13}(t_f-t)}
=
\frac{\text{prospects for eventual success after the move}}
{\text{prospects for eventual success before the move}}
\label{eq:prospect_ratio_private}
\end{equation}
which measures the factor by which the move $1\rightarrow3$ improves the walker's prospects.
Conditioning on the destination turns
the constant rate $k$ into a time-dependent rate determined by
the propagators to the target. 

These propagators are not a new calculation; they are a reinterpretation of Case 1's solution. Recall how that solution was obtained. We released the walker at node
1, that is, we solved the master equation with all of the probability placed on
node 1 at $t=0$, and we found the probability $p_j(t)$ of locating the walker at
node $j$ a time $t$ later. But a probability computed under the certainty that the
walker began at node 1 is already a conditional probability, the chance of being at
node $j$ given a start at node 1, and that is exactly the transition probability
$P_{1j}(t)$. The occupation probabilities of a walk released from a definite node
simply are the propagators out of that node. 
The Case 1 results
\begin{equation}
p_1(t)=P_{11}(t)=\frac13+\frac23 e^{-3kt},
\end{equation}
and
\begin{equation}
p_2(t)=p_3(t)=P_{12}(t)=P_{13}(t)=\frac13-\frac13 e^{-3kt},
\end{equation}
are therefore already the propagators carrying node 1 to nodes 1, 2, and 3, with
no further work.

Two features of the triangle let this single calculation supply every propagator
the reweighting needs. First, the unbiased rates do not change in time, so the
propagator depends only on how much time has elapsed, and we may read it over the
remaining time $\tau=t_f-t$. Second, under the unbiased dynamics the three nodes
are interchangeable, every node equivalent and every edge equal, so the propagator
can depend only on whether its start and end are the same node or two different
ones, never on their labels; releasing the walker instead from node 2 or node 3
would return these very same two curves. The whole propagator is thus built from
just two numbers, a same-node value $\tfrac13+\tfrac23 e^{-3k\tau}$ and a
different-node value $\tfrac13-\tfrac13 e^{-3k\tau}$. The bridge now singles out
node 3 as the target, and this is the only place the three nodes cease to stand on
an equal footing. The propagator to the target from the target itself is the
same-node value,
\begin{equation}
P_{33}(t_f-t)=\frac13+\frac23\,e^{-3k(t_f-t)},
\end{equation}
while the propagators to the target from the two non-target nodes are the
different-node value,
\begin{equation}
P_{13}(t_f-t)=P_{23}(t_f-t)=\frac13-\frac13\,e^{-3k(t_f-t)}.
\end{equation}
It is the demand that the walk end at node 3, and not any asymmetry of the
dynamics, that sets node 3 apart from nodes 1 and 2.

Introducing the shorthand $u=e^{-3k(t_f-t)}$ for the remaining-time factor, the
reweighted rates follow at once. The edge joining nodes 1 and 2 is left untouched,
since those two nodes carry the same propagator to the target and their ratio is
unity, so that
\begin{equation}
k_{12}^{\ast}=k_{21}^{\ast}=k.
\end{equation}
The two rates into the target are accelerated to
\begin{equation}
\lambda(t)\equiv k_{13}^{\ast}(t)=k_{23}^{\ast}(t)
=k\,\frac{P_{33}(t_f-t)}{P_{13}(t_f-t)}=k\,\frac{1+2u}{1-u},
\end{equation}
and the two rates out of the target are suppressed to
\begin{equation}
\mu(t)\equiv k_{31}^{\ast}(t)=k_{32}^{\ast}(t)
=k\,\frac{P_{13}(t_f-t)}{P_{33}(t_f-t)}=k\,\frac{1-u}{1+2u}.
\end{equation}

\begin{figure*}[t]
\centering
\includegraphics[width=\textwidth]{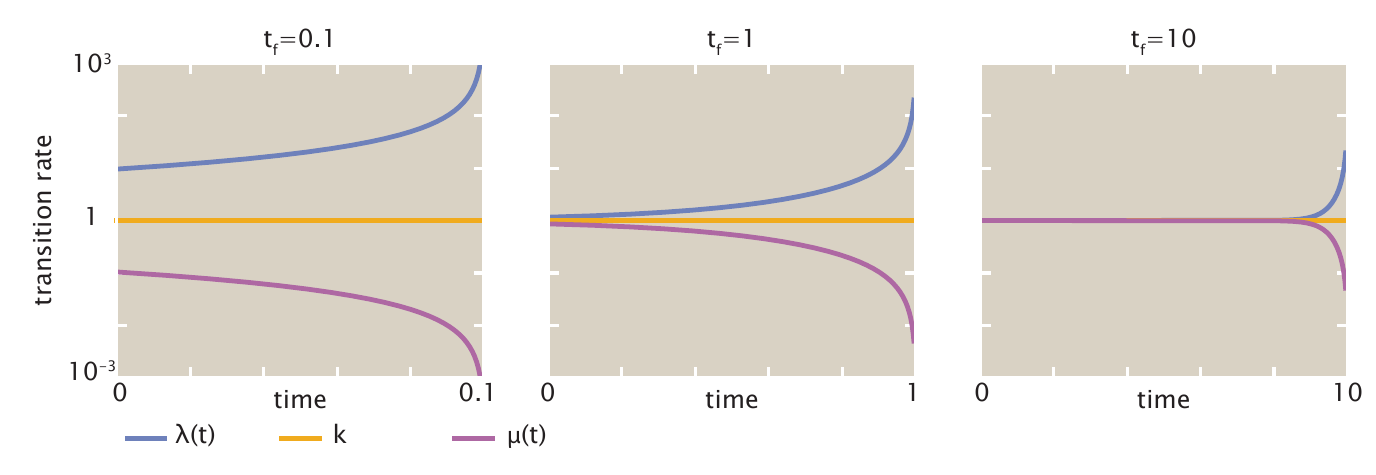}
\caption{\raggedright
Conditioned transition rates on the three-node graph for trajectories conditioned to start at node 1 at $t=0$ and arrive at node 3 at a prescribed final time $t_f$. Panels show three conditioning times, $t_f=0.1$, $1$, and $10$, with $k=1$, so that time is measured in units of $1/k$ and transition rates in units of $k$. The rate $\lambda(t)=k_{13}^*(t)=k_{23}^*(t)$ into the target node is shown in blue, the rate $\mu(t)=k_{31}^*(t)=k_{32}^*(t)$ out of the target node in purple, and the unchanged rate $k_{12}^*(t)=k_{21}^*(t)=k$ in orange. 
}
\label{fig:RenormalizeWeights}
\end{figure*}

As $t\to t_f$ the remaining time vanishes, $u\to 1$, the rate into the target diverges, and the rate out of it falls to zero. The walker is driven onto node 3 and held there, precisely as the bridge demands. When $u\ll1$, however, the
terms proportional to $u$ become negligible, and both $\lambda(t)$ and $\mu(t)$ approach the unconditioned rate $k$. Figure~\ref{fig:RenormalizeWeights} illustrates these induced rates for bridges of different durations: for long
bridges, the rates remain close to their unconditioned value $k$ for much of the trajectory before changing sharply as the prescribed final time approaches.

With these reweighted rates the master equation becomes
\begin{equation}
{d\over dt}
\left(
\begin{array}{c}
p_1\\
p_2\\
p_3
\end{array}
\right)
=
\left(
\begin{array}{ccc}
-(k+\lambda) & k & \mu\\
k & -(k+\lambda) & \mu\\
\lambda & \lambda & -2\mu
\end{array}
\right)
\left(
\begin{array}{c}
p_1\\
p_2\\
p_3
\end{array}
\right),
\end{equation}
with the time-dependent rates $\lambda(t)$ and $\mu(t)$ given above and the
initial condition ${\bf p}(0)=(1,0,0)$.

Although the rates depend on time, the same Bayesian logic that fixed them
delivers the solution as a product of propagators. The probability of finding the
conditioned walker at node $i$ at time $t$ is the propagator that carries it from
the start to node $i$, times the propagator that carries it from node $i$ on to
the target, divided by the propagator for the whole trip,
\begin{equation}
p_i(t)=\frac{P_{1i}(t)\,P_{i3}(t_f-t)}{P_{13}(t_f)}.
\end{equation}
It is striking that this solution carries no trace of the reweighted rates
$\lambda$ and $\mu$, being assembled entirely from the unbiased Case 1 propagators;
the altered rates and this product of bare propagators are two faces of the same
conditioning, the one describing it step by step and the other all at once.

We also note that this expression is the exact graph analogue of the Brownian bridge
formula
\begin{equation}
p_{\rm bb}(x,t)
=
\frac{
p_0(x,t\mid x_i,0)\,
p_0(x_f,t_f\mid x,t)
}
{
p_0(x_f,t_f\mid x_i,0)
}.
\end{equation}
In both cases the conditioned probability is obtained by multiplying a
forward propagator from the initial condition by a backward propagator to
the target and dividing by the propagator for the entire journey. The
forward propagator is the Case 1 result released from node 1, namely
$P_{11}(t)=\tfrac13+\tfrac23 e^{-3kt}$ and
$P_{12}(t)=P_{13}(t)=\tfrac13-\tfrac13 e^{-3kt}$. Introducing the elapsed-time and
remaining-time factors
\begin{equation}
w=e^{-3kt},\qquad u=e^{-3k(t_f-t)},\qquad U=e^{-3kt_f}=wu,
\end{equation}
the three occupation probabilities are
\begin{equation}
p_1(t)=\frac{(1+2w)(1-u)}{3(1-U)},
\label{eq:p1threeNodeConditional}
\end{equation}
\begin{equation}
p_2(t)=\frac{(1-w)(1-u)}{3(1-U)},
\end{equation}
and
\begin{equation}
p_3(t)=\frac{(1-w)(1+2u)}{3(1-U)}.
\label{eq:p3threeNodeConditional}
\end{equation}
Equivalently, we can return from the compact variables $w$, $u$, and $U$ to the original time variables $t$ and $t_f$. This gives the conditioned occupation probabilities directly as sums of elementary time-dependent terms,
{\small
\begin{align}
p_1(t)
&=
\frac{1-2e^{-3kt_f}}
     {3\left(1-e^{-3kt_f}\right)}
+
\frac{2}
     {3\left(1-e^{-3kt_f}\right)}
e^{-3kt}
-
\frac{1}
     {3\left(1-e^{-3kt_f}\right)}
e^{-3k(t_f-t)},
\\[6pt]
p_2(t)
&=
\frac{1+e^{-3kt_f}}
     {3\left(1-e^{-3kt_f}\right)}
-
\frac{1}
     {3\left(1-e^{-3kt_f}\right)}
e^{-3kt}
-
\frac{1}
     {3\left(1-e^{-3kt_f}\right)}
e^{-3k(t_f-t)},
\\[6pt]
p_3(t)
&=
\frac{1-2e^{-3kt_f}}
     {3\left(1-e^{-3kt_f}\right)}
-
\frac{1}
     {3\left(1-e^{-3kt_f}\right)}
e^{-3kt}
+
\frac{2}
     {3\left(1-e^{-3kt_f}\right)}
e^{-3k(t_f-t)}.
\end{align}}
This form separates the contribution of the two endpoint constraints. Each probability contains a constant term, a term proportional to $e^{-3kt}$, and a term proportional to $e^{-3k(t_f-t)}$. The factor $e^{-3kt}$ is largest near the initial time and decays as the system relaxes away from the initial condition. By contrast, $e^{-3k(t_f-t)}$ increases as $t$ approaches $t_f$, reflecting the growing influence of the final-time conditioning. Between these two boundary layers in time (for large $t_f$), the constant terms give the intermediate-time plateau values of the conditioned probabilities.

\begin{figure*}[t]
\centering
\includegraphics[width=\textwidth]{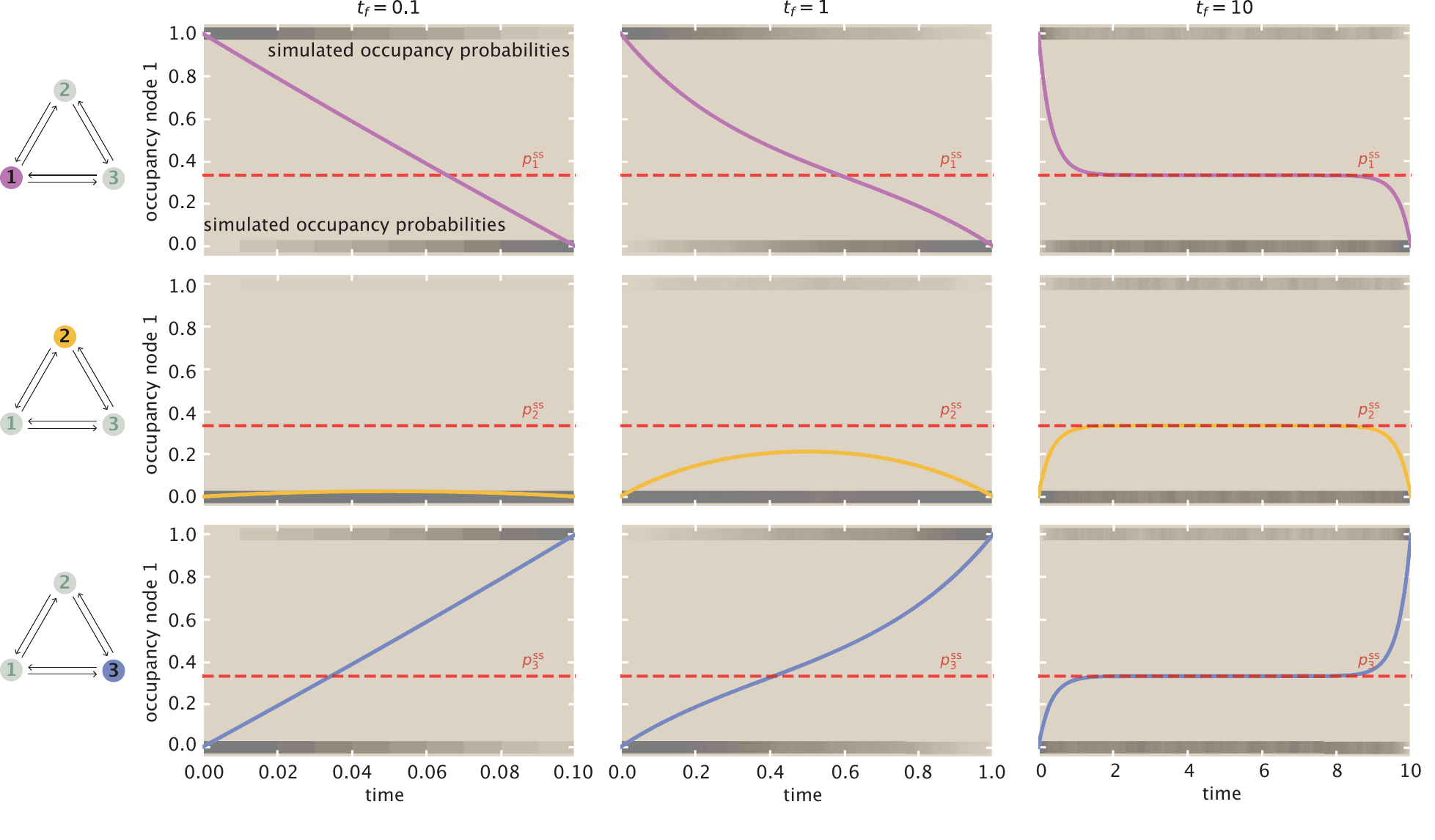}
\caption{\raggedright
Occupancy probabilities for trajectories on the three-node graph conditioned to start at node 1 at $t=0$ and arrive at node 3 at a prescribed final time $t_f$. Rows correspond to nodes 1, 2, and 3, while columns show three conditioning times, $t_f=0.1$, $1$, and $10$, with time measured in units of $1/k$ (equivalently, $k=1$). Solid curves denote the conditional occupancy probabilities, and dashed red lines indicate the corresponding steady-state occupancies $p_i^{\mathrm{ss}}$ of the unconditioned process. Gray bands show simulated occupancy probabilities from conditioned trajectories.}
\label{fig:threeNodeBridgeOccupancies}
\end{figure*}

As shown in Figure~\ref{fig:threeNodeBridgeOccupancies}, these occupation probabilities satisfy both ends of the bridge. At $t=0$ we have $w=1$, so that $p_1=1$ and
$p_2=p_3=0$, implying that the walker is released at node 1. At $t=t_f$ we have $u=1$, so that
$p_3=1$ and $p_1=p_2=0$, implying in turn that the walker is delivered with certainty to node 3. Between
these endpoints the probability drains out of node 1, passes in part through the
bystander node 2, and gathers at node 3, shepherded the whole way by the
reweighted rates that the conditioning has induced. Although we have constructed this bridge using the unconditioned, unbiased dynamics of Case 1, nothing in the conditioning procedure requires this particular choice. The same construction can be applied to other underlying dynamics, including driven nonequilibrium dynamics such as those of Case 2. We now make this generalization explicit.

\subsection{General Markovian description, conditioned on starting and ending states at specified times}
The triangular graph is a special example of a continuous time, discrete state Markov chain with $N=3$ nodes. How do the renormalized dynamics exposed by conditioning on arrival to a final state, or more demandingly doing so at a final time, lift to more general graphs? To address this question, consider a general Markov chain with states $i$ between 1 and $N$, where the rate  at which state $i$ jumps to state $j$ is $k_{ij}$. Let the probability of occupying these states at a time $t$ be stored in a vector $\mathbf{p}(t) = [p_1(t), \ldots, p_N(t)]^\top$, and populate the transition rates into a Laplacian matrix $\mathbf{L}$, whose $(j,i)$th entry stores the transition rate $k_{ij}$ from state $i$ to $j$, and whose diagonal entries give the total exit rate from each state, namely $L_{ii} = -\displaystyle \sum_{j\neq i} k_{ij}$. Then these probabilities evolve over time as,
\begin{align}
 \frac{d\mathbf{p}}{dt} = \mathbf{L}\mathbf{p}(t).
    \label{eq:generalMarkovian}
\end{align}
The brevity of eqn.~\ref{eq:generalMarkovian} belies the variety and complexity of the dynamics admitting such a description; such systems are sometimes called kinetic schemes or chemical master equations. Technically, the general solutions to such dynamics like eqn.~\ref{eq:generalMarkovian} (starting from an initial condition $\mathbf{p}(t=0) = \mathbf{p}_0)$  can be written compactly as a solution involving a matrix exponential, namely,
\begin{align}
    \mathbf{p}(t) & = \exp\left[t \mathbf{L}\right] \mathbf{p}_0
    \label{eq:matrixExponentialForMarkovDynamics_asConcise}
    \\
        & \equiv \left(\mathbf{I} + t \mathbf{L} + \frac{1}{2}t \mathbf{L}\mathbf{L} + \cdots \frac{1}{m!}t^m \mathbf{L}^m + \cdots \right) \mathbf{p}_0,
        \label{eq:matrixExponentialForMarkovDynamics}
\end{align}
where the second line uses the power series expansion representation of the matrix exponential; the $\mathbf{I}$ is the $N \times N$ identity matrix; and $\mathbf{L}^m$ represents the product of $m$ factors of $\mathbf{L}$. We hastily stress that in our view, the ability to write dynamics in the symbolic form of eqn.~\ref{eq:matrixExponentialForMarkovDynamics} does not yet represent true intellectual progress, up to a few very coarse advantages of mechanical manipulation. Rather, to the contrary, this representation should not obscure that the true conceptual, mathematical, and biological functionality and flexibility of a physical system obeying eqn.~\ref{eq:generalMarkovian} still remains essentially uninterrogated by eqn.~\ref{eq:matrixExponentialForMarkovDynamics}; this equation is essentially just a notational repackaging of the original problem. And indeed, in our view, new tools for thinking about these facilities are urgently needed. For instance, we have previously explored some useful tools for conceptually understanding the flexibility of such systems in or out of equilibrium using practical techniques from graph theory \cite{Mahdavi2024}, which replace arduous linear algebraic calculation with drawing of key diagrams. Yet we strongly argue that much more creative thinking on these axes is critically needed.

This said, at least one virtue of eqn.~\ref{eq:matrixExponentialForMarkovDynamics_asConcise} is that this representation furnishes a very concrete input-output map, or \emph{propagator}, from an initial condition $\mathbf{p}_0$ to a final state $\mathbf{p}(t)$. That is, the term $\exp[\mathbf{L}t]$ represents a matrix transformation specifying the transition probability of occupying a state $i$ at time $t=0$ and ending at any other state $j$ at any desired time $t$. Specifically, eqn.~\ref{eq:matrixExponentialForMarkovDynamics_asConcise} says that the probability $p_j(t)$ of being in state $j$ at time $t$ is given exactly by the $j$th element of $\exp[\mathbf{L}t] \mathbf{p}_0 $, computed as $\displaystyle\sum_{j}\left(\exp[\mathbf{L}t]\right)_{ji} (\mathbf{p}_0)_i$. So, if the system begins at state $i$ at time $0$ (namely with the initial condition $\mathbf{p}_0 = [\underbrace{0, \ldots, 1}_i \ldots, 0]^\top$), then this matrix multiplication shows that the output probability of being in some different state $j$ is exactly reported by $\left(\exp[\mathbf{L}t]\right)_{ji}$. In other words, 
\begin{align}
    p(\text{state}~ j ~\text{at time t}|\text{state $i$ at $t=0$}) = \left(\exp[\mathbf{L}t]\right)_{ji}.
\end{align}

\begin{figure}[h!]
    \centering
    \includegraphics[width=\linewidth]{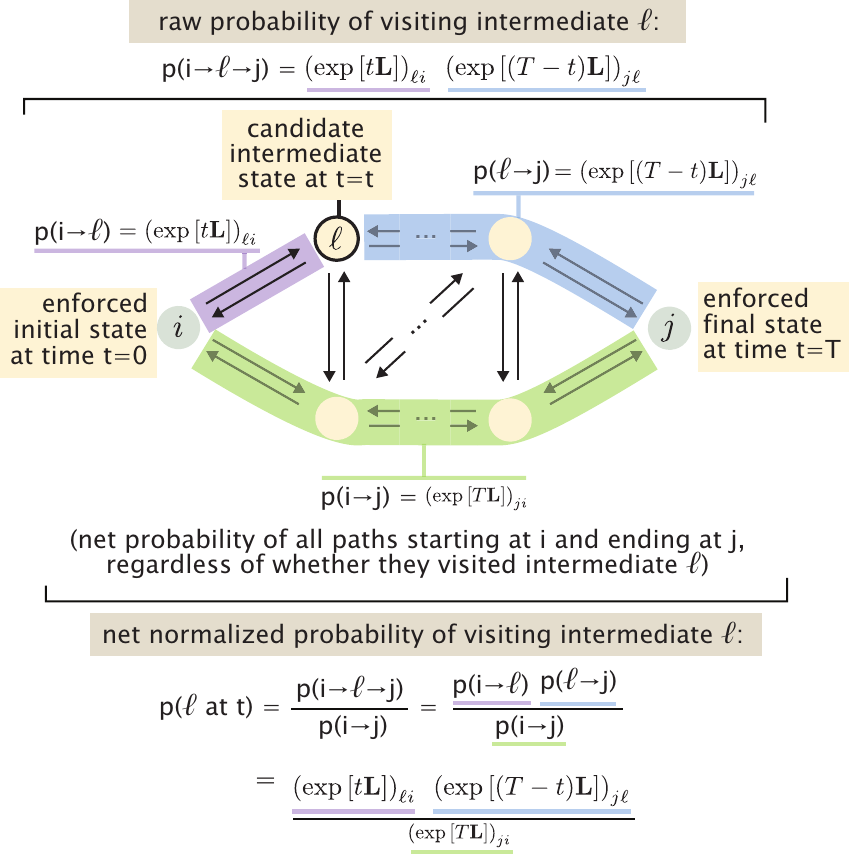}
    \caption{\raggedright Probability $p(\ell,t)$ of visiting an intermediate state $\ell$ at time $t$, once trajectories are conditioned on beginning at state $i$ at time $t=0$ and ending at state $j$ at time $T$.}
    \label{fig:explainGeneralPathProb}
\end{figure}

As shown in Fig.~\ref{fig:explainGeneralPathProb}, since the matrix exponential gives a general notation for the transition probabilities between specified starting and ending times, we may use it to express the probability of being at some intermediate state $\ell$ at time $t$, given that the system began at state $i$ at time $t=0$, and ends at state $j$ at time $t=T$. 
In particular, $\left(\exp\left[t\mathbf{L}\right]\right)_{\ell i}$ expresses the probability of starting at state $i$ and arriving at an intermediate state $\ell$ at time $t$. In the remaining time $(T-t)$ before trajectories must terminate at state $j$ at final time $T$, the probability that trajectories can do so while beginning anew at $\ell$ is $\left(\exp\left[(T-t)\mathbf{L}\right]\right)_{j \ell}$ (the fact that we need not consult any prior history (for instance from $i$) exploits the Markov property of trajectories).  Last, we must normalize this against the net probability that any path (regardless of whether $\ell$ is visited at the proper time) takes a time $T$ from $i$ to $j$, which is $\left(\exp\left[T\mathbf{L}\right]\right)_{ji}$. 
These components allow us to write the probability of occupying any state $\ell$ in the system, given the enforced conditioning, as
\begin{align*}
p\left(\ell ~\text{at }t\mid i\text{ at } t=0, j\text{ at }t=T \right)
\end{align*}
\begin{align}
&=
\frac{
p\left(\ell ~\text{at }t\mid i\text{ at } t=0 \right)
p\left(j ~\text{at }T\mid \ell \text{ at } t \right)
}{
p\left(j\text{ at }t=T\mid i\text{ at }t=0\right)
}
\\
&=
\frac{
\left(\exp\left[t\mathbf{L}\right]\right)_{\ell i}
\left(\exp\left[(T-t)\mathbf{L}\right]\right)_{j\ell}
}{
\left(\exp\left[T\mathbf{L}\right]\right)_{ji}
}.
\label{eq:matrixExpIntermediateMarginalProb}
\end{align}
This expression gives the generalizations of eqns.~\ref{eq:p1threeNodeConditional}-\ref{eq:p3threeNodeConditional} for generic state space graphs. 

\begin{figure}[h!]
    \centering
    \includegraphics[width=\linewidth]{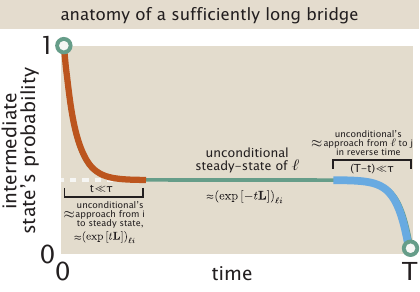}
    \caption{\raggedright Typical anatomy of a long bridge involves dwelling at the unconditional steady-state distribution for intermediate times bookended by transitions before and after resembling other unconditional transients.}
    \label{fig:typicalBridgeAnatomy}
\end{figure}

How do these conditional probability distributions tend to behave over time, compared to their unconditional counterparts? The behavior of eqns.~\ref{eq:p1threeNodeConditional}-\ref{eq:p3threeNodeConditional} explain how central aspects of the interesting anatomy seen in the earlier section's three-nodeded graph example in Fig.~\ref{fig:threeNodeBridgeOccupancies} may be expected to be very general. In particular, Fig.~\ref{fig:threeNodeBridgeOccupancies} hints that these bridge conditional probabilities ``procrastinate'' in satisfying their endpoint constraints, if they can afford to via a sufficiently long conditioning time $T$. This procrastination is most visible for the case of $T=10$ (the right-most column of plots) in Fig.~\ref{fig:threeNodeBridgeOccupancies}: sufficiently long bridges appear to reach an intermediate steady-state that matches the unconditioned distribution $\mathbf{p}^{\text{ss}}$ over states, bookended with transient boundary layers that lead from the start state to this intermediate steady-state, and then from the intermediate steady-state to the final conditioned state. To understand this general behavior, consider the literal algebraic contrast of the conditional dynamics $\hat{p}(\ell,t)$ of being in state $\ell$ at time $t$ conditioned on both start and end states, as specified by eqn.~\ref{eq:matrixExpIntermediateMarginalProb} (as shown in Fig.~\ref{fig:explainGeneralPathProb}) with the unconditional dynamics of trajectories merely starting at state $i$ at $t=0$: the latter unconditional proability of being in state $\ell$ at time $t$ is merely $p(\ell,t)=\exp\left(t\mathbf{L} \right)_{i\ell}$. Interestingly, notice that the conditional dynamics $\hat{p}(\ell,t)$ depart from the unconditional dynamics $p(\ell, t)$ only by a ratiometric factor, namely,
\begin{align}
    \frac{\hat{p}(\ell,t)}{p(\ell, t)} & = \frac{\left(\exp\left[(T-t)\mathbf{L}\right]\right)_{j \ell}}{\left(\exp\left[T\mathbf{L}\right]\right)_{j i}}.
    \label{eq:ratioInmdtProb}
\end{align}
This ratiometric factor exposes three epochs in how a bridge probability changes in time. Early in time, highlighted as the orange small $t$ trace in Fig.~\ref{fig:typicalBridgeAnatomy}, the ratiometric factor of eqn.~\ref{eq:ratioInmdtProb} approaches unity. This happens because both the numerator and denominator of eqn.~\ref{eq:ratioInmdtProb} are well approximated by the steady-state probability of the terminal state $j$: that is, regardless of whether the system starts at an intermediate state $\ell$ or the initial state $i$, the probability of being in state $j$ given by the matrix propagator should approach the steady-state value of the chain, $p^{\text{ss}}(j)$. As a result, the conditional probability $\hat{p}(\ell,t)$ closely mimics the unconditional relaxation from the starting state $i$ to the relevant intermediate state $\ell$. What time $t$ is sufficiently early depends on the rate at which the original unconditional Markov chain of states relaxes to its equilibrium; this characteristic relaxation timescale $\tau$ is captured by the second smallest-size eigenvalue of the Laplacian transition rate matrix $\mathbf{L}$. 

At intermediate times $t\gg \tau$ but $(T-t)\gg \tau$, the conditional probability still approximates the unconditional probability, $\hat{p}(\ell,t)\approx p(\ell,t)$; but now sufficient time has passed (compared to the relaxation time) that the unconditional system has plateaued to near its steady-state value  $p(\ell,\infty)$, highlighted as the flat teal region in Fig.~\ref{fig:typicalBridgeAnatomy}. At late times, namely $t\gg \tau$, $(T-t) <\tau$, a boundary layer traced schematically in blue in Fig.~\ref{fig:typicalBridgeAnatomy} manifests, in order to satisfy the endpoint constraint at last. Thus the coarse anatomy seen in the traces of the particular triangular cyclic graph of Fig.~\ref{fig:threeNodeBridgeOccupancies} are in fact generic features in relevant conditioning time limits.

More generally, these expressions of eqns.~\ref{eq:p1threeNodeConditional}-\ref{eq:p3threeNodeConditional} describing conditional distributions can be readily generalized in two important senses. First, one might pursue trajectories that only satisfy some softer, or more macroscopic, constraint than deterministically starting and terminating in specified states. Specifically, let the microscopic state of a system $x(t)$ specify some biological activity of interest $g[x(t)]$, where the functional bracket reminds us of a potential dependence on the whole trajectory; for instance, $g[x(t)] \equiv \alpha_i x_i(t)$ could encode the biochemical production or signaling rate of state $i$ given by a proportionality $\alpha_i$ at a specific time $t$; or $g[x(t)] \equiv \int_0^T dt~ \alpha_i x_i(t)$ would encode the accumulated such activity over the whole trajectory of length $T$. One could focus only on those trajectories that match a stipulated value $\langle g\rangle$ of this activity $g$ on average. Whether this constraint is demanded at a particular time (like the start or end), or aggregated over the whole trajectory length (making this a functional path constraint), this conditioning induces a new probability distribution over satisfactory trajectories (that departs from the unconditional distribution over paths). In fact, there are many possible probability distributions over such surviving trajectories. One logical principle to choose a particular one is to demand the distribution over surviving trajectories that satisfies the constraint but is otherwise as close as possible to the original unconditional probability distribution over paths $P[x(t)]$, for some informationally prudent notion of ``close.'' Said mathematically, the new probability distribution $\hat{P}[x(t)]$ (over full trajectories $x(t)$) should be the one that has the smallest Kullback-Leibler divergence (relative entropy) with respect to the original unconditional probability distribution while still satisfying the constraints, namely $\hat{P}[x(t)] \equiv \text{argmin}_{\hat{P}\text{ matching } \langle g \rangle} ~ \text{KL}\left( \left. P[x(t)] \right| \hat{P}[x(t)]\right)$. This general prescription is a maximum (relative) caliber principle (with respect to the original unconditional probability distribution), along the foundational ideas of Jaynes \cite{jaynes1980minimum} and many subsequent practitioners \cite{ghosh2006teaching, wu2009, presse2010dynamical, ghosh2020maximum, presse2013principles, dixit2018perspective}. This recipe clearly accommodates the fixed deterministic endpoint constraint discussed earlier as a special case, for instance considering all trajectories starting at $i$ with $g[x(t)] \equiv (x_j(T)=1)$; it also smoothly specifies a recipe for a more gently-constrained probability distribution over paths that are found in state $j$ with some desired softer probability $\alpha \leq 1$. 

A second generalization obtainable from the same prescription considers paths that start from some distribution $\mathbf{p}_0$ and end at some distribution $\mathbf{p}_T$ over states, instead of deterministically starting from particular (singleton) start and end states. This distributional problem was originally posed and analyzed by Schr{\"o}dinger \cite{Chetrite_2021} and the ensuing mathematical forms are called Schr{\"o}dinger bridges; L{\'e}onard provides a superb,  mathematically rigorous discussion of these ideas \cite{leonard2013surveyschrodingerproblemconnections}. In summary, all of these examples show the key idea that a new distribution over acceptable trajectories is induced by start and endpoint constraints, while being as close as possible to an original unconditioned distribution over paths (in an informational sense). 

Conditioning in this sense does not only induce intermediate probability distributions over states of the system; it also formally imposes effective new  transition rates among the states to comply with the starting and ending conditions. In brief, though a full description of these dynamics is beyond the scope of the present paper, one may see that conditioning induces a new probability distribution $\hat{\mathbf{p}}(t)$ with a time-dependent set of effective transition rates  $\mathbf{L}(t)$ such that $\frac{d\hat{\mathbf{p}}}{dt} = \hat{\mathbf{L}}(t)\hat{\mathbf{p}}$. In particular, such renormalized transition rates adopt the form of $\hat{L}_{\ell i}(t) = L_{\ell i} \frac{\left(\exp[(T-t) \mathbf{L}] \right)_{j\ell}}{\left(\exp[(T-t) \mathbf{L}] \right)_{j i}}$. The precise mechanics here are much less important than the concrete enunciation of the  more general point: conditioning effectively accomplishes new forces atop the unconditional dynamics. 

Taken together, these constructions provide two complementary ways of viewing conditioning. Retrospectively, conditioning selects from the original ensemble those trajectories that satisfy a prescribed outcome. Dynamically, the same conditioning can be represented by new effective transition rates that generate the conditioned ensemble directly.

\section{Discussion}
The eagle casts for prey; the poet and scientist cast for word and idea. Neurons wander through tissue, holding out for just the right synaptic partners to make us, \emph{us}. Many of the feats that most animate life are ultimately accomplished by exploratory trial and error. Organisms, cells, and molecules---in stark contrast to the granular deterministic control algorithms that dominate machines and computers---are often not controlled by microscopic instructions that steer their instantaneous trajectories. Instead, a different mechanism commonly organizes such systems: individuals sample variation by exploring randomly and widely, until fortunate trajectories are stabilized by selection and environmental forces. This picture of exploratory dynamics, visible recursively on all scales of biological function, invites the understanding that the magic of compounded variation and selection is available not only to Darwinian evolution on generational timescales, but also to physiological processes on much shorter timescales. In such cases, biological function can be verified once reached rather than micromanaged at every step of a trajectory \cite{Ravasio2026b}, a principle also alive in algorithmic complexity theory \cite{winfree2019chemical}. 
This view casts the variability in molecular and organismal activity pervading biology as a potential computational resource, and the energy it consumes not as a cost, but rather as a mechanism.

The biological diversity, architectural complexity, and new ingredients of these exploratory dynamics demand fresh quantitative thinking. In this work, we explored how common mathematics can unify a wide variety of biological settings steered by such selective, not instructive, rules. Using the typical time until success as a central measure, a common mathematical protocol describes how biochemical turnovers or resets rescue (or retard) the success of biological searching. Counterintuitively, these results explain (with fresh transparency) how continually erasing progress in a search, even when that progress is nearing triumph, can sometimes be of crucial help on average. This analysis is motivated by myriad examples where biological components are dynamically created and destroyed or subjected to aborted trajectories, from dynamic microtubule instability to kinetic proofreading to misfolded protein degradation. We also explored how the presence of local cues or drift velocities in the seeker's motion accelerates search and introduces new physics to the (im)prudence of resets.
Next, we showed how checkpoint mechanisms (allowing retention of partial progress) can reduce the average time to success from exponential in the distance to a target to subexponential or even linear in this distance (see Table~\ref{tab:ScalingSummary}). These sundry results are usefully organized by the P{\'e}clet number, which partitions at least three surprising regimes in searcher behavior, as explored by Fig.~\ref{fig:CueOptimization}. For instance, counterintuitively, our modeling reveals the surprise that faster resetting is more prudent than the optimal resetting rate without such cue. These results inspire provocative questions and beg much future work. For instance, do biological systems guided by cues bear sensitivity to such rich physical regimes and their transitions? 

We tackled how mathematics for exploration smoothly generalizes to the case of multiple exploratory searchers or multiple required targets. We showed how intraflagellar transport gives a surprising manifestation of search under reset to shape biochemicals in space, focusing in that case on the concentration field created by multiple walkers.
Such searches that are only permitted to explore within a typical reset or turnover time before restarting reflect a specific type of conditioning on ventured trajectories. In fact, broader types of macroscopic conditioning can also impose microscopic order on trajectories. 
To explore this more general idea, we first investigated trajectories conditioned to start at one position and end at another at a specified time, yielding a Brownian bridge. We explored how this concept enlarges to describe the ensemble of trajectories that end at a specified position specifically for the first time at a particular end time.  Armed with these continuous random walk results, we showed how these concepts can be generalized to random walks on graphs conditioned on starting and ending states at particular times, thus generating what we might call a Schr{\"o}dinger bridge on a graph. These graph bridges enjoy informational interpretations that quantify how much conditioning on a final state narrows an ensemble of realizable trajectories. Collectively, these results articulate how simple macroscopic conditioning can convert simple search dynamics bearing rich signatures of nonlocal forces compelled by compatibility with future conditions).

A vast and lovely primary literature exists on many of the themes we explore in this work. We enthusiastically refer the reader to sources such as \cite{evans2011, RoldanGrill2016, Evans2020a, eliazar2023entropy, pal2019first, PhysRevE.111.034114} on stochastic reset, \cite{Hachmo2023} on conditional probability, and \cite{tang2026foundations} for an excellent discussion of Schr{\"o}dinger bridges and their ilk. Throughout this paper, our intention was to be concrete, careful, expository, and intuitive complements to these formative works. To our knowledge, we also obtain several new results, including surprises about optimal reset policies under a variety of cue characteristics and explicit constrained random walk propagators on small concrete graphs. 

Some useful mathematical procedures recur throughout such analyses. A common mathematical protocol begins with the Markovian property of a search trajectory, which leads to a recurrence relation for its typical time to first reach a target state (Fig.~\ref{fig:explainSimpleHitTimeRecurse}). A continuum approximation turns this recurrence into a differential equation equipped with the boundary conditions appropriate to the problem. Its analytical, asymptotic, or numerical solutions then reveal how biophysical control knobs such as reset, catastrophic rescue, partial progress saving, and cues shape the typical time to biological success.

To accommodate biological complications such as multiple searchers or targets, or checkpoints, this protocol can be complemented by quantities such as the survival probability that an individual searcher trajectory has not yet succeeded by some time. For independent searchers, this immediately gives the probability that at least one has succeeded and hence the first success time of the group. 

Precise descriptions and predictions about the typical timing of successful exploratory trajectories, like central results in this work, should give empirical leverage to interpret experiments. For instance, consider the magnificent case study of synaptogenesis. How can only a few tens of thousands ($\text{few}\times 10^4$) of human genes give rise to almost one hundred billion ($\approx \text{9}\times 10^{10}$) neurons, which form up to $10^{15}$ synaptic connections? The specificity of such connections is a staggering puzzle: in neurodevelopment, growing neurons (microns in width) can crawl past extraordinary reaches of tissue---up to tens of centimeters---to find a particular sensorimotor synaptic partner cell \cite{gerhart2007}. One key to this extraordinary wiring or healing of nervous connections is a search and stabilize process, paved by specific chemical cues and stabilizing feedbacks. This occurs along the ideas of a so-called \emph{chemoaffinity hypothesis}, articulated by Roger Sperry in the 1940's and elaborated by many modern investigators. In a series of literally incisive experiments, Sperry and colleagues cut nerves attaching to the eye of fish and reconnected them in scrambled orientations \cite{sperry1963chemoaffinity}. They found that neural cells grew to reconnect to their original partner tissue cells and regions, ignoring wide swaths of possible alternative synaptic partners along the way \cite{meyer1998roger, sperry1963chemoaffinity}. The fact that not only particular synaptic destinies, but the paths by which optic fibers reconnected to their partner regions, was so stereotyped implicates a strong, putatively chemically specified encoding of many of these connections. Subsequent mechanistic work, including molecular-resolution imaging of neuronal connectivity over time, has sharpened this picture \cite{li2021cellular,lyu2025dimensionality, Lyu2025.03.01.640986, xu2024molecular}. Yet, the central exploratory character of cellular attachment trajectories, stabilized by suitable and impressively specific tissue cues, is ratified by many sources of evidence. Throughout this intellectual history, mathematics has served important guiding roles in sharpening empirical questions about the details of the establishment of such neural maps \cite{goodhill2007contributions}. However, importantly, most such models adjudicate possible mechanisms purely on the basis of the feasible structures of the ultimate input-output maps between potential neuronal partners, and their compatibility with wild-type and perturbation experiments. We propose---in light of the type of statistical mechanics of trajectories this present paper gives simple examples of---that candidate models might also be discriminated on additional empirical bases. Both the typical trajectory \emph{dynamics} and \emph{timescales} in setting up such developmental maps should vary dramatically among proposed mechanisms, and thus could be high-leverage tools to help distinguish between otherwise competitively plausible biological universes. We speculate that careful framing of theoretical knobs powering exploratory dynamics, and temporal experimental statistics in developmental wiring, can be a rewarding further direction.

Our analyses of simplified biological searchers often probed the optimal resetting, checkpoint, multiplicity, or cue strategy to speed up the typical time to biological success. The presence of optimal routines under some conditions raises the tantalizing question about whether various biological systems might have tuned their presiding parameters to pursue the vicinity of such optima, as in a longer philosophical tradition of biological thinking \cite{mora2011biological, perez2009structure, loeb2012optimal}. However, both naturalism and mathematics encourage caution and humility with prejudice in pursuing this question. The objective function that organisms follow in their searches is not straightforwardly only the expected first time to success, but very plausibly includes other dynamical features of trajectories. For instance, many (though not all!) of the disparate biological mechanisms that accomplish resets or effective turnovers are dissipative in character---a famous and very literal example being the hydrolysis of GTP driving the dynamic collapse of microtubules \cite{gerhart2007}. Clearly, the prudence of a search strategy might attend to both time and energy. Even armed with better understanding, living organisms probably do not care so monotonically or unidimensionally about speed or energy cost as computer science paradigms might formulate most emphatically. The multivariate objective functions truly attending different organisms, including perhaps preventing that ultimate adsorbing boundary condition of death, merit respect and future inquiry.

This said, the sweep of parametric variation revealed by theoretical predictions is valuable not only to inquire about point optima per se. Even if biological systems are more guided by a mechanics of tinkering \cite{jacob1977evolution}, and not necessarily optimal in a global sense,  mechanisms like reset and ingenious conditionality can powerfully facilitate solutions that are ``good enough,'' or enable new control knobs for their regulation.

Studies of exploratory dynamics invite many urgent conceptual, mathematical, biological, and experimental extensions. Our analyses of random walk under drift or with exponential cues factorized away a crucial layer of physics. Specifically, these descriptions baldly coarse-grained the means by which chemical feedbacks are literally transduced into differential motion towards a target. That transduction encompasses the magnificent area of biophysics we call chemotaxis, clearly with hugely rich behaviors. An integrated synthesis of both organismal sensitivities to cues, and the downstream search processes we explore, is surely required to understand the whole space of functionality relevant to organisms.

A related tradition asks how organisms might handle only sporadic information about targets, as opposed to continuous gradients about where targets are to be found relative to a seeker's current position. One beautiful entry in this tradition is the suggestion that simple ``infotaxis'' algorithms can confront the inherently sporadic information delivered by e.g. turbulent transport of biochemical bearings. This situates an interesting conceptual spectrum in connection with our work. At one end, an informative continuous cue is more or less always available to a biological agent---this is the domain of classical bacterial chemotaxis studies. At the other end, no cue is available to a biological searcher at all locally \emph{until} they arrive at the terminal condition. This setting is perhaps well typified by the capture of kinetochores by microtubules, or conceivably by the establishment of Ran protein patterns \cite{oh2016spatial}; relevant motions in these systems execute totally blind search until a terminal condition, where simple gradient descent or infotaxis would fail. In the middle lie situations where cues are available, but weakly or sporadically, and strategies such as infotactic search should gain purchase and perhaps explanatory relevance. We suspect dissecting these distinctions and their serious consequences will be a rewarding program in further work.

Interestingly, different biological contexts may encode the parameters and algorithmic structure of search processes at different scales. For example, a bacterium's pursuit of sugar molecules might be largely steered by its internal molecular architecture. Yet the way a bee pursues a fragrant flower is not only sculpted by its own decisions, but actively architected by the flower as well: both searcher and target are under strong, and deeply reciprocal, selection. A mathematics which better accommodates the duality of both searcher and target adaptiveness is a rich further direction.

We showed how multiple independent, statistically-identical, searchers can accelerate various search processes. Biology probably exploits parallel searchers in yet stronger and deeper ways. For example, experiments tracking the migration of T cells in zebrafish,
 exploratory search dynamics that support responses to antigens and infections, report that T cells show a wide spectrum of speeds, directional persistence, and turning rates \cite{jerison2020heterogeneous}.One hypothesis is that this heterogeneity is crucial for pursuing targets at many characteristic distances in disordered positions. Another  conceptual tradition points out that when step sizes of individual random walks vary widely say according to a power law, in a so called L{\'e}vy flight), search can be much more efficient, as studied extensively in ecology, including birds hunting prey and sporadic food \cite{viswanathan2011physics}. How resets, checkpoints, and other architectural decorations to search might complement these effects will be a rich future direction. %

Many of our calculations idealize the triumphant cessation of a biological trajectory by the first arrival to a special target state of biological interest. This is a reasonable abstraction for biological events in which downstream activity is much faster than the search itself, but is not a universally appropriate description. For instance, in the earlier example of synaptogenesis, many neurons initially attach to some tentative partner cells, then execute competition and feedbacked effects to accomplish the ultimate pruning to singular partnership \cite{gerhart2007, gerhart1997cells}. Thus, at least some of the functional magic of exploratory dynamics lies not only in variation and selection, but also the third force of \emph{competition}, or \emph{feedback}, or precise executions of \emph{imperfect} selection. Describing the mathematics of such graded activations in some unified way is a worthy future enterprise. 

Our last examples show how macroscopic conditioning is a much more general organizing concept than just restricting random walk excursions to be at most a certain typical length (as in resetting). Brownian bridges and Schr{\"o}dinger bridges give exact results for how distributions over trajectories are constrained by modest macroscopic criteria. They show sometimes strange and counterintuitive behaviors, including nonmonotonic changes in trajectory variances over time and nonlocal forcing from compatibility with their future boundary conditions, as in fictitious forces in noninertial reference frames in classical mechanics. One way to view their behaviors is as precise \emph{epistemic} statements, namely how just knowing one thing about a macroscopic constraint gives you further gifts of microscopic knowledge about the character of trajectories. This is a view close in spirit to Jaynes's union of statistical mechanics and information theory \cite{jaynes1957information, Jaynes2003ProbabilityTheory}, and allows the collapse of trajectory diversity under conditioning to be quantified directly in informational terms. Whether or how biological systems execute conditionality of the sorts communicated by these abstractions more literally and mechanistically---for example, constraining trajectories of cell states in between pinned time endpoints in a cell cycle is an exciting frontier for future work. In a separate domain, very recent and stimulating work shows that the power of conditionality is particularly vivid in examples of competing self-replicating populations \cite{Ravasio2026b} and error correcting polymer duplication mechanisms \cite{ravasio2026evolution}. These settings invoke rich new physics, such as any functional benefits in stereotyped trajectories; consequences when the number of searchers change in time; and how time delay penalties might correlate with errors. 

Finally, we note that another dynamic infusing exploratory biology awaits elegant and unifying mathematical exposition: time-dependent protocols or aging effects. Take the charismatic example of a phalanx of hundreds of thousands of ants, spreading out in massively parallel search of food. At first, no pheromones significantly organize or complicate the search, and the exploratory statistics are close to a random walk with resets to a home base. When enough trails have occurred, however, individual ants respond to a chemical patina of history. After a food source has been found and a feedback accelerates a highway of entrainment towards that source, the search dynamic is close to random seeking with cues. Eventually, even these cues dissipate. These typify a sort of epochal transition in the character of the search problem, its environmental inputs, and its transition rates. Smoothly and tractably accommodating the many feedbacks and hysteresis truly attending such dynamics will be a science question minting discoveries for long to come.

\section{Acknowledgements}
We dedicate this work to the contributions and example of Prof. Rudi Podgornik. We are grateful to the NIH for support through award numbers DP1OD000217 (Director's Pioneer Award) and NIH MIRA 1R35 GM118043-01 to RP.  
MK acknowledges support from the NIH under grants R35-GM145248 and R01 AG07334.
JK was supported by the Simons Foundation and NSF grants DMR-1610737 and MRSEC DMR-2011846. 
GLS thanks Riccardo Ravasio, Dongyang Li, Fangzhou Xiao, and John Marken for stimulating discussions. GLS gratefully acknowledges the NSF-Simons National Institute for Theory and Mathematics in Biology (NITMB), via a Fellowship (to GLS) and funding (to GLS and MM) supported via grants from the NSF (DMS-2235451) and Simons Foundation (MP-TMPS-00005320). SDM thanks Nishanth Chavourkar, Philip-David Medows, Amin Tajik, Heun Jin Lee, Sujit Datta, Matt Thomson, and James Wang for helpful and stimulating discussions. We are deeply grateful to the CZI Theory Institute Without Walls for the support that made our initial exploratory studies of exploratory dynamics possible. We thank Phil Nelson, Arvind Murugan, Itai Yanai, Erik Winfree, Xingbo Yang, Peter Foster, Tim Mitchison, Natalia Orlovsky, Lena Koslover, Yuhai Tu, Milo Lin, Ariel Amir, and Ana Duarte for further questions and insights.

\section*{AUTHOR DECLARATIONS}

\subsection*{Conflict of Interest}

The authors have no conflicts to disclose.
\subsection*{Author Contributions}

Sara D. Mahdavi: Formal analysis; Methodology; Software; Visualization; Writing -- original draft; Writing -- review \& editing. Gabriel L. Salmon: Formal analysis; Methodology; Software; Visualization; Writing -- original draft; Writing -- review \& editing. Minakshi Ashok: Formal analysis; Methodology; Software; Visualization; Writing -- original draft; Writing -- review \& editing.
Madhav Mani: Supervision; Writing -- review \& editing. Marc Kirschner: Conceptualization; Supervision. Jane Kondev:  Conceptualization; Formal analysis; Methodology; Supervision;
Writing -- review \& editing. Rob Phillips: Conceptualization; Formal analysis; Methodology; Software; Supervision; Visualization; Writing -- original draft; Writing -- review \& editing.

\section*{DATA AVAILABILITY}
The data that support the findings of this study will be openly available at \url{https://github.com/RPGroup-PBoC/2026_math_exploratory_dynamics}.




\section*{References}
\bibliography{ref}

@article{chandrasekhar1943stochastic,
  title={Stochastic problems in physics and astronomy},
  author={Chandrasekhar, Subrahmanyan},
  journal={Reviews of Modern Physics},
  volume={15},
  number={1},
  pages={1},
  year={1943},
  publisher={APS}
}

@article{levitan1991kinetics,
  title={Kinetics of fertilization in the sea urchin Strongylocentrotus franciscanus: interaction of gamete dilution, age, and contact time},
  author={Levitan, Don R and Sewell, Mary A and Chia, Fu-Shiang},
  journal={The Biological Bulletin},
  volume={181},
  number={3},
  pages={371--378},
  year={1991},
  publisher={Marine Biological Laboratory}
}

@article{levitan1993importance,
  title={The importance of sperm limitation to the evolution of egg size in marine invertebrates},
  author={Levitan, Don R},
  journal={The American Naturalist},
  volume={141},
  number={4},
  pages={517--536},
  year={1993},
  publisher={University of Chicago Press}
}

@article{Podgornik1989,
  author        = {Podgornik, R. and Rau, D. C. and Parsegian, V. A.},
  title         = {The action of interhelical forces on the organization of {DNA} double helices: Fluctuation-enhanced decay of electrostatic double-layer and hydration forces},
  journal       = {Macromolecules},
  volume        = {22},
  pages         = {1780--1786},
  year          = {1989}
}

@article{Podgornik1995,
  author        = {Podgornik, R. and Strey, H. H. and Rau, D. C. and Parsegian, V. A.},
  title         = {Watching molecules crowd: {DNA} double helices under osmotic stress},
  journal       = {Biophys. Chem.},
  volume        = {57},
  pages         = {111--121},
  year          = {1995}
}

@article{Strey1999,
  author        = {Strey, H. H. and Parsegian, V. A. and Podgornik, R.},
  title         = {Equation of state for polymer liquid crystals: Theory and experiment},
  journal       = {Phys. Rev. E},
  volume        = {59},
  pages         = {999--1008},
  year          = {1999}
}

@article{Zandi2020,
  author        = {Zandi, R. and Dragnea, B. and Travesset, A. and Podgornik, R.},
  title         = {On virus growth and form},
  journal       = {Phys. Rep.},
  volume        = {847},
  pages         = {1--102},
  year          = {2020},
  doi           = {10.1016/j.physrep.2019.12.005}
}

@article{Peterlin1953,
  author        = {Peterlin, A.},
  title         = {Light scattering by very stiff chain molecules},
  journal       = {Nature},
  volume        = {171},
  pages         = {259--260},
  year          = {1953}
}

@book{newton1687,
  author        = {Newton, I.},
  title         = {Philosophiae Naturalis Principia Mathematica},
  publisher     = {Royal Society},
  address       = {London},
  year          = {1687}
}

@book{Chandrasekhar1995,
  author        = {Chandrasekhar, S.},
  title         = {{Newton's Principia for the Common Reader}},
  publisher     = {Oxford University Press},
  address       = {Oxford, UK},
  year          = {1995}
}

@book{Pask2013,
  author        = {Pask, C.},
  title         = {{Magnificent Principia: Exploring Isaac Newton's Masterpiece}},
  publisher     = {Prometheus Books},
  address       = {Amherst, NY},
  year          = {2013}
}

@book{arnold1978,
  author        = {Arnold, V. I.},
  title         = {Ordinary Differential Equations},
  publisher     = {MIT Press},
  address       = {Cambridge, MA},
  year          = {1978}
}

@book{lin1988,
  author        = {Lin, C. C. and Segel, L. A.},
  title         = {Mathematics Applied to Deterministic Problems in the Natural Sciences},
  publisher     = {Society for Industrial and Applied Mathematics},
  address       = {Philadelphia, PA},
  year          = {1988},
  doi           = {10.1137/1.9781611971347}
}

@book{strogatz2015,
  author        = {Strogatz, S. H.},
  title         = {Nonlinear Dynamics and Chaos: With Applications to Physics, Biology, Chemistry, and Engineering},
  edition       = {2nd},
  publisher     = {Westview Press},
  address       = {Boulder, CO},
  year          = {2015}
}

@article{Bennison2018,
  author        = {Bennison, A. and Bearhop, S. and Bodey, T. W. and Votier, S. C. and Grecian, W. J. and Wakefield, E. D. and Hamer, K. C. and Jessopp, M.},
  title         = {Search and foraging behaviors from movement data: A comparison of methods},
  journal       = {Ecol. Evol.},
  volume        = {8},
  number        = {1},
  pages         = {13--24},
  year          = {2018},
  doi           = {10.1002/ece3.3593}
}

@article{victora2012,
  author        = {Victora, G. D. and Nussenzweig, M. C.},
  title         = {Germinal centers},
  journal       = {Annu. Rev. Immunol.},
  volume        = {30},
  pages         = {429--457},
  year          = {2012},
  doi           = {10.1146/annurev-immunol-020711-075032}
}

@article{mitchison1984,
  author        = {Mitchison, T. and Kirschner, M.},
  title         = {Dynamic instability of microtubule growth},
  journal       = {Nature},
  volume        = {312},
  number        = {5991},
  pages         = {237--242},
  year          = {1984},
  doi           = {10.1038/312237a0}
}

@article{lowery2009,
  author        = {Lowery, L. A. and Van Vactor, D.},
  title         = {The trip of the tip: understanding the growth cone machinery},
  journal       = {Nat. Rev. Mol. Cell Biol.},
  volume        = {10},
  number        = {5},
  pages         = {332--343},
  year          = {2009},
  doi           = {10.1038/nrm2679}
}

@article{gerhardt2003,
  author        = {Gerhardt, H. and Golding, M. and Fruttiger, M. and Ruhrberg, C. and Lundkvist, A. and Abramsson, A. and Jeltsch, M. and Mitchell, C. and Alitalo, K. and Shima, D. and Betsholtz, C.},
  title         = {{VEGF} guides angiogenic sprouting utilizing endothelial tip cell filopodia},
  journal       = {J. Cell Biol.},
  volume        = {161},
  number        = {6},
  pages         = {1163--1177},
  year          = {2003}
}

@article{Martins2011,
  author        = {Martins, B. M. C. and Swain, P. S.},
  title         = {Trade-Offs and Constraints in Allosteric Sensing},
  journal       = {PLoS Comput. Biol.},
  volume        = {7},
  number        = {11},
  pages         = {e1002261},
  year          = {2011},
  doi           = {10.1371/journal.pcbi.1002261}
}

@book{Phillips2020,
  author        = {R. Phillips},
  title         = {{The Molecular Switch, Allostery and Signaling}},
  publisher     = {Princeton University Press},
  address       = {Princeton},
  year          = {2020},
  note          = {(Illustrated by N. Orme)}
}

@incollection{monod1974chance,
  author        = {Monod, Jacques},
  title         = {On chance and necessity},
  booktitle     = {Studies in the Philosophy of Biology: Reduction and Related Problems},
  publisher     = {Springer},
  pages         = {357--375},
  year          = {1974}
}

@article{Jacob1961,
  author        = {Jacob, F. and Monod, J.},
  title         = {Genetic regulatory mechanisms in the synthesis of proteins},
  journal       = {J. Mol. Biol.},
  volume        = {3},
  pages         = {318--356},
  year          = {1961},
  doi           = {10.1016/S0022-2836(61)80072-7}
}

@article{Monod1963,
  author        = {J. Monod and J. P. Changeux and F. Jacob},
  title         = {Allosteric proteins and cellular control systems},
  journal       = {J. Mol. Biol.},
  volume        = {6},
  pages         = {306--329},
  year          = {1963}
}

@article{Monod1965,
  author        = {Monod, J. and Wyman, J. and Changeux, J. P.},
  title         = {On the Nature of Allosteric Transitions: A Plausible Model},
  journal       = {J. Mol. Biol.},
  volume        = {12},
  pages         = {88--118},
  year          = {1965}
}

@book{Muller-Hill1996,
  author        = {M\"{u}ller-Hill, B.},
  title         = {{The lac Operon: a short history of a genetic paradigm}},
  publisher     = {Walter de Gruyter},
  address       = {Berlin, New York},
  year          = {1996}
}

@article{Ullmann2011,
  author        = {Ullmann, A.},
  title         = {In Memoriam: Jacques Monod (1910-1976)},
  journal       = {Genome Biol. Evol.},
  volume        = {3},
  pages         = {1025--1033},
  year          = {2011}
}

@article{pais1979einstein,
  author        = {Pais, Abraham},
  title         = {Einstein and the quantum theory},
  journal       = {Rev. Mod. Phys.},
  volume        = {51},
  number        = {4},
  pages         = {863},
  year          = {1979}
}

@book{Rogers2005,
  author        = {Rogers, Donald W.},
  title         = {{Einstein's Other Theory: The Planck-Bose-Einstein Theory of Heat Capacity}},
  publisher     = {Princeton University Press},
  address       = {Princeton, NJ},
  year          = {2005},
  isbn          = {9780691118260}
}

@book{Rigden2005,
  author        = {Rigden, John S.},
  title         = {{Einstein 1905: The Standard of Greatness}},
  publisher     = {Harvard University Press},
  address       = {Cambridge, MA},
  year          = {2005},
  isbn          = {9780674015449}
}

@book{Rigden2002,
  author        = {J. S. Rigden},
  title         = {{Hydrogen The Essential Element}},
  publisher     = {Harvard University Press},
  address       = {Cambridge, Mass},
  year          = {2002}
}

@book{waddington1973or,
  author        = {Waddington, C. H.},
  title         = {{O.R. in World War 2}: Operational Research Against the {U-Boat}},
  series        = {Histories of Science Series},
  publisher     = {Elek Science},
  address       = {London},
  year          = {1973},
  isbn          = {978-0-236-15463-0}
}

@article{Holy1994,
  author        = {Holy, T. E. and Leibler, S.},
  title         = {Dynamic instability of microtubules as an efficient way to search in space},
  journal       = {Proc. Natl. Acad. Sci. U.S.A.},
  volume        = {91},
  number        = {12},
  pages         = {5682--5685},
  year          = {1994}
}

@article{Wollman2005,
  author        = {Wollman, R. and Cytrynbaum, E. N. and Jones, J. T. and Meyer, T. and Scholey, J. M. and Mogilner, A.},
  title         = {Efficient chromosome capture requires a bias in the `search-and-capture' process during mitotic-spindle assembly},
  journal       = {Curr. Biol.},
  volume        = {15},
  number        = {9},
  pages         = {828--832},
  year          = {2005}
}

@article{kondev2025,
  author        = {Kondev, J. and Kirschner, M. and Garcia, H. G. and Salmon, G. L. and Phillips, R.},
  title         = {Biological processes as exploratory dynamics},
  journal       = {Biophys. J.},
  volume        = {125},
  number        = {12},
  pages         = {2812--2837},
  year          = {2025},
  doi           = {10.1016/j.bpj.2025.09.009}
}

@article{Hachmo2023,
  author        = {Hachmo, O. and Amir, A.},
  title         = {Conditional probability as found in nature: Facilitated diffusion},
  journal       = {Am. J. Phys.},
  volume        = {91},
  number        = {8},
  pages         = {653--658},
  year          = {2023}
}

@article{mitchison1985,
  author        = {Mitchison, T. J. and Kirschner, M. W.},
  title         = {Properties of the kinetochore in vitro. {II}. Microtubule capture and {ATP}-dependent translocation},
  journal       = {J. Cell Biol.},
  volume        = {101},
  number        = {3},
  pages         = {766--777},
  year          = {1985}
}

@article{Gundersen2002,
  author        = {Gundersen, G. G.},
  title         = {Evolutionary conservation of microtubule-capture mechanisms},
  journal       = {Nat. Rev. Mol. Cell Biol.},
  volume        = {3},
  number        = {4},
  pages         = {296--304},
  year          = {2002}
}

@article{Heald2015,
  author        = {Heald, R. and Khodjakov, A.},
  title         = {Thirty Years of Search and Capture: The Complex Simplicity of Mitotic Spindle Assembly},
  journal       = {J. Cell Biol.},
  volume        = {211},
  number        = {6},
  pages         = {1103--1111},
  year          = {2015},
  doi           = {10.1083/jcb.201510015}
}

@incollection{vonHippel1979,
  author        = {von Hippel, P. H.},
  title         = {On the molecular bases of the specificity of interaction of transcriptional proteins with genome {DNA}},
  booktitle     = {{Biological Regulation and Development: Gene Expression}},
  publisher     = {Springer US},
  address       = {Boston, MA},
  pages         = {279--347},
  year          = {1979},
  doi           = {10.1007/978-1-4684-3417-0_8},
  type          = {Book Section}
}

@article{Winter1981a,
  author        = {Winter, R. B. and von Hippel, P. H.},
  title         = {Diffusion-driven mechanisms of protein translocation on nucleic acids. 2. {The} {\it {Escherichia} coli} repressor--operator interaction: equilibrium measurements},
  journal       = {Biochemistry},
  volume        = {20},
  number        = {24},
  pages         = {6948--6960},
  year          = {1981}
}

@article{Winter1981b,
  author        = {Winter, R. B. and Berg, O. G. and von Hippel, P. H.},
  title         = {Diffusion-driven mechanisms of protein translocation on nucleic acids. 3. {The} {\it {Escherichia} coli lac} repressor--operator interaction: kinetic measurements and conclusions},
  journal       = {Biochemistry},
  volume        = {20},
  number        = {24},
  pages         = {6961--6977},
  year          = {1981}
}

@article{Hu2006,
  author        = {T. Hu and A. Y. Grosberg and B. I. Shklovskii},
  title         = {How proteins search for their specific sites on {DNA}: the role of {DNA} conformation},
  journal       = {Biophys. J.},
  volume        = {90},
  number        = {8},
  pages         = {2731--2744},
  year          = {2006}
}

@book{Blitzstein2019,
  author        = {Blitzstein, J. K. and Hwang, J.},
  title         = {{Introduction to Probability}},
  publisher     = {Chapman and Hall/CRC Press},
  address       = {Boca Raton, FL},
  year          = {2019}
}

@article{evans2011,
  author        = {Evans, M. R. and Majumdar, S. N.},
  title         = {Diffusion with stochastic resetting},
  journal       = {Phys. Rev. Lett.},
  volume        = {106},
  number        = {16},
  pages         = {160601},
  year          = {2011},
  doi           = {10.1103/PhysRevLett.106.160601}
}

@article{Reuveni2016,
  author        = {Reuveni, S.},
  title         = {Optimal stochastic restart renders fluctuations in first passage times universal},
  journal       = {Phys. Rev. Lett.},
  volume        = {116},
  number        = {17},
  pages         = {170601},
  year          = {2016},
  doi           = {10.1103/PhysRevLett.116.170601}
}

@article{Evans2020a,
  author        = {Evans, M. R. and Majumdar, S. N. and Schehr, G.},
  title         = {Stochastic resetting and applications},
  journal       = {J. Phys. A: Math. Theor.},
  volume        = {53},
  number        = {19},
  pages         = {193001},
  year          = {2020},
  doi           = {10.1088/1751-8121/ab7cfe}
}

@article{Bressloff2020a,
  author        = {Bressloff, P. C.},
  title         = {Search processes with stochastic resetting and multiple targets},
  journal       = {Phys. Rev. E},
  volume        = {102},
  number        = {2},
  pages         = {022115},
  year          = {2020},
  doi           = {10.1103/PhysRevE.102.022115}
}

@article{Bressloff2020b,
  author        = {Bressloff, P. C.},
  title         = {Queueing theory of search processes with stochastic resetting},
  journal       = {Phys. Rev. E},
  volume        = {102},
  number        = {3},
  pages         = {032109},
  year          = {2020},
  doi           = {10.1103/PhysRevE.102.032109}
}

@book{Norris1998,
  author        = {Norris, J. R.},
  title         = {{Markov Chains}},
  publisher     = {Cambridge University Press},
  address       = {Cambridge, UK},
  year          = {1998}
}

@article{dynkin1989kolmogorov,
  author        = {Dynkin, E.B.},
  title         = {{Kolmogorov} and the theory of Markov processes},
  journal       = {Ann. Probab.},
  volume        = {17},
  number        = {3},
  pages         = {822--832},
  year          = {1989}
}

@book{kolmogorov2019selected,
  author        = {Kolmogorov, Andrei N and Shiryaev, Albert N},
  title         = {Selected works II: probability theory and mathematical statistics},
  publisher     = {Springer},
  year          = {2019}
}

@article{HUANG2024129389,
  author        = {Feng Huang and Hanshuang Chen},
  title         = {Extremal statistics for a one-dimensional {Brownian} motion with a reflective boundary},
  journal       = {Physica A},
  volume        = {633},
  pages         = {129389},
  year          = {2024},
  doi           = {10.1016/j.physa.2023.129389}
}

@article{chechkin2018random,
  author        = {Chechkin, Aleksei and Sokolov, Igor M},
  title         = {Random search with resetting: a unified renewal approach},
  journal       = {Phys. Rev. Lett.},
  volume        = {121},
  number        = {5},
  pages         = {050601},
  year          = {2018}
}

@article{evans2025stochastic,
  author        = {Evans, Martin R and Ray, Somrita},
  title         = {Stochastic resetting prevails over sharp restart for broad target distributions},
  journal       = {Phys. Rev. Lett.},
  volume        = {134},
  number        = {24},
  pages         = {247102},
  year          = {2025}
}

@article{naoz2008protein,
  author        = {Naoz, M. and
Manor, U. and
Sakaguchi, H. and
Kachar, B. and
Gov, N. S.},
  title         = {Protein localization by actin treadmilling and molecular motors regulates stereocilia shape and treadmilling rate},
  journal       = {Biophys. J.},
  volume        = {95},
  number        = {12},
  pages         = {5706--5718},
  year          = {2008}
}

@article{van2020ciliary,
  author        = {van der Burght, S. N. and
Rademakers, S. and
Johnson, J.-L. and
Li, C. and
Kremers, G.-J. and
Houtsmuller, A. B. and
Leroux, M. R. and
Jansen, G.},
  title         = {Ciliary tip signaling compartment is formed and maintained by intraflagellar transport},
  journal       = {Curr. Biol.},
  volume        = {30},
  number        = {21},
  pages         = {4299--4306},
  year          = {2020}
}

@article{dawson2010life,
  author        = {Dawson, Scott C and House, Susan A},
  title         = {Life with eight flagella: flagellar assembly and division in {Giardia}},
  journal       = {Curr. Opin. Microbiol.},
  volume        = {13},
  number        = {4},
  pages         = {480--490},
  year          = {2010}
}

@article{ott2025design,
  author        = {Ott, Carolyn M and Lippincott-Schwartz, Jennifer},
  title         = {Design principles of ciliary signaling},
  journal       = {J. Cell Sci.},
  volume        = {138},
  number        = {20},
  pages         = {jcs264325},
  year          = {2025}
}

@article{McInally2019,
  author        = {McInally, S. G. and Kondev, J. and Dawson, S. C.},
  title         = {Length-dependent disassembly maintains four different flagellar lengths in {Giardia}},
  journal       = {eLife},
  volume        = {8},
  pages         = {e48694},
  year          = {2019},
  doi           = {10.7554/eLife.48694}
}

@article{datta2022assemble,
  author        = {Datta, Arnab and Ghosh, Sagnik and Kondev, Jane},
  title         = {How to assemble a scale-invariant gradient},
  journal       = {eLife},
  volume        = {11},
  pages         = {e71365},
  year          = {2022}
}

@article{chien2017dynamics,
  author        = {Chien, Alexander and Shih, Sheng Min and Bower, Raqual and Tritschler, Douglas and Porter, Mary E and Yildiz, Ahmet},
  title         = {Dynamics of the {IFT} machinery at the ciliary tip},
  journal       = {eLife},
  volume        = {6},
  pages         = {e28606},
  year          = {2017}
}

@book{alberts2008molecular,
  author        = {Alberts, Bruce and Johnson, Alexander and Lewis, Julian and
               Raff, Martin and Roberts, Keith and Walter, Peter},
  title         = {Molecular Biology of the Cell},
  edition       = {5},
  publisher     = {Garland Science},
  address       = {New York},
  year          = {2008},
  isbn          = {978-0-8153-4105-5}
}

@article{kelch2015organ,
  author        = {Kelch, Inken D and Bogle, Gib and Sands, Gregory B and Phillips, Anthony RJ and LeGrice, Ian J and Rod Dunbar, P},
  title         = {Organ-wide {3D}-imaging and topological analysis of the continuous microvascular network in a murine lymph node},
  journal       = {Sci. Rep.},
  volume        = {5},
  number        = {1},
  pages         = {16534},
  year          = {2015}
}

@article{Kennedy1994,
  author        = {Kennedy, T. E. and Serafini, T. and de la Torre, J. R. and Tessier-Lavigne, M.},
  title         = {Netrins are diffusible chemotropic factors for commissural axons in the embryonic spinal cord},
  journal       = {Cell},
  volume        = {78},
  number        = {3},
  pages         = {425--435},
  year          = {1994}
}

@article{Berg1972,
  author        = {Berg, H. C. and Brown, D. A.},
  title         = {Chemotaxis in {\it Escherichia coli} analysed by three-dimensional tracking},
  journal       = {Nature},
  volume        = {239},
  number        = {5374},
  pages         = {500--504},
  year          = {1972}
}

@article{Ahmad2019,
  author        = {Ahmad, S. and Nayak, I. and Bansal, A. and Nandi, A. and Das, D.},
  title         = {First passage of a particle in a potential under stochastic resetting: A vanishing transition of optimal resetting rate},
  journal       = {Phys. Rev. E},
  volume        = {99},
  number        = {2},
  pages         = {022130},
  year          = {2019}
}

@article{Ray2021,
  author        = {Ray, S. and Reuveni, S.},
  title         = {Resetting transition is governed by an interplay between thermal and potential energy},
  journal       = {J. Chem. Phys.},
  volume        = {154},
  number        = {17},
  pages         = {171103},
  year          = {2021}
}

@article{chaffiol2005,
  author        = {Chaffiol, Antoine and Laloi, David and Pham-Del{\`e}gue, Minh-H{\`a}},
  title         = {Prior classical olfactory conditioning improves odour-cued flight orientation of honey bees in a wind tunnel},
  journal       = {J. Exp. Biol.},
  volume        = {208},
  number        = {19},
  pages         = {3731--3737},
  year          = {2005}
}

@article{sprayberry2018prevalence,
  author        = {Sprayberry, Jordanna DH},
  title         = {The prevalence of olfactory-versus visual-signal encounter by searching bumblebees},
  journal       = {Sci. Rep.},
  volume        = {8},
  number        = {1},
  pages         = {14590},
  year          = {2018}
}

@article{dong2023social,
  author        = {Dong, Shihao and Lin, Tao and Nieh, James C and Tan, Ken},
  title         = {Social signal learning of the waggle dance in honey bees},
  journal       = {Science},
  volume        = {379},
  number        = {6636},
  pages         = {1015--1018},
  year          = {2023}
}

@article{spaethe2001visual,
  author        = {Spaethe, J and Tautz, J and Chittka, L},
  title         = {Visual constraints in foraging bumblebees: flower size and color affect search time and flight behavior},
  journal       = {Proc. Natl. Acad. Sci. U.S.A.},
  volume        = {98},
  number        = {7},
  pages         = {3898--3903},
  year          = {2001}
}

@article{riley2005flight,
  author        = {Riley, Joe R and Greggers, Uwe and Smith, Alan D and Reynolds, Don R and Menzel, Randolf},
  title         = {The flight paths of honeybees recruited by the waggle dance},
  journal       = {Nature},
  volume        = {435},
  number        = {7039},
  pages         = {205--207},
  year          = {2005}
}

@article{Ravasio2026b,
  author        = {Ravasio, R. and Husain, K. and Evans, C. G. and Phillips, R. and Ribezzi-Crivellari, M. and Szostak, J. W. and Murugan, A.},
  title         = {Conditioning as a route to stereotyped behavior in growing populations},
  journal       = {arXiv},
  year          = {2026},
  note          = {arXiv:2605.13009}
}

@article{Simon1992,
  author        = {Simon, S. M. and Peskin, C. S. and Oster, G. F.},
  title         = {What drives the translocation of proteins?},
  journal       = {Proc. Natl. Acad. Sci. U.S.A.},
  volume        = {89},
  number        = {9},
  pages         = {3770--3774},
  year          = {1992},
  doi           = {10.1073/pnas.89.9.3770}
}

@article{Peskin1993,
  author        = {Peskin, C. S. and Odell, G. M. and Oster, G. F.},
  title         = {Cellular motions and thermal fluctuations: the {Brownian} ratchet},
  journal       = {Biophys. J.},
  volume        = {65},
  number        = {1},
  pages         = {316--324},
  year          = {1993},
  doi           = {10.1016/S0006-3495(93)81035-X}
}

@article{Lubensky1999,
  author        = {Lubensky, D. K. and Nelson, D. R.},
  title         = {Driven Polymer Translocation Through a Narrow Pore},
  journal       = {Biophys. J.},
  volume        = {77},
  number        = {4},
  pages         = {1824--1838},
  year          = {1999},
  doi           = {10.1016/S0006-3495(99)77027-X}
}

@book{Muthukumar2011,
  author        = {Muthukumar, M.},
  title         = {{Polymer Translocation}},
  publisher     = {CRC Press},
  address       = {Boca Raton},
  year          = {2011},
  doi           = {10.1201/b10901}
}

@article{Zandi2003,
  author        = {Zandi, R. and Reguera, D. and Rudnick, J. and Gelbart, W. M.},
  title         = {What drives the translocation of stiff chains?},
  journal       = {Proc. Natl. Acad. Sci. U.S.A.},
  volume        = {100},
  number        = {15},
  pages         = {8649--8653},
  year          = {2003},
  doi           = {10.1073/pnas.1533334100}
}

@article{Evilevitch2003,
  author        = {Evilevitch, A. and Lavelle, L. and Knobler, C. M. and Raspaud, E. and Gelbart, W. M.},
  title         = {Osmotic pressure inhibition of {DNA} ejection from phage},
  journal       = {Proc. Natl. Acad. Sci. U.S.A.},
  volume        = {100},
  number        = {16},
  pages         = {9292--9295},
  year          = {2003},
  doi           = {10.1073/pnas.1233721100}
}

@article{Inamdar2006,
  author        = {Inamdar, M. M. and Gelbart, W. M. and Phillips, R.},
  title         = {Dynamics of {DNA} ejection from bacteriophage},
  journal       = {Biophys. J.},
  volume        = {91},
  number        = {2},
  pages         = {411--420},
  year          = {2006},
  doi           = {10.1529/biophysj.105.070532}
}

@article{Hepp2016,
  author        = {Hepp, C. and Maier, B.},
  title         = {Kinetics of {DNA} uptake during transformation provide evidence for a translocation ratchet mechanism},
  journal       = {Proc. Natl. Acad. Sci. U.S.A.},
  volume        = {113},
  number        = {44},
  pages         = {12467--12472},
  year          = {2016},
  doi           = {10.1073/pnas.1608110113}
}

@article{sokolov2005target,
  author        = {Sokolov, Igor M and Metzler, Ralf and Pant, Kiran and Williams, Mark C},
  title         = {Target search of {N} sliding proteins on a {DNA}},
  journal       = {Biophys. J.},
  volume        = {89},
  number        = {2},
  pages         = {895--902},
  year          = {2005}
}

@article{bornschlogl2013filopodia,
  author        = {Bornschl{\"o}gl, Thomas},
  title         = {How filopodia pull: what we know about the mechanics and dynamics of filopodia},
  journal       = {Cytoskeleton},
  volume        = {70},
  number        = {10},
  pages         = {590--603},
  year          = {2013}
}

@article{weiss1983order,
  author        = {Weiss, George H and Shuler, Kurt E and Lindenberg, Katja},
  title         = {Order statistics for first passage times in diffusion processes},
  journal       = {J. Stat. Phys.},
  volume        = {31},
  number        = {2},
  pages         = {255--278},
  year          = {1983}
}

@article{yuste2001order,
  author        = {Yuste, Santos B and Acedo, L and Lindenberg, Katja},
  title         = {Order statistics for d-dimensional diffusion processes},
  journal       = {Phys. Rev. E},
  volume        = {64},
  number        = {5},
  pages         = {052102},
  year          = {2001}
}

@article{lawley2020universal,
  author        = {Lawley, Sean D},
  title         = {Universal formula for extreme first passage statistics of diffusion},
  journal       = {Phys. Rev. E},
  volume        = {101},
  number        = {1},
  pages         = {012413},
  year          = {2020}
}

@article{pal2019local,
  author        = {Pal, Arnab and Chatterjee, Rakesh and Reuveni, Shlomi and Kundu, Anupam},
  title         = {Local time of diffusion with stochastic resetting},
  journal       = {J. Phys. A: Math. Theor.},
  volume        = {52},
  number        = {26},
  pages         = {264002},
  year          = {2019}
}

@article{yuste1996order,
  author        = {Yuste, S Bravo and Lindenberg, Katja},
  title         = {Order statistics for first passage times in one-dimensional diffusion processes},
  journal       = {J. Stat. Phys.},
  volume        = {85},
  number        = {3},
  pages         = {501--512},
  year          = {1996}
}

@article{biroli2023critical,
  author        = {Biroli, Marco and Majumdar, Satya N and Schehr, Gr{\'e}gory},
  title         = {Critical number of walkers for diffusive search processes with resetting},
  journal       = {Phys. Rev. E},
  volume        = {107},
  number        = {6},
  pages         = {064141},
  year          = {2023}
}

@article{meyer2025optimal,
  author        = {Meyer, Hugues and Rieger, Heiko},
  title         = {Optimal number of agents in a collective search and when to launch them},
  journal       = {Phys. Rev. E},
  volume        = {111},
  number        = {6},
  pages         = {064112},
  year          = {2025}
}

@article{ha2024measuring,
  author        = {Ha, Gloria and Dieterle, Paul and Shen, Hao and Amir, Ariel and Needleman, Daniel J},
  title         = {Measuring and modeling the dynamics of mitotic error correction},
  journal       = {Proc. Natl. Acad. Sci. U.S.A.},
  volume        = {121},
  number        = {25},
  pages         = {e2323009121},
  year          = {2024}
}

@article{winey1995three,
  author        = {Winey, Mark and Mamay, Cynthia L and O'Toole, Eileen T and Mastronarde, David N and Giddings Jr, Thomas H and McDonald, Kent L and McIntosh, J Richard},
  title         = {Three-dimensional ultrastructural analysis of the {Saccharomyces} cerevisiae mitotic spindle.},
  journal       = {J. Cell Biol.},
  volume        = {129},
  number        = {6},
  pages         = {1601--1615},
  year          = {1995}
}

@article{joglekar2008molecular,
  author        = {Joglekar, Ajit P and Bouck, David and Finley, Ken and Liu, Xingkun and Wan, Yakun and Berman, Judith and He, Xiangwei and Salmon, ED and Bloom, Kerry S},
  title         = {Molecular architecture of the kinetochore-microtubule attachment site is conserved between point and regional centromeres},
  journal       = {J. Cell Biol.},
  volume        = {181},
  number        = {4},
  pages         = {587--594},
  year          = {2008}
}

@article{long2019mammalian,
  author        = {Long, Alexandra F and Kuhn, Jonathan and Dumont, Sophie},
  title         = {The mammalian kinetochore--microtubule interface: robust mechanics and computation with many microtubules},
  journal       = {Curr. Opin. Cell Biol.},
  volume        = {60},
  pages         = {60--67},
  year          = {2019}
}

@article{lawley2023slowest,
  author        = {Lawley, Sean D and Johnson, Joshua},
  title         = {Slowest first passage times, redundancy, and menopause timing},
  journal       = {J. Math. Biol.},
  volume        = {86},
  number        = {6},
  pages         = {90},
  year          = {2023},
  doi           = {10.1007/s00285-023-01921-9}
}

@book{Jaynes2003ProbabilityTheory,
  author        = {Jaynes, E. T.},
  editor        = {Bretthorst, G. Larry},
  title         = {Probability Theory: The Logic of Science},
  publisher     = {Cambridge University Press},
  address       = {Cambridge},
  year          = {2003},
  isbn          = {9780521592710}
}

@book{Amir2021ThinkingProbabilistically,
  author        = {Amir, Ariel},
  title         = {Thinking Probabilistically: Stochastic Processes, Disordered Systems, and Their Applications},
  publisher     = {Cambridge University Press},
  address       = {Cambridge},
  year          = {2021}
}

@article{schrodinger1931,
  author        = {Schr{\"o}dinger, E.},
  title         = {{{\"U}ber die Umkehrung der Naturgesetze}},
  journal       = {Sitzungsberichte der Preussischen Akademie der Wissenschaften, Physikalisch-mathematische Klasse},
  pages         = {144--153},
  year          = {1931}
}

@article{Chetrite2021,
  author        = {Chetrite, R. and Muratore-Ginanneschi, P. and Schwieger, K.},
  title         = {E. {Schr{\"o}dinger}’s 1931 paper ``On the Reversal of the Laws of Nature"},
  journal       = {Eur. Phys. J. H},
  volume        = {46},
  number        = {1},
  pages         = {28},
  year          = {2021}
}

@article{Bertoin2003,
  author        = {Bertoin, J. and Chaumont, L. and Pitman, J.},
  title         = {Path transformations of first passage bridges},
  journal       = {Electron. Commun. Probab.},
  volume        = {8},
  pages         = {155--166},
  year          = {2003},
  doi           = {10.1214/ECP.v8-1096}
}

@book{Feynman2010,
  author        = {Feynman, R. P. and Hibbs, A. R. and Styer, D. F.},
  title         = {{Quantum Mechanics and Path Integrals}},
  publisher     = {Dover},
  address       = {Mineola},
  year          = {2010}
}

@book{Wiegel1986,
  author        = {Wiegel, F. W.},
  title         = {{Introduction to Path-Integral Methods in Physics and Polymer Science}},
  publisher     = {World Scientific},
  address       = {Singapore},
  year          = {1986}
}

@book{redner2001,
  author        = {Redner, S.},
  title         = {A Guide to First-Passage Processes},
  publisher     = {Cambridge University Press},
  address       = {Cambridge},
  year          = {2001},
  doi           = {10.1017/CBO9780511606014}
}

@article{Redner2022,
  author        = {Redner, S.},
  title         = {A First Look at First-Passage Processes},
  journal       = {arXiv},
  year          = {2022},
  eprint        = {2201.10048},
  archiveprefix = {arXiv},
  primaryclass  = {cond-mat.stat-mech}
}

@book{Borodin2012,
  author        = {Borodin, A. N. and Salminen, P.},
  title         = {Handbook of Brownian motion-facts and formulae},
  publisher     = {Birkh{\"a}user},
  year          = {2012}
}

@book{Pitman2006,
  author        = {Pitman, J.},
  title         = {{Combinatorial Stochastic Processes}},
  series        = {Lecture Notes in Mathematics},
  volume        = {1875},
  publisher     = {Springer},
  address       = {Berlin, Heidelberg},
  year          = {2006}
}

@article{Gunawardena2012,
  author        = {Gunawardena, J.},
  title         = {A Linear Framework for Time-Scale Separation in Nonlinear Biochemical Systems},
  journal       = {PLoS One},
  volume        = {7},
  number        = {5},
  pages         = {e36321},
  year          = {2012},
  doi           = {10.1371/journal.pone.0036321}
}

@article{Gunawardena2013,
  author        = {Mirzaev, I. and Gunawardena, J.},
  title         = {{Laplacian} Dynamics on General Graphs},
  journal       = {Bull. Math. Biol.},
  volume        = {75},
  number        = {11},
  pages         = {2118--2149},
  year          = {2013}
}

@article{Mahdavi2024,
  author        = {Mahdavi, S. D. and Salmon, G. L. and Daghlian, P. and Garcia, H. G. and Phillips, R.},
  title         = {Flexibility and sensitivity in gene regulation out of equilibrium},
  journal       = {Proc. Natl. Acad. Sci. U.S.A.},
  volume        = {121},
  number        = {46},
  pages         = {e2411395121},
  year          = {2024}
}

@article{jaynes1980minimum,
  author        = {Jaynes, Edwin T},
  title         = {The minimum entropy production principle},
  journal       = {Annu. Rev. Phys. Chem.},
  volume        = {31},
  number        = {1},
  pages         = {579--601},
  year          = {1980}
}

@article{ghosh2006teaching,
  author        = {Ghosh, Kingshuk and Dill, Ken A and Inamdar, Mandar M and Seitaridou, Effrosyni and Phillips, Rob},
  title         = {Teaching the principles of statistical dynamics},
  journal       = {Am. J. Phys.},
  volume        = {74},
  number        = {2},
  pages         = {123--133},
  year          = {2006}
}

@article{wu2009,
  author        = {Wu, David and Ghosh, Kingshuk and Inamdar, Mandar and Lee, Heun Jin and Fraser, Scott and Dill, Ken and Phillips, Rob},
  title         = {Trajectory Approach to Two-State Kinetics of Single Particles on Sculpted Energy Landscapes},
  journal       = {Phys. Rev. Lett.},
  volume        = {103},
  number        = {5},
  pages         = {050603},
  year          = {2009},
  doi           = {10.1103/PhysRevLett.103.050603}
}

@article{presse2010dynamical,
  author        = {Press{\'e}, S and Ghosh, K and Phillips, R and Dill, KA},
  title         = {Dynamical fluctuations in biochemical reactions and cycles},
  journal       = {Phys. Rev. E},
  volume        = {82},
  number        = {3},
  pages         = {031905},
  year          = {2010}
}

@article{ghosh2020maximum,
  author        = {Ghosh, Kingshuk and Dixit, Purushottam D and Agozzino, Luca and Dill, Ken A},
  title         = {The maximum caliber variational principle for nonequilibria},
  journal       = {Annu. Rev. Phys. Chem.},
  volume        = {71},
  number        = {1},
  pages         = {213--238},
  year          = {2020}
}

@article{presse2013principles,
  author        = {Press{\'e}, Steve and Ghosh, Kingshuk and Lee, Julian and Dill, Ken A},
  title         = {Principles of maximum entropy and maximum caliber in statistical physics},
  journal       = {Rev. Mod. Phys.},
  volume        = {85},
  number        = {3},
  pages         = {1115--1141},
  year          = {2013}
}

@article{dixit2018perspective,
  author        = {Dixit, Purushottam D and Wagoner, Jason and Weistuch, Corey and Press{\'e}, Steve and Ghosh, Kingshuk and Dill, Ken A},
  title         = {Perspective: Maximum caliber is a general variational principle for dynamical systems},
  journal       = {J. Chem. Phys.},
  volume        = {148},
  number        = {1},
  pages         = {010901},
  year          = {2018},
  doi           = {10.1063/1.5012990}
}

@article{Chetrite_2021,
  author        = {Schrodinger, E. and Chetrite, Raphaël and Muratore-Ginanneschi, Paolo and Schwieger, Kay},
  title         = {E. {Schr{\"o}dinger}’s 1931 paper ``On the Reversal of the Laws of Nature'' [``{\"U}ber die Umkehrung der Naturgesetze'', Sitzungsberichte der preussischen Akademie der Wissenschaften, physikalisch-mathematische Klasse, 8 N9 144–153]},
  journal       = {Eur. Phys. J. H},
  volume        = {46},
  number        = {1},
  pages         = {28},
  year          = {2021},
  doi           = {10.1140/epjh/s13129-021-00032-7}
}

@misc{leonard2013surveyschrodingerproblemconnections,
  author        = {Christian Léonard},
  title         = {A survey of the Schr\"odinger problem and some of its connections with optimal transport},
  year          = {2013},
  eprint        = {1308.0215},
  archiveprefix = {arXiv},
  primaryclass  = {math.PR},
  url           = {https://arxiv.org/abs/1308.0215}
}

@inproceedings{winfree2019chemical,
  author        = {Winfree, Erik},
  title         = {Chemical reaction networks and stochastic local search},
  booktitle     = {International Conference on DNA Computing and Molecular Programming},
  organization  = {Springer},
  pages         = {1--20},
  year          = {2019}
}

@article{RoldanGrill2016,
  author        = {Rold\'an, \'Edgar and Lisica, Ana and S\'anchez-Taltavull, Daniel and Grill, Stephan W.},
  title         = {Stochastic resetting in backtrack recovery by {RNA} polymerases},
  journal       = {Phys. Rev. E},
  volume        = {93},
  number        = {6},
  pages         = {062411},
  year          = {2016},
  doi           = {10.1103/PhysRevE.93.062411}
}

@article{eliazar2023entropy,
  author        = {Eliazar, Iddo and Reuveni, Shlomi},
  title         = {Entropy of sharp restart},
  journal       = {J. Phys. A: Math. Theor.},
  volume        = {56},
  number        = {2},
  pages         = {024002},
  year          = {2023}
}

@article{pal2019first,
  author        = {Pal, Arnab and Eliazar, Iddo and Reuveni, Shlomi},
  title         = {First passage under restart with branching},
  journal       = {Phys. Rev. Lett.},
  volume        = {122},
  number        = {2},
  pages         = {020602},
  year          = {2019}
}

@article{PhysRevE.111.034114,
  author        = {Alston, Henry and Bertrand, Thibault},
  title         = {Boosting macroscopic diffusion with local resetting},
  journal       = {Phys. Rev. E},
  volume        = {111},
  number        = {3},
  pages         = {034114},
  year          = {2025},
  doi           = {10.1103/PhysRevE.111.034114}
}

@article{tang2026foundations,
  author        = {Tang, Sophia},
  title         = {Foundations of {Schr{\"o}dinger} bridges for generative modeling},
  journal       = {arXiv},
  year          = {2026},
  note          = {arXiv:2603.18992}
}

@book{gerhart2007,
  author        = {Kirschner, M. and Gerhart, J.},
  title         = {The Plausibility of Life: Resolving {Darwin's} Dilemma},
  publisher     = {Yale University Press},
  address       = {New Haven, CT},
  year          = {2005}
}

@article{sperry1963chemoaffinity,
  author        = {Sperry, Roger W},
  title         = {Chemoaffinity in the orderly growth of nerve fiber patterns and connections},
  journal       = {Proc. Natl. Acad. Sci. U.S.A.},
  volume        = {50},
  number        = {4},
  pages         = {703--710},
  year          = {1963}
}

@article{meyer1998roger,
  author        = {Meyer, Ronald L},
  title         = {Roger {Sperry} and his chemoaffinity hypothesis},
  journal       = {Neuropsychologia},
  volume        = {36},
  number        = {10},
  pages         = {957--980},
  year          = {1998}
}

@article{li2021cellular,
  author        = {Li, Tongchao and Fu, Tian-Ming and Wong, Kenneth Kin Lam and Li, Hongjie and Xie, Qijing and Luginbuhl, David J and Wagner, Mark J and Betzig, Eric and Luo, Liqun},
  title         = {Cellular bases of olfactory circuit assembly revealed by systematic time-lapse imaging},
  journal       = {Cell},
  volume        = {184},
  number        = {20},
  pages         = {5107--5121},
  year          = {2021}
}

@article{lyu2025dimensionality,
  author        = {Lyu, Cheng and Li, Zhuoran and Xu, Chuanyun and Wong, Kenneth Kin Lam and Luginbuhl, David J and McLaughlin, Colleen N and Xie, Qijing and Li, Tongchao and Li, Hongjie and Luo, Liqun},
  title         = {Dimensionality reduction simplifies synaptic partner matching in an olfactory circuit},
  journal       = {Science},
  volume        = {388},
  number        = {6746},
  pages         = {538--544},
  year          = {2025}
}

@article{Lyu2025.03.01.640986,
  author        = {Lyu, Cheng and Li, Zhuoran and Xu, Chuanyun and Kalai, Jordan and Luo, Liqun},
  title         = {Rewiring an olfactory circuit by altering the combinatorial code of cell-surface proteins},
  journal       = {bioRxiv},
  year          = {2025},
  doi           = {10.1101/2025.03.01.640986}
}

@article{xu2024molecular,
  author        = {Xu, Chuanyun and Li, Zhuoran and Lyu, Cheng and Hu, Yixin and McLaughlin, Colleen N and Wong, Kenneth Kin Lam and Xie, Qijing and Luginbuhl, David J and Li, Hongjie and Udeshi, Namrata D and others},
  title         = {Molecular and cellular mechanisms of teneurin signaling in synaptic partner matching},
  journal       = {Cell},
  volume        = {187},
  number        = {18},
  pages         = {5081--5101},
  year          = {2024}
}

@article{goodhill2007contributions,
  author        = {Goodhill, Geoffrey J},
  title         = {Contributions of theoretical modeling to the understanding of neural map development},
  journal       = {Neuron},
  volume        = {56},
  number        = {2},
  pages         = {301--311},
  year          = {2007}
}

@article{mora2011biological,
  author        = {Mora, Thierry and Bialek, William},
  title         = {Are biological systems poised at criticality?},
  journal       = {J. Stat. Phys.},
  volume        = {144},
  number        = {2},
  pages         = {268--302},
  year          = {2011}
}

@article{perez2009structure,
  author        = {P{\'e}rez-Escudero, Alfonso and Rivera-Alba, Marta and de Polavieja, Gonzalo G},
  title         = {Structure of deviations from optimality in biological systems},
  journal       = {Proc. Natl. Acad. Sci. U.S.A.},
  volume        = {106},
  number        = {48},
  pages         = {20544--20549},
  year          = {2009}
}

@article{loeb2012optimal,
  author        = {Loeb, Gerald E},
  title         = {Optimal isn’t good enough},
  journal       = {Biol. Cybern.},
  volume        = {106},
  number        = {11},
  pages         = {757--765},
  year          = {2012}
}

@article{jacob1977evolution,
  author        = {Jacob, Fran{\c{c}}ois},
  title         = {Evolution and tinkering},
  journal       = {Science},
  volume        = {196},
  number        = {4295},
  pages         = {1161--1166},
  year          = {1977}
}

@article{oh2016spatial,
  author        = {Oh, Doogie and Yu, Che-Hang and Needleman, Daniel J},
  title         = {Spatial organization of the {Ran} pathway by microtubules in mitosis},
  journal       = {Proc. Natl. Acad. Sci. U.S.A.},
  volume        = {113},
  number        = {31},
  pages         = {8729--8734},
  year          = {2016}
}

@article{jerison2020heterogeneous,
  author        = {Jerison, Elizabeth R and Quake, Stephen R},
  title         = {Heterogeneous {T} cell motility behaviors emerge from a coupling between speed and turning in vivo},
  journal       = {eLife},
  volume        = {9},
  pages         = {e53933},
  year          = {2020}
}

@book{viswanathan2011physics,
  author        = {Viswanathan, Gandhimohan M and Da Luz, Marcos GE and Raposo, Ernesto P and Stanley, H Eugene},
  title         = {The physics of foraging: an introduction to random searches and biological encounters},
  publisher     = {Cambridge University Press},
  year          = {2011}
}

@book{gerhart1997cells,
  author        = {Gerhart, J. and Kirschner, M.W. and Kirschner, M.},
  title         = {Cells, Embryos and Evolution},
  publisher     = {John Wiley \& Sons, Incorporated},
  year          = {1997},
  isbn          = {9780632043965},
  url           = {https://books.google.com/books?id=e7ZfkQEACAAJ}
}

@article{jaynes1957information,
  author        = {Jaynes, Edwin T},
  title         = {Information theory and statistical mechanics},
  journal       = {Phys. Rev.},
  volume        = {106},
  number        = {4},
  pages         = {620},
  year          = {1957}
}

@article{ravasio2026evolution,
  author        = {Ravasio, Riccardo and Husain, Kabir and Evans, Constantine G and Phillips, Rob and Ribezzi-Crivellari, Marco and Szostak, Jack W and Murugan, Arvind},
  title         = {Evolution of error correction through a need for speed},
  journal       = {Science},
  volume        = {391},
  number        = {6787},
  pages         = {818--824},
  year          = {2026}
}

@misc{lalley2016randomwalks,
  author        = {Lalley, Steven P.},
  title         = {{One-Dimensional Random Walks}},
  howpublished  = {Lecture notes for Statistics 312: Stochastic Processes, University of Chicago, Autumn 2016},
  year          = {2016},
  note          = {Available at \url{https://galton.uchicago.edu/~lalley/Courses/312/RW.pdf}; accessed 2026-06-17}
}

@book{lawler2010random,
  author        = {Lawler, Gregory F and Limic, Vlada},
  title         = {Random walk: a modern introduction},
  volume        = {123},
  publisher     = {Cambridge University Press},
  year          = {2010}
}

@article{polya1921aufgabe,
  author        = {P{\'o}lya, Georg},
  title         = {{\"U}ber eine Aufgabe der Wahrscheinlichkeitsrechnung betreffend die Irrfahrt im Stra{\ss}ennetz},
  journal       = {Math. Ann.},
  volume        = {84},
  number        = {1},
  pages         = {149--160},
  year          = {1921}
}

@article{baumler2023recurrence,
  author        = {B{\"a}umler, Johannes},
  title         = {Recurrence and transience of symmetric random walks with long-range jumps},
  journal       = {Electron. J. Probab.},
  volume        = {28},
  pages         = {1--24},
  year          = {2023}
}

@book{durrett2019probability,
  author        = {Durrett, Rick},
  title         = {Probability: Theory and Examples},
  edition       = {5},
  volume        = {49},
  publisher     = {Cambridge University Press},
  year          = {2019},
  page          = {250}
}

@misc{weisstein_polyas_random_walk_constants,
  author        = {Weisstein, Eric W.},
  title         = {{P{\'o}lya's Random Walk Constants}},
  howpublished  = {\url{https://mathworld.wolfram.com/PolyasRandomWalkConstants.html}},
  year          = {2026},
  note          = {From MathWorld---A Wolfram Resource. Accessed 17 June 2026}
}

@book{polya1984collected,
  author        = {Pólya, George},
  editor        = {Rota, Gian-Carlo},
  title         = {George Pólya: Collected Papers: Probability, Combinatorics, Teaching and Learning in Mathematics},
  series        = {Mathematicians of Our Time},
  volume        = {4},
  publisher     = {The MIT Press},
  year          = {1984},
  isbn          = {978-0-262-16097-1},
  url           = {https://mitpress.mit.edu/9780262160971/george-polya-collected-papers-volume-4/},
  month         = {sep},
  urldate       = {2026-07-31}
}

@article{krishnapur2004recurrent,
  author        = {Krishnapur, Manjunath and Peres, Yuval},
  title         = {Recurrent graphs where two independent random walks collide finitely often},
  journal       = {Electron. Commun. Probab.},
  volume        = {9},
  pages         = {72--81},
  year          = {2004},
  url           = {https://projecteuclid.org/journals/electronic-communications-in-probability/volume-9/issue-none/Recurrent-Graphs-where-Two-Independent-Random-Walks-Collide-Finitely-Often/10.1214/ECP.v9-1111.pdf}
}

@article{murtagh2026randomwalk,
  author        = {Murtagh, Jack},
  title         = {Socially Awkward Math},
  journal       = {Sci. Am.},
  volume        = {334},
  number        = {3},
  pages         = {72},
  year          = {2026},
  doi           = {10.1038/scientificamerican032026-1y79vdxh3SUBuPBq5qUTnU},
  note          = {Published online February 17, 2026}
}

@incollection{chungRandomWalksOfPolya,
  author        = {Chung, K.L.},
  title         = {Appendix 1: P\`{o}lya's Work in Probability},
  booktitle     = {The Random Walks of George P\`{o}lya},
  editor        = {Gerald L. Alexanderson},
  publisher     = {Mathematical Association of America},
  address       = {Washington, DC},
  pages         = {193--200},
  year          = {2000}
}

@book{doyle1984random,
  author        = {Doyle, Peter G and Snell, J Laurie},
  title         = {Random walks and electric networks},
  volume        = {22},
  publisher     = {American Mathematical Soc.},
  year          = {1984}
}

@article{hartmann2025diffusion,
  author        = {Hartmann, Alexander K and Majumdar, Satya N},
  title         = {Diffusion with stochastic resetting on a lattice},
  journal       = {Phys. Rev. E},
  volume        = {112},
  number        = {3},
  pages         = {034102},
  year          = {2025}
}

@book{koshy2008catalan,
  author        = {Koshy, Thomas},
  title         = {Catalan numbers with applications},
  publisher     = {Oxford University Press},
  year          = {2008}
}

@book{stanley2015catalan,
  author        = {Stanley, Richard P},
  title         = {Catalan numbers},
  publisher     = {Cambridge University Press},
  year          = {2015}
}

@article{PhysRevResearch.2.032029,
  author        = {Besga, Benjamin and Bovon, Alfred and Petrosyan, Artyom and Majumdar, Satya N. and Ciliberto, Sergio},
  title         = {Optimal mean first-passage time for a {Brownian} searcher subjected to resetting: Experimental and theoretical results},
  journal       = {Phys. Rev. Res.},
  volume        = {2},
  number        = {3},
  pages         = {032029(R)},
  year          = {2020},
  doi           = {10.1103/PhysRevResearch.2.032029}
}

@article{bhat2016stochastic,
  author        = {Bhat, Uttam and De Bacco, Caterina and Redner, S},
  title         = {Stochastic search with {Poisson} and deterministic resetting},
  journal       = {J. Stat. Mech.: Theory Exp.},
  volume        = {2016},
  number        = {8},
  pages         = {083401},
  year          = {2016}
}

@article{Evans_2013,
  author        = {Evans, Martin R and Majumdar, Satya N and Mallick, Kirone},
  title         = {Optimal diffusive search: nonequilibrium resetting versus equilibrium dynamics},
  journal       = {J. Phys. A: Math. Theor.},
  volume        = {46},
  number        = {18},
  pages         = {185001},
  year          = {2013},
  doi           = {10.1088/1751-8113/46/18/185001}
}

@misc{abramowitz1966handbook,
  author        = {Abramowitz, Milton and Stegun, Irene A and Romain, Jacques E},
  title         = {Handbook of mathematical functions, with formulas, graphs, and mathematical tables},
  publisher     = {American Institute of Physics},
  year          = {1966},
  note          = {Chapter 13: Confluent Hypergeometric Functions}
}

@article{DeBruyne2022,
  author        = {De Bruyne, Benjamin and Majumdar, Satya N and Schehr, Gr{\'e}gory},
  title         = {Optimal resetting {Brownian} bridges via enhanced fluctuations},
  journal       = {Phys. Rev. Lett.},
  volume        = {128},
  number        = {20},
  pages         = {200603},
  year          = {2022}
}

@article{NgGellerIntegralErrorFunction,
  title = {A table of integrals of the error functions.},
  author = {Edward W. Ng and Murray Geller},
  journal = {Journal of Research of the National Bureau of Standards, Section B: Mathematical Sciences},
  year = {1969},
  pages = {1},
  url = {https://api.semanticscholar.org/CorpusID:121798145},
}

@article{blackwell_waldequation,
  title = {On an equation of Wald},
  author = {Blackwell, David},
  journal = {The Annals of Mathematical Statistics},
  volume = {17},
  number = {1},
  pages = {84--87},
  year = {1946},
  publisher = {JSTOR},
}

@article{Ray2019Peclet,
  author  = {Ray, S. and Mondal, D. and Reuveni, S.},
  title   = {P{\'e}clet number governs transition to acceleratory restart in drift-diffusion},
  journal = {J. Phys. A: Math. Theor.},
  volume  = {52},
  pages   = {255002},
  year    = {2019}
}


\newpage
\clearpage
\onecolumngrid
\setcounter{page}{1}

\appendix
\renewcommand{\appendixname}{}

\setcounter{figure}{0}
\setcounter{table}{0}
\setcounter{equation}{0}

\renewcommand{\thefigure}{S\arabic{figure}}
\renewcommand{\thetable}{S\arabic{table}}
\renewcommand{\theequation}{S\arabic{equation}}

\startcontents[appendix]
\noindent\textbf{Supplemental Material for \emph{The Trajectory Statistics of Biological Exploratory Dynamics}}
\par\medskip
\printcontents[appendix]{}{1}{}

\startconstantfooter{\textcolor{gray}{The Trajectory Statistics of Biological Exploratory Dynamics: Mahdavi*, Salmon*, Ashok, Mani, Kirschner, Kondev, and Phillips}}


\newcounter{SIsavedequation}

\let\originalsection\section

\RenewDocumentCommand{\section}{s o m}{%

    \setcounter{SIsavedequation}{\value{equation}}%

    \IfBooleanTF{#1}{%
        \originalsection*{#3}%
    }{%
        \IfNoValueTF{#2}{%
            \originalsection{#3}%
        }{%
            \originalsection[#2]{#3}%
        }%
    }%

    \setcounter{equation}{\value{SIsavedequation}}%
}

\section{Quixotic exploration through random walks}
\label{section:BareRandomWalks}

In principle, organisms, cells, and molecules \emph{could} navigate spaces in pursuit of target states by the simplest algorithm possible: at each moment in time, flip an independent coin that decides where to step next, assembling an unguided random walk that diffuses searchers throughout their physical or behavioral spaces. Indeed, many chemical reactions in cells proceed dependably enough by diffusive, mass-action encounters in the crowded compartments of cells or organisms.

However, for some search processes, this type of unstructured diffusive motion is often \emph{not} the means by which biological systems explore or sample physiological variation. Instead, as prospectively hinted at in Fig.~\ref{fig:DifferentMath} and Fig.~\ref{fig:BiologicalCaseStudies} of the main text (\S \ref{section:RandomWalkReset}), many examples of exploratory dynamics are decorated by additional mechanisms (such as the surprising presence of ``resets'' aborting even imminently successful trajectories; rescues from these resets via checkpoints; parallelism; directional drift; signaling and sensing via cues; and beyond). Why? What makes these modifications to the most primordial search algorithm necessary or useful, given that these modifications often cost energy or impose architectural complexity? 

\begin{table}[h!]
    \centering
    \footnotesize
    \setlength{\tabcolsep}{2pt}

    \begin{tabular}{ccc}
        \parbox{0.13\columnwidth}{\centering\textbf{dimension}} &
        \parbox{0.30\columnwidth}{\centering\textbf{probability of ever hitting target}} &
        \parbox{0.22\columnwidth}{\centering\textbf{mean time to hit target}} \\
        \hline

        \parbox{0.13\columnwidth}{\centering $d=1$} &
        \parbox{0.30\columnwidth}{\centering $p(\mathrm{hit\ target})=1$} &
        \parbox{0.22\columnwidth}{\centering $\langle T\rangle = \infty$} \\
        \parbox{0.13\columnwidth}{\centering $d=2$} &
        \parbox{0.30\columnwidth}{\centering $p(\mathrm{hit\ target})=1$} &
        \parbox{0.22\columnwidth}{\centering $\langle T\rangle = \infty$} \\
        \parbox{0.13\columnwidth}{\centering $d\geq 3$} &
        \parbox{0.30\columnwidth}{\centering $p(\mathrm{hit\ target}) < 1$} &
        \parbox{0.22\columnwidth}{\centering $\langle T\rangle = \infty$} 
    \end{tabular}

    \caption{\raggedright Properties of ordinary random walks. See (for instance) Refs.~\cite{lalley2016randomwalks, lawler2010random, polya1921aufgabe}.}
    \label{table:randomWalkOrdinaryProps}
\end{table}
\FloatBarrier

In short, one hypothesis is that such ordinary random walks can be too slow to rely on to find targets, or even fail to do so completely. One explicit biological example where unmodified random walk searches might have biologically forbidding slow speeds, or failure rates, is the concrete setting of the capture of chromosomes by microtubules in cell division \cite{Holy1994}. Specifically, the centrosome must capture all chromosomes within a time acceptable to the pace of mitosis. Since the dynamic instability of microtubules underlying this process requires constant injection of biochemical energy, the time to accomplish these searches has biological and bioenergetic implications. 
To appreciate the stakes (and motivate our journey through various mechanisms in the  escapades of this paper), we now tour briskly some mathematical facts to this effect. 
First, consider the typical time that trajectories take to reach a target. As summarized by the third column of Table~\ref{table:randomWalkOrdinaryProps},
even in one or two dimensions, the distribution of first-passage times is
sufficiently broad that the mean time to hit a target $\langle T\rangle$
is \emph{infinite}, at least in an unbounded region.

This latter chronological catastrophe may be appreciated from many angles; one is to directly calculate the probability of a walker arriving at a target in one dimension by exact lattice counts, as discussed explicitly and pedagogically in Appendix \ref{app:ordinary_random_walks}. The resulting probability $f(t) dt$ of first arriving at a target near time $t$ can be described in a continuum approximation as scaling with $f(t) \thicksim t^{-3/2}$ in one dimension (see the Appendix,  eqn.~\ref{eq:continuumApprox_f(t)ForFirstHit1DrandomWalk}), anticipating a divergent mean first arrival time to any target.

Next, consider the probability of ever arriving at a particular target location in spaces of increasing dimension, as summarized by the second column of Table~\ref{table:randomWalkOrdinaryProps}. In one or two dimensions, an ordinary random walk will always eventually arrive at a target at any finite separation from a starting point. However, in spaces of three or more dimensions, the situation is spectacularly worse. Trajectories of such walks may never reach a specific target's location at all (namely, they show a hitting or recurrence probability of less than one)! 

\begin{figure}[t!]
    \centering
    \includegraphics[width=0.5\linewidth]{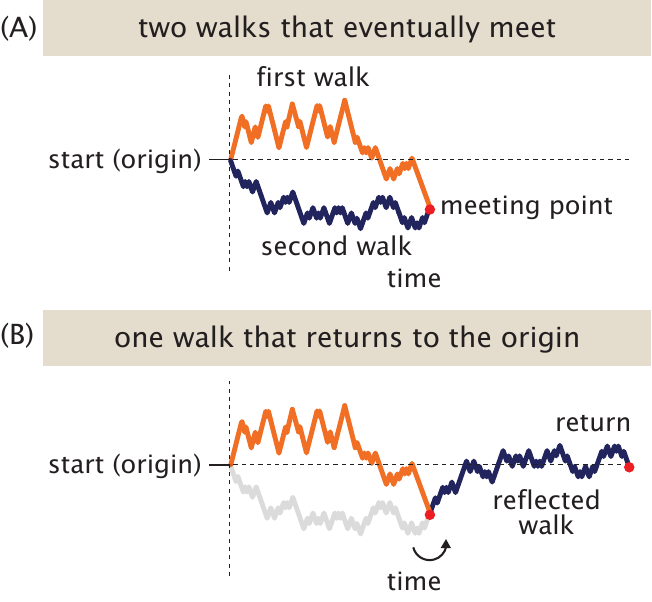}
    \caption{\raggedright Equivalence of the event that two separate random walks ever meet with the event that one random walk ever returns to the origin. (A) A pair of random walks (respectively, orange and navy) that meet at some point (in red). (B) Taking one of the random walks, swapping all positive steps for negative steps and vice versa, and appending to the other walk at the meeting point yields a new longer walk that accomplishes a return to the starting point.}
    \label{fig:whyRecurrenceIsSameAsTwoMeeting}
\end{figure}

This remarkable dimensionality-dependent behavior motivates the memorable quip, ``A drunk man will eventually find his way home, but a drunk bird may get lost forever,'' as limned by Shizuo Kakutani \cite{baumler2023recurrence, durrett2019probability}. 
(Here inebriation is a proxy for genuinely random wandering in either two or three dimensions.) P\`{o}lya proved this subtle and counterintuitive fact in a celebrated recurrence theorem  \cite{polya1921aufgabe, weisstein_polyas_random_walk_constants}. Amusingly (relatably?), P\`{o}lya was originally motivated to consider the mathematical problem by force of social embarrassment. Going for a (two-dimensional) walk in his  town of Zurich, he happened to encounter a student with their lover, then---his chagrin growing---met them again and again over the same walk. P\`{o}lya wondered whether this property of two random walks meeting could be expected to be generic \cite[``Two Incidents'' p. 582–585]{polya1984collected}; see also Refs.~\cite{polya1921aufgabe, krishnapur2004recurrent, murtagh2026randomwalk}. First, he realized that the question of whether two random walks could be expected to meet eventually is equivalent to the question of whether a single walk starting at the origin can be expected to return there. To appreciate this equivalence, take two walks that \emph{do} encounter each other at some point; reverse the steps of one of the walks; and append it to the other walk at their meeting place. The new longer walk definitely returns to the origin, as illustrated by Fig.~\ref{fig:whyRecurrenceIsSameAsTwoMeeting}. This operation maps every pair of walks that encounter each other to some other walk that accomplishes a return to the origin \cite{chungRandomWalksOfPolya}. Empowered to focus on this question of recurrence, classic and modern sources \cite{lalley2016randomwalks, lawler2010random} (including P\`{o}lya's original arguments \cite{polya1921aufgabe}) typically establish the dimensional breakdown in the certainty of visiting a site using heavy technical formalisms such as the machinery of generating functions, or else ingenious, but intricate, analogies to electrical circuits \cite{doyle1984random}. 

Instead, we now briefly trace a heuristic argument, avoiding technical fussiness to display the fundamental essence, for why the probability of ever visiting some site located at a finite distance $\mathbf{x}_T$ (a distance $R$ away) is less than one for spaces of dimensionalities three or larger. We will argue this as follows. First, we recognize that the desired probability of ever visiting a site is at most the marginal probability that it visits the site at a particular time $t$, summed over all possible times from zero to infinity. (By marginal probability, we mean the probability of being at a certain position at a given time, regardless of whether or how many times this site has been visited before in a trajectory.) This is a useful wedge into the problem, because we are already armed with knowing how this marginal probability behaves. Intuitively, this decreases very steeply with the dimensionality $d$ of the space the walk is performed in, since there are many more possible sites that a walker could be present at in spaces of larger dimension. Specifically this decay of probability steepens with dimensionality according to a power law prefactor $\thicksim t^{-d/2}$. Next we recall the other identifying characteristic of random walks: visiting sites a distance $R$ away typically requires $\thicksim R^2$ timesteps, no matter the dimensionality; so that governing sum over all times essentially turns on only for times $t\gtrsim R^2$, meaning the whole sum can be approximated as something proportional to $\thicksim R^{2-d}$. 
This competition between the sites plausibly visited by random walks in any allotted time, and the (larger) $d$-dimensional volume of sites that are candidates to be visited, exposes the essence of why every site in spaces of dimensions higher than two is not certainly visited by random walks.

\begin{figure*}
    \centering
    \includegraphics[width=\linewidth]{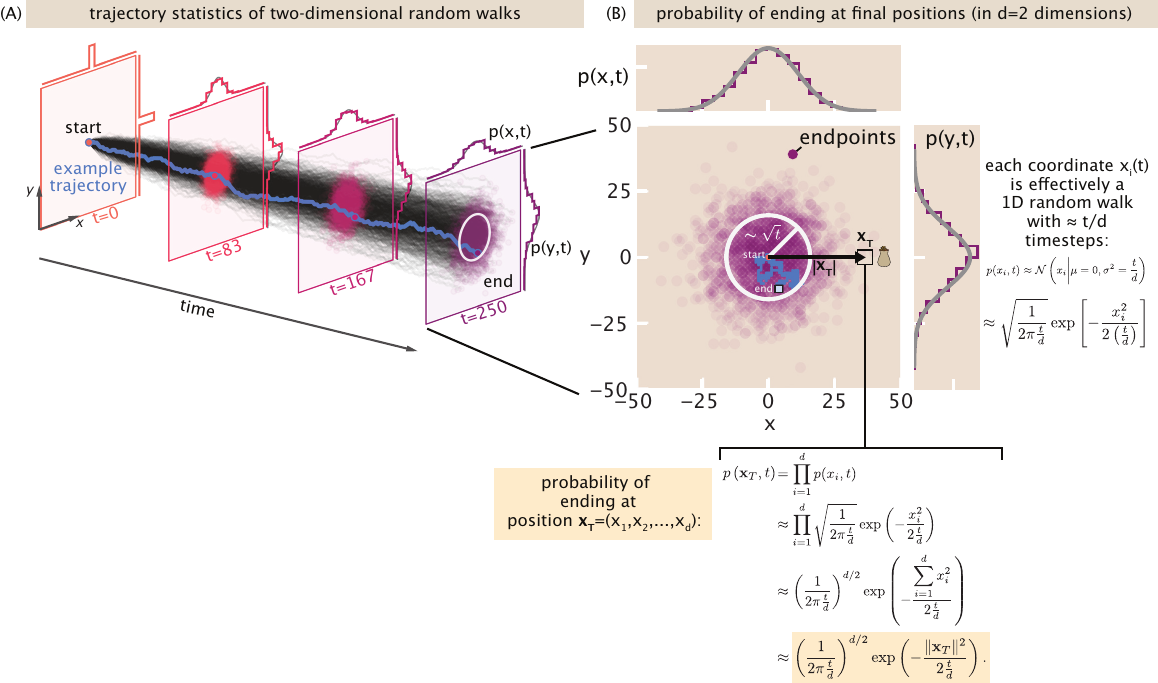}
    \caption{\raggedright Spreading of random walks over time and dimension. 
    (A) A few thousand individual symmetric random walks illustrated in $d=2$ dimensions visualized over time as black traces. A representative trajectory is shown in the pale blue line. The distributions of the positions of walks at a sequence of times are shown at the colored planes. (B) A representative temporal slice of the ensemble of random walks, corresponding to the final time in (A). Histograms (empirical in purple; theoretical, according to eqn.~\ref{eq:marginalCoordPos_2dwalks}, in grey) show how each spatial coordinate effectively receives a fraction $1/d$ of $t$ timesteps, thus effectively obeying an independent one dimensional walk's limiting normal distribution in position. Together these coordinates specify the distribution over $d$-dimensional positions.}
    \label{fig:dTwoDimCaliberWordlines}
\end{figure*}

In slightly more detail: let $\mathbf{x}_T = [x_1,\ldots,x_d]$ be the location of a target of interest in $d$-dimensional space, relative to start at the origin, a distance $R=\sqrt{x_1^2 + x_2^2 + \cdots +x_d^2}$ away. The desired probability $p_{\text{hit}}(\mathbf{x}_T)$ that a walker \emph{ever} hits $\mathbf{x}_T$ is complicated to think about. Yet, we already know many things about the marginal probability $p(\mathbf{x}_T,t)$ that a walk is found at the target $\mathbf{x}_T$ at some \emph{particular} time $t$. Specifically, the probability $p_{d=1}(x_T,t)$ that a random walk is found at a particular one-dimensional position $x_T$ is well approximated by a familiar normal distribution with mean position zero and variance $t$. (This behavior can be recalled simply by remembering that the walker steps left (giving a change in position $\Delta x=-1$) with equal probability as stepping right (giving $\Delta x=+1$), so the expected value of the squared update to position (yielding the variance) increases by $1$ at each time step, while the expected (mean) position of the walk remains unchanged as zero; see also Appendix \ref{app:ordinary_random_walks} for context for this behavior in terms of explicit lattice counts.) That is, we can write the probability density as
\begin{align}
    p_{d=1}(x_T,t) & \approx \mathcal{N}\left(x_T \bigg|0,t\right)\\
    & \approx \sqrt{\frac{1}{2\pi t}} \exp\left[-\frac{x_T^2}{2 t } \right].
\end{align}
This statement is approximate in the same sense that the central limit theorem is approximate; it holds increasingly exactly for long walks $t$ manifesting a continuous limit. 

\begin{figure*}
    \centering
    \includegraphics[width=0.7\linewidth]{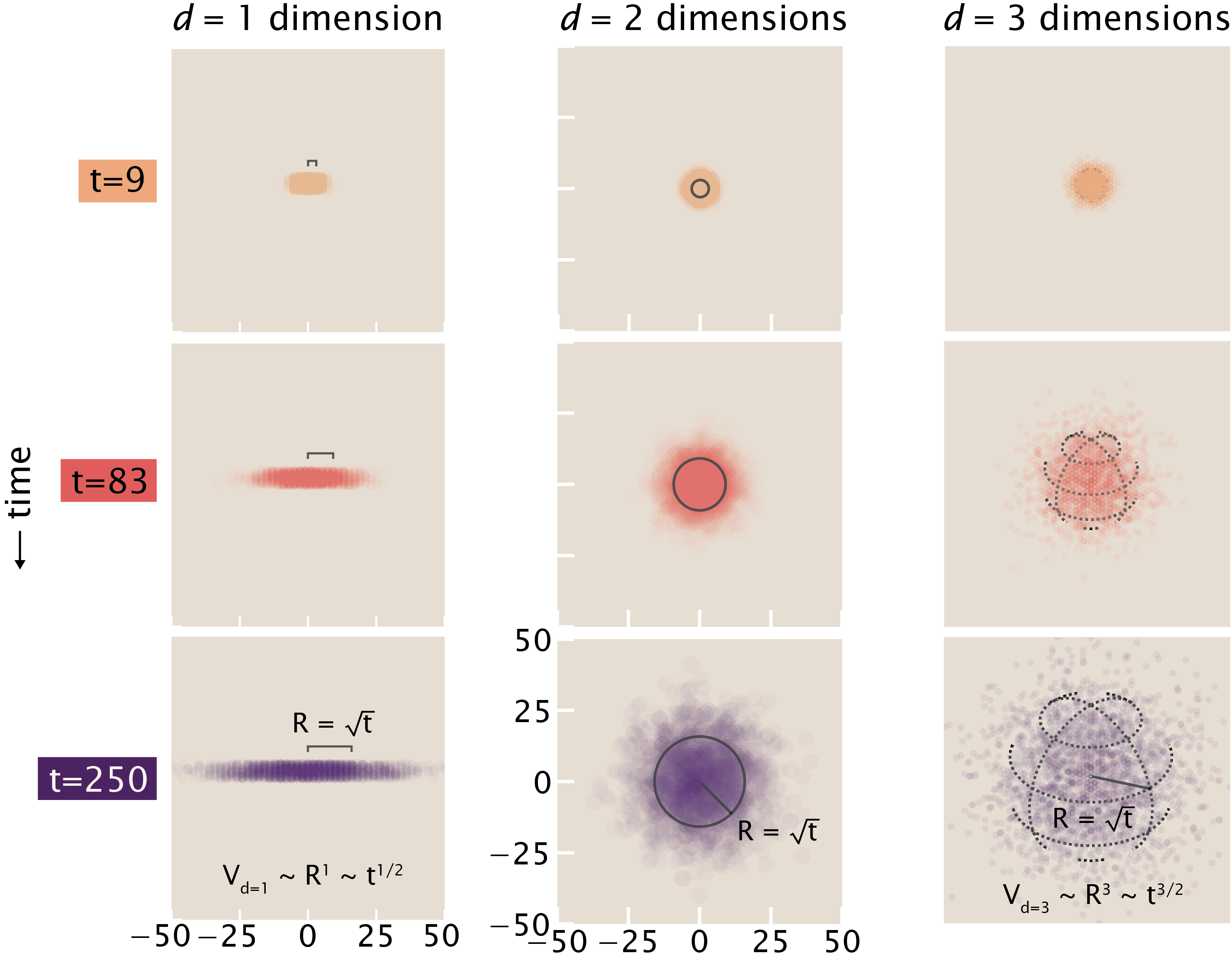}
    \caption{\raggedright Sampled distributions of positions of random walks in one, two, and three dimensional spaces over time.}
    \label{fig:filmStripOfOneTwoThreeDwalks}
\end{figure*}

 In spaces of higher dimensionality $d>1$, the time spent wandering is split among the $d$ possible spatial coordinates. This means that any individual coordinate's position $x_i$ at time $t$ is set by a one-dimensional random walk typically with only $\approx t/d$ timesteps. Accordingly, each coordinate is well described by a normal distribution with variance $t/d$, or
\begin{align}
   p(x_i,t) & \approx \mathcal{N}\left(x_i \bigg|\mu = 0,\sigma^2 = \frac{t}{d}\right)\\
    & \approx \sqrt{\frac{1}{2\pi \frac{t}{d}}} \exp\left[-\frac{x_i^2}{2 \frac{t}{d} } \right].
\label{eq:marginalCoordPos_2dwalks}
\end{align}
This description for the marginal distributions of coordinate positions is illustrated for the case of $d=2$ dimensions in  Fig.~\ref{fig:dTwoDimCaliberWordlines}. As a consequence, the vectorial position of walks in all $d$ dimensions spreads radially over time, with a typical mean squared displacement of $t$ or radius of $\sqrt{t}$. This is conspicuous in the widening ensemble of walks shown Fig.~\ref{fig:dTwoDimCaliberWordlines}(A) in two dimensions, or the spreads visualized in one, two, and three dimensions in Fig.~\ref{fig:filmStripOfOneTwoThreeDwalks}.

\begin{figure}[t!]
    \centering
    \includegraphics[width=0.75\linewidth]{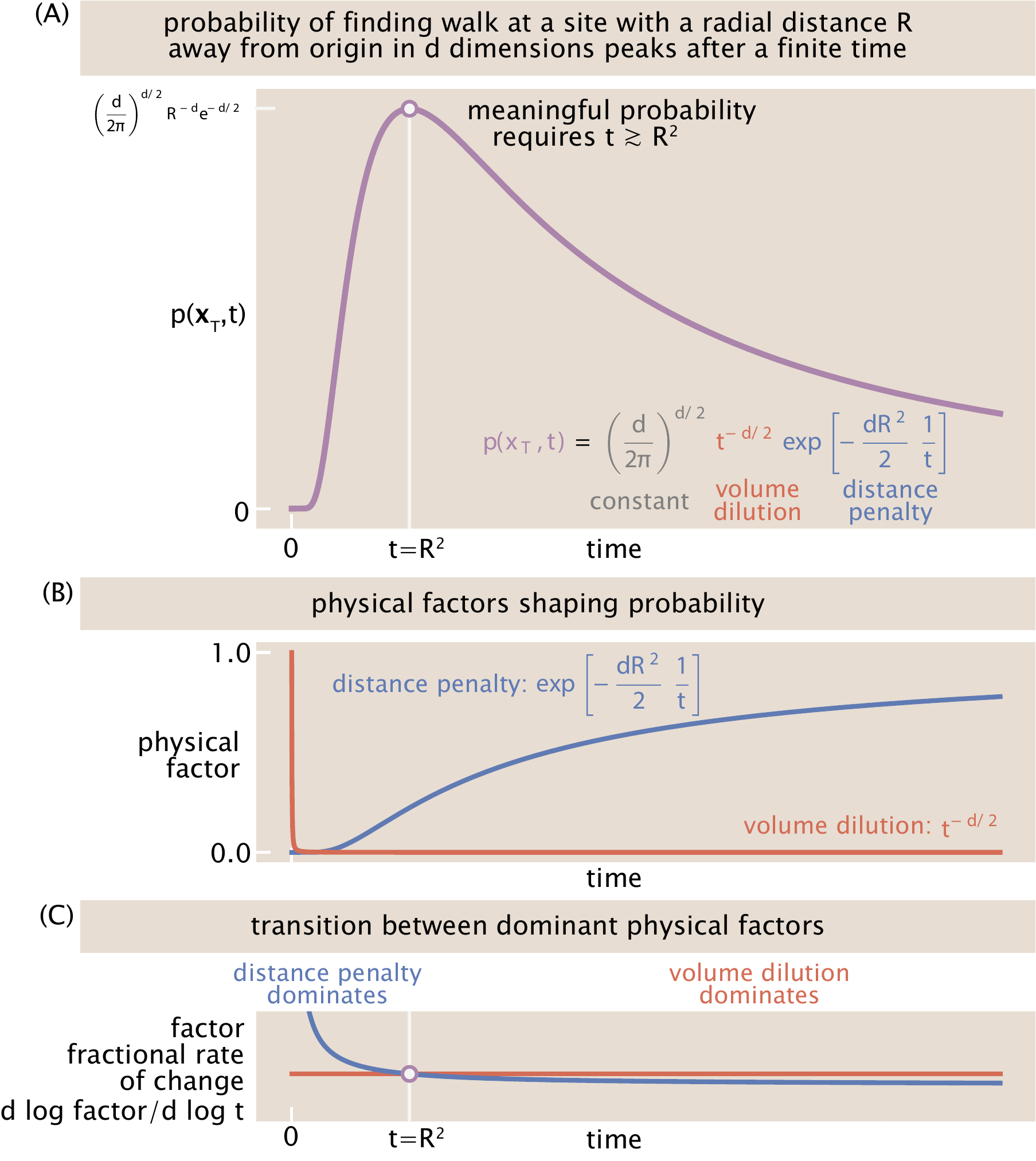}
    \caption{\raggedright The probability of a random walk occupying a site at a distance $R$ from the origin in $d$ dimensions. (A) The marginal probability $p(\mathbf{x}_T, t)$, as described by eqn.~\ref{eq:dDimProbability}, is strongly peaked in time; observing a walk at a radial position $R$ is small until a time $t\gtrsim R^2$, then declines again. (B) This probability is shaped by whether or not enough time has elapsed for walks to have a chance of reaching the radial position $R$, given by a distance penalty factor shown in blue, competing with the dilution over all possible sites walks could be found in the $d$-dimensional volume at time $t$, given by the volume dilution factor in red. (C) The dominance of these two effects switches at a peak time $t=R^2$; the fractional rates of change of these factors mark the crossover.}
    \label{fig:witnessUnderstandPeak}
\end{figure}

More algebraically, this behavior is explained by understanding that the probability of being at a vectorial position $\mathbf{x}_T$ is the product of the probabilities of finding each coordinate at their positions, over all $d$ dimensions. As shown by Fig.~\ref{fig:dTwoDimCaliberWordlines}(B), all this reasoning leads to the understanding that the marginal probability of finding a walk at position $\mathbf{x}_T$ at time $t$ in $d$ dimensions is,
\begin{align}
    p(\mathbf{x}_T,t) & \approx \left(\frac{1}{2\pi \frac{t}{d}}\right)^{d/2} \exp\left( -\frac{\lVert\mathbf{x_T}\rVert^2}{2\frac{t}{d}} \right)\\
    & =\left(\frac{d}{2\pi}\right)^{d/2}~ t^{-d/2} ~ \exp\left[-\frac{d R^2}{2}~\frac{1}{t} \right].
    \label{eq:dDimProbability}
\end{align}
Importantly, this probability of finding a walker at a site a distance $R$ from the origin is very small until times of order $t\gtrsim R^2$, after which the probability falls again, as witnessed by  Fig.~\ref{fig:witnessUnderstandPeak}(A). This behavior reflects two competing physical effects. When $t$ is too small, the target position is at too great a distance for walks to have meaningfully sampled; this effect is expressed by the dominance of the $\exp\left[-\frac{d R^2}{2}~\frac{1}{t} \right]$ factor of eqn.~\ref{eq:dDimProbability}, plotted in blue in Fig.~\ref{fig:witnessUnderstandPeak}(B). At long times, however, the fact that the walks have wandered over a burgeoning $d$-dimensional volume means that the probability that a walker is found at a particular site within this volume shrinks; this  volume dilution is reflected by a dominance of the $t^{-d/2}$ factor, plotted in red in Fig.~\ref{fig:witnessUnderstandPeak}(B). The probability is peaked when these factors have fractional changes in time that match in size, as tracked by crossover in the logarithmic derivatives shown in Fig.~\ref{fig:witnessUnderstandPeak}(C). 

\begin{figure}
    \centering
    \includegraphics[width=0.45\linewidth]{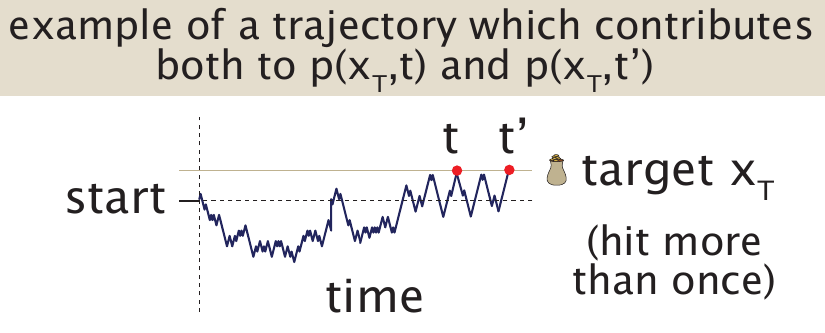}
    \caption{\raggedright Example of a trajectory which makes the union bound loose, namely, eqn.~\ref{eq:unionBoundForProbPhit} generally a strict inequality.}
    \label{fig:egRedundTrajUnionBound}
\end{figure}

Armed with this understanding given by eqn.~\ref{eq:dDimProbability}, we return to tackle our big question of finding the desired probability that a walk ever visits such a target site at any time, $p_\text{hit}(\mathbf{x}_T)$. These quantities are related in the following simple sense. If we add up the (now, known) probabilities $p(\textbf{x}_T,t)$ of the walk being at the target for all possible times $t$, we would get close to expressing the desired probability $p_\text{hit}(\mathbf{x}_T)$ of ever hitting the target. But we will have overcounted: for instance, this double-counts trajectories that are both present at the target at one time $t$ and also another time $t'$, which contribute to both the marginal probabilities $p(\textbf{x}_T,t)$ and $p(\textbf{x}_T,t')$ but actually contribute only once to $p_\text{hit}(\mathbf{x}_T)$, as visualized in the example of Fig.~\ref{fig:egRedundTrajUnionBound}. This just equips us to safely write the upper bound,
\begin{align}
     p_{\text{hit}}(\mathbf{x}_T) \leq \sum_{t=0}^\infty p(\mathbf{x}_T,t).
     \label{eq:unionBoundForProbPhit}
\end{align}
(More generally, this simple statement is just a particular instance of the general so-called ``union bound'' of probabilities, saying that the total probability $p\left(\displaystyle \cup_i A_i \right)$ of any of a set of $k$ events $\left\{ A_1, \ldots, A_k\right\}$ is at most the sum of their probabilities. For example, it is simplest to see this for the case of two events $\{A_1, A_2\}$: the probability of seeing either of these two events (their union) is $P(A_1 \cup A_2) = P(A_1) + P(A_2) - P(A_1 \cap A_2)$, or in pictures, \begin{tikzpicture}[baseline=-0.6ex] \fill[gray!25] (-0.14cm,0) circle[radius=0.32cm]; \fill[gray!25] (0.14cm,0) circle[radius=0.32cm]; \draw (-0.14cm,0) circle[radius=0.32cm]; \draw (0.14cm,0) circle[radius=0.32cm]; \node[ inner sep=0pt, font=\scriptsize ] at (0,0) {$A_1\cup A_2$}; \end{tikzpicture} \(\;=\;\) \begin{tikzpicture}[baseline=-0.6ex] \fill[gray!25] (-0.14cm,0) circle[radius=0.32cm]; \draw (-0.14cm,0) circle[radius=0.32cm]; \draw (0.14cm,0) circle[radius=0.32cm]; \node[ inner sep=0pt, font=\scriptsize ] at (-0.31cm,0) {$A_1$}; \end{tikzpicture} \(\;+\;\) \begin{tikzpicture}[baseline=-0.6ex] \fill[gray!25] (0.14cm,0) circle[radius=0.32cm]; \draw (-0.14cm,0) circle[radius=0.32cm]; \draw (0.14cm,0) circle[radius=0.32cm]; \node[ inner sep=0pt, font=\scriptsize ] at (0.31cm,0) {$A_2$}; \end{tikzpicture} \(\;-\;\) \begin{tikzpicture}[baseline=-0.6ex] \begin{scope} \clip (-0.14cm,0) circle[radius=0.32cm]; \fill[gray!25] (0.14cm,0) circle[radius=0.32cm]; \end{scope} \draw (-0.14cm,0) circle[radius=0.32cm]; \draw (0.14cm,0) circle[radius=0.32cm]; \node[inner sep=0pt] at (0,0) { \scalebox{0.62}{ \shortstack{ $A_1$\\[-0.55ex] $\cap$\\[-0.55ex] $A_2$ } } }; \end{tikzpicture}: neglecting the latter intersection piece $P(A_1 \cap A_2)$, which has positive probability, gives an upper bound. In our setting, the events $A_1$ and $A_2$ can mean a walk being found at time $t$ or $t'$, respectively.) 

Now we apply our understanding in eqn.~\ref{eq:dDimProbability} of how the marginal probabilities $p(\textbf{x}_T,t)$ of finding a walk at position $R$ behave over time: in effect, these are strongly suppressed for $t\ll R^2$, turn on for $t\gtrsim R^2$, and are modulated by a sharp power law in the dimensionality $d$. This emboldens us to more daringly approximate the bound of eqn.~\ref{eq:unionBoundForProbPhit} as,
\begin{align}
    p_{\text{hit}}(\mathbf{x}_T) & \leq \sum_{t=0}^\infty p(\mathbf{x}_T,t)\\
    &\leq \sum_{t=0}^\infty \left(\frac{d}{2\pi}\right)^{d/2}~ t^{-d/2} ~ \exp\left[-\frac{d R^2}{2}~\frac{1}{t} \right]\\
    & \lesssim C_d \sum_{t=R^2}^\infty t^{-d/2}.
     \label{eq:unionBoundForProbPhit}
\end{align}
The replacement in the last line imposes two core features we observed in Fig.~\ref{fig:witnessUnderstandPeak} and reflects a scaling behavior (not a pointwise relation). First, the exponential distance penalty $\exp\left[-\frac{d R^2}{2}~\frac{1}{t} \right]$ acts as a soft threshold function in time, making it very unlikely to find a walk at the radial position $R$ much before time $t\lesssim R^2$. Second, after this peak time, the exponential distance penalty's effect saturates and does not change the main scaling behavior (as shown in Fig.~\ref{fig:witnessUnderstandPeak}(B)). More precisely, we can squeeze it between two simple constants that depend only on the dimensionality, namely $\exp\left[-d/2 \right] \leq \exp\left[-\frac{d R^2}{2}~\frac{1}{t} \right] \leq 1$; so the behavior of the whole probability function is dominated by the $t^{-d/2}$ volume dilution behavior, up to some approximately constant factor $C_d$ close to one. 

\begin{figure}[h!]
    \centering
    \includegraphics[width=0.7\linewidth]{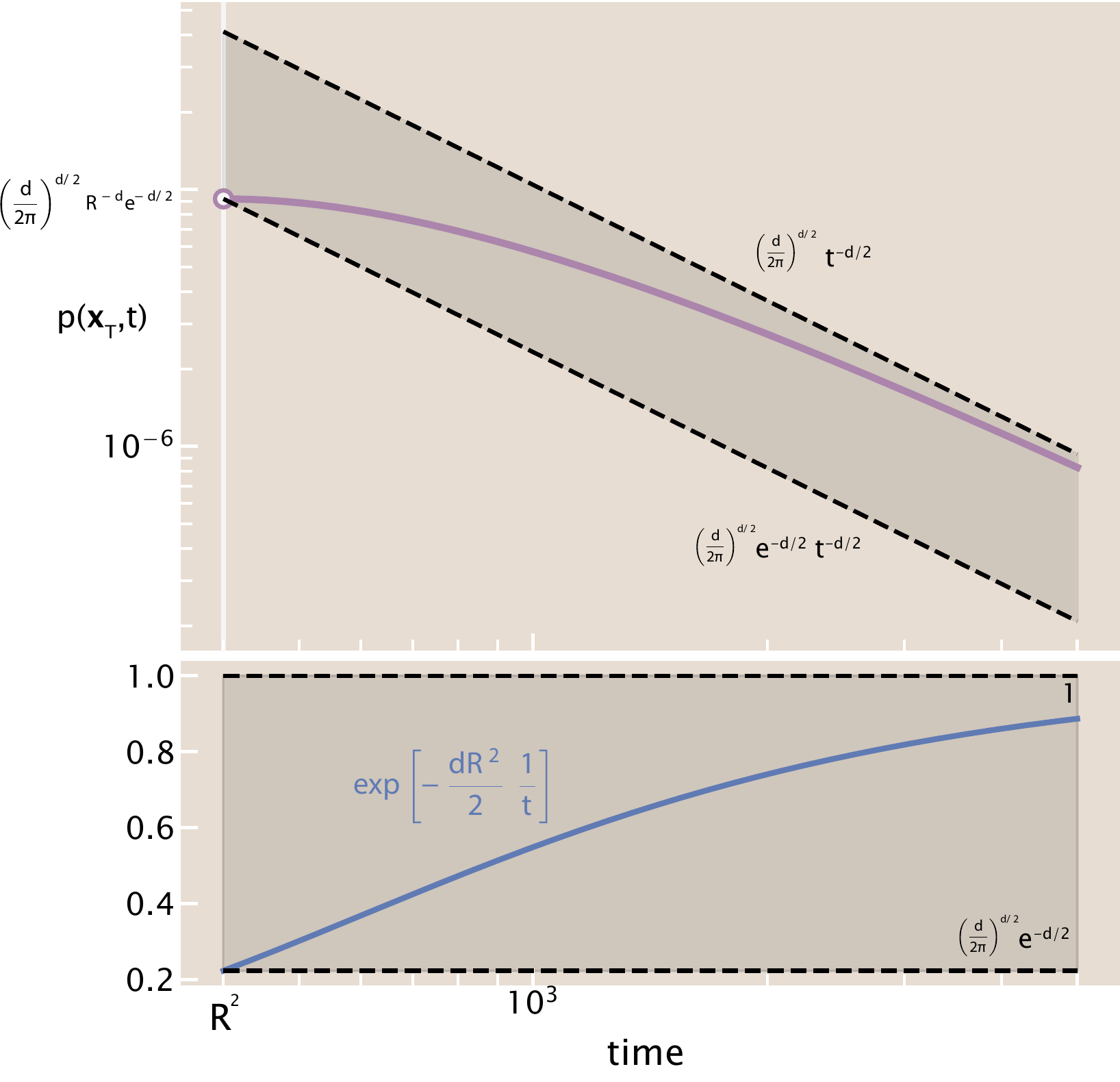}
    \caption{\raggedright Illustration of the bounds sandwiching the exponential distance penalty, $\exp\left[-d/2 \right] \leq \exp\left[-\frac{d R^2}{2}~\frac{1}{t} \right] \leq 1$. For illustration, this plot reflects the same purple and blue curves as in Fig.~\ref{fig:witnessUnderstandPeak}(A) and (B) of the main text except for times appearing just after the peak $t\geq R^2$ and with axes on a log-log basis to illustrate the agreement in power law exponents. Specifically, these plots reflect the case of a $d=3$ dimensional space at a target located at a site $R=25$ away from the origin.}
    \label{fig:squeezeBoundsForWalks}
\end{figure}

That is, the distance penalty $\exp\left[-\frac{d R^2}{2} \frac{1}{t}\right]$ is bounded from above and below by constants that are close to one, and just depend on the dimensionality $d$ of the space the random walk explores. These behaviors are visualized in Fig.~\ref{fig:squeezeBoundsForWalks}. The takeaway is that the remaining $t^{-d/2}$ behavior dominates the time dependence after the peak, $t>R^2$. 
\FloatBarrier

Next, to appreciate the core behavior of the summation in Eqn.~\ref{eq:unionBoundForProbPhit}, we realize it adopts the form of a Riemann sum and can be expressed as approaching an integral (particularly for distant targets with large $R$), giving,
\begin{align}
    p_{\text{hit}}(\mathbf{x}_T) & \lesssim C_d \int_{R^2}^\infty t^{-d/2}~dt\\
    & \lesssim C_d' ~R^{2-d},
    \label{eq:atLastTwoMinusD}
\end{align}
with $C_d'$ some suitable constant. 

At last, eqn.~\ref{eq:atLastTwoMinusD} exposes the critical dimensionality-dependence of recurrence or the probability of ever hitting a target. When $d>2$, the power law behavior $\thicksim R^{2-d}$ definitely shrinks to zero for sites with large enough $R$: this is enough to establish that these sites are not visited with certainty. (By itself, note that this expression eqn.~\ref{eq:atLastTwoMinusD} does \emph{not} yet comment forcefully that the optimistic converse is true for $d=1$ or $d=2$ dimensions, since the integral diverges (giving a vacuous bound on a hitting probability); additional theoretical ingredients are needed to appreciate that certain arrivals do happen in these lower dimensions.) 

\begin{figure*}
    \centering
    \includegraphics[width=0.8\linewidth]{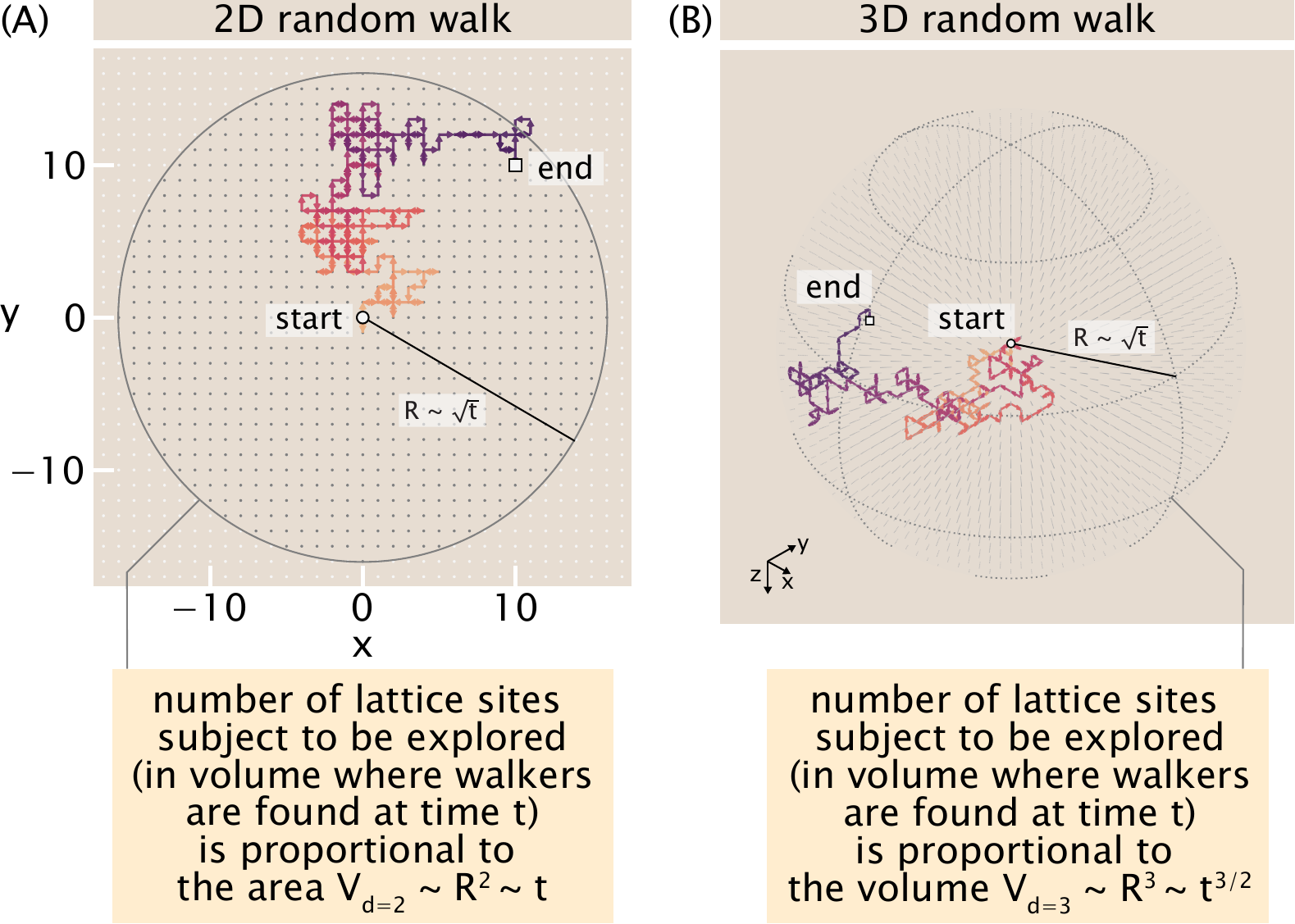}
    \caption{\raggedright The number of spatial locations standing to be explored at time $t$ explodes with dimensionality. This growing set of lattice sites standing to be explored are shown by gray dots found within a radius $R\thicksim \sqrt{t}$, describing the region where walkers are expected to reside.}
\label{fig:coverageCircleVersusSphereRandomWalks}
\end{figure*}

The foregoing analysis justifies, intuitively but concretely, why random walks explore so differently in one or two dimensional spaces, compared to higher dimensional spaces. Another vantage on this core transition in behavior with respect to dimensionality is to ask about the number of new sites actually  visited, relative to the number of sites that could feasibly be visited, up to time $t$, for random walks in various dimensions. To estimate the number of candidate sites to be explored, as we have argued, the typical region where a random walk is found has a characteristic radial size of $R\thicksim \sqrt{t}$. This means that random walks in dimensionality $d$ explore a volume \begin{align}
    V_d \thicksim R^d \thicksim t^{d/2}
\end{align} 
of possible candidate sites. As visualized for the humble cases of two and three dimensions in Fig.~\ref{fig:coverageCircleVersusSphereRandomWalks}, this volume of candidate sites to be explored grows very rapidly with dimensionality $d$. In stark contrast, the largest number of new sites a random walk can ever visit up to a time $t$ is at most $t$ (when every timestep leads to a new site). This competition between the limited rate of actually visiting fresh sites (optimistically linear in $t$) compared to the exploding volume of them that stands to be visited (a power law with exponent $d/2$ in dimension $d$) already anticipates that the latter can outpace the former near a marginal dimension around $d=2$. Making this point precise, define the expected density of visited sites in $d$-dimensions as $\rho_d(t)$: then,
\begin{align}
    \rho_d \leq \frac{t^1}{t^{d/2}} \thicksim t^{1-d/2},
\end{align}
which evinces that the density of visited sites decreases in time for $d\geq 3$. 

\begin{figure}
    \centering
    \includegraphics[width=0.5\linewidth]{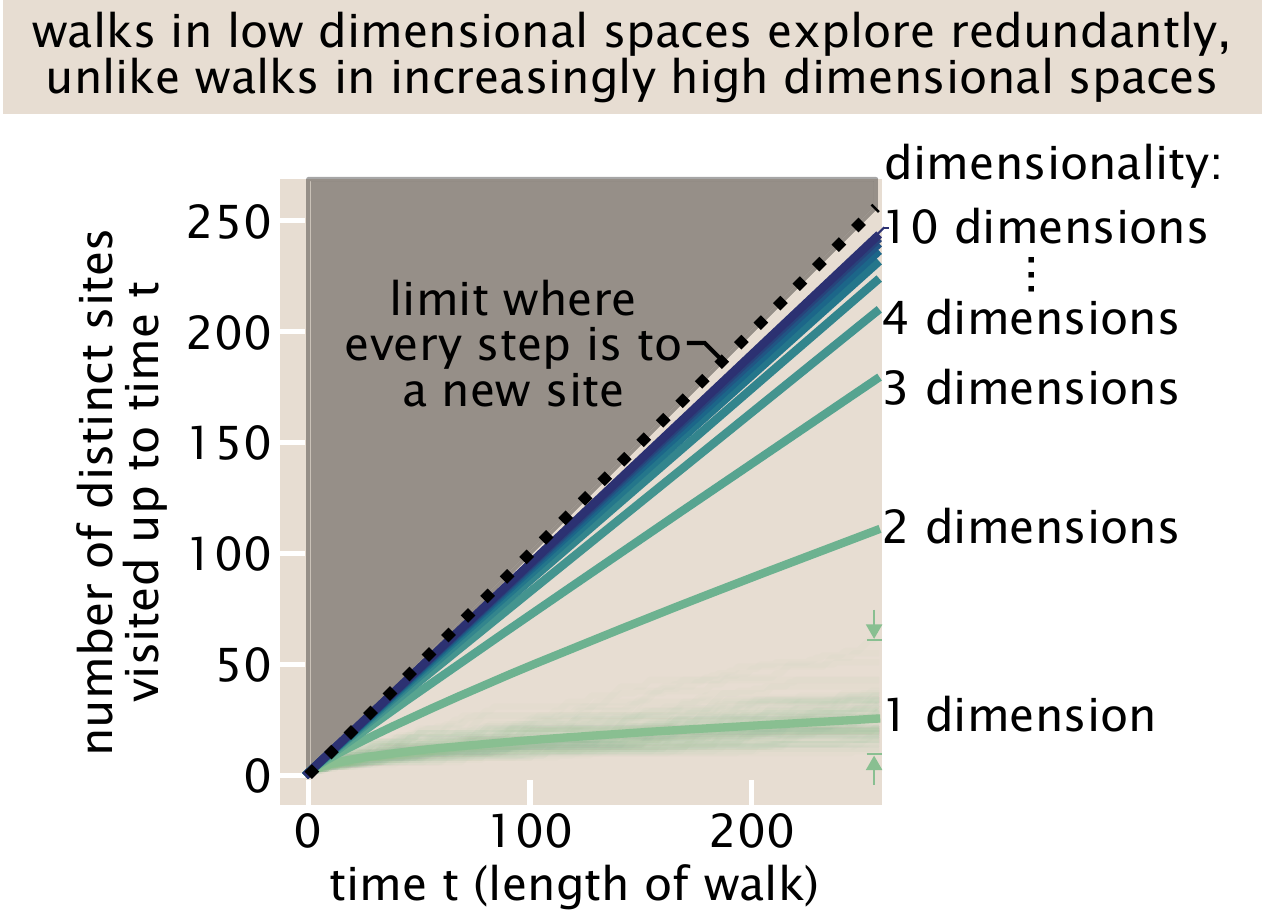}
    \caption{\raggedright Average coverage of new sites by random walks over time in spaces of varying dimensionalities. Specifically, for each dimensionality, 2400 random walks were performed; in each walk, the number of unique lattice sites visited up to time $t$ was counted, and the averages over these walks are reported as thick solid lines. (As a guide to the eye, the variability in this coverage is shown in the set of individual stochastic coverages in one dimension, shown as faint thin lines; arrows mark the extent of this variability.)   }
    \label{fig:dimensionalityBareWalkCoverage}
\end{figure}

In fact, the situation is more extreme than this picture captures: bare random walks do not sample fresh sites as fast as a linear rate in time. Instead, only a fraction of timesteps yield exposure to a new site not yet visited before, and this coverage rate itself depends heavily on dimension. This is shown explicitly by simulation in Fig.~\ref{fig:dimensionalityBareWalkCoverage}, where we perform walks in spaces up to ten dimensions, track the number of new locations sampled up to a given time $t$, and average over thousands of walks. Strikingly, particularly in spaces of low dimension such as $d=1$ or $d=2$, walks explore sites very thoroughly and redundantly as stressed by the starkly sublinear coverage curves. Only in spaces of more than a few dimensions does coverage begin to substantially resemble the ideal limit where every step leads to a fresh location. However, this coverage can still never exceed the speed limit of the $\thicksim t^1$ scaling shown as a black dotted line in Fig.~ \ref{fig:dimensionalityBareWalkCoverage}. The fundamental gulf between the fresh coverage a walk could ever expect to have, and the number of sites standing to be explored, expresses the essence of why bare random walks in high dimensional spaces may fundamentally fail to reach a target.

\FloatBarrier

\subsection{Impact of confinement and finite system sizes}
We just earned a feel for why unbounded random walks take an infinite time to reach a finite target, and why the probability of hitting a target becomes lower than one in spaces of sufficient dimensionality. However, an immediate, reasonable, and rich question arises. How does the effectiveness of random search change when we acknowledge that cells and organisms are not unbounded systems (for which we should be grateful), but rather have finite sizes? 

Of course, like many cells, the Zurich that P{\'o}lya walked in and found motivating was also finite. The original idealization to an infinite grid was very intentional: in P{\'o}lya's own words, ``If the question is conceived too narrowly, too realistically, \emph{au pied de la lettre}, with the actual data about the network of winding footpaths behind the [Zurich] hotel, it becomes unmanageably complicated and, moreover, uninteresting. Yes, the problem must be simplified. . .The boundary of such a regular network may be irregular---but let the city be very big, unlimited: this is the simplest case and the most interesting, too''\cite{polya1984collected}. Whether a cell or organism regards the math of infinite grids indeed ``most interesting,'' or instead feels more strongly finite-size effects imposed by confinement, depends on the degree of confinement. This surely varies richly across the spectrum of biology between systems that are small enough for random walks to search well enough, to those systems that are large or high-dimensional enough for further mechanisms to be helpful.

Technically, any such finite system size makes visiting a target site bound to happen eventually (giving $p_{\text{hit}}(\mathbf{x}_T)=1$) for any dimensionality, at least if given enough time, in contrast to the recurrence probability behaviors discussed earlier for searches in infinite spaces. Such confinement also makes the mean time to arrive at a target finite.

However, biological explorations must also occur within times acceptable to organisms. For instance, to elicit its function, a protein diffusing in a cell must encounter its destination or binding partner before being degraded (by active machinery or passive injury). Similarly, a starving bacterium or worm crawling through soil is in a race against time to encounter nutrients before their internal reserves run out. Growing neurons must encounter their synaptic partners in tolerable developmental times. Limited time also constrains decisions in higher dimensional space of cognitive behaviors. Accordingly, it is more revealing to ask about how finite physical sizes affect biological searches while acknowledging their accompanying finite time limits. 

First, consider the case of one dimension: how does confinement matter? 
Happily, this question is already analyzed by the main text's analysis of the heuristic that random walks with reset can be approximated as walking within a reflecting box. This led to eqn.~\ref{eq:mainHitTimeFromZeroIn1dReflectingBox} stating that in a one dimensional box where a reflecting boundary is located at $x=-L_R$, a target is located at $x=x_T$, and a random walk begins at the origin $x=0$, the expected time to first hit the target is 
\begin{align}
    T(0) &= \frac{x_T^2}{2D} + \frac{L_R x_T}{D} \\
    & \approx \frac{L_R x_T}{D}.
\end{align}

While the main text used the distance $L_R$ to reason about the effect of resets, here we can instead take this to mean the characteristic lengthscale of confinement we are interested in understanding the impact on. In summary, then, in one dimension, keeping the target location $x_T$ fixed but expanding the domain of confinement $L_R$ makes the average time to hit the target increase approximately linearly with $L_R$. Or else if the target's location is a fixed fraction of the system size $\thicksim L_R$ then as the system's size grows, the time to hit the target grows quadratically with the system size. Specifically, when the location of the target $x_T$ is some fixed fraction $\alpha$ of
the distance $L_R$ from the origin to the reflecting wall, namely
$x_T=\alpha L_R$, then, $
T(0)
=
\alpha\left(1+\frac{\alpha}{2}\right)\frac{L_R^2}{D}
\propto \frac{L_R^2}{D}$
that is, varying with the characteristic $L_R^2/D$ timescale one can anticipate just on dimensional grounds. 

Some biological systems enjoy a small enough characteristic size (and thus $L_R$) that the time to encounter targets is functionally acceptable: for instance, small molecules (diffusing at $D\gtrsim \text{few} ~\mu\text{m}^2/s$ in a cytosol) move across the length of a bacterium $L_R=1~\mu m$ in a few tenths of a second, or a typical mammalian tissue cell with $L_R=10 ~\mu \text{m}$ in a few tens of seconds. It takes over an hour for such a molecule to diffuse across a human egg cell with $L_R \approx 120~\mu$m. It would take almost a day to diffuse across a \emph{Drosophila} embryo of length $L_R = 500~\mu$m (comparable to the full time it actually takes for the larva to develop and hatch!), and if axons relied on diffusion alone to operate then transport of one molecule along their few centimeter lengths would take of order 10 years. 
In sum, even for one dimensional systems in physical space, some systems are small enough for bare random walks to be acceptable and others motivate the introduction of mechanisms such as drift (directed transport by motors), cues, and other contrivances.

Incidentally, this calculation also gives us a route to understand when resets can still be dynamically relevant when the biological space already enjoys confinement. Specifically, two length scales compete---the resets impose a lengthscale $L_{R,\text{reset}}$ and the system's size imposes a lengthscale $L_{R,\text{system}}$. If the system is already confined to a small enough $L_{R,\text{system}}$ then the resets may be slow enough (giving a long enough $L_{R,\text{reset}}$) not to matter significantly. Specifically, in this one-dimensional context, we can compare the average
time $T_{\text{reset}}$ to reach the target when the search is effectively confined by resetting with the time $T_{\text{system}}$ when it is instead confined by the finite size of the system. These times compare according
to the ratio,
\begin{align}
        \frac{T_{\text{reset}}} {T_{\text{system}}}& \approx \frac{\frac{L_{R,\text{reset}} x_T}{D}}{\frac{L_{R,\text{system}} x_T}{D}}\\
    & \approx \frac{L_{R,\text{reset}}}{L_{R,\text{system}}}\\
& \approx \frac{C \sqrt{DT_R}}{L_{R,\text{system}}}.
\end{align}
This makes clear that resetting matters in shortening the arrival time to the target ($\frac{T_{\text{reset}}} {T_{\text{system}}} \lesssim 1$) exactly when, 
\begin{align}
    T_R \lesssim \frac{L_{R,\text{system}}^2}{D}.
\end{align} 
Namely, in one dimension, reset can be relevant to help when the interval between resets is shorter than the characteristic time to diffuse across the system's finite length. 

\begin{figure}
    \centering
    \includegraphics[width=\linewidth]{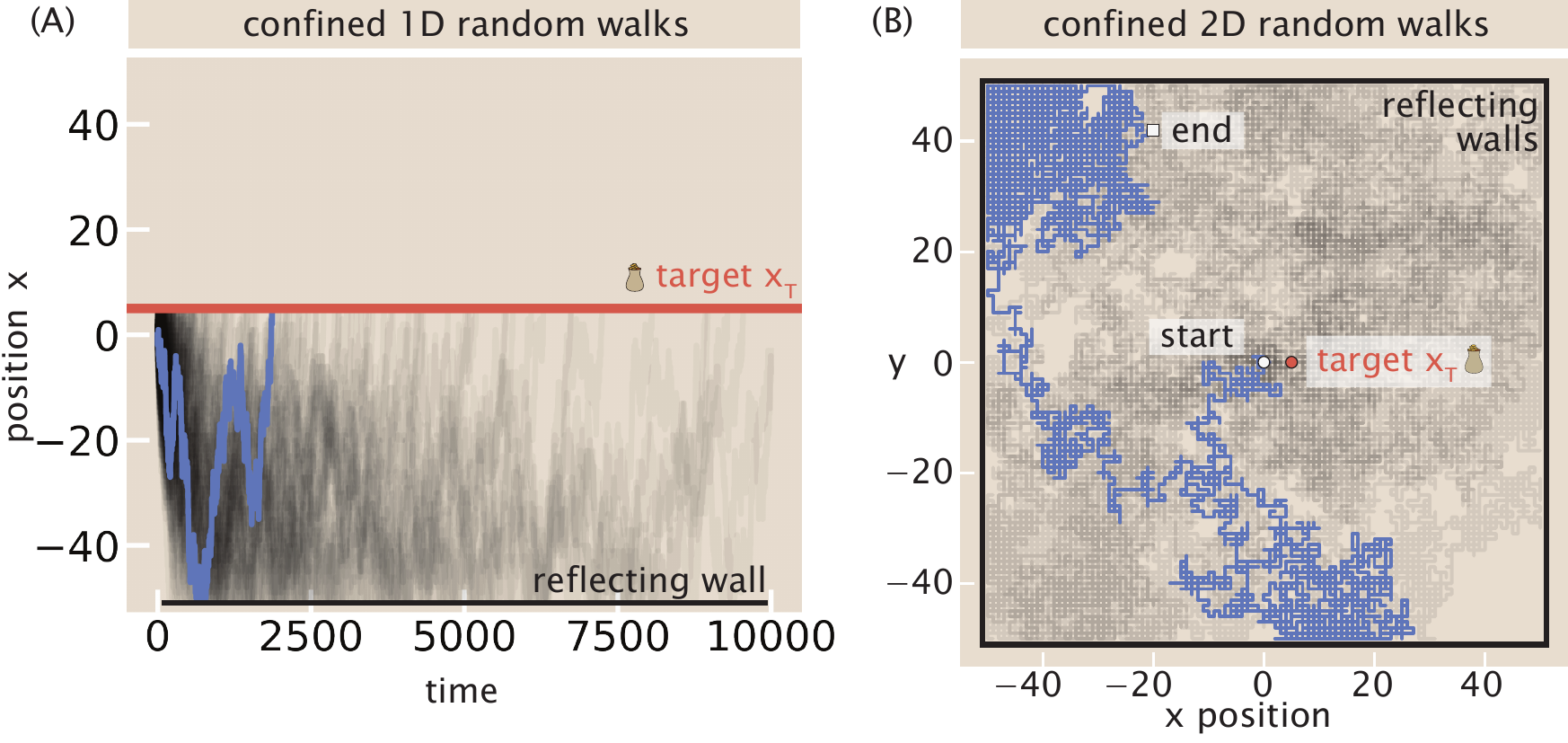}
    \caption{\raggedright Random walks in finite domains. This visualizes the confined case with $L_R=50$ in each coordinate dimension, small enough that walks can interact with the barrier before first arriving at the target. A target at $x_T=5$ in 1D (A), or at
$\mathbf{x}_T=[5,0]^\top$ in 2D (B), is shown as a red line or a red dot,
respectively. Every walk displayed is allowed up to $T=10000$ steps to find the target if it does so.}
    \label{fig:visualizeConstrainedWalks}
\end{figure}
\FloatBarrier

In higher dimensional spaces, such as two-dimensional or three-dimensional positions or larger state and behavioral spaces, the strong impact of dimensionality hinted at from our infinite-systems discussion persists in essence. To give a feel for such confined search processes, Fig.~\ref{fig:visualizeConstrainedWalks} illustrates simulated trajectories in boxes of size $L_R=50$ in one or two dimensions, with a target located only $x_T=5$ units away from the starting position of random walks. This system size is small enough for  walks to often interact with the reflecting boundaries before reaching the target, as visualized by two representative walks in blue in Fig.~\ref{fig:visualizeConstrainedWalks}(A) and (B). 

\begin{figure}[h!]
    \centering
    \includegraphics[width=0.7\linewidth]{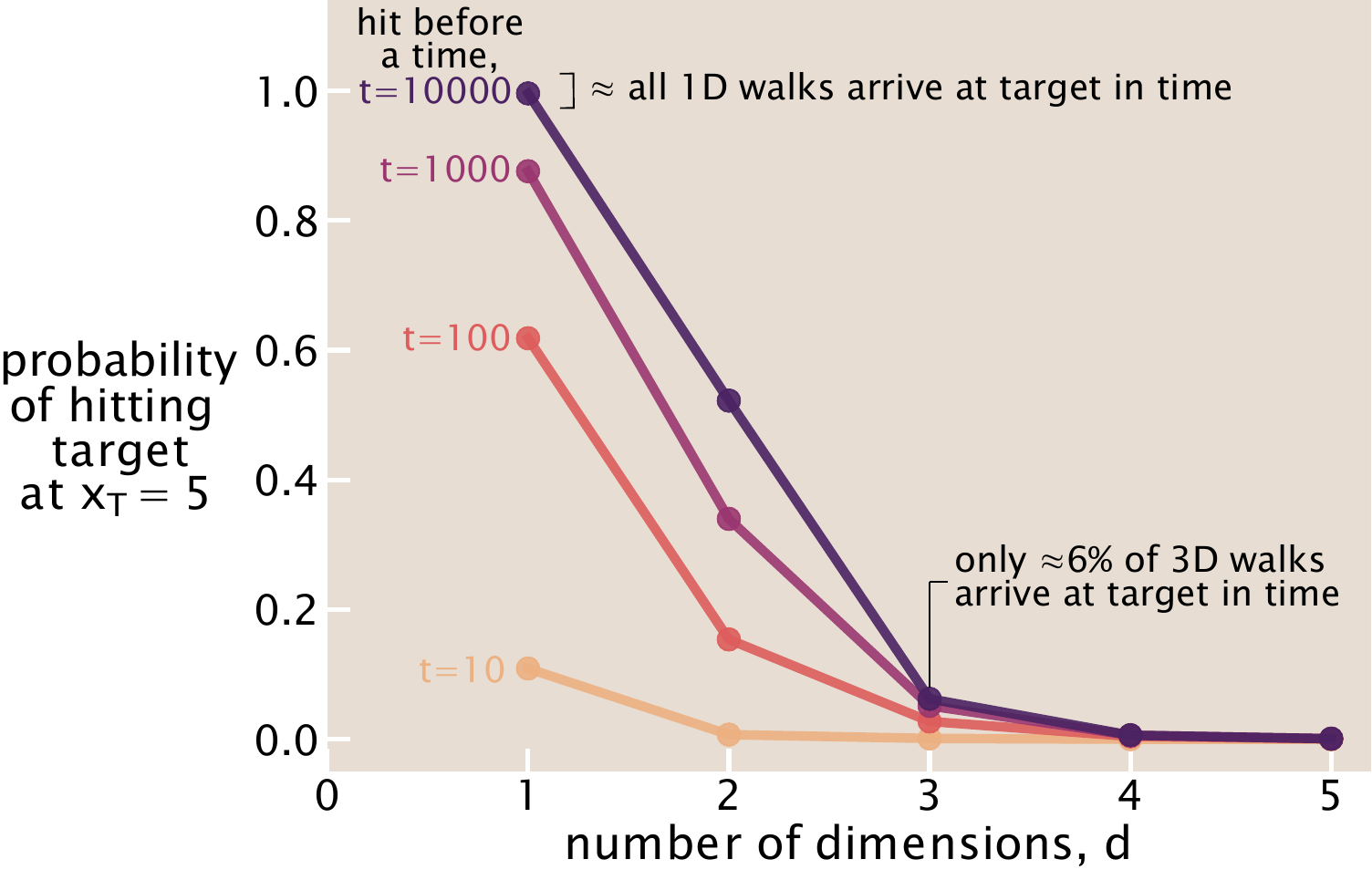}
    \caption{\raggedright Random walks in finite domains of varying dimensionalities. A target is localized at $\mathbf{x}_T = [5,0,0,\ldots, 0]$ in $d$-dimensional space. For each dimensionality $d$, an ensemble of 10000 random walks was performed, and the fraction of these walks that had hit the target by each allotted time is reported. The lines of increasingly orange color from top to bottom indicate the allowed times $t \in \{10,100,1000,10000\}$ steps.  (The width of the reflecting box was $L=50$ in each coordinate dimension.)}
    \label{fig:dimensionalityCollapseEvenInFiniteLattice}
\end{figure}

How many such walks reach the target five units away from the start in spaces of increasing dimensionality? Fig.~\ref{fig:dimensionalityCollapseEvenInFiniteLattice} reports the fraction of such fortunate walks under various budgets in allotted time. While in one dimension, nearly all walks arrive at the target within a generous time $t=10,000$ steps (as visualized by the dark purple line in Fig.~\ref{fig:dimensionalityCollapseEvenInFiniteLattice}), this drops to just above half in $d=2$ dimensions and only less than 6\% of such walks arrive at the target in three dimensions. More constrained time budgets (still large compared to the shortest possible time of $t=5$ required of the luckiest trajectories ballistically arriving at the target) drive this success probability dramatically down, as visualized by the lighter orange lines in Fig.~\ref{fig:dimensionalityCollapseEvenInFiniteLattice}. In summary, these exercises affirm that even in confined spaces, larger dimensionalities of spaces can punish the effectiveness of ordinary random walks.

\FloatBarrier

\subsection{Conclusions about bare random-walks}
Taken together, these stark mathematical behaviors could explain why biological systems often do not rely on ordinary random walks alone. Crucially, the behavioral and phase spaces that biological searchers explore are often much richer than literal (one, two, or three dimensional) physical positions. Cells express thousands of genes and so gene expression traverses spaces of thousands of dimensions; epigenetic modifications on polymers explore combinatorial spaces; the cytosolic soup bubbles with thousands of distinct chemical species (and adjusts their counts in regulatory spaces of comparable dimensionality). At a higher level, animal behaviors, let alone underlying cognition, are set by the exquisite coordination of myriad degrees of freedom. Accordingly, instead of such bare random walks, biological searchers may use superficially surprising modifications to this type of motion to accomplish searches much more effectively, as discussed with the case studies of resets, drift, cues, and checkpoints, as shown in the main text.

\subsection{Helpfulness of resets in random walks of higher dimensionality}
\label{sec:higherDmoreHelp}
The formidable influence of dimensionality on random walks also raises the significance of how resets might accelerate the time to encounter targets, as explored in the case of $d=1$ dimension in \S\ref{section:RandomWalkReset} in the main text. Since the failure of a seeker to ever find a target at all (or within a feasible time) is more likely in higher dimensional spaces, we might expect that the potential remedies of resets could give even starker benefits there than in lower dimensional spaces. Indeed,  Ref.~\cite{hartmann2025diffusion} insightfully analyzed a lattice model to find the optimal rates for reset in $d$ dimensions and found that more aggressive reset rates, approximately linearly scaling with $d$, manifest in higher dimensions. Much more thinking is warranted on such dynamics and policies, and whether richer effects unique to higher dimensional spaces emerge. But we expect these provocations are sure to be relevant to the large state, parameter, and behavioral spaces of biology.

\section{First-passage properties of ordinary random walks}
\label{app:ordinary_random_walks}


Here, we discuss transparently and explicitly 
how the probability of a random walker trajectory reaching a target for the first time can be calculated by an exact enumeration of the relevant lattice paths. Our spirit here is to be very concrete and explicit, and to provide details, illustrations, and remarks usually omitted or left implicit as homework for the reader in other wonderful foundational literature. Specifically, the approach shown here has an august history dating to Feller, Andre, Chandrasekhar, and other giants of probability. 
\begin{figure*}[h!]
    \centering
\includegraphics[width=\textwidth]{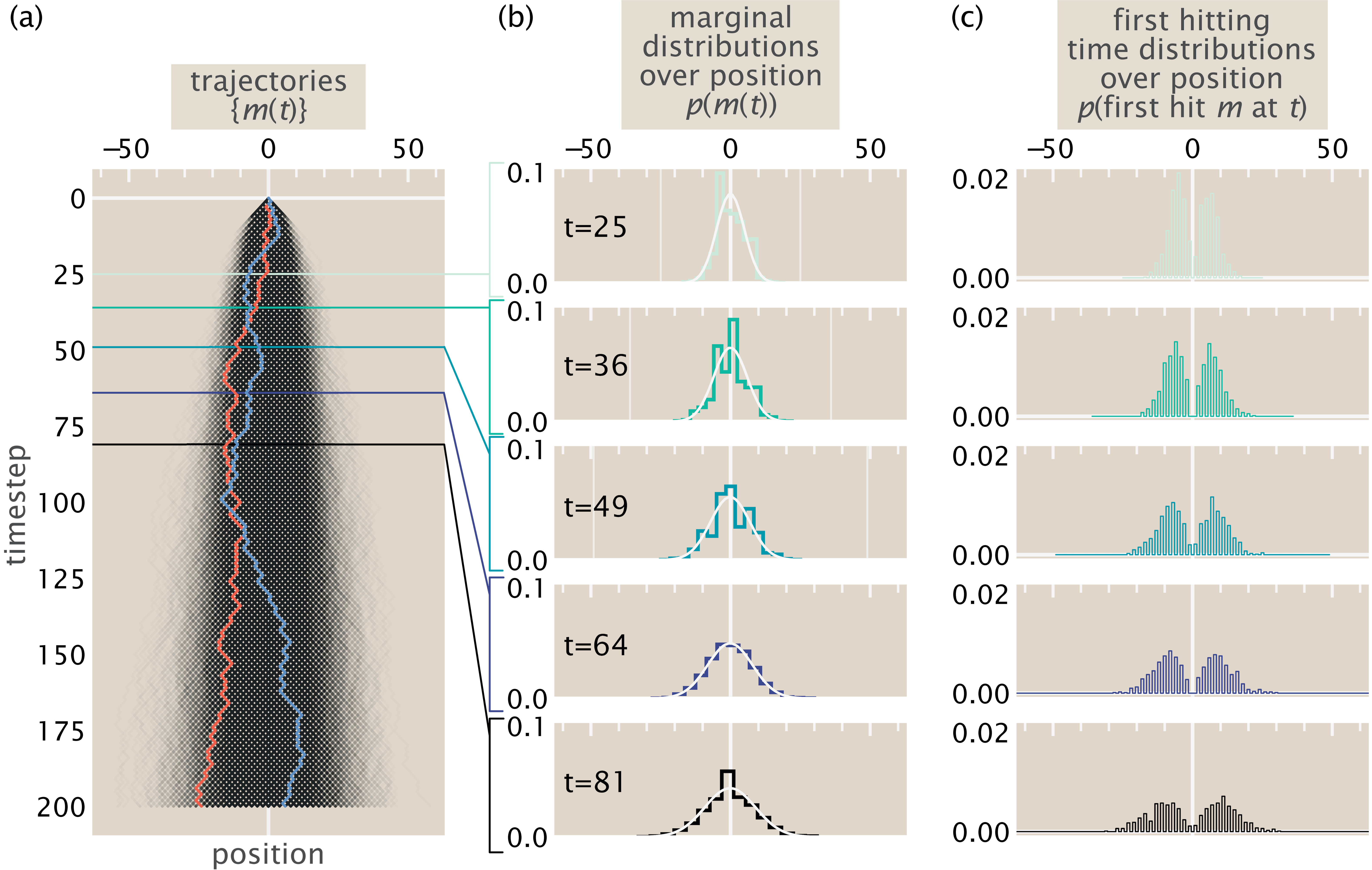}
    \caption{\raggedright Unbiased random walks in one dimension. (a) Trajectories of an ensemble of 8000 random walks over time (drawn in black, except for two arbitrary exemplary trajectories in blue or red, which respectively typify trajectories that succeed at visiting a target at $m=10$ or fail to do so within the indicated time window). (b) The marginal probability $p(m(t))$ of finding walkers at position $m(t)$ sliced at timepoints $t\in\{5^2, 6^2, 7^2, 8^2, 9^2\}$ (shown as increasingly deeply teal colored histograms corresponding to the horizontal lines in (a)); these timepoints are quadratically spaced to show how the variance scales approximately quadratically in time. The corresponding Gaussian density approximation with the corresponding mean and variance of the distribution is shown as a light white line for reference. (c) The empirical fraction of trajectories that hit position $m$ for the first time at time $t$ for these same example timepoints in (b). (Note the even-odd parity oscillation: if $m$ and $t$ have different parities, the probability of first hitting $m$ at time $t$ is zero.) }
\label{fig:ordinaryRandomWalks}
\end{figure*}
\FloatBarrier

What is the probability the walker is found at a position $m$ at time $t$? This net position is the difference between the number of rightwards steps $R$ accumulated up to time $t$, and the number of leftwards steps $L$ up to time $t$, namely $$m(t) = R - L;$$ 
and since every step must be right or left, 
$$t=R + L.$$ 
(Accordingly, note that $R=\frac{t+m}{2}$ and $L=\frac{t-m}{2}$: the latter expression requires that the total time $t$ and the net position $m$ must be both even or both odd, since their difference divided by two must be even because $L$ is an integer.)

The $R$ independent right steps have each occurred with probability $p$, and the left steps each with probability $1-p$; a particular sequence of these steps thus has probability $p^{R} (1-p)^{L} = p^{R} (1-p)^{t-R}$. However, there are many ways of arranging these total number of rightwards and leftwards steps; specifically there are ${t \choose R} = {t \choose L}$ ways, so the total probability of all ways to take $R$ steps to the right in a total trajectory of length $t$ is
\begin{align}
    \mathbb{P}\left(X(t) = m\right) &= \mathbb{P}(R - L = m) \\
    &= \mathbb{P}(R \text{ steps among } t) = {t \choose R} p^R (1-p)^{t-R} \\
    &= {t \choose \frac{t+m}{2}} p^{\frac{t+m}{2}} (1-p)^{\frac{t-m}{2}}
    \label{eq:binomial}
\end{align}
Naturally, this is the Binomial distribution with success parameter $p$ for $t$ trials; it has mean $\langle m(t) \rangle = t p$ and variance $\sigma_m^2 = \langle m^2 \rangle - \langle m \rangle^2 = tp(1-p)$.
For familiar reference, we plot an ensemble of unbiased (e.g. $p=1/2$) random walks in Figure \ref{fig:ordinaryRandomWalks}(A).

\subsection{First passage times by lattice counts}

Next we ask a different question than the probability of the walker's instantaneous position: what is the probability that the walker visits site $m$ \emph{for the first time} at time $t$? To simplify this discussion, we now specialize to the unbiased walk case $p=1/2$, which clarifies our problem as one of purely \emph{counting} desirable trajectories (that successfully terminate at $m$ but never have visited it before time $t$). 

To proceed, we choose to generalize the reasoning involved in a popular proof for the form of the Catalan numbers, commonly attributed to Andr{\'e} \cite[p. 276-278, ch. 9]{koshy2008catalan} and (separately) Feller \cite[p. 138, \S A2]{stanley2015catalan}. To lay the groundwork for our argument, as a clearer pedagogical example, we first give a very brief account of this tactic here.

\begin{tcolorbox}[breakable, parbox=false,  boxrule=0pt]
\emph{The Catalan numbers count special paths.}
The Catalan numbers are a sequence of numbers that show up with spooky regularity across counting problems that look completely different from each other (\cite{stanley2015catalan, koshy2008catalan}). The $n$th Catalan number is defined as,
\begin{align}
        C_n & \equiv \frac{1}{n+1}{2n\choose n} \label{eq:CatalanOrigDef}\\
        & = {2n \choose n} - {2n \choose n-1}. \label{eq:CatalanBetter}
    \end{align}
One beautiful combinatorial interpretation of the $n$th Catalan number is the number of lattice paths of length $2n$ connecting an origin $(0,0)$ to a point $(n,n)$ at an upper right corner of a two-dimensional grid, in such a way that the paths do not cross (but may touch) the $y=x$ diagonal, and where paths consist only of up $(0,+1)$ or right $(+1,0)$ steps. These path requirements are visualized in Fig. \ref{fig:catalanRequirement}, where an example of a good (valid) path not crossing the $y=x$ diagonal is shown in {\color[HTML]{AC85AC}purple} and a bad path that crosses this diagonal is shown in {\color[HTML]{EAC164}light orange}. \newline

{{\centering
\includegraphics[width=0.3\textwidth]{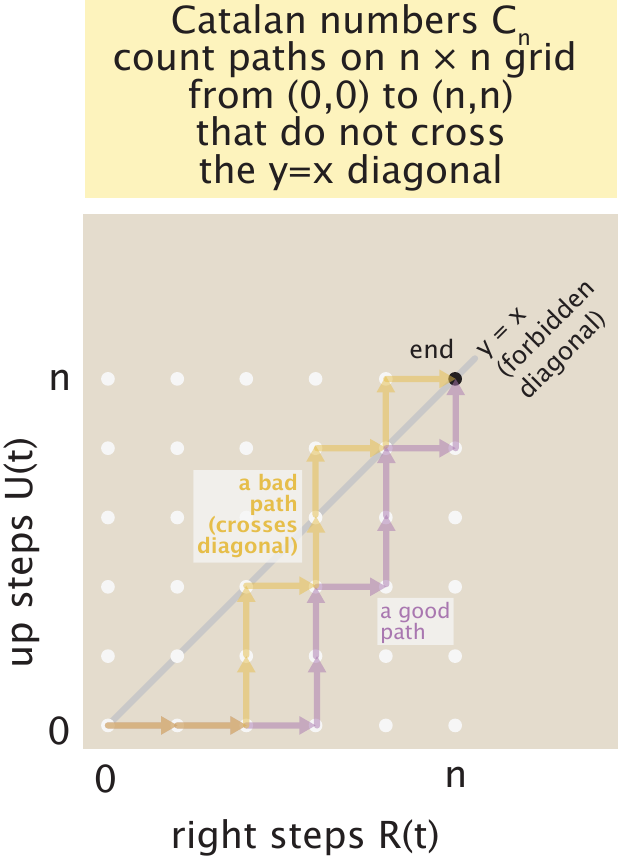} 
\captionof{figure}{Examples of good paths counted (and bad paths not counted) by the Catalan numbers.}
\label{fig:catalanRequirement}}}
\vspace{10pt}

These special lattice path objects seem like a very annoying set to count tidily; it is not obvious that they would somehow be accounted for by the concise algebraic definition in eqn.~\ref{eq:CatalanOrigDef}. However, the equivalent definition eqn.~\ref{eq:CatalanBetter} of the Catalan numbers gives clearer understanding. Specifically the subtraction in eqn.~\ref{eq:CatalanBetter} anticipates that the desired special objects can be accounted for by an easy count of a lager set of paths they belong to, minus some bad paths that violate their particular requirements. (This tactic will recur in later throughout our random walk analysis.)

To count these special permitted paths, the number of ``bad'' paths (that touch the diagonal prematurely at least once) can be subtracted from the total number of paths from $(0,0)$ to $(n,n)$ (good or bad). This latter total number of paths is straightforward to count, being simply the number of ways of choosing $n$  rightwards steps among the total number of steps $2n$, namely ${2n \choose n}$. The essential trickiness afflicts  the number of bad paths. Andr{\' e} (or Feller's) beautiful gambit, sometimes known as the \emph{reflection method}, is to count these bad paths by building a one-to-one mapping to \emph{another} set of paths that are much easier to count. We illustrate the important steps of this approach in Figure \ref{fig:catalanReflection}.
\begin{center}
\includegraphics[width=\textwidth]{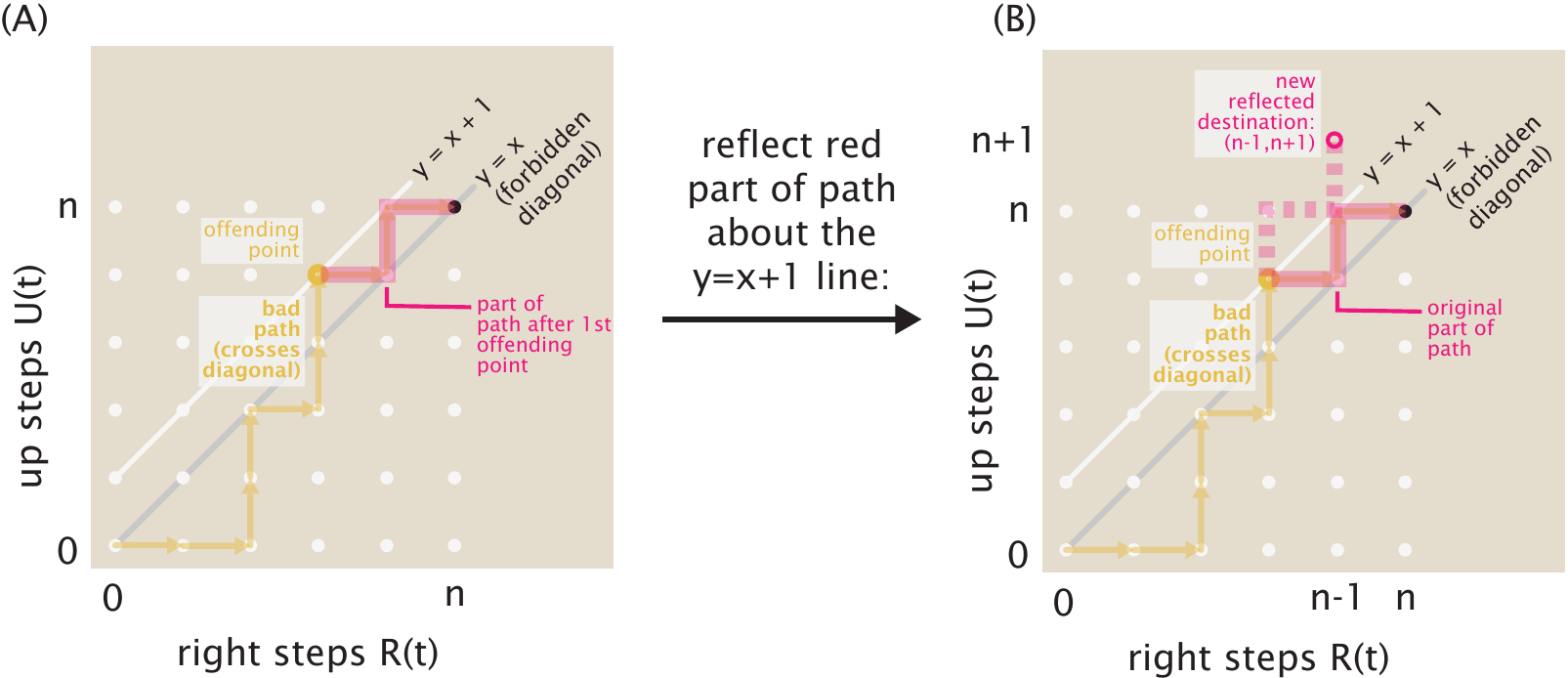} 
\captionof{figure}{Reflection procedure to count bad paths to exclude from Catalan counts. (a) Divide a bad path into parts before and after the point {\color[HTML]{EAC164}$\circ$} after it has first offendingly crossed the $y=x$ diagonal (the former subpath marked as an orange line {\color[HTML]{EAC164}$-$}, the latter emphasized as a thick pink line {\color[HTML]{DBABB8}\rule[0.35ex]{1em}{3pt}}). (b) Reflect the part of this path {\color[HTML]{DBABB8}%
 \raisebox{0.15ex}{%
   \rule{0.35em}{3pt}\kern0.3em
   \rule{0.35em}{3pt}\kern0.3em
   \rule{0.35em}{3pt}%
 }} after the offending point {\color[HTML]{EAC164}$\circ$} about the $y=x+1$ line. This reflected subpath (emphasized with a thick pink dashed line {\color[HTML]{DBABB8}%
 \raisebox{0.15ex}{%
   \rule{0.35em}{3pt}\kern0.3em
   \rule{0.35em}{3pt}\kern0.3em
   \rule{0.35em}{3pt}%
 }}), together with the other original part of the path {\color[HTML]{EAC164}$-$}, creates a new path from $(0,0)$ to a new point $(n-1, n+1)$, {\color[HTML]{EE337F}$\circ$}.}
\label{fig:catalanReflection}
\end{center}

First, as shown in Fig. \ref{fig:catalanReflection}(A), take a bad path and find the first premature point {\color[HTML]{EAC164}$\circ$}  where it has just naughtily crossed the diagonal $y=x$. This offending point {\color[HTML]{EAC164}$\circ$}  lies on the line $y=x+1$. Divide the bad path into two parts: as shown in Fig. \ref{fig:catalanReflection}(A), that subpath before this offending point, {\color[HTML]{EAC164}$-$}, and that subpath after, {\color[HTML]{DBABB8}\rule[0.35ex]{1em}{3pt}}. Then, reflect the {\color[HTML]{DBABB8}\rule[0.35ex]{1em}{3pt}} part of the path after the offending point about the $y=x+1$ diagonal that it is touching (by swapping every rightward move for an upward one): the outcome is visualized as a dashed pink line {\color[HTML]{DBABB8}%
 \raisebox{0.15ex}{%
   \rule{0.35em}{3pt}\kern0.3em
   \rule{0.35em}{3pt}\kern0.3em
   \rule{0.35em}{3pt}%
 }} in Fig.~\ref{fig:catalanReflection}(B). This new path has a different ultimate destination {\color[HTML]{EE337F}$\circ$} than the original terminal point $(n,n)$, because by virtue of having \emph{crossed} the diagonal, the offending point shows an imbalance of rightward and upward moves. Specifically, since each path is $2n$ steps long (e.g. is even), and the unreflected {\color[HTML]{EAC164}$-$} part of the path had one more upwards move than rightwards move (having arrived at the $y=x+1$ line), the reflected part of the path {\color[HTML]{DBABB8}%
 \raisebox{0.15ex}{%
   \rule{0.35em}{3pt}\kern0.3em
   \rule{0.35em}{3pt}\kern0.3em
   \rule{0.35em}{3pt}%
 }} reaches the new destination $(n-1,n+1)$, {\color[HTML]{EE337F}$\circ$}. By taking this reflected second part of the path (together with the unreflected initial part of the path), a new path from $(0,0)$ to $(n-1,n+1)$ has been created. These reflected paths from $(0,0)$ to $(n-1,n+1)$ (as an unrestricted set) are one-to-one with the original bad paths from $(0,0)$ to $(n,n)$ (since the reflection operation can be uniquely reversed); this bijection means that these sets of paths have the same count. There are ${2n \choose n-1}$ such unrestricted paths from $(0,0)$ to $(n-1,n+1)$, so the total number of inoffensive good paths is $\boxed{C_n = {2n \choose n} - {2n \choose n-1}}$.
\end{tcolorbox}
Now, we adapt the essence of this tactic to confront our present problem (counting the 1D random walk paths that arrive at a net position $m$ for the first time after exactly $t$ steps). The gambit is the same as above, save for some key details.

\begin{tcolorbox}[breakable, parbox=false,  boxrule=0pt]
    \emph{Generalized counting of arbitrary first arrival paths.} Counting the number of walk trajectories that terminate at a final position of $m = R(t)-L(t)$ in a total of $t=R(t)+L(t)$ steps, crucially \emph{without} visiting $m$ beforehand, shares the structure of the classic Catalan counting problem above, but with two key differences. First, the forbidden line that permissible trajectories must avoid is not the identity diagonal $R=L$ (e.g. $y=x$) line, but instead the $m= R - L$ line. Second, whereas the Catalan counting problem permitted touches to the diagonal line---and in particular identified all bad paths exactly by the property that they all crossed it---our setting demands we further exclude paths that even merely touch the $m=R-L$ line. Summarizing in other words, the Catalan problem counts all paths with $R>L$ as bad; our current problem (regarding the first passage of a random walk) counts all paths with $R\geq L+m$ as bad. 

    Accordingly, we change the procedure successful for the Catalan problem to this setting as follows. As before, we operate in the two-dimensional Cartesian lattice defined by the running number of rightward $R(t)$ and left $L(t)$ steps, where moves are depicted by right and upwards arrow moves in the $(R(t),L(t))$ lattice, respectively. Recall that in this lattice, descending diagonal lines of slope $-1$ together with $y$-intercept of $L(t)=t$ describe all endpoints of paths of total length $t$; conversely, ascending lines of slope $+1$ define the final positions $m = R(t) - L(t)$ of trajectories (as illustrated in Fig. \ref{fig:RLlatticeOrientation}). Thus, all valid trajectories of interest terminate at the intersection of these two lines $R(t) - L(t) =m$ and $R(t) + L(t) = t$, denoted by the black dot $\bullet$ in Fig.~\ref{fig:RLlatticeOrientation}.  
\begin{center}
\includegraphics[width=0.4\textwidth]{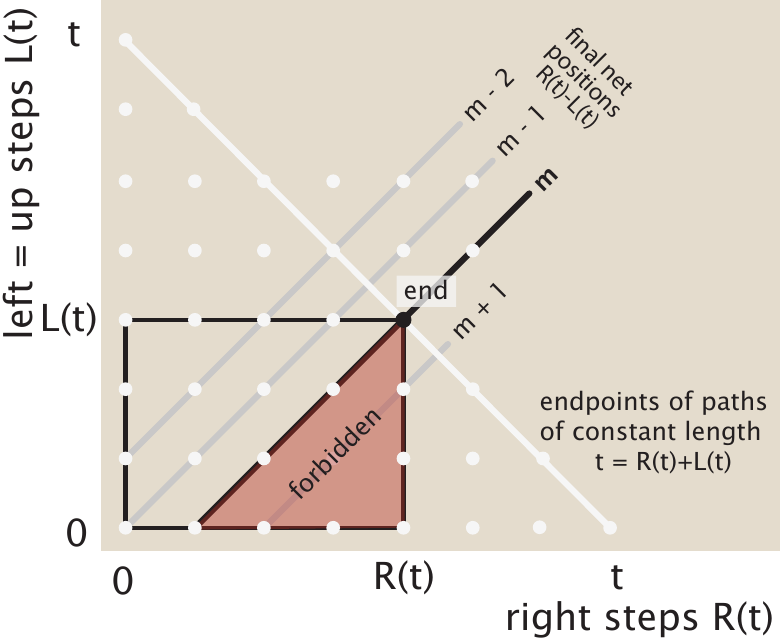}
\captionof{figure}{Properties of the lattice of the running number of rightward steps $R(t)$ against the running number of left steps $L(t)$ (plotted as upwards in the lattice). Descending slope-one lines characterize trajectories of different total duration $t=R(t)+L(t)$, and ascending slope-one lines give the possible net endpoints $m=R(t)-L(t)$ of paths. (For example, by way of illustration, here all trajectories terminating at the black dot $\bullet$ endpoint satisfy $R(t) = 4$, $L(t) =3$, $m=1$, and $t=7.$)}
\label{fig:RLlatticeOrientation}
\end{center}

To accommodate the new requirements, we first recognize that all good paths terminating in such a net position $m$ (let $m>0$ without loss of generality) necessarily end in a rightwards move, {\color{red}$\rightarrow$}, as shown in Fig.~\ref{fig:RLlatticePenultimateconceit}(A).\footnote{(Or will necessarily end in a leftwards move, if contrary to assumption, $m<0$.)}  (Specifically, note that obviously there are possible paths ending at the endpoint using a leftwards move; but by necessity all of these paths will have already transgressed with respect to the forbidden $m=R-L$ line at some earlier point, as shown by Fig.~\ref{fig:RLlatticePenultimateconceit}(B).) Thus, instead of counting paths terminating at $(R(t), L(t))$, we specialize instead to counting those paths that terminate at the new endpoint of $(R(t)-1, L(t))$ (shown as a magenta dot {\color[HTML]{A3509F}$\circ$} in Fig.~\ref{fig:RLlatticePenultimateconceit}(C)) and then deterministically take the required final rightwards step. 
So, our count is actually confined to a smaller sublattice from the origin $(0,0)$ to the effective new endpoint $(R(t)-1, L(t))=\left(\frac{t+m}{2}-1, \frac{t-m}{2} \right)$ (a region outlined as a magenta box in Fig.~\ref{fig:RLlatticePenultimateconceit}(C)). This leaves some remaining nontrivial bad paths to account for, as in Fig.~\ref{fig:RLlatticePenultimateconceit}(D). 

\begin{center}
\includegraphics[width=0.8\textwidth]{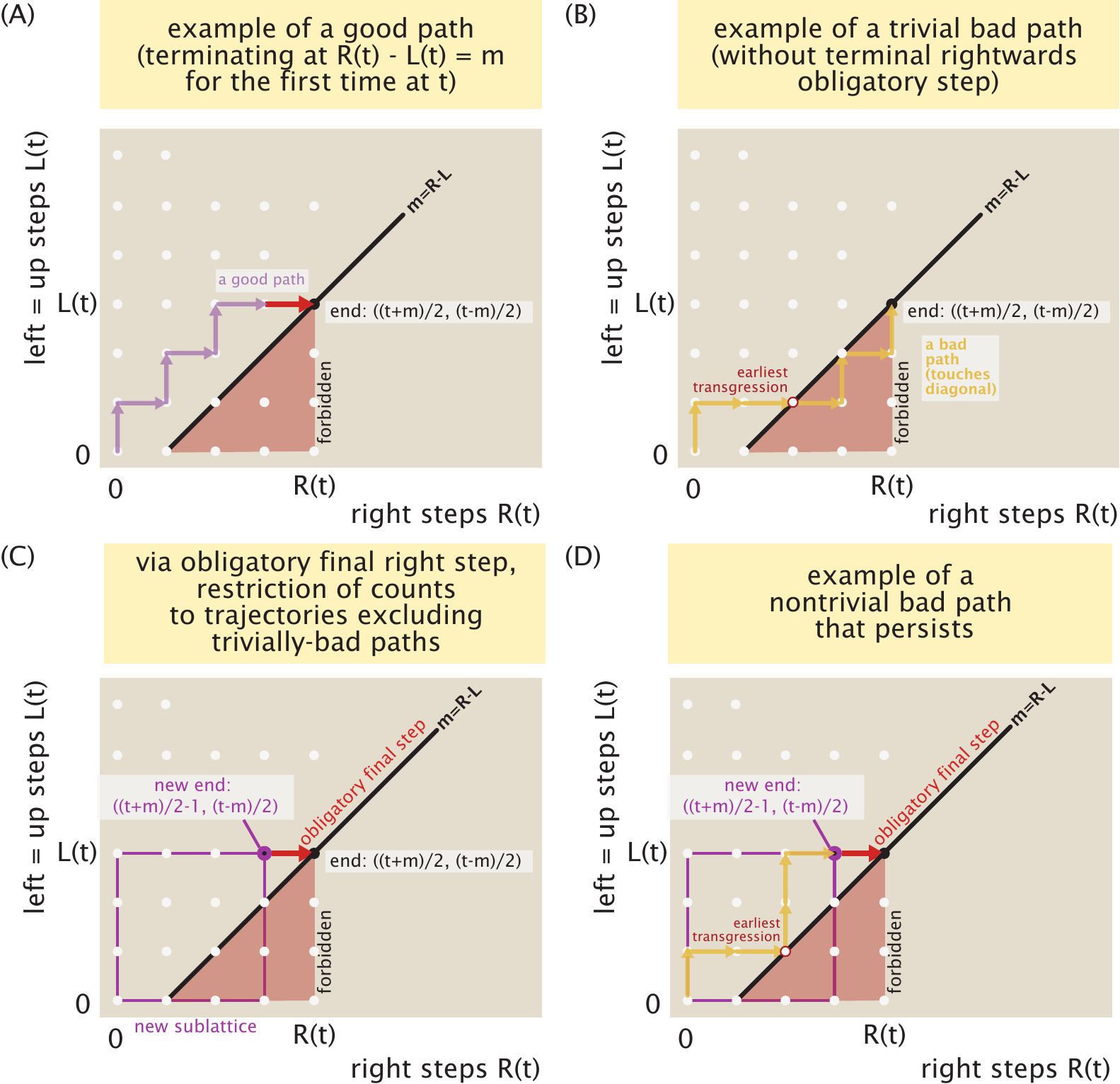}
\captionof{figure}{Structure of valid paths. (A) Example of good path showing the required final rightwards steps. (B) Example of a trivially bad path, which we can anticipate is automatically problematic since it lacks the required final rightwards move. (C) Counting valid paths to the destination $(R(t),L(t)) = \left(\frac{t+m}{2}, \frac{t-m}{2} \right)$, denoted with the black dot $\bullet$, can be narrowed to the count of valid paths to $(R(t)-1, L(t))=\left(\frac{t+m}{2}-1, \frac{t-m}{2} \right)$, {\color[HTML]{A3509F}$\circ$},  that then commit to the only feasible final rightwards step, {\color{red}$\rightarrow$}. (D) Example of a nontrivial bad path that manifests this required final rightwards move, but still earlier transgresses.}
\label{fig:RLlatticePenultimateconceit}
\end{center}

Now our gambit is largely similar to the Catalan numbers, up to the change in the offending diagonal we care about (avoiding crossing). Specifically, the total number of all paths (good or bad) are those ending at the new endpoint $(R(t)-1, L(t))=\left(\frac{t+m}{2}-1, \frac{t-m}{2} \right)$, {\color[HTML]{A3509F}$\circ$}, giving a total count ${t-1 \choose \frac{t-m}{2}}$, again counting all choices of where rightward steps appear among all $t-1$ timesteps of the relevant trajectory. 

We now count the nontrivial bad paths among these by establishing a bijection with easier-to-count paths, again by choosing a suitable reflection. As shown by the two examples in Fig.~\ref{fig:twoSuccessfulWalkerReflections}, first find the earliest point where a bad path offensively touches the relevant diagonal $R(t) - L(t) = m$ line. Consider all moves subsequent to this offending point (highlighted in thick pink {\color[HTML]{DBABB8}\rule[0.35ex]{1em}{3pt}} in Fig.~\ref{fig:twoSuccessfulWalkerReflections}), and for every rightwards move swap out a leftwords move, and vice versa, to accomplish a local rotation about the diagonal. The resulting path, highlighted in dashed pink {\color[HTML]{DBABB8}%
 \raisebox{0.15ex}{%
   \rule{0.35em}{3pt}\kern0.3em
   \rule{0.35em}{3pt}\kern0.3em
   \rule{0.35em}{3pt}%
 }} in  Fig.~\ref{fig:twoSuccessfulWalkerReflections}, will always map the previous terminal point {\color[HTML]{A3509F}$\circ$} at $\left(\frac{t+m}{2}-1, \frac{t-m}{2} \right)$ to a new reflected destination point $\left(\frac{t+m}{2}, \frac{t-m}{2}-1 \right)$, denoted as {\color[HTML]{EE337F}$\circ$} in Fig.~\ref{fig:twoSuccessfulWalkerReflections}. This new reflected destination does not vary over bad paths, as illustrated by Fig.~\ref{fig:twoSuccessfulWalkerReflections}. (This same destination point results because the former destination point $\left(\frac{t+m}{2}-1, \frac{t-m}{2} \right)$ definitionally has an extra leftwards move compared to all points on the diagonal $m=R-L$, so a reflection about this line decreases the number of leftwards moves and increases the number of rightwards moves by one.)

\begin{center}
\includegraphics[width=0.8\textwidth]{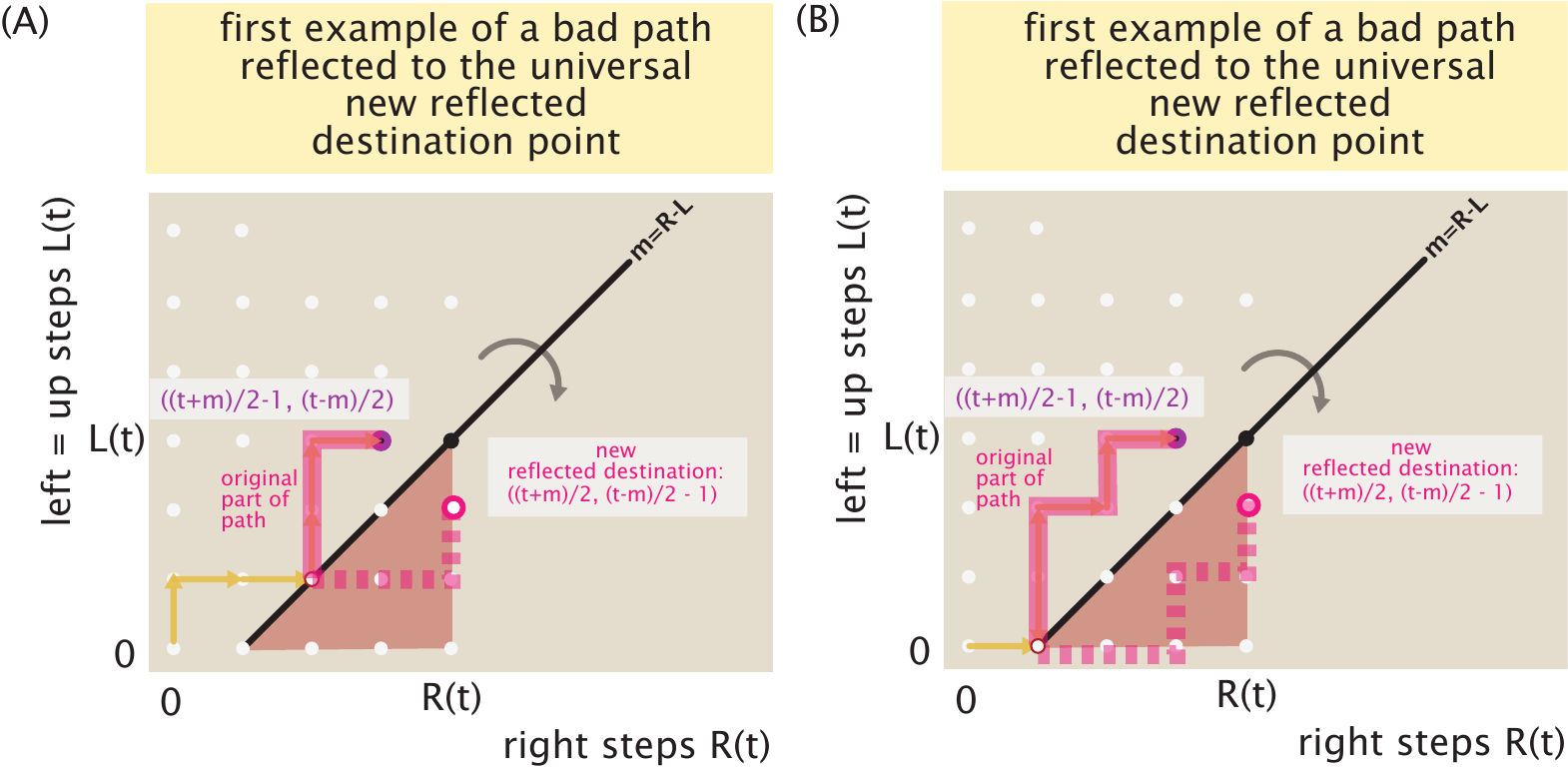}
\captionof{figure}{Two examples of applying a reflection to map distinct bad paths to the same reflected destination point, enabling counting. Given an example of a bad path as in (A) or (B), find the first point offending point where the path intersects the diagonal line $m=R(t) - L(t)$ prematurely, and retain the path's moves before this offending point, denoted by an orange line {\color[HTML]{EAC164}$-$}. Reflect the subsequent steps outlined in pink {\color[HTML]{DBABB8}\rule[0.35ex]{1em}{3pt}} with respect to the diagonal, yielding the resulting reflected subpath given by the dashed pink line {\color[HTML]{DBABB8}%
 \raisebox{0.15ex}{%
   \rule{0.35em}{3pt}\kern0.3em
   \rule{0.35em}{3pt}\kern0.3em
   \rule{0.35em}{3pt}%
 }}. No matter where the original bad paths (and in particular where their earliest transgressions specifically occur), the resulting destination of the reflected subpath arrives at the same new reflected destination point, always located at $\left(\frac{t+m}{2}, \frac{t-m}{2}-1 \right)$, denoted as {\color[HTML]{EE337F}$\circ$}.      }
\label{fig:twoSuccessfulWalkerReflections}
\end{center}

Since this reflection process is unique and reversible, we have found a correspondence the bad paths of interest to all lattice trajectories from the origin $(0,0)$ to the point $\left(\frac{t+m}{2}, \frac{t-m}{2}-1 \right)$. These lattice paths are still length $(t-1)$, and so there are ${t-1 \choose \frac{t+m}{2}}$ bad paths.

In sum, we just established that the number of paths that arrive at position $m = R(t) - L(t)$ for the first time exactly at time $t$ is
\begin{align}
    \#(m,t) &  = {t-1 \choose \frac{t-m}{2}} - {t-1 \choose \frac{t+m}{2}}\\
    &= \boxed{\frac{m}{t}{t \choose \frac{t-m}{2}}}. \label{eq:resultLatticePathCount}
\end{align}

\end{tcolorbox}

Note some special cases of this eqn.~\ref{eq:resultLatticePathCount}:
\begin{itemize}
    \item As noted early on, this relation is evaluated only for integer numbers of left or right steps, so we will regard this count as returning 0 when $t$ and $m$ differ in parity; 
    \item $t=m$ returns $\#(m,t) = 1 \times {t \choose 0} = 1$ path, as we intuitively require (the only way to get to position $m$ in exactly $t$ steps is to go there ballistically for $t$ e.g. consecutive rightwards steps). 
\end{itemize}
So for a random walk that steps right with probability $p$, the probability of first hitting $m$ at $t$ is just this degeneracy times the probability of taking the requisite number of rightwards steps,
\begin{align}
    \mathbb{P}(m,t) = \begin{cases}
        \frac{m}{t}{t \choose \frac{t-m}{2}}p^{(t+m)/2}(1-p)^{(t-m)/2} & \text{if $t \equiv m \mod 2$}, \label{eq:firstHitProbLatticeCountExact} \\
        0 & \text{else}.
    \end{cases}
\end{align}
or in the case of an unbiased random walk with $p=(1-p)=1/2$,
\begin{align}
    \mathbb{P}(m,t) = \begin{cases}
    \frac{m}{t}{t \choose \frac{t-m}{2}} \frac{1}{2^t}& \text{if $t \equiv m \mod 2$},\\
        0 & \text{else}.
    \end{cases}    
\end{align}

\subsection{Interpretations and continuum approximations}
To help interpret the first hitting probability specified by eqn.~\ref{eq:firstHitProbLatticeCountExact}, we note that the total marginal probability of visiting site $m$ at time $t$, regardless of whether it has been visited before, is just,
\begin{align}
    \mathbb{P}_{\text{marginal}}(m,t)={t \choose \frac{t-m}{2}}~p^{(t+m)/2}(1-p)^{(t-m)/2}.
\end{align}
(This follows from simply counting the paths of length $t$ with an excess of $m$ rightwards steps, without any further requirements.) This expression reveals an interesting connection: the first-hitting probability eqn.~\ref{eq:firstHitProbLatticeCountExact} is $m/t$ times the overall probability of the marginal probability, $$\mathbb{P}(m,t) = \frac{m}{t} \mathbb{P}_{\text{marginal}}(m,t).$$ 
This says that of all the paths that successfully end at site $m$ in time $t$, a fraction $m/t$ of them do so for the first time. 

The marginal distribution of a random walker's position $\mathbb{P}_{\text{marginal}}(m,t)$ famously approaches a Gaussian distribution in the small-timestep, large $t$ continuum limit. Specifically, for the unbiased $p=1/2$ case,
\begin{align}
    \mathbb{P}_{\text{marginal}}(m,t) \approx \frac{1}{\sqrt{2\pi t}} \exp\left[-\frac{m^2}{2t} \right].
\end{align}
This limit for the marginal distribution gives the analogous limit for the first-hitting distribution as,
\begin{align}
\label{eq:continuumApprox_f(t)ForFirstHit1DrandomWalk}
     \mathbb{P}(m,t) \approx \frac{m}{\sqrt{2\pi} t^{3/2}} \exp\left[-\frac{m^2}{2t} \right].
\end{align}

\subsection{Survival probability}
The probability that a walker has \emph{not} hit position $m$ by a time $t$ is equal to the chance it has not first hit $m$ at all preceding times $t'$, namely these nasty expressions,
\begin{align}
    \mathbb{P}(m,>t) & = 1- \sum_{t'=m}^t \mathbb{P}(m,t')\\
    & = 1- \sum_{t'=m}^t\frac{m}{t'}{t' \choose \frac{t'-m}{2}} \frac{1}{2^{t'}} \hspace{22pt}{\footnotesize\text{(now reindex: define $2\ell\equiv t-m$ as time exceeding the ballistic trajectory)}}\\
    & = 1-\frac{m}{2^m}\left( \sum_{\ell=0}^{\ell = \frac{t-m}{2}} \frac{1}{m+2\ell}\frac{1}{2^{2\ell}} {m+2\ell \choose \ell}\right)\\
   & = 1-\frac{m}{2^m}\left(\frac{1}{m} + \frac{1}{m+2}\frac{1}{2^2}(m+2) + \frac{1}{m+4}\frac{1}{2^4}\frac{(m+4)!}{(m+2)! 2!}+\cdots\right)\\
   & = 1-\frac{m}{2^m}\left(\frac{1}{m} + \frac{1}{2^2} + \cdots + \frac{1}{t-2} \frac{1}{2^{t-m-2}} {t-2 \choose \frac{t-m}{2}}+\frac{1}{t} \frac{1}{2^{t-m}} {t \choose \frac{t-m}{2}}\right)
   \label{eq:survivalProbRandomWalkerVeryExplicit}
\end{align}

These expressions smoothly capture the algebraic cases discussed in the main text, predicting the mean first passage time for the special cases of $r=m=5$, and $r=6=m+1$. There we discussed that in both cases, the survival probability is $31/32 = 1-(1/2)^5$. 
We can also see this from this equation eqn.~\ref{eq:survivalProbRandomWalkerVeryExplicit}: starting from $t=m$, the summation only contains one term, $\ell=0$, giving $P(m,>m) = 1-\frac{1}{2^m}$. For the case $t=m+1$, the probability is unchanged: this follows since nonzero terms in the summation advance by increments of two (as required by matching parity of $m$ and $t$), adding just one extra step does not change the survival probability, $\mathbb{P}(m, >t) = \mathbb{P}(m, >t+1)$.

\section{Analytics of random walks with deterministic reset}
\label{app:sophisticatedFunResetTheory}

Figure~\ref{fig:TimeToReachTarget_stochasticAndFits} of the main text compares the mean first passage times of random walks with different reset policies. Here, we present a brief exposition on the mean first passage time for a random walker subject to deterministic reset, which has previously been derived by Refs.~\cite{PhysRevResearch.2.032029, bhat2016stochastic}.  

Our general strategy is to first derive an analytical solution to the probability density $p(x, t)$ that the random walker will be at location $x$ at time $t$. Then, the integral $\int_{-\infty}^{\text{target}} p(x, t) dx$ gives the survival probability, from which the mean first passage time can be calculated.

\subsection{Probability distribution of random searcher at position $x$ at time $t$}
\label{app:analyticaldeterministicreset}
Let the random walker be on the one dimensional line, seeking a target located at $x = x_T$, periodically reset to the origin at every time period $T_R$. Between resets, the only dynamic at play is the diffusion of the random searcher, subject to an absorbing boundary condition at the target $x_T$. It is instructive to write the evolution of the probability distribution between resets iteratively. Consider then the time period $t = 0$ up to (but not including) $t = T_R$. The probability of a random searcher being at location $x$ and time $t$ under such a diffusion-absorption paradigm is given by 
\begin{align}
    p_0(x, t) = \frac{1}{\sqrt{4 \pi D t}} \left(e^{- \frac{x^2}{4 D t}} - e^{\frac{- \left(x - 2 x_T\right)^2}{4 D t}}\right) \,, \qquad 0 \leq t < T_R\,,
\end{align}
where the subscript $0$ in $p_0(x, t)$ reminds us that the expression is valid before the first reset occurs. 

At time $t = T_R$, the reset clock ticks. Any probability mass not absorbed at the target, given by $S(T_R) = \int_{-\infty}^{x_T} p_0(x, T_R) dx$, is reset back to the origin. The integral to obtain the survival probability evaluates to the closed form
\begin{align}
    \label{eqn:survival probability integral deterministic diffusion-absorption}
    S(T_R) \quad = \quad \int_{-\infty}^{x_T} p_0(x, T_R) dx \quad = \quad \text{erf}\left(\frac{x_T}{\sqrt{4 D T_R}}\right)\,,
\end{align}
where $\text{erf}(\cdot)$ is the error function. 

From the time interval $t = T_R$ upto (but not including) $t = 2 T_R$, the random walker is once again subjected to the diffusion-absorption paradigm, and the probability distribution of the searcher is given by 
\begin{align}
    p_1(x, t) = \underbrace{\text{erf}\left(\frac{x_T}{\sqrt{4 D T_R}}\right)}_{\text{survival probability of searcher at }t = T_R} \underbrace{\frac{1}{\sqrt{4 \pi D \left(t - T_R\right)}} \left(e^{- \frac{x^2}{4 D \left(t - T_R\right)}} - e^{\frac{- \left(x - 2 x_T\right)^2}{4 D \left(t - T_R\right)}}\right)}_{\text{diffusion-absorption during }T_R \leq t < 2 T_R} \,, \qquad T_R \leq t < 2 T_R\,,
\end{align}
where the subscript $1$ in $p_1(x, t)$ reminds us that the expression is valid only between the first and second resets. In the above expression, the survival probability $\text{erf} \left(\frac{x_T}{\sqrt{4 D T_R}}\right)$ accounts for the probability that the random searcher did \emph{not} find its target and was reset to the origin at $t = T_R$. The remaining terms account for the diffusion-absorption dynamics \emph{after} the first reset.

\begin{figure}[h!]
    \centering
    \includegraphics[width=\textwidth]{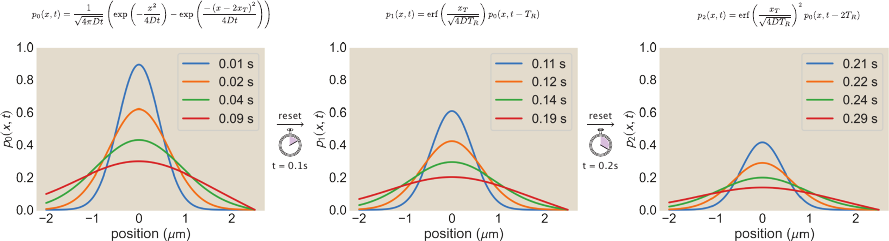}
\caption{\raggedright Probability distribution $p(x, t)$ of random walker subject to deterministic reset for a diffusion coefficient $D = 10 \, \mu \text{m}^2$/s and reset time period $T_R = 0.1$ seconds, with the target placed $2.5 \, \mu \text{m}$ to the right of the origin.}
    \label{fig:pxt_deterministic_reset}
\end{figure}

At time $t = 2 T_R$, the reset clock ticks once again. The probability mass that survived is given by 
\begin{align}
    \begin{split}
        S(2 T_R) \quad = \quad \int_{-\infty}^{x_T} p_1(x, 2T_R) dx \quad & = \quad \text{erf}\left(\frac{x_T}{\sqrt{4 D T_R}}\right) \int_{-\infty}^{x_T} \frac{1}{\sqrt{4 \pi D T_R}} \left(e^{- \frac{x^2}{4 D T_R}} - e^{\frac{- \left(x - 2 x_T\right)^2}{4 D T_R}}\right) dx,\\
        & = \text{erf}\left(\frac{x_T}{\sqrt{4 D T_R}}\right)^2,
    \end{split}
\end{align}
and once again, the random searcher that has survived thus far is subjected to the diffusion-absorption paradigm over the time period $t = 2 T_R$ upto (but not including) $t = 3 T_R$. The probability distribution over this time period is given by 
\begin{align}
    p_2(x, t) = \underbrace{\text{erf}\left(\frac{x_T}{\sqrt{4 D T_R}}\right)^2}_{\text{survival probability of searcher at }t = 2 T_R} \underbrace{\frac{1}{\sqrt{4 \pi D \left(t - 2 T_R\right)}} \left(e^{- \frac{x^2}{4 D \left(t - 2 T_R\right)}} - e^{\frac{- \left(x - 2 x_T\right)^2}{4 D \left(t - 2 T_R\right)}}\right)}_{\text{diffusion-absorption during }2 T_R \leq t < 3 T_R} \,, \qquad 2 T_R \leq t < 3 T_R\,.
\end{align}
Writing the solutions over the zeroth, first, and second time periods one on top of another, 
\begin{align}
    p_0(x, t) & = && && \quad \frac{1}{\sqrt{4 \pi D t}}  \left(e^{- \frac{x^2}{4 D t}} - e^{\frac{- \left(x - 2 x_T\right)^2}{4 D t}}\right) 
        \,, & \qquad 0 \leq t < T_R \label{eq:p0},\\[4pt]
    p_1(x, t) & = && \text{erf}\left(\frac{x_T}{\sqrt{4 D T_R}}\right) && \frac{1}{\sqrt{4 \pi D \left(t - T_R\right)}} \left(e^{- \frac{x^2}{4 D \left(t - T_R\right)}} - e^{\frac{- \left(x - 2 x_T\right)^2}{4 D \left(t - T_R\right)}}\right) 
        \,, & \qquad T_R \leq t < 2 T_R \label{eq:p1},\\[4pt]
    p_2(x, t) &= && \text{erf}\left(\frac{x_T}{\sqrt{4 D T_R}}\right)^2 && \frac{1}{\sqrt{4 \pi D \left(t - 2 T_R\right)}}  \left(e^{- \frac{x^2}{4 D \left(t - 2 T_R\right)}} - e^{\frac{- \left(x - 2 x_T\right)^2}{4 D \left(t - 2T_R\right)}}\right) 
        \,, & \qquad 2 T_R \leq t < 3 T_R \label{eq:p2}, 
\end{align}
and so on, we see that the solution can be generally written as 
\begin{align}
    p(x, t) = \text{erf}\left(\frac{x_T}{\sqrt{4 D T_R}}\right)^n p_0(x, t - n T_R) \,, & \qquad \qquad n T_R \leq t < \left( n + 1 \right)T_R.
\end{align}
Figure~\ref{fig:pxt_deterministic_reset} shows an example of the dynamics of the probability distribution of a random walker subject to deterministic reset.
\subsection{Survival probability of random searcher}
We are now interested in evaluating the survival probability $S(t)$ by integrating $p(x, t)$ from $x = -\infty$ to $x = x_T$, 
\begin{align}
    S(t) & = \int_{-\infty}^{x_T} p(x, t) dx \,,\\
    & = \int_{-\infty}^{x_T} \left(\text{erf}\left(\frac{x_T}{\sqrt{4D T_R}}\right)\right)^n p_0(x, t - n T_R) dx\,,\qquad n T_R \leq t < \left(n + 1\right) T_R\,.
\end{align}
Pulling the error function outside of the integral,
\begin{align}
    \label{eqn: s(t) integral diffusion-reset}
    S(t) & = \left( \text{erf} \left(\frac{x_T}{\sqrt{4 D T_R}}\right) \right)^n \int_{-\infty}^{x_T} p_0(x, t - n T_R) dx\,,\qquad n T_R \leq t < \left(n + 1\right) T_R\,.
\end{align}
Using eqn.~\ref{eqn:survival probability integral deterministic diffusion-absorption}, the survival probability evaluates to
\begin{align}
    \label{eqn: s(t) diffusion-reset}
    S(t) & = \left( \text{erf} \left(\frac{x_T}{\sqrt{4 D T_R}}\right) \right)^n \text{erf}\left(\frac{x_T}{\sqrt{4 D \left(t - n T_R\right)}}\right)\,,\qquad n T_R \leq t < \left(n + 1\right) T_R\,.
\end{align}
That is, the survival probability at any time $t$ is the survival probability at time $t = n T_R$ (after $n$ resets) multiplied by the survival probability after duration $t - n T_R$. 

\begin{figure}[h!]
    \centering
    \includegraphics[width=0.9\textwidth]{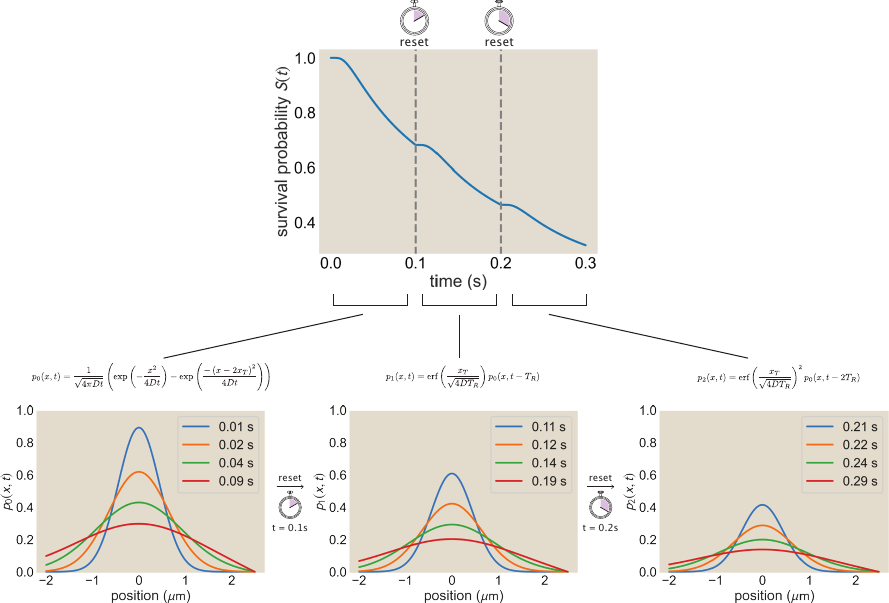}
\caption{\raggedright Survival probability and probability distribution of a random walker subject to deterministic reset for a diffusion coefficient $D = 10 \, \mu \text{m}^2$/s and reset time period $T_R = 0.1$ seconds, with the target placed $2.5 \, \mu \text{m}$ to the right of the origin.}
    \label{fig:survival_probability_deterministic_reset}
\end{figure}

Figure~\ref{fig:survival_probability_deterministic_reset} depicts the survival probability for the dynamics shown in Figure~\ref{fig:pxt_deterministic_reset}.

\subsection{Mean first passage time of random searcher}
\label{app:Mean first passage time of random searcher}
The evaluation of the mean first passage time requires some careful algebra. The probability of hitting the target $f(t)$ at time $t$ is related to the survival probability as
\begin{align}
    \label{eqn:f(t) for diffusion-reset}
    f(t) = \frac{d(1 - S)}{dt} = - \frac{dS}{dt}\,.
\end{align}
Then, the mean first passage time is given by 
\begin{align}
    T = \mathbb{E}[t] = \int_0^{\infty} t f(t) dt\,.
\end{align}
Substituting the expression for $f(t)$ from eqn.~\ref{eqn:f(t) for diffusion-reset}, 
\begin{align}
    T = - \int_0^{\infty} t \frac{dS}{dt} dt\,.
\end{align}
Performing integration by parts, we get 
\begin{align}
    T = - \left[ t S - \int S dt \right]_0^{\infty} = \int_0^{\infty} S dt\,,
\end{align}
where we have used the fact that eventually the survival probability is zero for $t \rightarrow \infty$. The above equation directly relates the survival probability to the mean first passage time.  

Substituting the expression for the survival probability for random walks with deterministic reset in eqn~\ref{eqn: s(t) diffusion-reset}, 
\begin{align}
T=
\sum_{n=0}^{\infty}
\int_{nT_R}^{(n+1)T_R}
\left(\text{erf}\left(\frac{x_T}{\sqrt{4DT_R}}\right)
\right)^n \text{erf}\left(\frac{x_T}{\sqrt{4D(t-nT_R)}} \right) dt.
\end{align}
Splitting up the integral over the time periods between resets, we get 
\begin{align}
    T = \int_0^{T_R} \text{erf}\left(\frac{x_T}{\sqrt{4 D t}}\right) dt + \int_{T_R}^{2 T_R} \text{erf} \left(\frac{x_T}{\sqrt{4 D T_R}}\right) \text{erf}\left(\frac{x_T}{\sqrt{4 D \left(t - T_R\right)}}\right) dt +  \int_{2 T_R}^{3 T_R} \left( \text{erf} \left(\frac{x_T}{\sqrt{4 D T_R}}\right) \right)^2 \text{erf}\left(\frac{x_T}{\sqrt{4 D \left(t - 2 T_R\right)}}\right) dt + \hdots \,.
\end{align}
Applying the transformation $\tau = t - n T_R$ for the $n^{th}$ integral, where $n = 0, 1, 2 \hdots$, 
\begin{align}
    T = \int_0^{T_R} \text{erf}\left(\frac{x_T}{\sqrt{4 D \tau }}\right) d\tau + \int_{0}^{T_R} \text{erf} \left(\frac{x_T}{\sqrt{4 D T_R}}\right) \text{erf}\left(\frac{x_T}{\sqrt{4 D \tau }}\right) d\tau  +  \int_{0}^{T_R} \left( \text{erf} \left(\frac{x_T}{\sqrt{4 D T_R}}\right) \right)^2 \text{erf}\left(\frac{x_T}{\sqrt{4 D \tau }}\right) d\tau + \hdots \,.
\end{align}
The integral can be factorized into an infinite sum of a geometric series multiplied by the integral of the error function over the time period between resets,
\begin{align}
    T = \left(1 + \text{erf} \left(\frac{x_T}{\sqrt{4 D T_R}}\right) + \left( \text{erf} \left(\frac{x_T}{\sqrt{4 D T_R}}\right) \right)^2  \hdots\right) \int_0^{T_R} \text{erf}\left(\frac{x_T}{\sqrt{4 D \tau }}\right) d\tau\,.
\end{align}
For a geometric series $1, r, r^2, r^3, \hdots$, the infinite sum is given by $1 / \left(1 - r\right)$, where the multiplicative factor $r$ satisfies the condition $r < 1$. Here, the multiplicative factor is $r = \text{erf} \left(\frac{x_T}{\sqrt{4 D T_R}}\right)$, whose value is strictly less than $1$. Thus, the mean first passage time becomes 
\begin{align}
    T = \frac{1}{1 - \text{erf} \left(\frac{x_T}{\sqrt{4 D T_R}}\right)}\int_0^{T_R} \text{erf}\left(\frac{x_T}{\sqrt{4 D \tau }}\right) d\tau\,.
\end{align}
It turns out that the fraction $1 / \left( 1 - \text{erf} \left(\frac{x_T}{\sqrt{4 D T_R}}\right) \right)$ measures the mean number of attempts taken to reach the target. The integral $\int_0^{T_R} \text{erf}\left(\frac{x_T}{\sqrt{4 D \tau }}\right) d\tau$ can be interpreted as the mean first passage time within the time interval between resets. That is to say, the mean first passage time is schematically given by 
\begin{align}
    \label{eqn:MFPT deterministic reset schematic}
    T \quad = \quad \text{average number of attempts} \quad  \times \quad \text{average time to reach target between two resets}\,.
\end{align}
The reader may recognize the parallels between the above expression and eqn.~\ref{eq:geometricMeanTime} of the main text. This is because random walks with deterministic resets can be approached via the language of the geometric distribution, which is explored further in Appendix~\ref{app:geometric}.

In fact, the structure of eqn.~\ref{eqn:MFPT deterministic reset schematic} can be generalized using Wald's identity, briefly explained below. 
\begin{tcolorbox}[breakable, parbox=false,  boxrule=0pt]
\emph{Wald's Identity\cite{blackwell_waldequation}.} Consider a sequence of independent and identically drawn real-valued random variables, $\left(X_1, X_2, X_3,\hdots\right)$, with a common expectation $\mathbb{E}[X]$. Let $N$ be an integer random variable with finite mean, such that an event occurring at discrete time $N = n$ is determined entirely by the history of the sequence, $\left(X_1, X_2, X_3,\hdots X_n\right)$, and does not depend on future values of the sequence, $X_{n+1}, X_{n+2}, \hdots$ and so forth. Such a random variable $N$ is called a stopping time.

Then, Wald's identity states
\begin{align}
    \mathbb{E}\left[X_1 + X_2 + X_3 + \hdots X_N\right] = \mathbb{E}\left[N\right] \mathbb{E}\left[X\right]\,.
\end{align}
\end{tcolorbox}
Note that Wald's identity generalizes eqn.~\ref{eq:geometricMeanTime} to stopping criteria that may depend on the entire history of the sequence.

Overlaying Wald's identity on the mean first passage time for random walks with deterministic reset, 
\begin{align}
    \label{eqn:determinstic_theoretical_MFPT}
    \begin{split}
        T = \underbrace{\frac{1}{1 - \text{erf} \left(x_T/\sqrt{4 D T_R}\right)}}_{\mathbb{E}\left[ N \right]} \underbrace{\int_0^{T_R} \text{erf}\left(\frac{x_T}{\sqrt{4 D \tau }}\right) d\tau}_{\mathbb{E}\left[\text{duration of one attempt}\right]}\,.
    \end{split}
\end{align}

\section{Average hitting times as explicit averages over trajectories}
\label{app:quickJaneKolmogorovNote}
\begin{figure}
    \centering
    \includegraphics[width=0.4\linewidth]{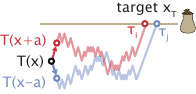}
    \caption{\raggedright One way to view the recursion relations for mean first passage time (for instance eqn.~\ref{eq:recursionBasic})---which are also called Kolmogorov backwards equations---is in terms of explicit averages over trajectories.}
    \label{fig:quickJaneSketchOfKolmogorovBackwardsEquations}
\end{figure}
\FloatBarrier
Here we briefly explain another vantage on the why the mean first hitting time for a target $T(x)$ starting at position $x$ obeys a recursion relation or local average (Kolmogorov's backwards equations) with those of its neighbors, which explicitly centers trajectories. Imagine that a set of $N$ random walks are performed starting at position $x$ until they hit the target. Call $\tau_i$ the time that the $i$th trajectory finally reaches the target. Then the empirical mean first hitting time of these is just the average of these $N$ hitting time values,
\begin{align}
    T(x) \approx \frac{1}{N}\sum_{i=1}^N \tau_i.
\end{align}
Consider the first step of each of these walks. As we are reminded by Fig.~\ref{fig:quickJaneSketchOfKolmogorovBackwardsEquations}, either the trajectory stepped to $x+a$ or else the trajectory stepped to $x-a$. This bins all the trajectories into these two classes, which means we may split the above average sum into two parts,
\begin{align}
    T(x) \approx \frac{1}{N}\left( {\color[HTML]{C74444}\sum_{i: \text{ step to $x+a$}}^N \tau_i} + {\color[HTML]{738FC1} \sum_{j: \text{ step to $x-a$}}^N \tau_j}  \right),
\end{align}
If ${\color[HTML]{C74444}N_+}$ of these trajectories first stepped to $x+a$, and ${\color[HTML]{738FC1}N_-}$ stepped to $x-a$, then we may further rewrite the average as,
\begin{align}
    T(x) \approx {\frac{{\color[HTML]{C74444}N_+}}{N} \left(\frac{1}{{\color[HTML]{C74444}N_+}} ~ {\color[HTML]{C74444}\sum_{i: \text{ step to $x+a$}}^N \tau_i}\right) + {\frac{{\color[HTML]{738FC1}N_-}}{N} \left(\frac{1}{{\color[HTML]{738FC1}N_-}} ~ {\color[HTML]{738FC1}\sum_{j: \text{ step to $x-a$}}^N \tau_j}\right)}}.
\end{align}
Notice that $\displaystyle \frac{{\color[HTML]{C74444}N_+}}{N}$ represents the realized probability of stepping to $x+a$, and analogously $\displaystyle \frac{{\color[HTML]{738FC1}N_-}}{N}$ represents the probability of stepping to $x-a$. Further, $\displaystyle \left(\frac{1}{{\color[HTML]{C74444}N_+}} ~ {\color[HTML]{C74444}\sum_{i: \text{ step to $x+a$}}^N \tau_i}\right)$ represents the average arrival time of the trajectory conditioned on it first stepping to $x+a$, which we may identify as ${\color[HTML]{C74444}T(x+a)}$, and analogously $\displaystyle \left(\frac{1}{{\color[HTML]{738FC1}N_-}} ~ {\color[HTML]{738FC1}\sum_{j: \text{ step to $x-a$}}^N \tau_j}\right)$ as ${\color[HTML]{738FC1}T(x-a)}$. For a symmetric random walk where stepping left or right has equal probability, this allows us to rewrite the average further as,
\begin{align}
    T(x) \approx \frac{1}{2k} + \frac{1}{2}{\color[HTML]{C74444}T(x+a)} + \frac{1}{2}{\color[HTML]{738FC1}T(x-a)},
\end{align}
where we acknowledged the time required for the first step. Over many trajectories $N\rightarrow \infty$, this perspective recovers the eqn.~\ref{eq:recursionBasic} of the main text reasoned about directly from conditional probability.

\section{Resets confine the search of the random walker}

When is a random walk with reset equivalent to a random walk with confinement? The answer depends on our notion of ``equivalence''. In Ref.~\cite {Evans_2013}, the authors show this equivalence in the sense that the steady-state distribution of a diffusing particle with Poissonian reset is equivalent to that of a diffusing particle subject to a confining potential \emph{without} reset, even though their steady-state probability currents are not equivalent. The steady-state distribution of a diffusing particle subject to a confining potential is equilibrium in nature, while that of a diffusing particle subject to Poissonian reset is non-equilibrium in nature.

In Appendix \S~\ref{app:deterministic reset confines between resets}, we discuss equivalence of resets to confinement in the sense that it places an upper bound on the spread of the random walker's probability distribution. In Appendix \S~\ref{app:deterministic reset is equivalent to confined random walker}, we discuss equivalence of resets to confinement in the sense that the asymptotic mean first passage time of a random walker with deterministic reset approaches that of a random walker under confinement, with the reflecting boundary being appropriately placed. Interestingly, for the same condition, the dynamics of the probability distributions for deterministic reset approaches that of confined diffusion.

\subsection{Resets place an upper bound on the spread of the distribution}
\label{app:deterministic reset confines between resets}
In Appendix \S~\ref{app:analyticaldeterministicreset}, we showed that the probability distribution of a random walker subjected to periodic reset is prescribed as
\begin{align}
    p(x, t) = \text{erf}\left(\frac{x_T}{\sqrt{4 D T_R}}\right)^n p_0(x, t - n T_R) \,, & \qquad \qquad n T_R \leq t < \left( n + 1 \right)T_R,
\end{align}
where $n$ is the number of resets that occurred by time $t$, and 
\begin{align}
    p_0(x, t - n T_R) = \frac{1}{\sqrt{4 \pi D \left(t - n T_R\right)}} \left(e^{- \frac{x^2}{4 D \left(t - n T_R\right)}} - e^{\frac{- \left(x - 2 x_T\right)^2}{4 D \left(t - n T_R\right)}}\right), \qquad n T_R \leq t < \left( n + 1 \right)T_R,
\end{align}
describes the probability distribution of a random walker subject to an absorbing boundary condition at $x = x_T$ after the $n^{th}$ reset occurs. Let us shift the time frame by $\tau = t - n T_R$, so that at $t = n T_R$, $\tau = 0$. The solution becomes
\begin{align}
    p(x, t) = \text{erf}\left(\frac{x_T}{\sqrt{4 D T_R}}\right)^n \frac{1}{\sqrt{4 \pi D \tau}}  \left(e^{- \frac{x^2}{4 D \tau}} - e^{\frac{- \left(x - 2 x_T\right)^2}{4 D \tau}}\right) \qquad \qquad 0 \leq \tau < T_R,
\end{align}
where $n$ still tracks the number of resets that have occurred by time $\tau = t - n T_R$. The error function $\text{erf}\left(\frac{x_T}{\sqrt{4 D T_R}}\right)^n$ quantifies the probability mass that survived up to the $n^{th}$ reset and does not contribute to any spatial variation. Factoring out the surviving probability mass by dividing both sides of the previous equation by $S(n T_R) = \text{erf}\left(\frac{x_T}{\sqrt{4 D T_R}}\right)^n$, 
\begin{align}
    \frac{p(x, n T_R + \tau)}{S\left(n T_R\right)} = \frac{1}{\sqrt{4 \pi D \tau}}  \left(e^{- \frac{x^2}{4 D \tau}} - e^{\frac{- \left(x - 2 x_T\right)^2}{4 D \tau}}\right) \qquad \qquad 0 \leq \tau < T_R,
\end{align}
we are left only with the dynamics of a random walker subject to an absorbing boundary condition at $x = x_T$ within the time period of a reset. Since the reset clock ticks at $\tau = T_R$, the spread of the above distribution is capped at $\sqrt{D T_R}$.

\begin{figure}[h!]
    \centering
    \includegraphics[width=0.4\textwidth]{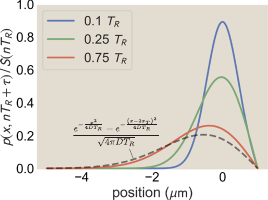}
\caption{\raggedright The probability distribution spread of a random walker between two resets is strictly lower than that of a random walker without reset at $t = T_R$. Here, diffusion coefficient $D = 10 \, \mu \text{m}^2$/s and reset time period $T_R = 0.1$ seconds, with the target placed $1 \, \mu \text{m}$ to the right of the origin.}
    \label{fig:pxt_between_resets_is_confined}
\end{figure}

\subsection{Equivalence of asymptotics of mean first passage times}
\label{app:deterministic reset is equivalent to confined random walker}
Thus far, we have discussed how resets endow an upper bound to the probability distribution spread of the random walker between any two resets. Under stronger conditions, before the first reset, the dynamics of a random walker with deterministic reset are seemingly equivalent to that of a confined walker without reset for all time. Intuitively, this is true if the resetting random walker (without confinement) speedily finds its target both before the first reset time $T_R$ and before the confined random walker (without reset) interacts with the boundary $L_R$. 

Let us quantify the above intuition. The mean first passage time for a random walker subject to reset was derived in Appendix~\ref{app:Mean first passage time of random searcher}, 
\begin{align}
    T = \frac{1}{1 - \text{erf} \left(\frac{x_T}{\sqrt{4 D T_R}}\right)}\int_0^{T_R} \text{erf}\left(\frac{x_T}{\sqrt{4 D \tau }}\right) d\tau\,.
\end{align}
The integral is of the form $\int_0^T \text{erf}\left(\sqrt{b / t}\right) dt$. With the change of variable $z = \sqrt{b / t}$, the integral becomes 
\begin{align}
    \int_0^T \text{erf}\left(\sqrt{b / t}\right) dt & = 2 b \int_{\sqrt{b/T}}^\infty \text{erf}\left(z\right) / z^3 dz,\\
    & = 2 b \left(\frac{\text{erf}\left(z \right)}{2 z^{2}} + \frac{1}{\sqrt{\pi}} \int_{\sqrt{b/T}}^\infty \frac{e^{-z^2}}{z^2} dz\right), \label{eqn:error function integral}\\
    & = 2 b \left(\frac{\text{erf}\left(z \right)}{2 z^{2}}  + \frac{e^{-b / T}}{\sqrt{\pi}\sqrt{b / T}} - 1 + \text{erf}\left(\sqrt{b / T}\right)\right)\label{eqn:exponential integral},\\
    & = T \left(\left(2 \frac{b}{T} + 1\right) \text{erf}\left(\sqrt{\frac{b}{T}}\right) + 2 \sqrt{\frac{b }{\pi T}} e^{-b / T} - 2\frac{b}{T}\right) 
    \label{eqn: rearrangement},
\end{align}
where eqn.~\ref{eqn:error function integral} arises from eqn.~14 of the tabulation\cite{NgGellerIntegralErrorFunction} (with $n = 3$ and $a = 1$), eqn.~\ref{eqn:error function integral} is obtained after calculating the integral, and eqn.~\ref{eqn: rearrangement} is obtained after a rearrangement of terms.
Using ~\ref{eqn: rearrangement}, the mean first passage time becomes
\begin{align}
T = \frac{T_R\left( \left(1 + 2 \left(\dfrac{x_T}{2 \sqrt{D T_R}}\right)^2\right)\text{erf}\left(\frac{x_T}{2 \sqrt{D T_R}}\right) + \dfrac{2}{\sqrt{\pi}}\left(\dfrac{x_T}{2 \sqrt{D T_R}}\right)\,\exp\left(-\left( \dfrac{x_T}{2 \sqrt{D T_R}} \right)^2\right) - 2 \left(\dfrac{x_T}{2 \sqrt{D T_R}}\right)^2 \right)}{1 - \text{erf}\left(\dfrac{x_T}{2\sqrt{D T_R}}\right)}\,,
\end{align}
In the limit of infinite reset time $T_R \rightarrow \infty$, for fixed target location $x_T$, the fraction $x_T / \sqrt{D T_R}$ is much smaller than $1$. In this limit, we have the approximations erf$\left(x_T / 2\sqrt{D T_R}\right) \sim \frac{2 x_T / 2 \sqrt{D T_R}}{ \sqrt{\pi}}$, $1 - \text{erf}\left(x_T / 2 \sqrt{D T_R}\right) \sim 1$, $\left(x_T / 2 \sqrt{D T_R}\right)^2 \sim 0$, and $\exp\left(-\left(x_T / 2\sqrt{D T_R}\right)^2\right) \sim 1$. The mean first passage time becomes
\begin{align}
    T \sim \dfrac{2}{\sqrt{\pi}} x_T\sqrt{\frac{T_R}{D}}\,.
\end{align}
Compare the above to eqn~\ref{eqn: MFPT under confinement if LR >> xT} of Section~\ref{sec:Approximate Description of The Simple Random Walk With Reset}, the mean first passage time for a random walker confined between the boundaries $x = - L_R$ and $x = x_T$ without any reset, 
\begin{align}
    T_{\text{confined}}(0) \sim \frac{x_T L_R}{D} =  \frac{2}{\sqrt{\pi}}x_T\sqrt{\frac{T_R}{D}} \text{  under confinement without reset}\,, \qquad T\sim \frac{2}{\sqrt{\pi}}x_T\sqrt{\frac{T_R}{D}}  \text{  under periodic reset}\,,
\end{align}
where the reflecting boundary was placed at $L_R =\frac{2}{\sqrt{\pi}} \sqrt{D T_R}$ in eqn~\ref{eqn: location of reflecting boundary following extremal value arguments} of Section~\ref{sec:Approximate Description of The Simple Random Walk With Reset} following arguments of extremal value statistics.

Thus, the mean first passage time under reset scales similarly to that of a confined random walker, provided the target is much closer to the initial condition than the lengthscale $L_R$. 

\begin{figure}[h!]
    \centering
    \includegraphics[width=0.7\textwidth]{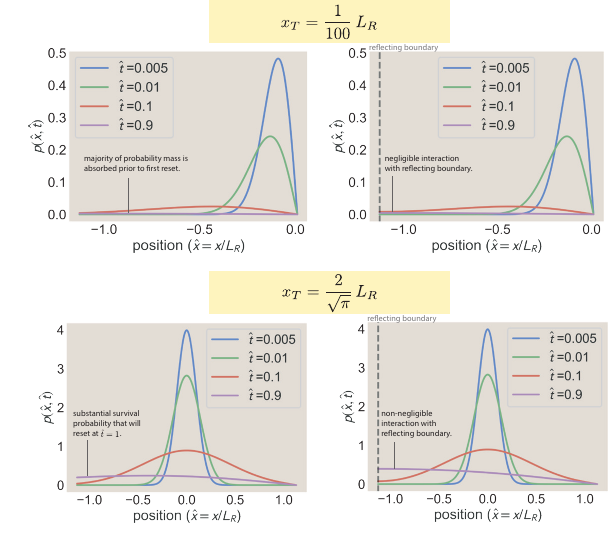}
\caption{\raggedright The probability distribution of a random walker prior to the first reset (left column) approaches that of a random walker with a reflecting boundary at $x=-L_R$ (right column), provided $x_T \ll L_R$, where $L_R = \frac{2}{\sqrt{\pi}} \sqrt{D T_R}$ is the effective confinement length. When $x_T = 0.01 L_R$, we see that the majority of the probability mass is absorbed at the target $x = x_T$ before the reset time $T_R$ \emph{and} before the probability mass substantially diffuses to the reflecting boundary. When $x_T = \pi L_R$, a non-negligible probability mass will be reset at time $T_R$, thus significantly deviating from the dynamics of diffusion without reset. Here, the diffusion coefficient is $D = 10 \, \mu \text{m}^2$/s, the reset time period is $T_R = 0.1$ seconds, and time has been normalized using $\hat{t} = \frac{t}{T_R}$.}
    \label{fig:pxt_is_equivalent_if_xT_much_less_than_LR}
\end{figure}

We see similar agreement of the dynamics of the probability distributions in Figure~\ref{fig:pxt_is_equivalent_if_xT_much_less_than_LR}. Here, the probability distribution for a random walker under confinement was approximated by the solution
\begin{align}
    p_{\text{confined}}(x, t) =  \frac{2}{L_R \left(1 + \frac{x_T}{L_R}\right)} \sum_{n=0}^{\infty}  \cos\left( \frac{(2n+1)\pi}{2} \frac{1 + \frac{x}{L_R}}{1 + \frac{x_T}{L_R}} \right) \cos\left( \frac{(2n+1)\pi}{2} \frac{1}{1 + \frac{x_T}{L_R}} \right) \exp\left(-\frac{D}{L_R^2}\left(\frac{(2n+1)\pi}{2\left(1 + \frac{x_T}{L_R}\right)}\right)^2t\right)\,,
\end{align}
which is the solution to the diffusion equation with the reflecting boundary condition $\frac{\partial}{\partial}p\biggl|_{x = - L_R} = 0$ and the absorbing boundary condition $p(x_T, t) = 0$ in the absence of reset. The plot corresponds to simulations of 500 modes.

\FloatBarrier

\section{Additional results for random walks with resetting}

\subsection{Additional sweep of behavior of random walk under resets from simulation}
\label{sec:additionalResetRandomWalkSweep}
In the main text, exploring how random walks with resets operate, we dissected the example of a particular target location $x_T=5$ using simulations that prompted theoretical understanding. Here we briefly display the joint landscape of how both target locations and reset rates determine the mean time to successfully reach targets.
\begin{figure}[h!]
    \centering
    \includegraphics[width=0.5\textwidth]{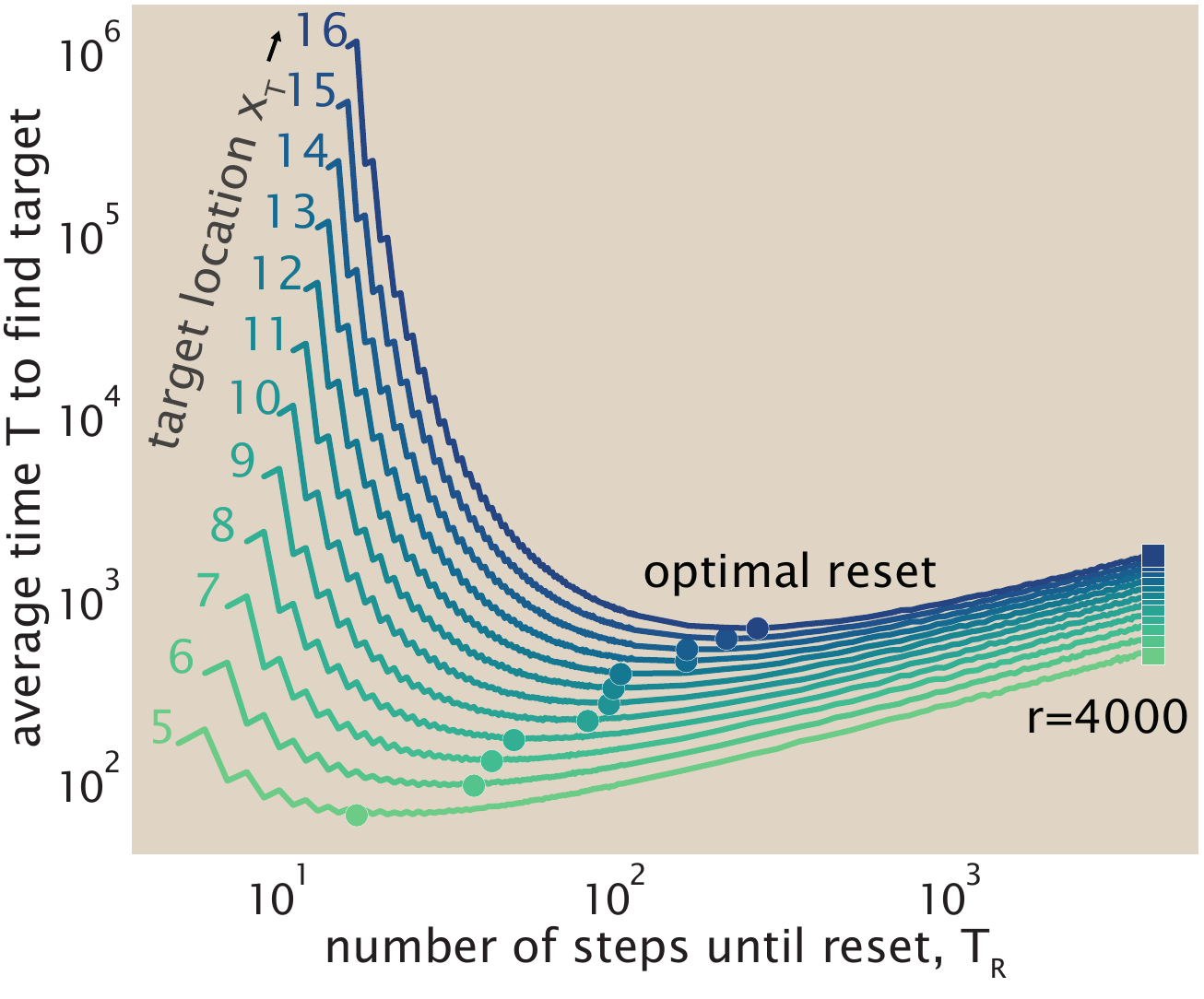}
\caption{\raggedright Average time for seekers to reach a target, subject to different deterministic reset periods $T_R$ (on the $x$-axis) and different target locations $x_T$ (shown as lines with different hues of sea green to dark blue), using numerical simulations.}
    \label{fig:TimeToReachTarget_sweep}
\end{figure}

\FloatBarrier

\subsection{The geometric distribution as a language of exploration with memoryless resets}
\label{app:geometric}

In the work of Kondev \emph{et al.} \cite{kondev2025}, the
first-passage problem with resetting is interpreted in terms of a geometric
distribution. When resetting returns the system to the same initial state,
any trajectory leading to success can be decomposed into a sequence of
independent attempts, consisting of several failed attempts followed by a
final successful one.

We denote by $T_0$ the first-passage time to the target in the absence of
resetting, with probability density $f_0(t)$. We denote by $R$ the time of
the first reset event after the beginning of an attempt. Since resets occur
as a Poisson process with constant rate $r$, $R$ is exponentially
distributed,
\begin{equation}
p_R(t)=r e^{-rt},
\end{equation}
and the probability that no reset has occurred by time $t$ is
\begin{equation}
\mathbb{P}(R>t)=e^{-rt}.
\end{equation}
An attempt is successful when the target is reached before the next reset.
We therefore define the success probability
\begin{equation}
q
=
\mathbb{P}(T_0<R).
\end{equation}
Conditioning on the value of the no-reset first-passage time gives
\begin{align}
q
&=
\int_0^\infty
\mathbb{P}(T_0<R\mid T_0=t)\,
f_0(t)\,dt
\\
&=
\int_0^\infty
\mathbb{P}(R>t)\,
f_0(t)\,dt
\\
&=
\int_0^\infty
e^{-rt}f_0(t)\,dt.
\end{align}
Thus,
\begin{equation}
q=\mathbb{E}\!\left[e^{-rT_0}\right],
\label{eq:q_laplace_transform}
\end{equation}
so that the probability of success during a single reset-delimited attempt
is simply the Laplace transform of the no-reset first-passage-time
distribution evaluated at the reset rate $r$.

We next characterize the durations of successful and failed attempts. The
duration of a successful attempt is
\begin{equation}
T_s
=
T_0
\,|\,
T_0<R,
\end{equation}
whereas the duration of a failed attempt is
\begin{equation}
T_f
=
R
\,|\,
R<T_0.
\end{equation}

For the successful attempts, Bayes' rule gives
\begin{align}
\mathbb{E}[T_s]
&=
\int_0^\infty
t\,
\mathbb{P}(T_0=t\mid T_0<R)\,dt
\\
&=
\frac{
\int_0^\infty
t\,
\mathbb{P}(T_0<R\mid T_0=t)\,
f_0(t)\,dt
}{
\mathbb{P}(T_0<R)
}.
\end{align}
Using $\mathbb{P}(R>t)=e^{-rt}$, we obtain
\begin{equation}
\mathbb{E}[T_s]
=
\frac{
\mathbb{E}\!\left[T_0e^{-rT_0}\right]
}{q}.
\label{eq:mean_successful_attempt}
\end{equation}

We proceed similarly for the duration of a failed attempt. By definition,
\begin{align}
\mathbb{E}[T_f]
&=
\int_0^\infty
t\,
\mathbb{P}(R=t\mid R<T_0)\,dt
\\
&=
\frac{
\int_0^\infty
t\,r e^{-rt}\,
\mathbb{P}(T_0>t)\,dt
}{
\mathbb{P}(R<T_0)
}.
\end{align}
Since the two possible outcomes of an attempt are success or failure,
\begin{equation}
\mathbb{P}(R<T_0)=1-q.
\end{equation}
Integrating the numerator by parts gives
\begin{equation}
\mathbb{E}[T_f]
=
\frac{1}{r}
-
\frac{
\mathbb{E}\!\left[T_0e^{-rT_0}\right]
}{
1-q
}.
\label{eq:mean_failed_attempt}
\end{equation}

Each attempt succeeds independently with probability $q$. Therefore, the
number $N_f$ of failed attempts before the first successful attempt is a
geometric random variable,
\begin{equation}
\mathbb{P}(N_f=n)
=
(1-q)^nq,
\qquad
n=0,1,2,\ldots,
\end{equation}
with mean
\begin{equation}
\mathbb{E}[N_f]
=
\frac{1-q}{q}.
\end{equation}

The total first-passage time is the sum of the durations of all failed
attempts and the duration of the final successful attempt. Its mean is
therefore
\begin{equation}
\langle T\rangle
=
\mathbb{E}[N_f]\,
\mathbb{E}[T_f]
+
\mathbb{E}[T_s].
\end{equation}
Substituting eqns.~\eqref{eq:mean_successful_attempt} and
\eqref{eq:mean_failed_attempt}, we obtain
\begin{align}
\langle T\rangle
&=
\frac{1-q}{q}
\left[
\frac{1}{r}
-
\frac{
\mathbb{E}\!\left[T_0e^{-rT_0}\right]
}{
1-q
}
\right]
+
\frac{
\mathbb{E}\!\left[T_0e^{-rT_0}\right]
}{q}
\\
&=
\frac{1-q}{rq}.
\label{eq:mfpt_geometric_reset}
\end{align}

Thus, a first-passage process with memoryless resetting can be coarse-grained
into a sequence of independent attempts governed by a geometric
distribution. All of the microscopic first-passage dynamics enter this
description through the single-attempt success probability
\begin{equation}
q=\mathbb{E}\!\left[e^{-rT_0}\right].
\end{equation}

\section{Dependence of the mean first-passage time on the reset rate $r$ in 1D with diffusion and drift}
\label{app:dep_T_r}

In \S\ref{subsec:RandomWalkResetsConstantDrift} (``Random Walk With Resets and Drift'') of the main text, we discussed how the prudence of resetting in the presence of a constant cue can be organized by the P\'eclet number. Here we derive this condition.

We recall that the mean first-passage time to reach $x_T>0$ starting from the origin in the presence of stochastic (``Poisson'') resetting at rate $r$ can be written as,
\begin{equation}
\langle T\rangle(x_T|0)=\frac{1}{r}\left(e^{x_T/x^*}-1\right),
\end{equation}
where the effective length scale is
\begin{equation}
x^*=\frac{2D/v}{\sqrt{1+\frac{4Dr}{v^2}}-1}.
\label{eq:length scale_diffusion_drift_app}
\end{equation}
We consider here $v>0$, corresponding to drift toward the target, and
determine when the mean first-passage time admits an optimum at a finite
reset rate.
It is convenient to introduce the dimensionless quantity
\begin{equation}
s(r)=\sqrt{1+\frac{4Dr}{v^2}}
\end{equation}
together with the additional dimensionless partner,
\begin{equation}
\chi=\frac{v x_T}{2D}.
\end{equation}
(Note that this variable already looks suspiciously like a P\'eclet number, as we will soon see.) 

Expressing eqn.~\eqref{eq:length scale_diffusion_drift_app} in terms of these definitions, one finds
\begin{equation}
\frac{x_T}{x^*}= \frac{v x_T}{2D}\,(s-1)=\chi(s-1).
\end{equation}

Since $s^2-1=\frac{4Dr}{v^2}$, the prefactor $1/r$ in the mean first passage time can be rewritten in terms of $s$, resulting in the expression
\begin{equation}
\langle T\rangle(x_T|0)=\frac{4D}{v^2}\,
\frac{e^{\chi(s-1)}-1}{s^2-1}.
\label{eq:T_of_s}
\end{equation}

Our goal is to ask whether the mean first passage time $\langle T\rangle$ has any minimum with respect to a finite value of the resetting rate $r$. A sufficient condition for this to be true is $\frac{d\langle T\rangle}{dr}=0$ for some $r>0$. This condition can be translated in terms of the natural variable $s$, since the map $r\mapsto s(r)$ is strictly increasing for $r\ge 0$, so extrema of $\langle T\rangle$ as a function of $r$ correspond one-to-one with extrema as a function of $s$. Thus, our question translates to whether there exists a solution $s>1$ to this equation,
\begin{equation}
\frac{d}{ds}\left(\frac{e^{\chi(s-1)}-1}{s^2-1}\right)=0.
\end{equation}
This equation itself reduces to the implicit stationarity condition,
\begin{equation}
\frac{\chi e^{\chi(s-1)}}{e^{\chi(s-1)}-1}=\frac{2s}{s^2-1},
\qquad s>1.
\label{eq:stationary_s}
\end{equation}
To help analyze when eqn.~\eqref{eq:stationary_s} admits a solution, we rewrite this condition in terms of yet another parameter, given by $t=s-1>0$ (not to be confused \sout{at all} with time).  This conveniently rewrites the condition as,
\begin{equation}
\frac{\chi}{1-e^{-\chi t}}=\frac{2(1+t)}{t(2+t)}.
\end{equation}
Multiplying both sides by $t$ yields
\begin{equation}
b(\chi t)=a(t),
\qquad t>0,
\label{eq:abt_equation}
\end{equation}
where we have defined
\begin{equation}
b(x)=\frac{x}{1-e^{-x}},
\qquad
a(t)=\frac{2(1+t)}{2+t}.
\end{equation}
It is convenient to recast eqn.~\eqref{eq:abt_equation} as a root-finding problem.  We therefore define
\begin{equation}
F(t)=b(\chi t)-a(t),
\qquad t\ge 0,
\label{eq:F_def}
\end{equation}
so that stationary points correspond to solutions of $F(t)=0$ with $t>0$. The function $a(t)$ is strictly increasing and bounded on $t\ge 0$,
\begin{equation}
a(0)=1,
\qquad
\lim_{t\to\infty}a(t)=2,
\qquad
a'(t)=\frac{2}{(2+t)^2}>0.
\label{eq:a_props}
\end{equation}
The function $b(x)$ is strictly increasing and unbounded on $x\ge 0$,
\begin{equation}
\lim_{x\to 0}b(x)=1,
\qquad
\lim_{x\to\infty}b(x)=\infty,
\label{eq:b_limits}
\end{equation}
and $b'(x)>0$ for all $x>0$ since
\begin{equation}
b'(x)=\frac{1-e^{-x}(1+x)}{(1-e^{-x})^2}>0,
\label{eq:bprime_pos}
\end{equation}
using $(1+x)e^{-x}\le 1$ for $x\ge 0$.
In addition, differentiating once more shows that $b$ is convex on $x\ge 0$,
\begin{equation}
b''(x)=\frac{e^{-x}\bigl[x-2+(x+2)e^{-x}\bigr]}{(1-e^{-x})^3}\ge 0,
\qquad x\ge 0,
\label{eq:bpp}
\end{equation}
with strict inequality for $x>0$.
Similarly, $a$ is concave on $t\ge 0$ since
\begin{equation}
a''(t)=-\frac{4}{(2+t)^3}<0.
\label{eq:app_concave}
\end{equation}
From eqn.~\eqref{eq:F_def} we have
\begin{equation}
F''(t)=\chi^2\,b''(\chi t)-a''(t).
\end{equation}
Using $b''(x)\ge 0$ for $x\ge 0$ and $-a''(t)=\frac{4}{(2+t)^3}>0$, we obtain
\begin{equation}
F''(t)\ge \frac{4}{(2+t)^3}>0,
\qquad t\ge 0.
\label{eq:F_convex}
\end{equation}
Thus $F$ is strictly convex on $[0,\infty)$ and $F'(t)$ is strictly increasing.
At $t=0$ we have $F(0)=b(0)-a(0)=1-1=0$.
The behavior near $t=0^+$ (which is to say, approaching $t=0$ starting from positive values to the right of zero) is controlled by the first-order expansion
\begin{equation}
b(\chi t)=1+\frac{\chi}{2}t+O(t^2),
\qquad
a(t)=1+\frac{1}{2}t+O(t^2),
\end{equation}
which implies
\begin{equation}
F(t)=\frac{\chi-1}{2}\,t+O(t^2),
\qquad t\to 0^+.
\label{eq:F_smallt}
\end{equation}
Hence $F(t)>0$ for all sufficiently small $t>0$ if $\chi>1$, whereas $F(t)<0$ for all sufficiently small $t>0$ if $\chi<1$.
For large $t$, eqn.~\eqref{eq:b_limits} gives $b(\chi t)\sim \chi t$, while eqn.~\eqref{eq:a_props} gives $a(t)\to 2$.
Therefore
\begin{equation}
\lim_{t\to\infty}F(t)=+\infty.
\label{eq:F_infty}
\end{equation}
Since $F''(t)>0$, the derivative $F'(t)$ is strictly increasing on
$[0,\infty)$. From the small-$t$ expansion in
eqn.~\eqref{eq:F_smallt}, we have
\begin{equation}
F'(0)=\frac{\chi-1}{2}.
\end{equation}

When $\chi\geq 1$, we have $F'(0)\geq0$. Since $F'$ is strictly
increasing, $F'(t)>0$ for all $t>0$. Together with $F(0)=0$, this
implies
\begin{equation}
F(t)>0,
\qquad t>0,
\end{equation}
and therefore no extremum exists at finite reset rate.

When $\chi<1$, we instead have $F'(0)<0$, so $F(t)<0$ for sufficiently
small positive $t$. On the other hand,
\begin{equation}
\lim_{t\to\infty}F(t)=+\infty.
\end{equation}
By continuity, $F$ must therefore cross zero at some $t^\star>0$.
Moreover, because $F'$ is strictly increasing, $F$ has at most one
minimum and can cross zero only once after leaving the origin.
Thus there exists a unique solution $t^\star>0$ when $\chi<1$.
In summary, a finite optimal reset rate exists if and only if
\begin{equation}
\chi=\frac{v x_T}{2D}<1,
\end{equation}
in which case the minimizer is unique. Since the P\'eclet number over
the target distance is
\begin{equation}
\mathrm{Pe}=\frac{v x_T}{D},
\end{equation}
this condition is equivalently
\begin{equation}
\mathrm{Pe}<2.
\end{equation}
Thus, a finite optimal reset rate exists only when diffusion is
sufficiently strong relative to drift.

Another route to appreciate that the P\'eclet number's value relative to 2 sets the prudence of reset can be seen more directly via the immediate expansion of eqn.~\ref{eq:pecletExpandClever} of the main text. The foregoing analysis is a more complete and rigorous articulation of this core behavior.

\section{Random walk with reset and exponential cue}
\subsection{Nondimensionalization of the exponential-cue backward equation}
\label{app:expo_cue_non_dim}
In \S \ref{subsec:exponentialCue} of the main text (``Random Seeker with an Exponential Cue''), we tamed a little of the complexity of learning how different reset policies vary with the properties of directional cues (whose strength also vary with position) by expressing the differential equation for the first passage time in terms of a smaller number of non-dimensional parameters. 

Here we explicitly nondimensionalize this eqn.~\ref{eq:backward_exp_cue} of the main text. We first measure position in units of the location of the target and time in units of the time to diffuse to the target, yielding the dimensionless positions and first passage times,
\begin{equation}
\tilde x=\frac{x}{x_T},
\qquad
\tilde T(\tilde x)=\frac{D}{x_T^2}T(x),
\end{equation}
or equivalently,
\begin{equation}
x=x_T\tilde x,
\qquad
T(x)=\frac{x_T^2}{D}\tilde T(\tilde x).
\end{equation}

We first remind ourselves how derivatives with respect to $x$ are related to
derivatives with respect to $\tilde x$. We have
\begin{equation}
\frac{dx}{d\tilde x}=x_T.
\end{equation}
Starting from
\begin{equation}
\tilde T(\tilde x)
=
\frac{D}{x_T^2}T(x_T\tilde x),
\end{equation}
we differentiate with respect to $\tilde x$. By the chain rule,
\begin{align}
\frac{d\tilde T}{d\tilde x}
&=
\frac{D}{x_T^2}
\frac{dT}{dx}
\frac{dx}{d\tilde x}
\\
&=
\frac{D}{x_T^2}
T'(x)x_T
\\
&=
\frac{D}{x_T}T'(x).
\end{align}
Therefore,
\begin{equation}
T'(x)
=
\frac{x_T}{D}
\tilde T'(\tilde x).
\label{eq:first_derivative_nondim_cue}
\end{equation}

We can apply the same procedure once more to obtain the second derivative.
Differentiating the previous relation with respect to $\tilde x$ gives
\begin{align}
\frac{d^2\tilde T}{d\tilde x^2}
&=
\frac{D}{x_T}
\frac{d}{d\tilde x}T'(x)
\\
&=
\frac{D}{x_T}
\frac{dT'}{dx}
\frac{dx}{d\tilde x}
\\
&=
\frac{D}{x_T}
T''(x)x_T
\\
&=
D T''(x).
\end{align}
Hence,
\begin{equation}
T''(x)
=
\frac{1}{D}
\tilde T''(\tilde x).
\label{eq:second_derivative_nondim_cue}
\end{equation}

Substituting these relations, together with
$\tilde\ell=\ell_c/x_T$, into eqn.~\ref{eq:backward_exp_cue} gives
\begin{equation}
0
=
1
+
\tilde T''(\tilde x)
+
\frac{v_*x_T}{D}
e^{-(1-\tilde x)/\tilde\ell}
\tilde T'(\tilde x)
-
\frac{rx_T^2}{D}\tilde T(\tilde x)
+
\frac{rx_T^2}{D}\tilde T(0).
\end{equation}
Using the dimensionless parameters,
\begin{equation}
\mathrm{Pe}
=
\frac{v_*x_T}{D},
\qquad
\tilde T_R
=
\frac{D}{rx_T^2},
\end{equation}
we obtain the dimensionless backward equation
\begin{equation}
0
=
1
+
\tilde T''(\tilde x)
+
\mathrm{Pe}\,
e^{-(1-\tilde x)/\tilde\ell}
\tilde T'(\tilde x)
-
\frac{\tilde T(\tilde x)}{\tilde T_R}
+
\frac{\tilde T(0)}{\tilde T_R}
.
\label{eq:dimensionless_backward_exp_cue}
\end{equation}
This is the desired dimensionless eqn.~\ref{eq:pecletDescriptionOfEq100} of the main text.

\subsection{Solving the backward equation for the mean first-passage time}
\label{app:derivation_hypergeometric_with_cues}

We now solve the backward equation associated with the exponentially localized cue. To make further progress understanding this equation, we now introduce an auxiliary function $\phi(\tilde x)$ that captures the
position-dependent part of the mean first-passage time, thus expressing the dynamics as
\begin{equation}
\tilde T(\tilde x)
=
\left[\tilde T(0)+\tilde T_R\right]
\left[1-\phi(\tilde x)\right].
\label{eq:T_phi_decomposition}
\end{equation}
Substituting this change of variables into
eqn.~\ref{eq:pecletDescriptionOfEq100}, the constant terms cancel and $\phi$
satisfies the homogeneous equation
\begin{equation}
\phi''(\tilde x)
+
\mathrm{Pe}
\,e^{-(1-\tilde x)/\tilde\ell}
\phi'(\tilde x)
-
\frac{1}{\tilde T_R}\phi(\tilde x)
=
0.
\label{eq:phi_exp_cue}
\end{equation}
The absorbing boundary condition $\tilde T(1)=0$ gives
\begin{equation}
\phi(1)=1.
\label{eq:phi1is1}
\end{equation}
Evaluating eqn.~\ref{eq:T_phi_decomposition} at the starting point
$\tilde x=0$ gives
\begin{equation}
\tilde T(0)
=
\tilde T_R
\left(
\frac{1}{\phi(0)}-1
\right).
\label{eq:T_phi_relation}
\end{equation}

As $\tilde x\rightarrow-\infty$, the cue vanishes exponentially and
eqn.~\ref{eq:phi_exp_cue} reduces to
\begin{equation}
\phi''-\frac{1}{\tilde T_R}\phi=0.
\label{eq:hitTimeUnderExpCueAt0}
\end{equation}
Requiring the solution to remain bounded as $\tilde x\rightarrow-\infty$
selects the decaying exponential branch,
\begin{equation}
\phi(\tilde x)
\propto
e^{\tilde x/\sqrt{\tilde T_R}}
\qquad
(\tilde x\rightarrow-\infty).
\label{eq:phi_far_field}
\end{equation}

When the cue is absent, this far-field equation holds everywhere, and the
normalization $\phi(1)=1$ of eqn.~\ref{eq:phi1is1} gives
\begin{equation}
\phi(\tilde x)
=
e^{(\tilde x-1)/\sqrt{\tilde T_R}}.
\end{equation}
Substituting $\phi(0)=e^{-1/\sqrt{\tilde T_R}}$ into
eqn.~\ref{eq:T_phi_relation} recovers the result obtained
previously for diffusion with resetting of eqn.~\ref{eqn:SearchTimeFinalDimensionless}. For the exponential cue, eqn.~\ref{eq:phi_exp_cue} admits a closed-form
solution for $\phi(\tilde{x})$ in terms of hypergeometric functions that we now derive.

We show here that this equation can be transformed into Kummer's equation and therefore solved in terms of confluent hypergeometric functions \cite{abramowitz1966handbook}.

We introduce the change of variables
\begin{equation}
z
=
\tilde{\ell} \, \mathrm{Pe}\,
e^{-(1-\tilde{x})/\tilde{\ell}} .
\label{eq:change_of_variable_z}
\end{equation}
For $\tilde{\ell}>0$ and $\mathrm{Pe}>0$, this function is smooth and monotone. Since
\begin{equation}
\frac{dz}{d\tilde{x}}
=
\frac{z}{\tilde{\ell}},
\end{equation}
we have
\begin{equation}
\frac{d\phi}{d\tilde{x}}
=
\frac{z}{\tilde{\ell}}
\frac{d\phi}{dz},
\end{equation}
and
\begin{equation}
\frac{d^2\phi}{d\tilde{x}^2}
=
\frac{z}{\tilde{\ell}^2}
\frac{d\phi}{dz}
+
\left(\frac{z}{\tilde{\ell}}\right)^2
\frac{d^2\phi}{dz^2}.
\end{equation}
Substituting these expressions into eqn.~\ref{eq:phi_exp_cue} gives
\begin{equation}
z^2 \phi''(z)
+
z(z+1)\phi'(z)
-
\nu^2 \phi(z)
=
0,
\label{eq:ode_on_phi_of_z}
\end{equation}
where
\begin{equation}
\nu
=
\frac{\tilde{\ell}}{\sqrt{\tilde{T}_{R}}}.
\end{equation}
To reduce this equation to the Kummer equation, we write
\begin{equation}
\phi(z)
=
z^\nu u(z).
\end{equation}
Then
\begin{equation}
\phi'(z)
=
\nu z^{\nu-1}u(z)
+
z^\nu u'(z),
\end{equation}
and
\begin{equation}
\phi''(z)
=
\nu(\nu-1)z^{\nu-2}u(z)
+
2\nu z^{\nu-1}u'(z)
+
z^\nu u''(z).
\end{equation}
Substitution into eqn.~\ref{eq:ode_on_phi_of_z} yields
\begin{equation}
z u''(z)
+
(2\nu+1+z)u'(z)
+
\nu u(z)
=
0.
\end{equation}
Finally, defining $s=-z$, we obtain
\begin{equation}
s u''(s)
+
(2\nu+1-s)u'(s)
-
\nu u(s)
=
0.
\end{equation}
This is exactly Kummer's equation \cite{abramowitz1966handbook}, defined as,
\begin{equation}
s u''(s)
+
(b-s)u'(s)
-
a u(s)
=
0,
\end{equation}
with parameters
\begin{equation}
a=\nu,
\qquad
b=2\nu+1.
\end{equation}
A standard solution is the confluent hypergeometric function of the first kind \cite{abramowitz1966handbook},
\begin{equation}
M(a,b,s)
=
\sum_{n=0}^{\infty}
\frac{(a)_n}{(b)_n}
\frac{s^n}{n!},
\end{equation}
where $(a)_0=1$ and
\begin{equation}
(a)_n
=
a(a+1)\cdots(a+n-1)
\end{equation}
is the rising factorial. A second linearly independent solution is given by Tricomi's confluent hypergeometric function $U(a,b,s)$ \cite{abramowitz1966handbook}. Thus the general solution can be written as
\begin{equation}
u(s)
=
C_{1}M(a,b,s)
+
C_{2}U(a,b,s).
\end{equation}

Returning to the original variable $\tilde{x}$, we find
\begin{align}
\phi(\tilde{x})
&=
D_{1}
e^{-(1-\tilde{x})/\sqrt{\tilde{T}_{R}}}
M\left(
\nu,
2\nu+1,
-\tilde{\ell},\mathrm{Pe},
e^{-(1-\tilde{x})/\tilde{\ell}}
\right)
\nonumber\\
&\quad
+
D_{2}
e^{-(1-\tilde{x})/\sqrt{\tilde{T}_{R}}}
U\left(
\nu,
2\nu+1,
-\tilde{\ell},\mathrm{Pe},
e^{-(1-\tilde{x})/\tilde{\ell}}
\right),
\end{align}
where $D_{1}$ and $D_{2}$ are constants.

The physically relevant solution is selected by considering the far-left limit $\tilde{x}\to -\infty$, where the cue vanishes. In this limit, the solution should reduce to the decaying no-cue form
\begin{equation}
\phi(\tilde{x})
\propto
e^{-(1-\tilde{x})/\sqrt{\tilde{T}_{R}}}.
\end{equation}
Since $M(\nu,2\nu+1,0)=1$, the $M$ branch has precisely this behavior. By contrast, the $U$ branch is singular as $\tilde{x}\to -\infty$ for $\nu>0$ and gives the growing solution. We therefore set $D_{2}=0$.

The remaining constant is fixed by the normalization condition $\phi(1)=1$. This gives
\begin{equation}
\phi(\tilde{x})
=
e^{-(1-\tilde{x})/\sqrt{\tilde{T}_{R}}}
\frac{
M\left(
\nu,
2\nu+1,
-\tilde{\ell},\mathrm{Pe},
e^{-(1-\tilde{x})/\tilde{\ell}}
\right)
}{
M\left(
\nu,
2\nu+1,
-\tilde{\ell},\mathrm{Pe}
\right)
}.
\label{eq:phi_solution_hypergeometric}
\end{equation}
In particular, evaluating this expression for an initial position at $\tilde{x}=0$ gives
\begin{equation}
\phi(0)
=
e^{-1/\sqrt{\tilde{T}_{R}}}
\frac{
M\left(
\nu,
2\nu+1,
-\tilde{\ell},\mathrm{Pe},
e^{-1/\tilde{\ell}}
\right)
}{
M\left(
\nu,
2\nu+1,
-\tilde{\ell},\mathrm{Pe}
\right)
}.
\label{eq:phi_zero_hypergeometric}
\end{equation}
The dimensionless mean first-passage time from the reset position is therefore finally,
\begin{equation}
\tilde{T}(0)
=
\tilde{T}_{R}
\left(
\frac{1}{\phi(0)}-1
\right).
\label{eq:mfpt_hypergeometric_final}
\end{equation}
This is eqn.~\ref{eq:T_phi_relation}.

Numerical analyses in the main text in the $({\rm Pe}, \ell)$ plane are found as follows. To appreciate the behavior of this larger physical problem, we
evaluate the solution of eqn.~\ref{eq:hitTimeUnderExpCueAt0} numerically at $\tilde{x}=0$ via $\tilde{\phi}(0)$ and then compute the
mean first-passage time using eqn.~\ref{eq:T_phi_relation}. 

\subsection{Asymptotic checks for the exponential cue solution}

\subsubsection{Large $\rm Pe$ for fixed $\tilde\ell$ and $\tilde T_{R}$ }
\label{app:asymptoticExponentialCue}

\begin{figure}[H]
    \centering
    \includegraphics[width=0.5\textwidth]{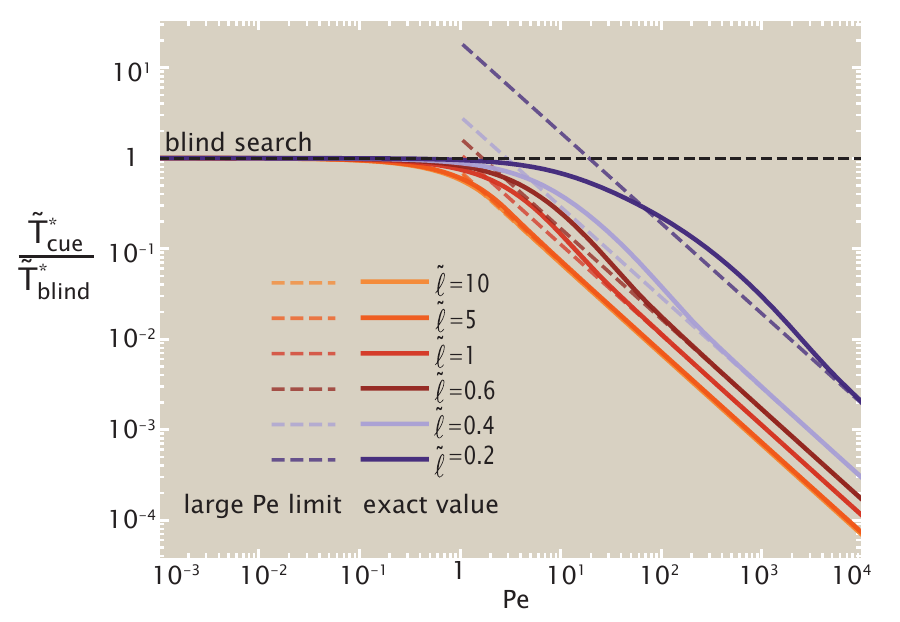}
    \caption{\raggedright
    Large-$\mathrm{Pe}$ behavior of the optimized search time.
    Solid curves show the numerical result for different cue lengths
    $\tilde\ell$, while dashed curves show the large-$\mathrm{Pe}$
    drift-time asymptote. The numerical solution approaches the predicted
    asymptotic behavior as $\mathrm{Pe}$ increases.
    }
    \label{fig:largePeExponentialCue}
\end{figure}

We now record a few useful asymptotic limits of the exact solution. We first look at the large Peclet limit. We first consider the limit of $\mathrm{Pe}$ going to infinity at fixed $\tilde\ell$ and fixed $\tilde T_{R}$. Using the large negative argument asymptotic behavior of Kummer's function \cite{abramowitz1966handbook},
\begin{equation}
    M(a,b,-y)
    \sim
    \frac{\Gamma(b)}{\Gamma(b-a)}y^{-a},
    \qquad
    y\to+\infty, 
\end{equation}
Using this asymptotic form, we find
\begin{equation}
\phi(0)
\to 1,
\qquad
\mathrm{Pe}\to\infty.
\end{equation}
Since the dimensionless mean first-passage time is
\begin{equation}
\tilde{T}(0)
=
\tilde{T}_{R}
\left(
\frac{1}{\phi(0)}-1
\right),
\end{equation}
this immediately implies
\begin{equation}
\tilde{T}(0)\to 0,
\qquad
\mathrm{Pe}\to\infty.
\end{equation}
As the Péclet number becomes large, the drift toward the target dominates diffusion and resetting. In the limit $\mathrm{Pe}\to\infty$, the walker is transported to the target infinitely rapidly, so the mean first-passage time vanishes. To understand better how the mean first passage time converges to 0 in the limit of large $\rm Pe$ number, we keep the first order correction (given by Mathematica),
\begin{equation}
    \phi(0)
    =
    1
    -
    \frac{\tilde\ell\left(e^{1/\tilde\ell}-1\right)}
    {\tilde T_{R}\mathrm{Pe}}
    +
    O\!\left(\mathrm{Pe}^{-2}\right).
\end{equation}
Therefore
\begin{equation}
    \tilde T(0)
    =
    \frac{\tilde\ell\left(e^{1/\tilde\ell}-1\right)}
    {\mathrm{Pe}}
    +
    O\!\left(\mathrm{Pe}^{-2}\right).
\end{equation}
This leading term is exactly the deterministic drift time from $\tilde x=0$ to $\tilde x=1$ in the velocity field
\begin{equation}
    \tilde v(\tilde x)
    =
    \mathrm{Pe}\,e^{-(1-\tilde x)/\tilde\ell}.
\end{equation}
Indeed,
\begin{equation}
    \tilde t_{\mathrm{drift}}
    =
    \int_0^1
    \frac{d\tilde x}
    {\mathrm{Pe}\,e^{-(1-\tilde x)/\tilde\ell}}
    =
    \frac{1}{\mathrm{Pe}}
    \int_0^1
    e^{(1-\tilde x)/\tilde\ell}
    d\tilde x
    =
    \frac{\tilde\ell\left(e^{1/\tilde\ell}-1\right)}
    {\mathrm{Pe}}.
\end{equation}
Thus, in the strong-cue limit, the mean first-passage time approaches the
drift-limited travel time. Figure~\ref{fig:largePeExponentialCue} shows that
the numerical solution converges to this asymptotic prediction at large
$\mathrm{Pe}$.

The deterministic drift time also suggests a natural dimensionless quantity for
characterizing when the cue substantially accelerates the search. We compare it
to the optimal blind-search time and define
\begin{equation}
\Theta_{\rm drift}
=
\frac{\tilde t_{\rm drift}}{\tilde T_{\rm blind}^*}
=
\frac{
\tilde\ell\left(e^{1/\tilde\ell}-1\right)
}{
\mathrm{Pe}\,\tilde T_{\rm blind}^*
}.
\end{equation}
When $\Theta_{\rm drift}\gg 1$, following the cue deterministically would be
slower than the characteristic blind-search time, whereas when
$\Theta_{\rm drift}\ll 1$, the cue provides a much faster route to the target.

\begin{figure}[H]
    \centering
    \includegraphics[width=0.5\textwidth]{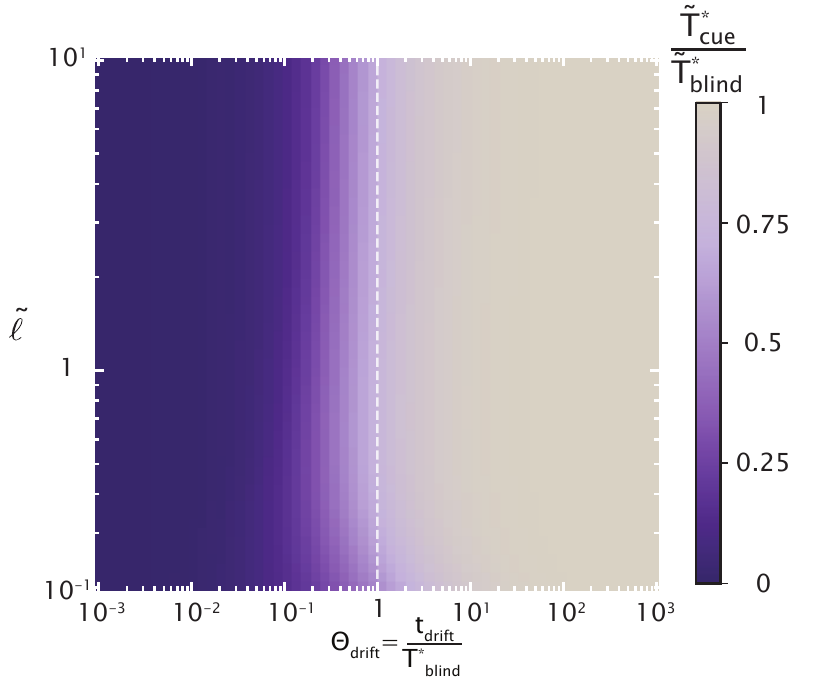}
    \caption{\raggedright
    Optimized search time expressed in terms of the normalized deterministic
    drift time $\Theta_{\rm drift}=\tilde t_{\rm drift}/\tilde T_{\rm blind}^*$.
    The color indicates the optimized search time in the presence of the cue,
    normalized by the optimal blind-search time,
    $\tilde T_{\rm cue}^*/\tilde T_{\rm blind}^*$.
    }
    \label{fig:thetaDriftCue}
\end{figure}

Figure~\ref{fig:thetaDriftCue} replots the optimized search-time ratio (previously shown in the $(\ell, {\rm Pe})$ plane in Fig.~\ref{fig:CueOptimization} of the main text in the $(\Theta_{\rm drift},\tilde\ell)$ plane instead. The nearly vertical contours show that
much of the dependence of the optimized search time on $\mathrm{Pe}$ and
$\tilde\ell$ is captured by the single combination $\Theta_{\rm drift}$.

\subsubsection{Large $\tilde T_{R}$ for fixed $\tilde\ell$ and $\rm Pe$ }
\label{app:largeTRFixedCue}
We next consider the limit of large reset time,
$\tilde T_R\to\infty$, corresponding to the no-reset regime.
For a purely diffusive walker in one dimension, removing resetting leads to
an infinite mean first-passage time. Here, however, the walker experiences a
drift toward the target. We therefore ask whether the exponentially localized
cue is sufficient to produce a finite mean first-passage time at fixed,
finite $\mathrm{Pe}$ and $\tilde\ell$.

To study this limit, we introduce the small parameter
\begin{equation}
\epsilon
=
\frac{1}{\sqrt{\tilde T_R}},
\end{equation}
so that $\epsilon\to0$ as $\tilde T_R\to\infty$. The equation for the
auxiliary function $\phi$ can then be written exactly as
\begin{equation}
\phi''(\tilde x)
+
\mathrm{Pe}\,
e^{-(1-\tilde x)/\tilde\ell}
\phi'(\tilde x)
-
\epsilon^2\phi(\tilde x)
=
0.
\label{eq:phi_large_TR_epsilon}
\end{equation}

The large-$\tilde T_R$ limit is slightly subtle. Although $\epsilon$
appears explicitly in eqn.~\ref{eq:phi_large_TR_epsilon} only through
$\epsilon^2$, the far-field boundary condition contains a contribution
already at order $\epsilon$. Indeed, as $\tilde x\to-\infty$, the cue
vanishes and eqn.~\ref{eq:phi_large_TR_epsilon} reduces to
\begin{equation}
\phi''-\epsilon^2\phi=0.
\end{equation}
Requiring the solution to remain bounded as $\tilde x\to-\infty$ gives
\begin{equation}
\phi(\tilde x)\propto e^{\epsilon\tilde x},
\end{equation}
and therefore
\begin{equation}
\frac{\phi'(\tilde x)}{\phi(\tilde x)}
\longrightarrow
\epsilon,
\qquad
\tilde x\to-\infty.
\label{eq:phi_log_derivative_far_field}
\end{equation}

At leading order, $\epsilon=0$, and eqn.~\ref{eq:phi_large_TR_epsilon}
reduces to
\begin{equation}
\phi_0''(\tilde x)
+
\mathrm{Pe}\,
e^{-(1-\tilde x)/\tilde\ell}
\phi_0'(\tilde x)
=
0.
\end{equation}
At the same order, the far-field condition
$\phi'/\phi\to\epsilon$ becomes
\begin{equation}
\phi_0'(-\infty)=0.
\end{equation}
Defining $q_0(\tilde x)=\phi_0'(\tilde x)$, we obtain
\begin{equation}
q_0'
+
\mathrm{Pe}\,
e^{-(1-\tilde x)/\tilde\ell}
q_0
=
0.
\end{equation}
Since $q_0(-\infty)=0$, the solution is $q_0(\tilde x)=0$ everywhere.
Therefore $\phi_0$ is constant. Finally, the normalization
$\phi_0(1)=1$ fixes this constant, giving
\begin{equation}
\phi_0(\tilde x)=1.
\end{equation}
To determine the order of the first correction, we expand
\begin{equation}
\phi(\tilde x)
=
\phi_0(\tilde x)
+
\epsilon\psi(\tilde x)
+
O(\epsilon^2).
\end{equation}
At leading order, $\phi_0(\tilde x)=1$. We then match this expansion to the
far-field condition eqn. ~\ref{eq:phi_log_derivative_far_field}
.
Substituting the expansion gives
\begin{equation}
\frac{\phi'}{\phi}
=
\epsilon\psi'(\tilde x)
+
O(\epsilon^2),
\end{equation}
and therefore matching at order $\epsilon$ requires
\begin{equation}
\psi'(-\infty)=1.
\end{equation}
Thus the coefficient of the order-$\epsilon$ correction is nonzero, and the
expansion must contain a term linear in $\epsilon$:
\begin{equation}
\phi(\tilde x)
=
1+\epsilon\psi(\tilde x)
+
O(\epsilon^2).
\end{equation}
\label{eq:phi_large_TR_expansion}

Substituting this expression
into eqn.~\ref{eq:phi_large_TR_epsilon}, we have
\begin{equation}
\epsilon\psi''(\tilde x)
+
\mathrm{Pe}\,
e^{-(1-\tilde x)/\tilde\ell}
\epsilon\psi'(\tilde x)
-
\epsilon^2
\left[
1+\epsilon\psi(\tilde x)
\right]
+
O(\epsilon^2)
=
0.
\end{equation}
Dividing by $\epsilon$ and keeping only the leading terms as
$\epsilon\to0$, we obtain
\begin{equation}
\psi''(\tilde x)
+
\mathrm{Pe}\,
e^{-(1-\tilde x)/\tilde\ell}
\psi'(\tilde x)
=
0.
\end{equation}
Introducing
\begin{equation}
q(\tilde x)=\psi'(\tilde x),
\end{equation}
we obtain the first-order equation
\begin{equation}
q'(\tilde x)
+
\mathrm{Pe}\,
e^{-(1-\tilde x)/\tilde\ell}
q(\tilde x)
=
0.
\end{equation}
Integrating gives
\begin{equation}
q(\tilde x)
=
C
\exp\left[
-\tilde\ell\,\mathrm{Pe}\,
e^{-(1-\tilde x)/\tilde\ell}
\right].
\end{equation}
Since $q(-\infty)=1$, the integration constant is $C=1$, and hence
\begin{equation}
\psi'(\tilde x)
=
\exp\left[
-\tilde\ell\,\mathrm{Pe}\,
e^{-(1-\tilde x)/\tilde\ell}
\right].
\end{equation}

Using $\psi(1)=0$, we then find
\begin{equation}
\psi(0)
=
-
\int_0^1
\exp\left[
-\tilde\ell\,\mathrm{Pe}\,
e^{-(1-\tilde x)/\tilde\ell}
\right]
d\tilde x.
\end{equation}
It is therefore convenient to define
\begin{equation}
A(\mathrm{Pe},\tilde\ell)
=
\int_0^1
\exp\left[
-\tilde\ell\,\mathrm{Pe}\,
e^{-(1-\tilde x)/\tilde\ell}
\right]
d\tilde x,
\label{eq:A_large_TR_integral}
\end{equation}
so that
\begin{equation}
\phi(0)
=
1-
\frac{A(\mathrm{Pe},\tilde\ell)}
{\sqrt{\tilde T_R}}
+
O\!\left(\tilde T_R^{-1}\right).
\end{equation}

Using
\begin{equation}
\tilde T(0)
=
\tilde T_R
\left(
\frac{1}{\phi(0)}-1
\right),
\end{equation}
we obtain
\begin{equation}
\frac{1}{\phi(0)}-1
=
\frac{A(\mathrm{Pe},\tilde\ell)}
{\sqrt{\tilde T_R}}
+
O\!\left(\tilde T_R^{-1}\right),
\end{equation}
and therefore
\begin{equation}
\tilde T(0)
\sim
A(\mathrm{Pe},\tilde\ell)
\sqrt{\tilde T_R}
\qquad
\tilde T_R\to\infty.
\label{eq:T_large_TR_asymptotic}
\end{equation}

Thus, at any fixed finite $\mathrm{Pe}$ and $\tilde\ell$, the exponentially
localized cue does not regularize the no-reset problem: the mean
first-passage time diverges as $\sqrt{\tilde T_R}$ as the reset time becomes
large.
The coefficient $ A(\mathrm{Pe},\tilde\ell)$ is positive for $\tilde\ell>0$ and $\mathrm{Pe}>0$. Therefore,
\begin{equation}
    \tilde T(0)\to\infty
    \qquad
    \text{as}
    \qquad
    \tilde T_{R}\to\infty.
\end{equation}
Thus, for finite $\mathrm{Pe}$ and finite $\tilde\ell$, taking the no-reset limit makes the mean first-passage time diverge. This confirms that the optimum reset time remains finite at finite cue strength.

We now consider the joint limit of large Péclet number and large nondimensional reset time. The motivation for this limit is that, at high Péclet number, the mean first-passage time approaches the deterministic drift-dominated travel time. In the formal limit $\mathrm{Pe}\to\infty$, this behavior holds for any reset time, since the drift is strong enough to bring the searcher to the target almost immediately. For large but finite $\mathrm{Pe}$, however, we expect the relevant reset time to be large but finite. Indeed, resets must be sufficiently rare to allow the drift to carry the searcher to the target, but the reset time cannot be taken to infinity at fixed $\mathrm{Pe}$, since we have shown that the mean first-passage time diverges in the no-reset limit. This motivates studying the combined asymptotic regime in which both $\mathrm{Pe}$ and $\tilde T_{R}$ are large.
Introducing the cue-scale Péclet number
\begin{equation}
\mathrm{Pe}_{\rm cue,0}
=
\tilde\ell\,\mathrm{Pe}\,e^{-1/\tilde\ell},
\end{equation}
the prefactor can be written as
\begin{equation}
A(\mathrm{Pe},\tilde\ell)
=
\int_0^1
\exp\left[
-\mathrm{Pe}_{\rm cue,0}
e^{\tilde x/\tilde\ell}
\right]
d\tilde x.
\end{equation}
For $\mathrm{Pe}_{\rm cue,0}\gg1$, this integral is dominated by
$\tilde x\simeq0$. Expanding
\begin{equation}
e^{\tilde x/\tilde\ell}
\simeq
1+\frac{\tilde x}{\tilde\ell},
\end{equation}
we obtain
\begin{align}
A(\mathrm{Pe},\tilde\ell)
&\sim
e^{-\mathrm{Pe}_{\rm cue,0}}
\int_0^\infty
\exp\left[
-\frac{\mathrm{Pe}_{\rm cue,0}}{\tilde\ell}\tilde x
\right]
d\tilde x
\\
&=
\frac{\tilde\ell}{\mathrm{Pe}_{\rm cue,0}}
e^{-\mathrm{Pe}_{\rm cue,0}}.
\end{align}
Equivalently,
\begin{equation}
A(\mathrm{Pe},\tilde\ell)
\sim
\frac{e^{1/\tilde\ell}}{\mathrm{Pe}}
\exp\left[
-\tilde\ell\mathrm{Pe}e^{-1/\tilde\ell}
\right].
\end{equation}
Therefore, in the large-$\tilde T_{R}$ and large-$\mathrm{Pe}$ regime,
\begin{equation}
    \tilde T(0)
    \sim
    \frac{e^{1/\tilde\ell}}{\mathrm{Pe}}
    \exp\!\left[
    -\tilde\ell \mathrm{Pe}\,e^{-1/\tilde\ell}
    \right]
    \sqrt{\tilde T_{R}}.
\end{equation}
This expression shows that, although $\tilde T(0)$ still diverges as $\tilde T_{R}\to\infty$ at any fixed $\mathrm{Pe}$, the prefactor of this divergence becomes exponentially small when the cue is strong at the reset position.

Finally, we get an estimate for the optimal reset time in the strong-cue regime. A simple estimate for the reset time at which the large-$\tilde T_{R}$ asymptotic behavior reaches this drift-limited scale is obtained by matching
\begin{equation}
    \frac{e^{1/\tilde\ell}}{\mathrm{Pe}}
    \exp\!\left[
    -\tilde\ell \mathrm{Pe}\,e^{-1/\tilde\ell}
    \right]
    \sqrt{\tilde T_{R}}
    \sim
        \tilde t_{\mathrm{drift}}
    =
    \frac{\tilde\ell\left(e^{1/\tilde\ell}-1\right)}
    {\mathrm{Pe}}.
\end{equation}
After canceling the common factor $1/\mathrm{Pe}$, this gives
\begin{equation}
    \sqrt{\tilde T_{R}}
    \sim
    \tilde\ell
    \left(1-e^{-1/\tilde\ell}\right)
    \exp\!\left[
    \tilde\ell \mathrm{Pe}\,e^{-1/\tilde\ell}
    \right].
\end{equation}
Thus the corresponding reset time is
\begin{equation}
    \tilde T_{R}
    \sim
    \tilde\ell^2
    \left(1-e^{-1/\tilde\ell}\right)^2
    \exp\!\left[
    2\tilde\ell \mathrm{Pe}\,e^{-1/\tilde\ell}
    \right].
\end{equation}
This scale grows exponentially with $\mathrm{Pe}$ at fixed $\tilde\ell$. Hence, in the strong-cue regime, the optimal reset time is pushed to very large values. Physically, this corresponds to the fact that the search is already close to drift-limited: resetting becomes less useful because it interrupts an almost deterministic motion toward the target. Thus this estimate gives the large-$\mathrm{Pe}$ scaling of the optimal reset time, rather than an exact finite-$\mathrm{Pe}$ formula.

\newpage
\subsection{Existence of a minimum for the optimal reset time}
\label{app:existOptResetTimeMin}

\begin{figure}[h!]
    \centering
\includegraphics[width=0.75\textwidth]{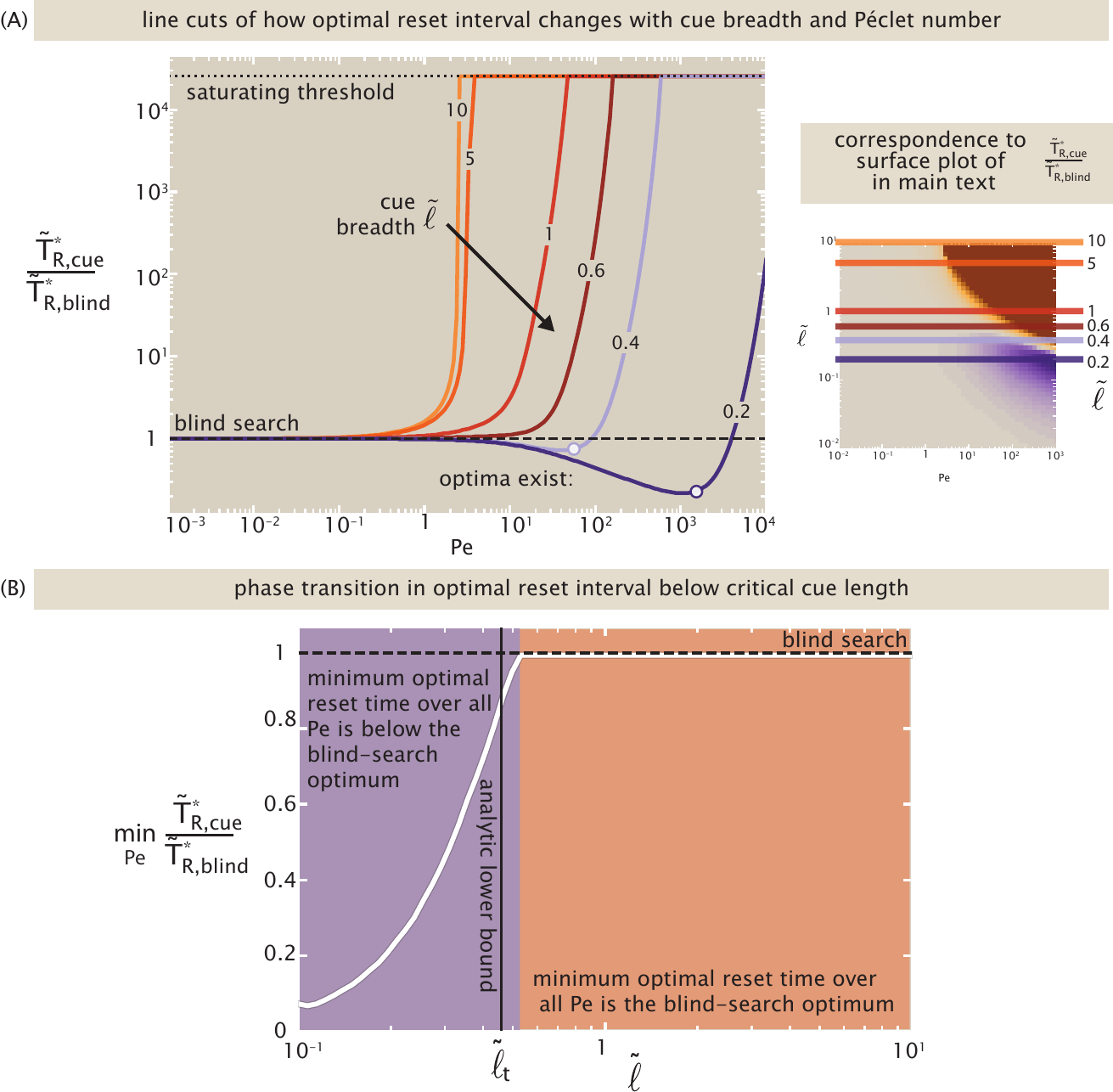}
    \caption{\raggedright
    Localized cues can favor more frequent resetting.
    (A) Optimal reset time relative to the blind-search optimum as a function
of $\mathrm{Pe}$ for several cue lengths $\tilde\ell$.
Values above $10^4$ are saturated for visualization, reflecting the
divergence of the optimal reset time in the strong-cue regime. (The heatmap of Fig.~\ref{fig:CueOptimization}(A) of the main text is repeated at right as a guide to which slices through the landscape the plot at left shows.)
    (B) Minimum optimal reset-time ratio over cue strength as a function of
    $\tilde\ell$. The vertical line marks the analytic lower bound derived
    from the weak-cue expansion.
    }
    \label{fig:localizedCueResetting}
\end{figure}
\FloatBarrier

Figure~\ref{fig:CueOptimization}(A) of the main text, plotting the optimal reset interval $\tilde{T}_{R,\text{cue}}^*$ (relative to the blind search value), revealed the onset of a surprising regime (visible as a purple island in the heatmap). Here, a moderately localized cue paired with a high-drift was shown to recommend resetting at a \emph{smaller} rate than in blind search. Thus, while broad and strong cues suppress resetting, a localized cue can instead favor more frequent resetting. Here we discuss the onset of this regime in more quantitative detail.

Figure~\ref{fig:localizedCueResetting}(A) inspects line cuts through the heatmap's surface at various values of the cue breadth $\ell$. Specifically, note that curves shown with $\ell\lesssim 0.4$ present minima in $\tilde{T}_{R,\text{cue}}^*/\tilde{T}_{R,\text{blind}}^*$ less than one, and are colored in purple. 
To characterize when this behavior occurs, for each cue length we consider
\begin{equation}
\min_{\mathrm{Pe}}
\frac{\tilde T_{R,\mathrm{cue}}^*}
{\tilde T_{R,\mathrm{blind}}^*},
\end{equation}
as shown in Fig.~\ref{fig:localizedCueResetting}(B). Note the explicit onset of a discrete transition in optimal reset policy. A value below one means that there exists some cue strength for which the
optimal resetting is more frequent than in blind search. Numerically, this
occurs only for sufficiently localized cues.

We can obtain an analytic lower bound on this threshold by examining the
response of the optimal reset time to an infinitesimally weak cue. It is
convenient to use the cue-scale Péclet number
\begin{equation}
\mathrm{Pe}_{\rm cue,0}
=
\tilde\ell\,\mathrm{Pe}\,e^{-1/\tilde\ell}.
\end{equation}
At fixed $\tilde\ell$, the weak-cue limit corresponds to
$\mathrm{Pe}_{\rm cue,0}\to0$.

We write
\begin{equation}
\alpha=\frac{1}{\sqrt{\tilde T_R}},
\qquad
\nu=\alpha\tilde\ell,
\end{equation}
so that $\tilde T_R=1/\alpha^2$. The mean first-passage time is
\begin{equation}
\tilde T(0;\alpha)
=
\frac{1}{\alpha^2}
\left(
\frac{1}{\phi(0)}-1
\right).
\end{equation}

We recall the hypergeometric solution,
\begin{equation}
\phi(0)
=
\frac{
\mathrm{Pe}_{\rm cue,0}^{\,\nu}
M(\nu,2\nu+1,-\mathrm{Pe}_{\rm cue,0})
}{
\left(\mathrm{Pe}_{\rm cue,0}e^{1/\tilde\ell}\right)^\nu
M\!\left(
\nu,2\nu+1,
-\mathrm{Pe}_{\rm cue,0}e^{1/\tilde\ell}
\right)
}.
\end{equation}
For $\mathrm{Pe}_{\rm cue,0}\ll1$, we use
\begin{equation}
M(a,b,-z)
=
1-\frac{a}{b}z+O(z^2).
\end{equation}
The prefactor in $\phi(0)$ simplifies as
\begin{equation}
\left(
\frac{\mathrm{Pe}_{\rm cue,0}}
{\mathrm{Pe}_{\rm cue,0}e^{1/\tilde\ell}}
\right)^\nu
=
e^{-\nu/\tilde\ell}
=
e^{-\alpha}.
\end{equation}
Therefore,
\begin{equation}
\phi(0)
=
e^{-\alpha}
\frac{
1-\dfrac{\nu}{2\nu+1}\mathrm{Pe}_{\rm cue,0}
+O(\mathrm{Pe}_{\rm cue,0}^2)
}{
1-\dfrac{\nu}{2\nu+1}
e^{1/\tilde\ell}\mathrm{Pe}_{\rm cue,0}
+O(\mathrm{Pe}_{\rm cue,0}^2)
}.
\end{equation}
Using
\begin{equation}
\frac{1-a z}{1-b z}
=
1+(b-a)z+O(z^2),
\end{equation}
we obtain
\begin{equation}
\phi(0)
=
e^{-\alpha}
\left[
1+
\frac{\nu}{2\nu+1}
\left(e^{1/\tilde\ell}-1\right)
\mathrm{Pe}_{\rm cue,0}
+
O(\mathrm{Pe}_{\rm cue,0}^2)
\right].
\end{equation}
It is useful to define
\begin{equation}
B_{\tilde\ell}(\alpha)
=
\frac{
\alpha\left(e^{1/\tilde\ell}-1\right)
}{
2\alpha+1/\tilde\ell
}.
\label{eq:B_ell_alpha}
\end{equation}
Indeed, since $\nu=\alpha\tilde\ell$, the previous expression can be written as
\begin{equation}
\phi(0)
=
e^{-\alpha}
\left[
1+
\mathrm{Pe}_{\rm cue,0}
B_{\tilde\ell}(\alpha)
+
O(\mathrm{Pe}_{\rm cue,0}^2)
\right].
\end{equation}
Therefore,
\begin{equation}
\frac{1}{\phi(0)}
=
e^\alpha
\left[
1-
\mathrm{Pe}_{\rm cue,0}
B_{\tilde\ell}(\alpha)
+
O(\mathrm{Pe}_{\rm cue,0}^2)
\right],
\end{equation}
and the mean first-passage time becomes
\begin{equation}
\tilde T(0;\alpha)
=
\frac{e^\alpha-1}{\alpha^2}
-
\mathrm{Pe}_{\rm cue,0}
\frac{e^\alpha B_{\tilde\ell}(\alpha)}{\alpha^2}
+
O(\mathrm{Pe}_{\rm cue,0}^2).
\label{eq:weak_cue_MFPT_expansion}
\end{equation}

The blind-search result corresponds to
$\mathrm{Pe}_{\rm cue,0}=0$:
\begin{equation}
\tilde T_{\rm blind}(\alpha)
=
\frac{e^\alpha-1}{\alpha^2}.
\end{equation}
The blind-search optimum is obtained by minimizing
$\tilde T_{\rm blind}(\alpha)$ with respect to $\alpha$. We denote the
minimizing value by $\alpha_0$, defined by
\begin{equation}
\left.
\frac{d\tilde T_{\rm blind}}{d\alpha}
\right|_{\alpha=\alpha_0}
=0.
\end{equation}
This gives
\begin{equation}
e^{\alpha_0}(\alpha_0-2)+2=0,
\end{equation}
with
\begin{equation}
\alpha_0\simeq 1.5936,
\qquad
\tilde T_{R,\rm blind}^*
=
\frac{1}{\alpha_0^2}
\simeq 0.394.
\end{equation}

We now ask how the optimal reset rate changes when
$\mathrm{Pe}_{\rm cue,0}$ is increased slightly away from zero. We write
\begin{equation}
\alpha^*(\mathrm{Pe}_{\rm cue,0})
=
\alpha_0
+
\mathrm{Pe}_{\rm cue,0}\alpha_1
+
O(\mathrm{Pe}_{\rm cue,0}^2).
\end{equation}
Using the expansion in eqn.~\ref{eq:weak_cue_MFPT_expansion}, the sign of
the first correction $\alpha_1$ is controlled by
\begin{equation}
\operatorname{sign}(\alpha_1)
=
\operatorname{sign}
\left[
1-\frac{1}{\alpha_0}
-
\frac{2}{2\alpha_0+1/\tilde\ell}
\right].
\label{eq:alpha1_sign}
\end{equation}
Since
\begin{equation}
\tilde T_R^*
=
\frac{1}{(\alpha^*)^2},
\end{equation}
the initial change in $\tilde T_R^*$ has the opposite sign to the initial
change in $\alpha^*$.

The sign change occurs when the bracket in
eqn.~\ref{eq:alpha1_sign} vanishes. This gives
\begin{equation}
\tilde\ell_{\rm lb}
=
\frac{\alpha_0-1}
{2\alpha_0(2-\alpha_0)}
\simeq 0.46.
\label{eq:ell_critical_reset_time}
\end{equation}
Thus, to leading order at small $\mathrm{Pe}_{\rm cue,0}$,
\begin{equation}
\tilde\ell>\tilde\ell_{\rm lb}
\quad\Longrightarrow\quad
\tilde T_R^*
\text{ initially increases with }
\mathrm{Pe}_{\rm cue,0},
\end{equation}
whereas
\begin{equation}
\tilde\ell<\tilde\ell_{\rm lb}
\quad\Longrightarrow\quad
\tilde T_R^*
\text{ initially decreases with }
\mathrm{Pe}_{\rm cue,0}.
\end{equation}

This explains the different behavior of the line cuts in
Fig.~\ref{fig:localizedCueResetting}(A). For broader cues,
$\tilde\ell>\tilde\ell_{\rm lb}$, increasing the cue strength initially
favors less frequent resetting. For more localized cues,
$\tilde\ell<\tilde\ell_{\rm lb}$, increasing the cue strength initially
favors more frequent resetting, producing a dip in $\tilde T_R^*$.

The value $\tilde\ell_{\rm lb}\simeq0.46$ therefore provides an analytic
lower bound on the threshold cue length shown in
Fig.~\ref{fig:localizedCueResetting}(B). Near the actual threshold, the
optimal reset time can first increase before dipping below the blind-search
value at finite cue strength, so the weak-cue calculation does not determine
the exact threshold.

\section{The Near-Target Limit for a Single Checkpoint}
\label{app:one_checkpoint_near_target}

The \S~\ref{section:RWCheckpoints} of the main text (``Random Walk with Resets to Checkpoints'') reported that the speedup in the time to reach a target with checkpoints is diluted from an exponential benefit to a linear benefit in $x_T/\lambda_D$, where $x_T$ is the target location and $\lambda_D$ is the typical diffusive lengthscale of the walk. Here we justify this claim.

For $x_T \ll \lambda_D$, we expand eqn.~\eqref{eq:ratio_one_checkpoint}
to leading order in $\Lambda = x_T/\lambda_D$,
\begin{equation}
\frac{\langle T_{\rm one\ checkpoint}\rangle}
{\langle T_{\rm no\ checkpoint}\rangle}
=
\frac{2\left(e^{\Lambda/2}-1\right)}
{e^{\Lambda}-1},
\end{equation}
using
\begin{equation}
e^\Lambda = 1+\Lambda+\frac{\Lambda^2}{2}+ O(\Lambda^3),
\qquad
e^{\Lambda/2}
=
1+\frac{\Lambda}{2}
+\frac{\Lambda^2}{8}
+O(\Lambda^3).
\end{equation}
It gives
\begin{equation}
\frac{\langle T_{\rm one\ checkpoint}\rangle}
{\langle T_{\rm no\ checkpoint}\rangle}
=
\frac{1+\Lambda/4+O(\Lambda^2)}
{1+\Lambda/2+O(\Lambda^2)}.
\end{equation}
Expanding the denominator to first order,
\begin{equation}
\frac{\langle T_{\rm one\ checkpoint}\rangle}
{\langle T_{\rm no\ checkpoint}\rangle}
=
\left(1+\frac{\Lambda}{4}\right)
\left(1-\frac{\Lambda}{2}\right)
+ O(\Lambda^2)
=
1-\frac{\Lambda}{4}
+ O(\Lambda^2).
\end{equation}
Therefore,
\begin{equation}
\frac{\langle T_{\rm one\ checkpoint}\rangle}
{\langle T_{\rm no\ checkpoint}\rangle}
=
1-\frac{x_T}{4\lambda_D}
+ O\!\left(
\frac{x_T^2}{\lambda_D^2}
\right),
\qquad x_T\ll\lambda_D.
\end{equation}
Thus, in the short-target-distance regime, the gain from introducing a checkpoint
is linear in $x_T/\lambda_D$ rather than exponential.
\section{Parallel search with multiple independent searchers}
\subsection{Mean earliest-arrival time from the collective survival probability}
\label{app:Tmin_parallel_search}
Here we explain the observation in the main text, \S~\ref{subsubsec:parallelSearchNindep}, that the integral of the survival probability $S(t)$ yields the mean first passage time $\langle T\rangle$, including for multiple independent searchers.
 
For $N$ independent searchers, the probability that none of the searchers has
reached the target by time $t$ is
\begin{equation}
\mathbb{P}(\tau_{\min}>t)
=
S(t)^N,
\end{equation}
where $S(t)$ is the survival probability of a single searcher and
\begin{equation}
\tau_{\min}
=
\min(\tau_1,\ldots,\tau_N),
\end{equation}
and $\tau_i$ is the first passage time of the $i$th searcher. 
The corresponding probability density for the earliest arrival is
\begin{equation}
f_{\min}(t)
=
-\frac{d}{dt}\mathbb{P}(\tau_{\min}>t)
=
-\frac{d}{dt}S(t)^N.
\end{equation}
The mean earliest-arrival time is then
\begin{equation}
T_{\min}
\equiv
\left\langle \tau_{\min}\right\rangle
=
\int_0^\infty
t\,f_{\min}(t)\,dt.
\end{equation}
Substituting the expression for $f_{\min}(t)$ gives
\begin{equation}
T_{\min}
=
-\int_0^\infty
t\,\frac{d}{dt}S(t)^N\,dt.
\end{equation}
Integrating by parts,
\begin{equation}
T_{\min}
=
-\left[tS(t)^N\right]_0^\infty
+
\int_0^\infty
S(t)^N\,dt.
\end{equation}
The boundary term vanishes at $t=0$ because of the prefactor $t$, and at
$t\to\infty$ because the probability of surviving without reaching the
absorbing target tends to zero. We obtain
\begin{equation}
T_{\min}
=
\int_0^\infty
S(t)^N\,dt.
\label{eq:Tmin_parallel_search_appendix}
\end{equation}
The mean earliest-arrival time for $N$ independent searchers is consequently
determined entirely by the single-searcher survival probability $S(t)$.

\subsection{Derivation of the backward equation for the survival probability}
\label{app:survival_backward_equation}

\begin{figure*}[h]
    \centering
    \includegraphics[width=0.7\linewidth]{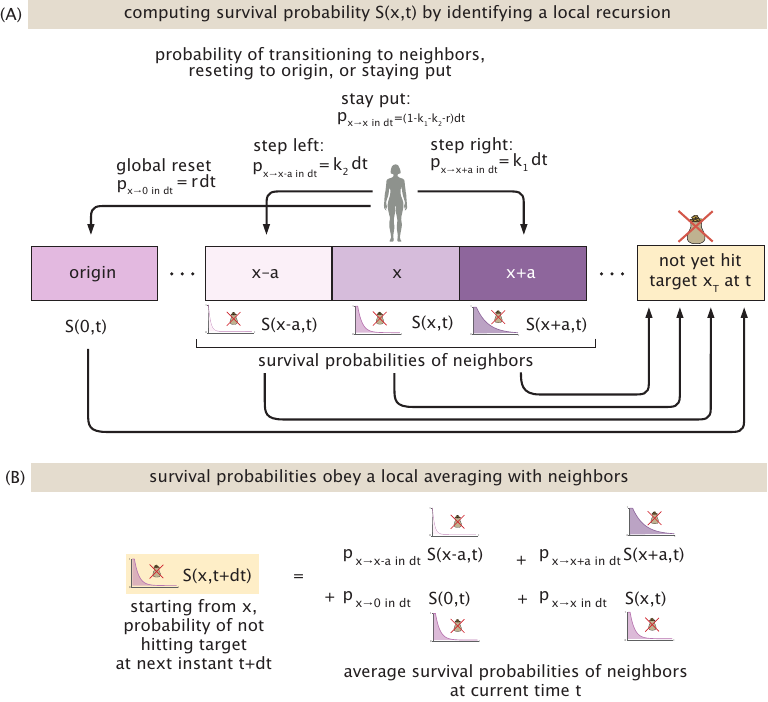}
\caption{\raggedright Origins of the local averaging property of a survival probability $S(x,t)$ that a seeker has not hit a target at time $t$. (A) The updated survival probability $S(x,t+dt)$ follows from available jumps in the time interval $dt$ from neighboring positions (weighted by the probabilities of such jumps), with their own survival probabilities at the previous time $t$. (Each site's purple shading reflects its local survival probability, stressing a coherence with those of neighboring sites.) (B) This averaging conditioned on possible jumps yields a self-consistent expression for the survival probability.} \label{fig:explainSimpleSurvProbRecurse}
\end{figure*}
\FloatBarrier

In \S\ref{sec:severalSearchers} of the main text, we study statistical properties of arrivals by multiple searchers. A key tool along the way is the survival probability $S(x,t)$ of a search not yet having found the target by time $t$ after having started at position $x$.
Here we derive eqn.~\ref{eq:backward_survival_drift_reset} for this
position-dependent survival probability $S(x,t)$ introduced in the main text, $\S$~\ref{subsubsec:parallelSearchNindep}.
We use the same backward-equation approach employed previously for the mean
first-passage time, now applying the local averaging argument to survival
probabilities rather than times. We consider a continuous-time random walk on a lattice with spacing $a$.
The walker steps to the right with rate $k+\Delta k$, to the left with rate
$k-\Delta k$, and resets to the origin with rate $r$.

The probability $S(x,t+dt)$ of surviving for a total duration $t+dt$ is
obtained by averaging over the possible events during the first interval
$dt$. Because the process is Markovian, after this first event the probability
of surviving for the remaining time $t$ depends only on the new position.
Therefore,
\begin{align}
S(x,t+dt)
&=
(k+\Delta k)dt\,S(x+a,t)
+
(k-\Delta k)dt\,S(x-a,t)
\nonumber\\
&\qquad
+
r dt\,S(0,t)
+
\left[1-(2k+r)dt\right]S(x,t).
\end{align}
Subtracting $S(x,t)$, dividing by $dt$, and taking the limit $dt\to0$ gives
\begin{align}
\frac{\partial S(x,t)}{\partial t}
&=
(k+\Delta k)\left[S(x+a,t)-S(x,t)\right]
\nonumber\\
&\qquad
+
(k-\Delta k)\left[S(x-a,t)-S(x,t)\right]
-rS(x,t)+rS(0,t).
\end{align}
Expanding $S(x\pm a,t)$ to second order in $a$ gives
\begin{equation}
\frac{\partial S(x,t)}{\partial t}
=
2a\Delta k\,\frac{\partial S(x,t)}{\partial x}
+
a^2k\,\frac{\partial^2 S(x,t)}{\partial x^2}
-rS(x,t)+rS(0,t).
\end{equation}
Using
\begin{equation}
v=2a\Delta k,
\qquad
D=a^2k,
\end{equation}
we recover
\begin{equation}
\frac{\partial S(x,t)}{\partial t}
=
v\,\frac{\partial S(x,t)}{\partial x}
+
D\,\frac{\partial^2 S(x,t)}{\partial x^2}
-rS(x,t)
+rS(0,t),
\end{equation}
which is eqn.~\ref{eq:backward_survival_drift_reset} of the main text.

\subsection{Characteristic earliest-arrival time and relation to the mean}
\label{app:largeN_characteristic_time}
In \S~\ref{subsubsec:parallelSearchNindep} of the main text, we discussed for instance the heuristic that the required number of searchers $N$ for earliest arrivals satisfies $NF(t) = O(1)$, where $F(t)$ is the probability that a single searcher has arrived by time $t$. Here we derive this statement, its consequences, and some related results.

To extract the large-$N$ behavior of the mean earliest-arrival time, it is
useful to introduce an auxiliary characteristic timescale. We define the
cumulative first-passage probability of a single searcher as
\begin{equation}
F(t)
=
1-S(t)
=
\mathbb{P}(\tau_i\leq t).
\end{equation}
For $N$ independent searchers, the collective survival probability is
\begin{equation}
\mathbb{P}(\tau_{\min}>t)
=
[1-F(t)]^N.
\end{equation}
We define a characteristic first-arrival time $t_N(q)$ by fixing the
collective survival probability to some value $0<q<1$,
\begin{equation}
\mathbb{P}(\tau_{\min}>t_N)
=
[1-F(t_N)]^N
=
q.
\label{eq:tN_q_definition_appendix}
\end{equation}
Solving for the corresponding single-searcher arrival probability gives
\begin{equation}
F(t_N)
=
1-q^{1/N}.
\end{equation}
For large $N$,
\begin{equation}
q^{1/N}
\approx
\exp\left(\frac{\ln q}{N}\right)
\approx
1+\frac{\ln q}{N}
+
O\left(\frac{1}{N^2}\right),
\end{equation}
and therefore
\begin{equation}
NF(t_N)
=
-\ln q
+
O\left(\frac{1}{N}\right).
\end{equation}
Hence any fixed choice of $0<q<1$,
\begin{equation}
NF(t_N)=O(1),
\qquad
F(t_N)=O\left(\frac{1}{N}\right).
\label{eq:NF_tN_general_appendix}
\end{equation}
This result can also be obtained directly from the collective survival
probability. For $F(t)\ll1$,
\begin{align}
\mathbb{P}(\tau_{\min}>t)
&=
\exp\left[N\ln(1-F(t))\right]
\\
&=
\exp\left[
-NF(t)
-\frac{N}{2}F(t)^2
+O\left(NF(t)^3\right)
\right].
\end{align}
On the earliest-arrival timescale, $F(t)=O(1/N)$, so all terms beyond
$NF(t)$ vanish as $N\to\infty$. Therefore,
\begin{equation}
\mathbb{P}(\tau_{\min}>t)
\sim
e^{-NF(t)}.
\label{eq:collective_survival_largeN_appendix}
\end{equation}
For convenience, we choose
\begin{equation}
q=e^{-1},
\end{equation}
for which
\begin{equation}
NF(t_N)\sim1.
\label{eq:NF_tN_appendix}
\end{equation}
With this convention,
\begin{equation}
\mathbb{P}(\tau_{\min}\leq t_N)
\sim
1-e^{-1}
\simeq0.63.
\end{equation}
The particular value of $q$ is not important for the leading asymptotic
behavior; this choice simply provides a convenient reference timescale.

Although $t_N$ is a characteristic quantile rather than the mean
earliest-arrival time, the two become asymptotically equivalent as
$N\to\infty$. To show this, we use the short-time first-passage probability
derived in Appendix~\S\ref{app:largeN_earliest_arrival_asymptotics},
\begin{equation}
F(t)
\sim
\frac{2\sqrt{Dt}}{x_T\sqrt{\pi}}
\exp\left[
-\frac{x_T^2}{4Dt}
+\frac{v x_T}{2D}
\right].
\label{eq:F_short_time_for_tN_appendix}
\end{equation}
It is convenient to introduce
\begin{equation}
y
=
\frac{x_T^2}{4Dt},
\qquad
y_N
=
\frac{x_T^2}{4Dt_N}.
\end{equation}
Since $NF(t_N)\sim1$, times close to $t_N$ may be written as
\begin{equation}
y=y_N+z.
\end{equation}
Using the short-time form above then gives, to leading order,
\begin{equation}
NF(t)
\sim
e^{-z}.
\end{equation}
Together with
eqn.~\ref{eq:collective_survival_largeN_appendix}, this gives
\begin{equation}
\mathbb{P}(\tau_{\min}>t)
\sim
\exp\left[-e^{-z}\right].
\end{equation}
Equivalently, the rescaled variable $z$ follows the standard Gumbel
distribution in the large-$N$ limit, whose mean is the Euler--Mascheroni
constant $\gamma_E$.

Since
\begin{equation}
t
=
\frac{x_T^2}{4D(y_N+z)}
=
t_N
\left[
1-\frac{z}{y_N}
+
O\left(\frac{1}{y_N^2}\right)
\right],
\end{equation}
averaging over the Gumbel fluctuations gives
\begin{equation}
T_{\min}
=
\left\langle\tau_{\min}\right\rangle
=
t_N
\left[
1-\frac{\gamma_E}{y_N}
+
O\left(\frac{1}{y_N^2}\right)
\right].
\label{eq:Tmin_tN_correction_appendix}
\end{equation}
Because $y_N\sim\ln N$,
\begin{equation}
\frac{T_{\min}}{t_N}
\longrightarrow
1,
\qquad
N\to\infty.
\label{eq:Tmin_tN_equivalence_appendix}
\end{equation}
Thus the auxiliary characteristic time $t_N$ and the mean earliest-arrival
time $T_{\min}$ have the same leading large-$N$ behavior. The characteristic
time is used only as an intermediate tool for extracting the extreme-value
asymptotics; the physical quantity considered in the main text remains
$T_{\min}$.

\subsection{Short-time and large-$N$ asymptotics of the earliest-arrival time}
\label{app:largeN_earliest_arrival_asymptotics}

Here we derive the large-$N$ behavior of the characteristic earliest-arrival
time introduced in the main text. In particular, we determine the short-time
behavior of the single-searcher cumulative first-passage probability $F(t)$
and use it to obtain both the leading large-$N$ scaling of $t_N$ and the first
drift-dependent correction.

For large $N$, the earliest successful trajectory occurs at progressively
shorter times. We therefore begin by determining the short-time behavior of
the single-searcher arrival probability. In the absence of resetting, the
first-passage density for a searcher starting at the origin, undergoing
diffusion with coefficient $D$ and constant drift $v$ toward a target at
$x_T>0$, is
\begin{equation}
f_0(t)
=
\frac{x_T}{\sqrt{4\pi D t^3}}
\exp\left[
-\frac{(x_T-vt)^2}{4Dt}
\right].
\label{eq:fpt_drift_no_reset_parallel_app}
\end{equation}
For this arrival to occur before the first reset, the searcher must also
experience no reset during the interval $[0,t]$. Since resets occur as a
Poisson process with rate $r$, the cumulative probability of reaching the
target by time $t$ without having reset is
\begin{equation}
F_{\rm nr}(t)
=
\int_0^t
e^{-rs}f_0(s)\,ds.
\label{eq:F_no_reset_parallel_app}
\end{equation}
Expanding the exponent in the integrand gives
\begin{align}
-\frac{(x_T-vs)^2}{4Ds}-rs
&=
-\frac{x_T^2}{4Ds}
+
\frac{v x_T}{2D}
-
\left(
\frac{v^2}{4D}+r
\right)s.
\end{align}
We define
\begin{equation}
A=\frac{x_T^2}{4D},
\qquad
B=\frac{v^2}{4D}+r,
\qquad
c=\frac{v x_T}{2D}.
\label{eq:ABC_parallel_app}
\end{equation}
Equation~\ref{eq:F_no_reset_parallel_app} can then be written as
\begin{equation}
F_{\rm nr}(t)
=
\frac{x_T}{\sqrt{4\pi D}}
e^{c}
\int_0^t
s^{-3/2}
e^{-A/s-Bs}\,ds.
\label{eq:F_nr_integral_app}
\end{equation}
To determine the small-$t$ equivalent of the remaining integral, define
\begin{equation}
W(t)
=
\int_0^t
s^{-3/2}e^{-A/s-Bs}\,ds,
\end{equation}
and consider
\begin{equation}
g(t)
=
\frac{\sqrt{t}}{A}e^{-A/t}.
\end{equation}
Both $W(t)$ and $g(t)$ vanish as $t\to0^+$. We may therefore use
L'H\^opital's rule: when both the numerator and denominator of a ratio vanish
in the limit, the limit of their ratio can be obtained from the ratio of
their derivatives, provided the latter limit exists. Here,
\begin{equation}
W'(t)
=
t^{-3/2}e^{-A/t-Bt},
\end{equation}
while
\begin{equation}
g'(t)
=
t^{-3/2}e^{-A/t}
\left(1+\frac{t}{2A}\right).
\end{equation}
It follows that
\begin{align}
\lim_{t\to0^+}
\frac{W(t)}{g(t)}
&=
\lim_{t\to0^+}
\frac{W'(t)}{g'(t)}
\\
&=
\lim_{t\to0^+}
\frac{e^{-Bt}}
{1+t/(2A)}
\\
&=
1.
\end{align}
Hence
\begin{equation}
W(t)
\sim
\frac{\sqrt{t}}{A}e^{-A/t},
\qquad
t\to0^+.
\end{equation}
Substituting this result into eqn.~\ref{eq:F_nr_integral_app} gives
\begin{equation}
F_{\rm nr}(t)
\sim
\frac{2\sqrt{Dt}}{x_T\sqrt{\pi}}
\exp\left[
-\frac{x_T^2}{4Dt}
+
\frac{v x_T}{2D}
\right],
\qquad
t\to0^+.
\label{eq:F_nr_short_time_parallel_app}
\end{equation}

We next verify that the full first-passage probability $F(t)$, including
trajectories that may reset, has the same leading short-time behavior.
Conditioning on the time $\tau$ of the first reset gives
\begin{equation}
F(t)
=
F_{\rm nr}(t)
+
\int_0^t
r e^{-r\tau}
S_0(\tau)
F(t-\tau)\,d\tau,
\label{eq:F_reset_renewal_parallel_app}
\end{equation}
where $S_0(\tau)$ is the survival probability in the absence of resetting.
Since $S_0(\tau)\leq1$ and $F(t-\tau)\leq F(t)$,
\begin{equation}
0
\leq
F(t)-F_{\rm nr}(t)
\leq
\left(1-e^{-rt}\right)F(t).
\end{equation}
This implies
\begin{equation}
F_{\rm nr}(t)
\leq
F(t)
\leq
e^{rt}F_{\rm nr}(t).
\end{equation}
Since $e^{rt}\to1$ as $t\to0^+$,
\begin{equation}
F(t)
\sim
F_{\rm nr}(t).
\end{equation}
The full single-searcher arrival probability therefore satisfies
\begin{equation}
F(t)
\sim
\frac{2\sqrt{Dt}}{x_T\sqrt{\pi}}
\exp\left[
-\frac{x_T^2}{4Dt}
+
\frac{v x_T}{2D}
\right],
\qquad
t\to0^+.
\label{eq:F_short_time_parallel_app}
\end{equation}

We now use this short-time form to determine the large-$N$ behavior of the
characteristic earliest-arrival time. For the convenient choice $q=e^{-1}$
introduced in the main text,
\begin{equation}
N F(t_N)\sim1.
\label{eq:NF_tN_asymptotic_app}
\end{equation}
Substituting eqn.~\ref{eq:F_short_time_parallel_app} gives
\begin{equation}
N
\frac{2\sqrt{Dt_N}}{x_T\sqrt{\pi}}
\exp\left[
-\frac{x_T^2}{4Dt_N}
+
\frac{v x_T}{2D}
\right] \sim 1.
\label{eq:tN_before_y_app}
\end{equation}
Using $A$ and $c$ defined in eqn.~\ref{eq:ABC_parallel_app}, we introduce
\begin{equation}
y_N
=
\frac{A}{t_N}.
\end{equation}
Since $t_N=A/y_N$ and
\begin{equation}
\frac{2\sqrt{DA}}{x_T}=1,
\end{equation}
eqn.~\ref{eq:tN_before_y_app} becomes
\begin{equation}
\frac{N}{\sqrt{\pi y_N}}
e^{-y_N+c} \sim 1.
\end{equation}
Taking the logarithm gives
\begin{equation}
y_N
+
\frac{1}{2}\ln y_N
=
\ln N
+
c
-
\frac{1}{2}\ln\pi
+
o(1).
\label{eq:yN_largeN_app}
\end{equation}
At leading order,
\begin{equation}
y_N\sim\ln N,
\end{equation}
and hence
\begin{equation}
t_N
\sim
\frac{x_T^2}{4D\ln N},
\qquad
N\to\infty.
\label{eq:tN_leading_largeN_app}
\end{equation}
To obtain the first correction, we write
\begin{equation}
y_N
=
\ln N+\delta_N,
\end{equation}
where $\delta_N/\ln N\to0$. Substituting this form into
eqn.~\ref{eq:yN_largeN_app} gives
\begin{equation}
\ln N+\delta_N
+
\frac{1}{2}
\ln\left(\ln N+\delta_N\right)
=
\ln N+c-\frac{1}{2}\ln\pi+o(1).
\end{equation}
Since
\begin{equation}
\ln\left(\ln N+\delta_N\right)
=
\ln\ln N+o(1),
\end{equation}
we find
\begin{equation}
\delta_N
=
-\frac{1}{2}\ln\ln N
+
c
-\frac{1}{2}\ln\pi
+
o(1).
\end{equation}
Consequently,
\begin{equation}
y_N
=
\ln N
-\frac{1}{2}\ln\ln N
+
\frac{v x_T}{2D}
-\frac{1}{2}\ln\pi
+
o(1),
\end{equation}
and
\begin{equation}
t_N
=
\frac{x_T^2}{4D}
\left[
\ln N
-\frac{1}{2}\ln\ln N
+\frac{v x_T}{2D}
-\frac{1}{2}\ln\pi
+o(1)
\right]^{-1}.
\label{eq:tN_largeN_correction_app}
\end{equation}
Expanding around the leading large-$N$ result gives the equivalent form
\begin{equation}
t_N
=
\frac{x_T^2}{4D\ln N}
\left[
1+
\frac{
\frac{1}{2}\ln\ln N
+\frac{1}{2}\ln\pi
-\frac{v x_T}{2D}
}{
\ln N
}
+
o\left(\frac{1}{\ln N}\right)
\right].
\label{eq:tN_largeN_expanded_app}
\end{equation}

We finally convert this characteristic timescale into the mean
earliest-arrival time used in the main text. From
Appendix~\S\ref{app:largeN_characteristic_time},
\begin{equation}
T_{\min}
=
t_N
\left[
1-\frac{\gamma_E}{y_N}
+
O\left(\frac{1}{y_N^2}\right)
\right],
\end{equation}
where
\begin{equation}
y_N
=
\frac{x_T^2}{4Dt_N}
\sim
\ln N.
\end{equation}
Using eqn.~\ref{eq:tN_largeN_expanded_app} therefore gives
\begin{equation}
T_{\min}
=
\frac{x_T^2}{4D\ln N}
+
\frac{x_T^2}{8D}
\frac{\ln\ln N}{(\ln N)^2}
+
\frac{x_T^2}{4D(\ln N)^2}
\left[
\frac{1}{2}\ln\pi
-\gamma_E
-\frac{v x_T}{2D}
\right]
+
o\left(\frac{1}{(\ln N)^2}\right).
\label{eq:Tmin_largeN_full_app}
\end{equation}
This gives the decomposition used in the main text,
\begin{equation}
T_{\min}
=
T_{\min}^{(0)}
+
T_{\min}^{(1)}
+
T_{\min}^{(2)}
+\cdots,
\end{equation}
with
\begin{equation}
T_{\min}^{(0)}
=
\frac{x_T^2}{4D\ln N},
\qquad
T_{\min}^{(1)}
=
\frac{x_T^2}{8D}
\frac{\ln\ln N}{(\ln N)^2},
\end{equation}
and
\begin{equation}
T_{\min}^{(2)}
=
\frac{x_T^2}{4D(\ln N)^2}
\left[
\frac{1}{2}\ln\pi
-\gamma_E
-\frac{v x_T}{2D}
\right].
\end{equation}
The leading behavior is therefore
\begin{equation}
T_{\min}
\sim
\frac{x_T^2}{4D\ln N},
\qquad
N\to\infty.
\end{equation}
The first correction is independent of drift, while drift first appears
in $T_{\min}^{(2)}$. Resetting does not appear at this order because the
winning trajectory occurs on a timescale much shorter than the mean reset
time.

\subsection{Approach to the reset-independent large-$N$ regime}\label{app:Nc_parallel_search}

We now derive the characteristic population size $N_c$ introduced in the main text.
For a large population of independent searchers, the first arrival occurs
when
\begin{equation}
N F(t)=O(1).
\end{equation}
We define $N_c$ by evaluating this condition at a time
$t=\varepsilon/r$, where $0<\varepsilon\ll1$. At this time, the first
arrival occurs on a timescale much shorter than the mean reset time $1/r$,
so that resetting has little opportunity to influence the winning trajectory.

Using the short-time result
\begin{equation}
F(t)
\sim
\frac{2\sqrt{Dt}}{x_T\sqrt{\pi}}
\exp\left[
-\frac{x_T^2}{4Dt}
+\frac{v x_T}{2D}
\right],
\end{equation}
we obtain
\begin{equation}
F\left(\frac{\varepsilon}{r}\right)
\sim
\sqrt{\frac{\varepsilon}{\pi\rho}}
\exp\left[
-\frac{\rho}{\varepsilon}
+\frac{v x_T}{2D}
\right],
\end{equation}
where
\begin{equation}
\rho
=
\frac{r x_T^2}{4D}.
\end{equation}
The large-$N$ condition
\begin{equation}
N_c
F\left(\frac{\varepsilon}{r}\right)
\sim
1
\end{equation}
therefore gives
\begin{equation}
N_c(\varepsilon)
\sim
\sqrt{\frac{\pi\rho}{\varepsilon}}
\exp\left[
\frac{\rho}{\varepsilon}
-\frac{v x_T}{2D}
\right],
\qquad
\varepsilon\to0.
\label{eq:Nc_reset_correction_app}
\end{equation}

The dominant dependence is exponential in $\rho/\varepsilon$. Increasing
the reset rate therefore shifts the onset of the reset-independent regime
to larger populations, whereas positive drift toward the target lowers
$N_c$.

For the numerical comparison in
Fig.~\ref{fig:Nc_multiple_searchers}, we determine $N_c$ numerically from
\begin{equation}
T_{\min}(N_c)=\frac{\varepsilon}{r}.
\end{equation}
Since the mean earliest-arrival time and the characteristic
large-$N$ first-arrival timescale are asymptotically equivalent, this
numerical definition approaches the asymptotic prediction above when
$N_c$ is sufficiently large.

\subsection{Optimal number of searchers in the presence of a resource cost}
\label{app:optimal_number_parallel_searchers}

Parallel search reduces the mean time to first success, but deploying
additional searchers may itself carry a cost, as discussed in the main text, $\S$\ref{subsubsec:parallelSearchNindep}. To illustrate this tradeoff,
we assign an effective cost $\gamma$ to each searcher, expressed in units
of time, and consider
\begin{equation}
\mathcal C(N)
=
T_{\min}(N)+\gamma N.
\label{eq:parallel_search_cost_app}
\end{equation}
Using the leading large-$N$ behavior
\begin{equation}
T_{\min}(N)
\sim
\frac{x_T^2}{4D\ln N},
\end{equation}
we obtain
\begin{equation}
\mathcal C(N)
\simeq
\frac{x_T^2}{4D\ln N}
+
\gamma N.
\label{eq:parallel_search_cost_asymptotic_app}
\end{equation}
For the purpose of optimization, we treat $N$ as a continuous variable. The
derivative of eqn.~\ref{eq:parallel_search_cost_asymptotic_app} is
\begin{equation}
\frac{d\mathcal C}{dN}
=
-\frac{x_T^2}{4D}
\frac{1}{N(\ln N)^2}
+
\gamma.
\end{equation}
The optimal population size $N^*$ therefore satisfies
\begin{equation}
\gamma
=
\frac{x_T^2}{4D}
\frac{1}{N^*(\ln N^*)^2}.
\label{eq:optimal_N_condition_app}
\end{equation}

The left-hand side is the fixed marginal cost of adding one searcher, while
the right-hand side is the magnitude of the marginal reduction in the
earliest-arrival time. The optimum is reached when these two contributions
balance.

It is also possible to write the optimal number of searchers $N^*$ explicitly. Defining
\begin{equation}
A
=
\frac{x_T^2}{4D},
\end{equation}
eqn.~\ref{eq:optimal_N_condition_app} becomes
\begin{equation}
N^*(\ln N^*)^2
=
\frac{A}{\gamma}.
\end{equation}
Writing
\begin{equation}
x=\ln N^*,
\qquad
N^*=e^x,
\end{equation}
gives
\begin{equation}
x^2 e^x
=
\frac{A}{\gamma}.
\end{equation}
Taking the positive square root,
\begin{equation}
x e^{x/2}
=
\sqrt{\frac{A}{\gamma}}.
\end{equation}
Introducing $u=x/2$ gives
\begin{equation}
u e^u
=
\frac{1}{2}
\sqrt{\frac{A}{\gamma}}.
\end{equation}
The solution can therefore be expressed in terms of the Lambert $W$ function,
defined by $W(z)e^{W(z)}=z$,
\begin{equation}
x
=
2W\left(
\frac{1}{2}
\sqrt{\frac{A}{\gamma}}
\right).
\end{equation}
Hence
\begin{equation}
N^*
=
\exp\left[
2W\left(
\frac{1}{2}
\sqrt{\frac{A}{\gamma}}
\right)
\right],
\qquad
A=\frac{x_T^2}{4D}.
\label{eq:optimal_N_lambert_app}
\end{equation}

The second derivative,
\begin{equation}
\frac{d^2\mathcal C}{dN^2}
=
\frac{A}{N^2}
\left[
\frac{1}{(\ln N)^2}
+
\frac{2}{(\ln N)^3}
\right],
\end{equation}
is positive for $N>1$, confirming that this stationary point is a minimum
within the continuous approximation. Since the actual population size is
discrete, the physical optimum is obtained by comparing the neighboring
integers around the value predicted by
eqn.~\ref{eq:optimal_N_lambert_app}.

The origin of the finite optimum is the diminishing return of parallelization:
the characteristic earliest-arrival time decreases only as $1/\ln N$, so the
marginal improvement from each additional searcher scales as
$1/[N(\ln N)^2]$, while the marginal cost remains equal to $\gamma$.

\section{Additional results for parallel search}
\label{app:parallel_search}
\subsection{Large-$N$ behavior of the $M$th arrival}
\label{app:MN_largeN}

As discussed in the main text, $\S$\ref{subsubsec:mofnsuccesses}, the $M$th arrival occurs when the number of
successful searchers becomes of order $M$. Since a single searcher has
reached the target by time $t$ with probability
\begin{equation}
F(t)=1-S(t),
\end{equation}
the expected number of successful searchers is $NF(t)$. For $N\gg M$ with
$M$ fixed, the characteristic time of the $M$th arrival therefore satisfies
\begin{equation}
NF(t)=O(M),
\end{equation}
or equivalently
\begin{equation}
F(t)=O\left(\frac{M}{N}\right).
\label{eq:FMN_scaling}
\end{equation}

For large $N$, this corresponds to the short-time tail of the
single-searcher first-passage distribution. As derived above,
\begin{equation}
F(t)
\sim
\frac{2\sqrt{Dt}}{x_T\sqrt{\pi}}
\exp\left[
-\frac{x_T^2}{4Dt}
+
\frac{v x_T}{2D}
\right].
\label{eq:short_time_F_MN_appendix}
\end{equation}
Resetting does not appear at this order because these very early arrivals
typically reach the target before a reset occurs.

To determine the leading dependence on $N$, we define
\begin{equation}
z=\frac{x_T^2}{4Dt}.
\end{equation}
The short-time first-passage probability can then be written as
\begin{equation}
F(t)
\sim
\frac{1}{\sqrt{\pi z}}
\exp\left[
-z+\frac{v x_T}{2D}
\right].
\end{equation}
Using $F(t)=O(M/N)$ and taking the logarithm gives
\begin{equation}
z+\frac{1}{2}\ln z
=
\ln\left(\frac{N}{M}\right)
+O(1).
\label{eq:z_MN_scaling}
\end{equation}
For $N\gg M$, the logarithm on the right-hand side is large, while
$\ln z$ and the remaining terms contribute only subleading corrections.
Therefore,
\begin{equation}
z
\sim
\ln\left(\frac{N}{M}\right).
\end{equation}
Using $z=x_T^2/(4Dt)$ gives the leading large-$N$ behavior
\begin{equation}
T_{M:N}
\sim
\frac{x_T^2}
{4D\ln(N/M)},
\qquad
N\to\infty
\quad\text{with fixed }M.
\label{eq:TMN_effective_parallelization}
\end{equation}

For $M=1$, this reduces to the earliest-arrival result
$T_{1:N}\sim x_T^2/(4D\ln N)$. More generally, the appearance of $N/M$
has a simple interpretation: waiting for $M$ successes among $N$ searchers
has, at leading order, the same short-time scaling as waiting for one
success among roughly $N/M$ searchers.

\subsection{Approach to the reset-independent regime for the $M$th arrival}
\label{app:Nc_M_parallel_search}

We next generalize this criterion for the onset of the reset-independent regime to the $M$th arrival.
The basic idea is that resetting becomes irrelevant to the successful
trajectory once the required number of arrivals typically occurs on a
timescale much shorter than the reset time $1/r$.

As in the main text, we introduce a small dimensionless number
$\varepsilon$ and consider the time
\begin{equation}
t_\varepsilon
=
\frac{\varepsilon}{r}.
\end{equation}
For $M$ successes to have occurred by this time, the expected number of
successful searchers must be of order $M$,
\begin{equation}
N_c F(t_\varepsilon)
=
O(M).
\end{equation}
We define the characteristic population size $N_c$ by the condition
\begin{equation}
N_c F(t_\varepsilon)
\simeq
M.
\label{eq:NcM_condition}
\end{equation}

Using the short-time form in
eqn.~\ref{eq:short_time_F_MN_appendix}, we have
\begin{equation}
F\left(\frac{\varepsilon}{r}\right)
\sim
\frac{2}{x_T\sqrt{\pi}}
\sqrt{\frac{D\varepsilon}{r}}
\exp\left[
-\frac{r x_T^2}{4D\varepsilon}
+
\frac{v x_T}{2D}
\right].
\end{equation}
Introducing
\begin{equation}
\rho
=
\frac{r x_T^2}{4D},
\label{eq:rho_NcM}
\end{equation}
this becomes
\begin{equation}
F\left(\frac{\varepsilon}{r}\right)
\sim
\sqrt{\frac{\varepsilon}{\pi\rho}}
\exp\left[
-\frac{\rho}{\varepsilon}
+
\frac{v x_T}{2D}
\right].
\label{eq:F_epsilon_over_r}
\end{equation}

Substituting eqn.~\ref{eq:F_epsilon_over_r} into
eqn.~\ref{eq:NcM_condition} gives
\begin{equation}
N_c(M,\varepsilon)
\sim
M
\sqrt{\frac{\pi\rho}{\varepsilon}}
\exp\left[
\frac{\rho}{\varepsilon}
-
\frac{v x_T}{2D}
\right],
\label{eq:NcM_appendix}
\end{equation}
which is eqn.~\ref{eq:Nc_M} of the main text.

In particular,
\begin{equation}
N_c(M,\varepsilon)
\sim
M\,N_c(1,\varepsilon),
\end{equation}
so requiring $M$ successful arrivals increases the characteristic number
of searchers needed to outrun resetting approximately linearly with $M$.
The precise numerical value of $N_c$ depends on the operational
criterion used to define when resetting has become negligible, but the
linear dependence on $M$ and the exponential dependence on
$\rho/\varepsilon$ follow directly from the short-time first-passage
probability.

\subsection{Long-time survival and the last arrival}
\label{app:last_arrival_largeN}

To determine the behavior of the last arrival, we first need the long-time
survival probability of a single searcher. We follow the approach of
Pal \emph{et al.}~\cite{pal2019local}, who considered a drift-diffusion process with
Poisson resetting in the presence of an absorbing boundary. Their absorbing
boundary is placed at the origin; here, instead, the searcher resets to
$x=0$ and the absorbing target is located at $x=x_T$.

Let $S(x,t)$ denote the probability that a searcher starting at position
$x$ has not yet reached the target by time $t$. As derived above, it obeys
the backward equation
\begin{equation}
\frac{\partial S}{\partial t}
=
D\frac{\partial^2 S}{\partial x^2}
+
v\frac{\partial S}{\partial x}
-rS(x,t)
+rS(0,t),
\label{eq:last_arrival_backward}
\end{equation}
with
\begin{equation}
S(x_T,t)=0,
\qquad
S(x,0)=1.
\end{equation}

Following Ref.~\cite{pal2019local}, we take the Laplace transform in time,
\begin{equation}
\widetilde S(x,s)
=
\int_0^\infty e^{-st}S(x,t)\,dt.
\end{equation}
Equation~\ref{eq:last_arrival_backward} then becomes
\begin{equation}
D\frac{d^2\widetilde S}{dx^2}
+
v\frac{d\widetilde S}{dx}
-(r+s)\widetilde S(x,s)
=
-1-r\widetilde S(0,s).
\label{eq:last_arrival_laplace_ODE}
\end{equation}

Solving this equation while requiring the solution to remain bounded away
from the target and imposing $\widetilde S(x_T,s)=0$ gives, for a searcher
starting from the reset position,
\begin{equation}
\widetilde S(0,s)
=
\frac{
1-e^{-\kappa(s)x_T}
}{
s+r e^{-\kappa(s)x_T}
},
\label{eq:survival_laplace_reset}
\end{equation}
where
\begin{equation}
\kappa(s)
=
\frac{
\sqrt{v^2+4D(r+s)}-v
}{
2D
}.
\end{equation}

We now use this result to determine the long-time behavior. Recall that an
exponential decay $e^{-\alpha t}$ has Laplace transform $1/(s+\alpha)$.
We therefore look for the slowest exponential contribution to
eqn.~\ref{eq:survival_laplace_reset}, which is set by a zero of its
denominator. If we denote this value by $s_*$, it satisfies
\begin{equation}
s_*
+
r
\exp\left[
-\frac{x_T}{2D}
\left(
\sqrt{v^2+4D(r+s_*)}-v
\right)
\right]
=0.
\end{equation}
Since the second term is positive, any such solution necessarily has
$s_*<0$. We therefore write $s_*=-\alpha$, with $\alpha>0$, giving
\begin{equation}
\alpha
=
r
\exp\left[
-\frac{x_T}{2D}
\left(
\sqrt{v^2+4D(r-\alpha)}-v
\right)
\right].
\label{eq:alpha_last_arrival}
\end{equation}
This is a transcendental equation for $\alpha$, and in general does not
admit a closed-form solution; instead, it implicitly determines the
long-time decay rate from $D$, $v$, $r$, and $x_T$. The single-searcher
survival probability therefore behaves at long times as
\begin{equation}
S(t)
\sim
C e^{-\alpha t},
\qquad
t\to\infty,
\label{eq:long_time_survival_alpha}
\end{equation}
where $C$ is a constant.
We can now return to the last of $N$ independent searchers. The probability
that all $N$ searchers have arrived by time $t$ is
\begin{equation}
\mathbb P(\tau_{N:N}\leq t)
=
[1-S(t)]^N.
\end{equation}
At the late times relevant for the final arrival, $S(t)\ll1$, and therefore
\begin{equation}
[1-S(t)]^N
\simeq
e^{-NS(t)}.
\end{equation}
The last arrival occurs when the expected number of searchers that remain
unsuccessful becomes of order one,
\begin{equation}
NS(t)=O(1).
\end{equation}
Using the long-time form $S(t)\sim C e^{-\alpha t}$ gives
\begin{equation}
NCe^{-\alpha t}=O(1),
\end{equation}
so that, to leading order for $N\gg1$,
\begin{equation}
T_{N:N}
\sim
\frac{\ln N}{\alpha}.
\label{eq:last_arrival_mean_leading_appendix}
\end{equation}

\section{Derivation of the first-passage Brownian bridge statistics}
\label{app:FPbb_derivation}

In this appendix, we derive the conditional probability density, mean trajectory, and variance of the first-passage Brownian bridge, and analyze the monotonicity and maximal excursion of its mean trajectory.

\subsection{Conditional probability density}

We set $t_i=0$ and $x_i=0$, and introduce the variable
\begin{equation}
z=x_T-x,
\end{equation}
which represents the remaining distance to the target.
In terms of $z$, the absorbing-boundary propagator becomes
{\small 
\begin{equation}
p_{\mathrm{abs}}(x_T-z,t \mid 0,0;x_T)
=
\frac{1}{\sqrt{4\pi D t}}
\left[
e^{-(z-x_T)^2/(4Dt)}
-
e^{-(z+x_T)^2/(4Dt)}
\right].
\end{equation} }

The first-passage density in the denominator is
\begin{equation}
f_{\mathrm{FP}}(t_f \mid 0,0;x_T)
=
\frac{x_T}{\sqrt{4\pi D t_f^3}}
e^{-x_T^2/(4Dt_f)},
\end{equation}
while the numerator contains
\begin{equation}
f_{\mathrm{FP}}(t_f-t \mid x_T-z,t;x_T)
=
\frac{z}{\sqrt{4\pi D (t_f-t)^3}}
e^{-z^2/[4D(t_f-t)]}.
\end{equation}
Substituting these expressions into the conditional probability gives
exponentials of the form
\begin{equation}
-\frac{(z\pm x_T)^2}{4Dt}
-\frac{z^2}{4D(t_f-t)}
+\frac{x_T^2}{4Dt_f}.
\end{equation}
Collecting powers of $z$ and completing the square yields
\begin{equation}
-\frac{t_f}{4Dt(t_f-t)}
\left(
z\pm\frac{x_T(t_f-t)}{t_f}
\right)^2.
\end{equation}
We therefore identify
\begin{equation}
\mu(t)=\frac{x_T(t_f-t)}{t_f},
\qquad
\sigma(t)^2=\frac{2Dt(t_f-t)}{t_f}.
\end{equation}
The probability density becomes
{\small 
\begin{equation}
p_{\mathrm{FPbb}}(x_T-z,t \mid 0,0;x_T,t_f)
=
\frac{z}{\mu(t)\sqrt{2\pi\sigma(t)^2}}
\left[
e^{-\frac{(z-\mu(t))^2}{2\sigma(t)^2}}
-
e^{-\frac{(z+\mu(t))^2}{2\sigma(t)^2}}
\right].
\end{equation} }

\subsection{Mean and variance}

The mean remaining distance to the target is
\begin{equation}
\langle z(t)\rangle
=
\int_0^\infty
z\,p_{\mathrm{FPbb}}(x_T-z,t \mid 0,0;x_T,t_f)\,dz.
\end{equation}
Substituting the probability density gives
\begin{equation}
\langle z(t)\rangle
=
\frac{1}{\mu(t)}
\int_0^\infty z^2
\left[
\phi(z;\mu(t),\sigma(t)^2)
-
\phi(z;-\mu(t),\sigma(t)^2)
\right]dz,
\end{equation}
where $\phi(z;\mu,\sigma^2)$ is the Gaussian density with mean $\mu$
and variance $\sigma^2$.
Using the change of variables
\begin{equation}
u=\frac{z\mp\mu}{\sqrt{2}\sigma},
\end{equation}
together with the definition of the error function,
\begin{equation}
\operatorname{erf}(x)
=
\frac{2}{\sqrt{\pi}}
\int_0^x e^{-u^2}\,du,
\end{equation}
the Gaussian integrals can be evaluated explicitly. After expanding the
square and integrating term by term, we obtain
\begin{equation}
\langle z(t)\rangle
=
\left(\mu(t)+\frac{\sigma(t)^2}{\mu(t)}\right)
\operatorname{erf}\left(
\frac{\mu(t)}{\sqrt{2}\sigma(t)}
\right)
+
\sqrt{\frac{2}{\pi}}
\sigma(t)
e^{-\frac{\mu(t)^2}{2\sigma(t)^2}}.
\end{equation}
Since $z=x_T-x$, the average trajectory is
\begin{equation}
\langle x(t)\rangle
=
x_T-\langle z(t)\rangle.
\end{equation}

To compute the variance, we evaluate the second moment
\begin{equation}
\langle z^2(t)\rangle
=
\frac{1}{\mu(t)}
\int_0^\infty z^3
\left[
\phi(z;\mu(t),\sigma(t)^2)
-
\phi(z;-\mu(t),\sigma(t)^2)
\right]dz.
\end{equation}
Using the same Gaussian integration procedure gives
\begin{equation}
\langle z^2(t)\rangle
=
\mu(t)^2+3\sigma(t)^2.
\end{equation}
The variance is therefore
\begin{equation}
\mathrm{Var}[z(t)]
=
\langle z^2(t)\rangle
-
\langle z(t)\rangle^2.
\end{equation}
Since $z=x_T-x$, we have $\mathrm{Var}[x(t)]=\mathrm{Var}[z(t)]$.
\subsection{Monotonicity and maximal excursion}

To understand the shape of the average trajectory, it is useful to write
\begin{equation}
\langle x(t)\rangle
=
x_T-\mu(t)h(y(t)),
\end{equation}
where
\begin{equation}
h(y)
=
(1+y^2)\operatorname{erf}\left(\frac{1}{\sqrt{2}y}\right)
+
\sqrt{\frac{2}{\pi}}y e^{-\frac{1}{2y^2}},
\end{equation}
and
\begin{equation}
y(t)=\frac{\sigma(t)}{\mu(t)}
=
\alpha\sqrt{\frac{t}{t_f-t}},
\end{equation}
with
\begin{equation}
\alpha=\frac{\sqrt{2Dt_f}}{x_T}.
\end{equation}
Differentiating the average trajectory gives
\begin{equation}
\frac{d}{dt}\langle x(t)\rangle
=
\frac{x_T}{t_f}s(y),
\end{equation}
where
\begin{equation}
s(y)
=
(1-\alpha^2)
\operatorname{erf}\left(\frac{1}{\sqrt{2}y}\right)
+
\sqrt{\frac{2}{\pi}}y e^{-\frac{1}{2y^2}}.
\end{equation}
If $\alpha\leq 1$, then
\begin{equation}
s(y)>0
\end{equation}
for all $y>0$, so the average trajectory is monotonic. If
$\alpha>1$, the sign of $s(y)$ changes during the trajectory, implying
non-monotonic behavior.

The turning point is obtained from
\begin{equation}
s(y_*)=0.
\end{equation}
For large $\alpha$, the solution satisfies
\begin{equation}
y_*=\alpha-\frac{1}{3\alpha}
+
O\left(\frac{1}{\alpha^3}\right),
\end{equation}
so that
\begin{equation}
t_* \simeq \frac{t_f}{2}.
\end{equation}
To obtain this asymptotic form, we start from the equation defining
$y_*$,
\begin{equation}
(\alpha^2-1)
\operatorname{erf}\left(\frac{1}{\sqrt{2}y_*}\right)
=
\sqrt{\frac{2}{\pi}}y_* e^{-\frac{1}{2y_*^2}}.
\end{equation}
For large $\alpha$, the solution has large $y_*$. We therefore define
\begin{equation}
u=\frac{1}{\sqrt{2}y_*},
\end{equation}
so that $u\ll1$. The equation becomes
\begin{equation}
\alpha^2-1
=
\frac{e^{-u^2}}
{\sqrt{\pi}u\operatorname{erf}(u)}.
\end{equation}
Using the small-$u$ expansions
\begin{equation}
\operatorname{erf}(u)
=
\frac{2}{\sqrt{\pi}}
\left(
u-\frac{u^3}{3}+O(u^5)
\right),
\end{equation}
and
\begin{equation}
e^{-u^2}=1-u^2+O(u^4),
\end{equation}
we obtain
\begin{align}
\alpha^2-1
&=
\frac{1-u^2+O(u^4)}
{2u^2\left(1-\frac{u^2}{3}+O(u^4)\right)}
\\
&=
\frac{1}{2u^2}
\left[
1-\frac{2u^2}{3}+O(u^4)
\right]
\\
&=
\frac{1}{2u^2}
-\frac{1}{3}
+
O(u^2).
\end{align}
Since $1/(2u^2)=y_*^2$, this gives
\begin{equation}
\alpha^2-1
=
y_*^2-\frac{1}{3}
+
O\left(\frac{1}{y_*^2}\right).
\end{equation}
Therefore,
\begin{equation}
y_*^2
=
\alpha^2-\frac{2}{3}
+
O\left(\frac{1}{\alpha^2}\right).
\end{equation}
Taking the square root gives
\begin{equation}
y_*
=
\alpha
\left[
1-\frac{2}{3\alpha^2}
+
O\left(\frac{1}{\alpha^4}\right)
\right]^{1/2}.
\end{equation}
Expanding the square root finally yields
\begin{equation}
y_*=
\alpha-\frac{1}{3\alpha}
+
O\left(\frac{1}{\alpha^3}\right).
\end{equation}
At the turning point, the average remaining distance to the target is
\begin{equation}
\langle z(t_*)\rangle
=
\mu(t_*)h(y_*).
\end{equation}
Using $t_*\simeq t_f/2$, we obtain
\begin{equation}
\mu(t_*)
\simeq
\frac{x_T}{2}.
\end{equation}
In the large-$\alpha$ limit,
\begin{equation}
h(\alpha)
\simeq
2\sqrt{\frac{2}{\pi}}\alpha,
\end{equation}
so that
\begin{equation}
\langle z(t_*)\rangle
\simeq
x_T\sqrt{\frac{2}{\pi}}\alpha.
\end{equation}
Using
\begin{equation}
\alpha=\frac{\sqrt{2Dt_f}}{x_T},
\end{equation}
we finally obtain
\begin{equation}
\langle z(t_*)\rangle
\simeq
2\sqrt{\frac{Dt_f}{\pi}}.
\end{equation}
Thus, in the regime $\sqrt{2Dt_f}\gg x_T$, the maximal excursion is
controlled entirely by the diffusive length scale $\sqrt{Dt_f}$.

\section{First-passage Brownian bridges with stochastic resetting}
\label{app:FPbb_resetting}

\subsection{Conditional probability density and local transition kernel}

We now generalize the framework developed above to diffusive processes
with stochastic resetting. While Brownian bridges with resetting have
previously been studied~\cite{DeBruyne2022}, to our knowledge the
corresponding first-passage Brownian bridge with stochastic resetting has
not previously been characterized. Here we derive its conditioned
dynamics and examine the organization of successful trajectories into
reset-terminated failed excursions followed by a final successful
first-passage event.
We consider a Brownian particle that diffuses with
diffusion coefficient $D$ and resets to the origin with
Poissonian rate $r$, with the initial position taken to be $x_i=0$.
During a small time interval $\Delta t$, the particle
either diffuses normally with probability $1-r\Delta t$, or resets to the
origin with probability $r\Delta t$.

We first ask for the conditional probability density of finding the
particle at position $x$ at time $t$, given that the trajectory first
reaches the target position $x_T$ at the prescribed final time $t_f$.
We write
\begin{equation}
p_{\mathrm{FPbb}}^{(r)}(x,t \mid x_i,t_i;x_T,t_f).
\end{equation}
Using Bayes' theorem and the Markov property, we get
\begin{equation}
p_{\mathrm{FPbb}}^{(r)}(x,t\mid x_i,t_i;x_T,t_f)
=
\frac{
p_{\mathrm{surv}}^{(r)}(x,t\mid x_i,t_i;x_T)
f_{\mathrm{FP}}^{(r)}(t_f-t\mid x;x_T)
}{
f_{\mathrm{FP}}^{(r)}(t_f-t_i\mid x_i;x_T)
},
\end{equation}
for $x<x_T$, where $f_{\mathrm{FP}}^{(r)}(T\mid x;x_T)$ denotes the
first-passage-time density to $x_T$ for a particle starting at $x$, with
all resets returning the particle to the origin.

Because the trajectory is conditioned on first arrival at $x_T$ at time
$t_f$, it must not have reached $x_T$ before the intermediate time $t$.
Therefore, the appropriate probability density for the first part of the
trajectory is the reset survival propagator,
\begin{equation}
p_{\mathrm{surv}}^{(r)}(x,t\mid x_i,t_i;x_T),
\end{equation}
which gives the probability density that the particle is at position $x$
at time $t$, having started from $x_i$ at time $t_i$, while remaining
below the target position $x_T$ up to time $t$ in the presence of
resetting. This quantity satisfies the renewal equation
\begin{align}
p_{\mathrm{surv}}^{(r)}(x,t\mid x_i,t_i;x_T)
&=
e^{-r(t-t_i)}
p_{\mathrm{abs}}(x,t\mid x_i,t_i;x_T)
\nonumber\\
&\quad+
r\int_{t_i}^{t}
e^{-r(\tau-t_i)}
S_0(\tau-t_i\mid x_i;x_T)\,
p_{\mathrm{surv}}^{(r)}(x,t\mid 0,\tau;x_T)
\,d\tau,
\end{align}
where $p_{\mathrm{abs}}$ is the absorbing propagator without resetting
and $S_0(\tau-t_i\mid x_i;x_T)$ is the corresponding no-reset survival
probability.

This expression is the reset analog of the first-passage Brownian bridge
density. The numerator counts all reset-diffusion microtrajectories that
start at $x_i$ at time $t_i$, survive below the target up to time $t$,
pass through position $x$ at time $t$, and then first reach $x_T$
exactly at time $t_f$. The denominator counts all reset-diffusion
microtrajectories that start at $x_i$ at time $t_i$ and first reach
$x_T$ at time $t_f$. The conditional probability is therefore the ratio
between these two ensembles.

The reset-modified first-passage density
$f_{\mathrm{FP}}^{(r)}$ can be defined through a renewal equation. Let
$S_r(T\mid x;x_T)$ denote the survival probability over a time $T$ for a
particle starting at $x$ and resetting to the origin, and let
$S_0(T\mid x;x_T)$ denote the corresponding survival probability without
resetting. A trajectory that survives until time $T$ either experiences
no reset during the entire interval, or experiences a first reset at time
$\tau$ and then starts the process again from the origin. This gives
\begin{equation}
S_r(T\mid x;x_T)
=
e^{-rT}S_0(T\mid x;x_T)
+
r\int_0^T
e^{-r\tau}
S_0(\tau\mid x;x_T)
S_r(T-\tau\mid 0;x_T)
\,d\tau.
\end{equation}

\begin{figure*}[t]
    \centering
    \includegraphics[width=\linewidth]{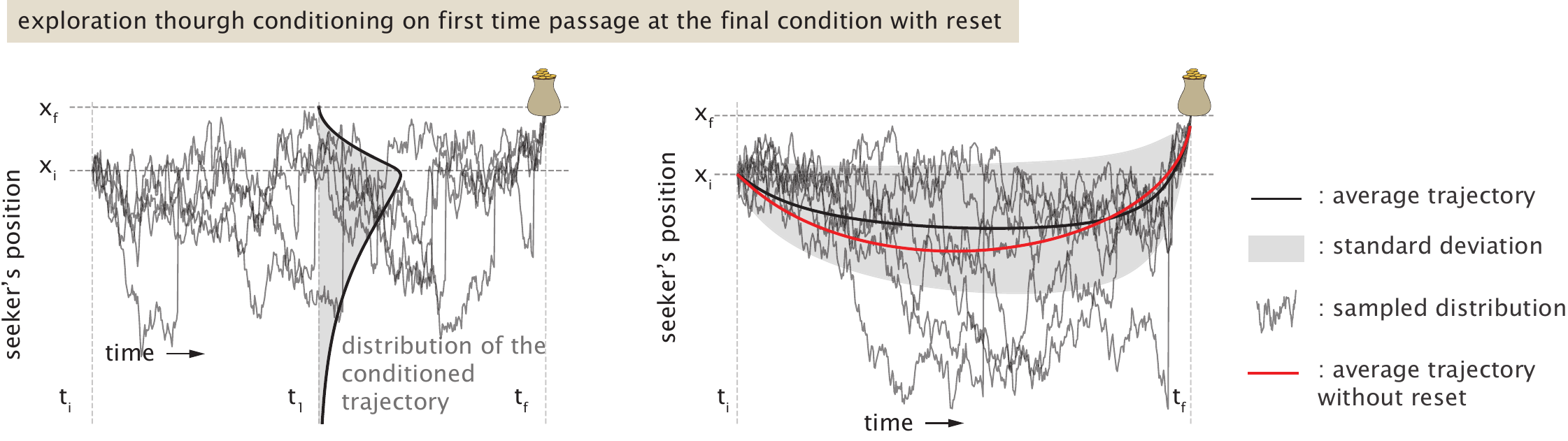}
    \caption{\raggedright 
    Conditioned first-passage trajectories with stochastic resetting.
    Left: sample trajectories conditioned to first reach the target position
    $x_T$ at the final time $t_f$. The shaded profile at the intermediate
    time $t_1$ represents the distribution of conditioned trajectories,
    obtained from the analytical first-passage bridge distribution with reset.
    Right: schematic illustration of the conditioned ensemble. The black curve
    denotes the mean conditioned trajectory with reset, while the shaded region
    indicates the corresponding standard deviation. The red curve shows the
    mean first-passage bridge trajectory in the absence of resetting.
    }
    \label{fig:conditioned_bridge_with_reset}
\end{figure*}
\FloatBarrier

The corresponding reset-modified first-passage-time density is then
obtained from
\begin{equation}
f_{\mathrm{FP}}^{(r)}(T\mid x;x_T)
=
-\frac{\partial}{\partial T}
S_r(T\mid x;x_T).
\end{equation}

This renewal equation therefore determines the first-passage statistics
needed to construct the conditional density above. By analogy with the
no-reset case, we can also write a local transition kernel, which is
useful for simulating conditioned trajectories step by step. The
one-step survival propagator in the presence of resetting is, to leading
order in $\Delta t$,
{\small
\begin{equation}
p_{\mathrm{surv}}^{(r)}(y,t+\Delta t\mid x,t;x_T)
=
(1-r\Delta t)
p_{\mathrm{abs}}(y,t+\Delta t\mid x,t;x_T)
+
r\Delta t\,\delta(y),
\end{equation}}
for $y<x_T$.

Conditioning this one-step kernel on first arrival at $x_T$ at time
$t_f$ gives
\begin{equation}
p_{\mathrm{FPbb}}^{(r)}(y,t+\Delta t\mid x,t;x_T,t_f)
=
p_{\mathrm{surv}}^{(r)}(y,t+\Delta t\mid x,t;x_T)
\frac{
f_{\mathrm{FP}}^{(r)}(t_f-t-\Delta t\mid y;x_T)
}{
f_{\mathrm{FP}}^{(r)}(t_f-t\mid x;x_T)
},
\end{equation}
for $y<x_T$.

This local kernel gives a direct numerical prescription for generating
reset-diffusion trajectories conditioned on first arrival at the target
at the final time. It also lets us ask how successful trajectories are
organized when unsuccessful excursions are physically terminated by
resetting.

\subsection{Decomposition into failed and successful excursions}

 The renewal structure of the reset process suggests a decomposition of a
successful trajectory into a sequence of failed excursions terminated by
resets, followed by one final successful excursion that first reaches the
target. If the trajectory experiences $n$ resets before the final
first-passage event, we denote by $\Delta_i$ the duration between the
$(i-1)$th and $i$th resets. The contribution to the first-passage-time
density at $t_f$ from trajectories with exactly $n$ resets is
\begin{equation}
P_n(t_f)
=
\int_0^{t_f} d\Delta_1
\int_0^{t_f-\Delta_1} d\Delta_2
\cdots
\int_0^{t_f-\sum_{i=1}^{n-1}\Delta_i}
d\Delta_n\,
g(\Delta_1,\ldots,\Delta_n).
\end{equation}
where
\begin{align}
g(\Delta_1,\ldots,\Delta_n)
&=
\prod_{i=1}^n
\left[
r e^{-r\Delta_i}
S_0(\Delta_i)
\right]
e^{-r\left(t_f-\sum_{i=1}^n\Delta_i\right)}
f_{\mathrm{FP}}
\left(
t_f-\sum_{i=1}^n\Delta_i
\right)\\
&=
r^n e^{-r t_f}
\left[
\prod_{i=1}^n S_0(\Delta_i)
\right]
f_{\mathrm{FP}}
\left(
t_f-\sum_{i=1}^n\Delta_i
\right).
\end{align}
Here $S_0(\Delta)$ is the no-reset survival probability and
$f_{\mathrm{FP}}(s)$ is the no-reset first-passage-time density. The factor
\begin{equation}
r e^{-r\Delta_i}S_0(\Delta_i)
\end{equation}
is the probability density that the $i$th failed excursion survives below
the target for a duration $\Delta_i$ and is then terminated by a reset,
while
\begin{equation}
e^{-r\left(t_f-\sum_{i=1}^n\Delta_i\right)}
f_{\mathrm{FP}}
\left(
t_f-\sum_{i=1}^n\Delta_i
\right)
\end{equation}
is the probability density that the final excursion experiences no reset
and first reaches the target after the remaining time
\begin{equation}
s=t_f-\sum_{i=1}^n\Delta_i.
\end{equation}

The most probable number of resets $n^*$ is obtained by maximizing
$P_n(t_f)$ over integer $n$. For a symmetric stationary point of the
joint distribution of excursion durations, the failed excursions have a
common duration,
\begin{equation}
\Delta_1=\cdots=\Delta_{n^*}=\Delta^*,
\end{equation}
so that
\begin{equation}
t_f=n^*\Delta^*+s^*.
\end{equation}
The corresponding stationary durations $\Delta^*$ and $s^*$ satisfy
\begin{equation}
\frac{d}{d\Delta}
\left[
n^*\log S_0(\Delta)
+
\log f_{\mathrm{FP}}(t_f-n^*\Delta)
\right]
=0.
\end{equation}
This construction identifies the dominant reset-conditioned trajectories
in terms of repeated failed excursions followed by a final successful
first-passage bridge. These dominant trajectories can be compared with
the mean conditioned trajectories discussed above, although the most
probable and mean trajectories need not coincide.

\subsection{Mean temporal organization of reset-conditioned trajectories}

The weight $P_n(t_f)$ also provides access to the statistics of the
reset-conditioned ensemble. In particular, the mean number of resets is
given by
\begin{equation}
\langle n\rangle
=
\frac{
\sum_{n=0}^{\infty}
n\,P_n(t_f)
}{
\sum_{n=0}^{\infty}
P_n(t_f)
}.
\end{equation}
For trajectories that undergo at least one reset, the mean duration of a
failed excursion can be computed from
\begin{equation}
\langle \Delta\rangle
=
\frac{
\sum_{n=1}^{\infty}
\int d\Delta_1\cdots d\Delta_n\,
\left(
\frac{1}{n}
\sum_{i=1}^n \Delta_i
\right)
g(\Delta_1,\ldots,\Delta_n)
}{
\sum_{n=1}^{\infty}
P_n(t_f)
}.
\end{equation}
Similarly, the mean duration of the final successful excursion is
\begin{equation}
\langle s\rangle
=
\frac{
\sum_{n=0}^{\infty}
\int d\Delta_1\cdots d\Delta_n\,
\left(
t_f-\sum_{i=1}^n\Delta_i
\right)
g(\Delta_1,\ldots,\Delta_n)
}{
\sum_{n=0}^{\infty}
P_n(t_f)
}.
\end{equation}
These averaged quantities characterize the temporal organization of
reset-conditioned trajectories and can be compared with the dominant
trajectory obtained from the saddle-point construction above.

For small diffusion constant, successful trajectories are dominated by a
single long successful excursion, so that
\begin{equation}
\langle s\rangle \approx t_f,
\qquad
\langle n\rangle \approx 0.
\end{equation}
In this regime, resets are strongly suppressed because a reset leaves
less time for the particle to reach the target within the prescribed
final time. Failed excursions, when they occur, are therefore pushed
toward short durations. As the diffusion constant increases,
trajectories can accommodate more failed excursions while still reaching
the target at time $t_f$, leading initially to an increase in the mean
number of resets. At still larger diffusion constant, the behavior
crosses over toward the finite large-$D$ limit derived below.

The relative sizes of $\langle \Delta\rangle$ and $\langle s\rangle$ are
not fixed across parameter space. Instead, they reflect how the fixed
total duration $t_f$ is partitioned between failed excursions and the
final successful excursion. For small diffusion constant, the conditioned
trajectory is dominated by the final successful excursion, so
$\langle s\rangle$ is close to $t_f$ while failed excursions are rare
and short. As the diffusion constant increases, the final successful
excursion can reach the target in a shorter time, leaving more of the
conditioned trajectory available for reset-terminated failed excursions.
In this regime, $\langle \Delta\rangle$ can become comparable to, or even
larger than, $\langle s\rangle$.

\subsection{Mean number of resets in the conditioned first-passage ensemble}

The decomposition above also gives a useful way to compute the mean
number of resets directly from the reset-modified first-passage-time
density. Summing over all possible numbers of resets gives
\begin{equation}
f_{\mathrm{FP}}^{(r)}(t_f)
=
\sum_{n=0}^{\infty}P_n(t_f).
\end{equation}
Writing
\begin{equation}
P_n(t_f)
=
r^n e^{-r t_f} A_n(t_f),
\end{equation}
where
\begin{equation}
A_n(t_f)
=
\int d\Delta_1\cdots d\Delta_n\,
\left[
\prod_{i=1}^n S_0(\Delta_i)
\right]
f_{\mathrm{FP}}
\left(
t_f-\sum_{i=1}^n\Delta_i
\right),
\end{equation}
with the integrations restricted to
$\sum_{i=1}^n\Delta_i\leq t_f$, we obtain
\begin{equation}
f_{\mathrm{FP}}^{(r)}(t_f)
=
e^{-r t_f}
\sum_{n=0}^{\infty}
r^n A_n(t_f).
\end{equation}
Since the reset rate appears as a factor $r^n$ for trajectories
containing $n$ reset events, differentiation with respect to $r$ gives
\begin{equation}
\langle n\rangle
=
r\frac{\partial}{\partial r}
\log
\left[
e^{r t_f}f_{\mathrm{FP}}^{(r)}(t_f)
\right],
\end{equation}
or equivalently
\begin{equation}
\langle n\rangle
=
r\frac{\partial}{\partial r}
\log f_{\mathrm{FP}}^{(r)}(t_f)
+
r t_f.
\label{eq:firstPassageBrownBridgeWithResets_numResetsTypical}
\end{equation}
This expression relates the mean number of resets in the conditioned
ensemble directly to the sensitivity of the first-passage-time density
to the reset rate.

The limiting behavior can be understood directly from the reset-sector
weights. In the small-$D$ limit, a trajectory that reaches the target at
the prescribed time $t_f$ is strongly penalized if a finite portion of
the available time is spent in failed excursions. Indeed, writing
\begin{equation}
s=t_f-\sum_{i=1}^n\Delta_i,
\end{equation}
the ratio of no-reset first-passage densities is
\begin{equation}
\frac{
f_{\mathrm{FP}}(s)
}{
f_{\mathrm{FP}}(t_f)
}
=
\left(
\frac{t_f}{s}
\right)^{3/2}
\exp\left[
-\frac{L^2}{4D}
\left(
\frac{1}{s}
-
\frac{1}{t_f}
\right)
\right],
\qquad
L=x_T-x_i.
\end{equation}
As $D\to0$, the dominant reset-containing histories therefore have a
total failed-excursion time that shrinks to zero. For fixed $n$ this
gives
\begin{equation}
\frac{P_n(t_f)}{P_0(t_f)}
\sim
\left(
\frac{4rDt_f^2}{L^2}
\right)^n,
\end{equation}
and hence
\begin{equation}
\langle n\rangle\to0
\qquad
(D\to0).
\end{equation}

The large-$D$ limit is different. Although each no-reset survival factor
becomes small,
\begin{equation}
S_0(\Delta)
\sim
\frac{L}{\sqrt{\pi D\Delta}},
\end{equation}
the one-reset contribution remains of the same order in $D$ as the
zero-reset contribution. In particular,
\begin{equation}
P_0(t_f)
\sim
e^{-rt_f}
\frac{L}{2\sqrt{\pi D}\,t_f^{3/2}},
\end{equation}
while
\begin{equation}
P_1(t_f)
\sim
e^{-rt_f}r
\frac{L}{\sqrt{\pi D t_f}}.
\end{equation}
Therefore,
\begin{equation}
\frac{P_1(t_f)}{P_0(t_f)}
\to
2rt_f.
\end{equation}
All sectors with $n\geq2$ are suppressed relative to these leading
contributions, so that
\begin{equation}
\langle n\rangle
\to
\frac{2rt_f}{1+2rt_f}
\qquad
(D\to\infty).
\end{equation}
Thus the mean number of resets vanishes in the small-diffusion limit but
approaches a finite value in the large-diffusion limit.

\newpage
\clearpage

\stopcontents[appendix]



\end{document}